\newif\ifdraft
\draftfalse

\ifdefined\DRAFT\drafttrue\fi

\ifdraft
   \documentclass[twocolumn,tighten,times,linenumbers,trackchanges,twocolappendix]{aastex631}
\else
   \documentclass[twocolumn,tighten,times,twocolappendix]{aastex631}
\fi

\usepackage{amsmath}
\let\tablenum\relax

\usepackage{afterpage}

\usepackage{siunitx}
\DeclareSIUnit{\parsec}{pc}
\DeclareSIUnit{\Mpc}{\mega\parsec}
\DeclareSIUnit{\year}{yr}

\usepackage{textgreek}

\usepackage[utf8]{inputenc}
\usepackage{xspace}
\usepackage{acronym}
\usepackage{booktabs}

\usepackage{color, colortbl}

\usepackage{graphicx}
\usepackage[caption=false]{subfig}

\usepackage{tabularx}
\usepackage{multirow}

\usepackage{makecell}

\newcolumntype{Y}{>{\centering\arraybackslash}X}
\newcolumntype{Z}{>{\raggedright\arraybackslash}X}
\newcolumntype{K}{>{\raggedleft\arraybackslash}X}
\newcolumntype{U}{>{\hsize=1.01\hsize}Y}
\newcolumntype{V}{>{\hsize=1.2\hsize}Y}
\newcolumntype{W}{>{\hsize=0.71\hsize}Y}

\usepackage{scrextend}
\usepackage{textgreek}

\usepackage{microtype}

\AtBeginDocument{}

\newcommand{\checkbar}{\checkmark\kern-1.1ex\raisebox{.7ex}{\rotatebox[origin=c]{135}{--}}}
\newcommand{\checkminus}{\sout{\checkmark}}
\newcommand{\checkprevext}{+}
\newcommand{\checkprevlvk}{++}
\newcommand{\checknodata}{\!\!\nodata\!\!}

\newcommand{\tigerinsp}{\mathrm{I}}
\newcommand{\tigerpostinsp}{\mathrm{PI}}

\newcommand{\tablefootnote}[1]{}

\usepackage{makerobust}

\renewcommand{\today}{\number\day\space\ifcase\month\or
  January\or February\or March\or April\or May\or June\or
  July\or August\or September\or October\or November\or December\fi
  \space\number\year}

\usepackage{float}

\definecolor{NOTECOLOR}{rgb}{0.4, 0.2, 0.1}
\definecolor{notecolor}{rgb}{0.4, 0.2, 0.1}
\definecolor{danger-red}{rgb}{0.8, 0.4, 0.0}
\definecolor{ok-green}{rgb}{0.0, 0.6, 0.5}

\newcommand{\reviewed}[1]{{#1}}

\newcommand{\macro}[1]{#1}

\definecolor{mgreen}{rgb}{0.1,0.7,0.1}

\newcommand{\subpapersection}[1]{\section{#1}}
\newcommand{\subpapersubsection}[1]{\subsection{#1}}

\newcommand{\paperscommonname}{GWTC-4.0: Tests of General Relativity. }

\newtoggle{fullauthorlist}
\toggletrue{fullauthorlist}

\newtoggle{endauthorlist}
\toggletrue{endauthorlist}

\newcommand{\LVKCollabAuthors}{The LIGO Scientific Collaboration, the Virgo Collaboration, and the KAGRA Collaboration}

\newcommand{\LVKCorrespondence}{The full author list is given at the end.\\
 For correspondence: LSC P\&P Committee, for LVK Publications, via}

\newcommand\gwtc[1][?]{\mbox{GWTC\if#1?\else-#1\fi}}
\newcommand\thisgwtcversionmajor{4}
\newcommand\thisgwtcversionminor{0}
\newcommand\thisgwtcversionfull{\thisgwtcversionmajor.\thisgwtcversionminor}
\newcommand\thisgwtcversion\thisgwtcversionfull

\newcommand{\soft}[1]{\textsc{#1}}

\newcommand{\GSTLAL}{\soft{GstLAL}\xspace}

\newcommand{\CWB}{\soft{cWB}\xspace}

\newcommand{\PYCBC}{\soft{PyCBC}\xspace}
\newcommand{\MBTA}{\soft{MBTA}\xspace}

\newcommand{\BAYESWAVE}{\soft{BayesWave}\xspace}
\newcommand{\BILBY}{\soft{Bilby}\xspace}

\newcommand{\LALSUITE}{\soft{LALSuite}\xspace}

\newcommand{\PBILBY}{\soft{ParallelBilby}\xspace}
\newcommand{\ASIMOV}{\soft{Asimov}\xspace}
\newcommand{\PESUMMARY}{\soft{PESummary}\xspace}

\newcommand{\NUMPY}{\soft{NumPy}\xspace}
\newcommand{\SCIPY}{\soft{SciPy}\xspace}

\newcommand{\SEABORN}{\soft{seaborn}\xspace}
\newcommand{\GWPY}{\soft{GWpy}\xspace}
\newcommand{\DYNESTY}{\soft{Dynesty}\xspace}

\newcommand{\IMRPhenomPVTWO}{\soft{IMRPhenomPv2}\xspace}

\newcommand{\IMRPhenomPVTHREEHM}{\soft{IMRPhenomPv3HM}\xspace}
\newcommand{\IMRPhenomXAS}{\soft{IMRPhenomXAS}\xspace}

\newcommand{\IMRPhenomXP}{\soft{IMRPhenomXP}\xspace}
\newcommand{\IMRPhenomXPHM}{\soft{IMRPhenomXPHM}\xspace}
\newcommand{\IMRPhenomXPHMST}{\soft{IMRPhenomXPHM\_SpinTaylor}\xspace}

\newcommand{\SEOBNRFOURROM}{\soft{SEOBNRv4\_ROM}\xspace}

\newcommand{\SEOBNRFOURHMROM}{\soft{SEOBNRv4HM\_ROM}\xspace}

\newcommand{\SEOBNRFIVEHMROM}{\soft{SEOBNRv5HM\_ROM}\xspace}

\newcommand{\IMRPhenomXPNRTidalTWO}{\soft{IMRPhenomXP\_NRTidalv2}\xspace}

\newcommand{\IMRPhenomNSBH}{\soft{IMRPhenomNSBH}\xspace}

\newcommand{\SEOBNRFOURNRTidalTWO}{\soft{SEOBNRv4\_ROM\_NRTidalv2}\xspace}
\newcommand{\SEOBNRFOURNRtidalTWONSBH}{\soft{SEOBNRv4\_ROM\_NRTidalv2\_NSBH}\xspace}

\newcommand{\SURSEVENDQFOUR}{\soft{NRSur7dq4}\xspace}

\DeclareSIUnit\parsec{pc}
\DeclareSIUnit\Mpc{\mega\parsec}
\DeclareSIUnit\yr{yr}
\DeclareSIUnit\GpcCubedYear{\giga\parsec\cubed\yr}

\newcommand{\Msun}{\ensuremath{\mathit{M_\odot}}}

\newcommand{\ip}[2]{\ensuremath{\langle #1 | #2 \rangle}}

\newcommand\PEpdfp{\ensuremath{p}}

\newcommand{\PEparameter}{\ensuremath{\boldsymbol{\theta}}}%

\newcommand\PEpdf[2][?]{\ensuremath{\PEpdfp({#2}\ifx#1?\else | {#1}\fi)}}

\newcommand\PEpriorpdfpi{\ensuremath{\pi}}
\newcommand\PEpdfprior[1]{\ensuremath{\PEpriorpdfpi({#1})}}
\newcommand\PEprior[1][\PEparameter]{\PEpdfprior{#1}}
\newcommand\PEpriorpe[1][\PEparameter]{{\let\keepPEpriorpdfpi\PEpriorpdfpi\def\PEpriorpdfpi{\keepPEpriorpdfpi_{\text{PE}}}\PEprior[#1]\let\PEpriorpdfpi\keepPEpriorpdfpi}}

\newcommand{\PN}[0]{\ac{PN}\xspace}
\newcommand{\BBH}[0]{\ac{BBH}\xspace}
\newcommand{\BNS}[0]{\ac{BNS}\xspace}
\newcommand{\NSBH}[0]{\ac{NSBH}\xspace}
\newcommand{\BH}[0]{\ac{BH}\xspace}
\newcommand{\NR}[0]{\ac{NR}\xspace}
\newcommand{\SNR}[0]{\ac{SNR}\xspace}

\newcommand{\GR}[0]{\ac{GR}\xspace}

\usepackage{amstext}

\makeatletter
\@ifpackageloaded{cleveref}{
  \AtBeginDocument{
    \def\ltx@label#1{\cref@label{#1}}%
    \def\label@in@display@noarg#1{\cref@old@label@in@display{#1}}%
    \def\label@in@mmeasure@noarg#1{%
      \begingroup%
        \measuring@false%
        \cref@old@label@in@display{#1}%
      \endgroup}%
  }
}{}
\makeatother

\protected\def\protectedacused{\acused}

\acrodef{LIGO}[LIGO]{Laser Interferometer Gravitational-Wave Observatory}
\acrodef{LHO}[LHO]{\ac{LIGO} Hanford Observatory}
\acrodef{LLO}[LLO]{\ac{LIGO} Livingston Observatory}
\acrodef{KAGRA}[KAGRA]{KAGRA}\acused{KAGRA}
\acrodef{iKAGRA}[iKAGRA]{initial-phase \ac{KAGRA}}
\acrodef{bKAGRA}[bKAGRA]{baseline-design \ac{KAGRA}}
\acrodef{GEO}[GEO]{GEO\,600 \ac{GW} detector}
\acrodef{aLIGO}{Advanced \ac{LIGO}}
\acrodef{A+}{Advanced+ \ac{LIGO}}
\acrodef{Asharp}[\ensuremath{\text{A}^\sharp}]{\ac{LIGO} \acs{Asharp}}
\acrodef{AdV}{Advanced \acl{Virgo}}
\acrodef{AdV+}{Advanced \acl{Virgo}+}
\acrodef{Virgo}{Virgo}\acused{Virgo}
\acrodef{VirgoNEXT}[Virgo\_nEXT]{Virgo\_nEXT}\acused{VirgoNEXT}

\acrodef{LSC}[LSC]{\acs{LIGO} Scientific Collaboration}
\acrodef{LV}[LV]{\acs{LIGO}--\acs{Virgo} Collaboration\protect\protectedacused{LVC}}
\acrodef{LVC}[LV]{\acs{LIGO}--\acs{Virgo} Collaboration\protect\protectedacused{LV}}
\acrodef{LVK}[LVK]{\acs{LIGO}--\ac{Virgo}--\ac{KAGRA} Collaboration}
\acrodef{IGWN}[IGWN]{International \ac{GWH} Observatory Network}

\acrodef{O1}[O1]{first observing run}
\acrodef{O2}[O2]{second observing run}
\acrodef{O3}[O3]{third observing run}
\acrodef{O3a}[O3a]{first half of the third observing run}
\acrodef{O3b}[O3b]{second half of the third observing run}
\acrodef{O3GK}[O3GK]{observing run}
\acrodef{O4}[O4]{fourth observing run}
\acrodef{O4a}[O4a]{first part of the fourth observing run}
\acrodef{O4b}[O4b]{second part of the fourth observing run}
\acrodef{O4c}[O4c]{third part of the fourth observing run}
\acrodef{O5}[O5]{fifth observing run}

\acrodef{BH}[BH]{black hole}
\acrodef{BBH}[BBH]{binary \ac{BH}}
\acrodef{BNS}[BNS]{binary \ac{NS}}
\acrodef{IMBH}[IMBH]{intermediate-mass \ac{BH}}
\acrodef{NS}[NS]{neutron star}
\acrodef{BHNS}[BHNS]{\ac{BH}--\ac{NS} binary}
\acrodefplural{BHNS}[BHNSs]{\ac{BH}--\ac{NS} binaries}
\acrodef{NSBH}[NSBH]{\ac{NS}--\ac{BH} binary}
\acrodefplural{NSBH}[NSBHs]{\ac{NS}--\ac{BH} binaries}
\acrodef{PBH}[PBH]{primordial \ac{BH}}
\acrodef{CBC}[CBC]{compact binary coalescence}

\acrodef{GW}[GW]{gravitational wave\protect\protectedacused{GWH}}
\acrodef{GWH}[GW]{gravitational-wave\protect\protectedacused{GW}}
\acrodef{IFO}[IFO]{interferometer}
\acrodef{SNR}[SNR]{signal-to-noise ratio}
\acrodef{FAR}[FAR]{false-alarm rate}
\acrodef{IFAR}[IFAR]{inverse false-alarm rate}
\acrodef{FAP}[FAP]{false alarm probability}
\acrodef{PSD}[PSD]{power spectral density}
\acrodef{GR}[GR]{general relativity}
\acrodef{NR}[NR]{numerical relativity}
\acrodef{PN}[PN]{post-Newtonian}
\acrodef{EOB}[EOB]{effective-one-body}
\acrodef{ROM}[ROM]{reduced-order model}
\acrodef{IMR}[IMR]{inspiral--merger--ringdown}
\acrodef{PDF}[pdf]{probability density function}
\acrodef{PE}[PE]{parameter estimation}
\acrodef{CI}[CI]{credible interval}
\acrodef{CL}[CL]{credible level}
\acrodef{EOS}[EoS]{equation of state}
\acrodef{KLD}[KLD]{Kullback--Leibler divergence}
\acrodef{JSD}[JSD]{Jensen--Shannon divergence}
\acrodef{GCN}[GCN]{General Coordinates Network}
\acrodef{GWTC}[GWTC]{Gravitational-Wave Transient Catalog}
\acrodef{GWOSC}[GWOSC]{Gravitational Wave Open Science Center}
\acrodef{WDM}[WDM]{Wilson--Debauchies--Meyer}

\acrodef{CWB}[cWB]{coherent WaveBurst}
\acrodef{LAL}[LAL]{\ac{LIGO} algorithm library}

\acrodef{CHRoCC}{central heating radius of curvature correction}
\acrodef{NonSENS}{non-stationary estimation and noise subtraction}

\acrodef{PTA}{Pulsar Timing Array}

\input{common__event_macros}

\newcommand{\BILBYTGR}{\soft{BilbyTGR}\xspace}

\newcommand{\IMRPhenomXPHMMSA}{\soft{IMRPhenomXPHM\_MSA}\xspace}
\newcommand{\IMRPhenomXPMSA}{\soft{IMRPhenomXP\_MSA}\xspace}

\newcommand{\TGRFARTHRESH}{\reviewed{$\le \qty{e-3}{\yr^{-1}}$}\xspace}

\newcommand{\TGRNUMEVENTSPREV}{\reviewed{49}\xspace}
\newcommand{\TGRNUMEVENTSPREVPLUSOFOURA}{\reviewed{91}\xspace}  

\newcommand{\TGRFigureWidth}{3.375in}
\newcommand{\TGRFigureWidthPage}{\textwidth}

\DeclareRobustCommand{\COMMONNAME}[1]{\IfEqCase{#1}{{GW230529}{\MINIMALNAME{GW230529_181500}{}}{GW230814other}{\MINIMALNAME{GW230814_061920}{}}{GW230814single}{\MINIMALNAME{GW230814_230901}{}}{GW231123}{\MINIMALNAME{GW231123_135430}{}}}[\textcolor{red}{???}]}

\DeclareRobustCommand{\TGRImrctGWTCFOURResults}[1]{\IfEqCase{#1}{{DMFGWTC4PHENOM}{\reviewed{\ensuremath{-0.00^{+0.05}_{-0.05}}}}{DCHIFGWTC4PHENOM}{\reviewed{\ensuremath{-0.04^{+0.07}_{-0.06}}}}{GRQUANTGWTC4}{\reviewed{\ensuremath{89.1}}}{GRQUANTGWTC3}{\ensuremath{79.6}}}}

\DeclareRobustCommand{\TGRImrctGWTCFOURResultsWOGWOneNineZeroEightOneFour}[1]{\IfEqCase{#1}{{DMFGWTC4PHENOM}{\reviewed{\ensuremath{0.01^{+0.05}_{-0.05}}}}{DCHIFGWTC4PHENOM}{\reviewed{\ensuremath{-0.03^{+0.06}_{-0.06}}}}{GRQUANTGWTC4}{\reviewed{\ensuremath{86.0}}}}}

\DeclareRobustCommand{\TGRImrctHierPopGWTCFOURResults}[1]{\IfEqCase{#1}{{DMFGWTC4HIERPOP}{\reviewed{\ensuremath{0.00^{+0.07}_{-0.06}}}}{DCHIFGWTC4HIERPOP}{\reviewed{\ensuremath{-0.05^{+0.11}_{-0.11}}}}{GRQUANTDevParamsTwoDGWTC4HIERPOP}{\reviewed{\ensuremath{51.7}}}{GRQUANTMeansTwoDGWTC4HIERPOP}{\reviewed{\ensuremath{94.2}}}{GRQUANTHyperParamsFourDGWTC4HIERPOP}{\reviewed{\ensuremath{73.1}}}}}

\DeclareRobustCommand{\TGRImrctHierPopGWTCFOURResultsWOGWOneNineZeroEightOneFour}[1]{\IfEqCase{#1}{{DMFGWTC4HIERPOP}{\reviewed{\ensuremath{0.02^{+0.06}_{-0.06}}}}{DCHIFGWTC4HIERPOP}{\reviewed{\ensuremath{-0.02^{+0.07}_{-0.07}}}}{GRQUANTDevParamsTwoDGWTC4HIERPOP}{\reviewed{\ensuremath{44.8}}}{GRQUANTMeansTwoDGWTC4HIERPOP}{\reviewed{\ensuremath{78.7}}}{GRQUANTHyperParamsFourDGWTC4HIERPOP}{\reviewed{\ensuremath{38.6}}}}}

\DeclareRobustCommand{\TGRPOLlogBImprovement}[1]{\IfEqCase{#1}{
{S}{\reviewed{10.48}}
{V}{\reviewed{3.63}}
{TS}{\reviewed{0.40}}
{TV}{\reviewed{0.21}}
{VS}{\reviewed{3.48}}
{TVS}{\reviewed{0.23}}
{min}{\reviewed{0.21}}
{max}{\reviewed{10.48}}
}}

\DeclareRobustCommand{\TGRPOLlogBResults}[1]{\IfEqCase{#1}{
{S}{\reviewed{-14.72}}
{Serr}{\reviewed{0.59}}
{V}{\reviewed{-5.33}}
{Verr}{\reviewed{0.58}}
{TS}{\reviewed{-0.20}}
{TSerr}{\reviewed{0.57}}
{TV}{\reviewed{0.10}}
{TVerr}{\reviewed{0.57}}
{VS}{\reviewed{-5.21}}
{VSerr}{\reviewed{0.58}}
{TVS}{\reviewed{-0.31}}
{TVSerr}{\reviewed{0.57}}
{min}{\reviewed{-14.72}}
{minerr}{\reviewed{0.59}}
{max}{\reviewed{0.10}}
{maxerr}{\reviewed{0.57}}
}}

\DeclareRobustCommand{\TGRTIGERBound}[1]{\IfEqCase{#1}{
{dchiminus2}{\reviewed{\num{6.2e-04}}}
{dchi0}{\reviewed{\num{8.2e-02}}}
{dchi1}{\reviewed{0.18}}
{dchi2}{\reviewed{0.11}}
{dchi3}{\reviewed{\num{5.7e-02}}}
{dchi4}{\reviewed{0.52}}
{dchi5l}{\reviewed{0.20}}
{dchi6}{\reviewed{0.30}}
{dchi6l}{\reviewed{1.07}}
{dchi7}{\reviewed{0.80}}
{min}{\reviewed{\num{6.2e-04}}}
{max}{\reviewed{1.1}}
}}

\DeclareRobustCommand{\TGRTIGERBoundPI}[1]{\IfEqCase{#1}{
{db1}{\reviewed{\num{2.9e-02}}}
{db2}{\reviewed{\num{9.1e-03}}}
{db3}{\reviewed{\num{8.3e-03}}}
{db4}{\reviewed{\num{1.6e-02}}}
{dc1}{\reviewed{0.12}}
{dc2}{\reviewed{\num{4.2e-02}}}
{dc4}{\reviewed{0.13}}
{dcl}{\reviewed{0.30}}
{min}{\reviewed{\num{8.3e-03}}}
{max}{\reviewed{0.30}}
}}

\DeclareRobustCommand{\TGRTIGERBoundImprovement}[1]{\IfEqCase{#1}{
{dchiminus2}{\reviewed{2.6}}
{dchi0}{\reviewed{1.7}}
{dchi1}{\reviewed{1.6}}
{dchi2}{\reviewed{1.5}}
{dchi3}{\reviewed{1.7}}
{dchi4}{\reviewed{1.5}}
{dchi5l}{\reviewed{1.3}}
{dchi6}{\reviewed{1.6}}
{dchi6l}{\reviewed{1.3}}
{dchi7}{\reviewed{1.6}}
{min}{\reviewed{1.3}}
{max}{\reviewed{2.6}}
}}

\DeclareRobustCommand{\TGRTIGERBoundImprovementPI}[1]{\IfEqCase{#1}{
{db1}{\reviewed{1.2}}
{db2}{\reviewed{1.2}}
{db3}{\reviewed{1.3}}
{db4}{\reviewed{1.6}}
{dc1}{\reviewed{1.3}}
{dc2}{\reviewed{1.3}}
{dc4}{\reviewed{1.1}}
{dcl}{\reviewed{1.4}}
{min}{\reviewed{1.1}}
{max}{\reviewed{1.6}}
}}

\DeclareRobustCommand{\TGRFTIBound}[1]{\IfEqCase{#1}{
{dchiMinus2}{\reviewed{\num{1.6e-03}}}
{dchi0}{\reviewed{\num{9.5e-02}}}
{dchi1}{\reviewed{0.26}}
{dchi2}{\reviewed{0.14}}
{dchi3NS}{\reviewed{\num{8.0e-02}}}
{dchi4NS}{\reviewed{0.46}}
{dchi5lNS}{\reviewed{0.14}}
{dchi6NS}{\reviewed{0.19}}
{dchi6l}{\reviewed{1.52}}
{dchi7NS}{\reviewed{0.34}}
{min}{\reviewed{\num{1.6e-03}}}
{max}{\reviewed{1.5}}
}}

\DeclareRobustCommand{\TGRFTIBoundImprovement}[1]{\IfEqCase{#1}{
{dchiMinus2}{\reviewed{1.2}}
{dchi0}{\reviewed{1.4}}
{dchi1}{\reviewed{1.2}}
{dchi2}{\reviewed{1.5}}
{dchi3NS}{\reviewed{1.8}}
{dchi4NS}{\reviewed{1.5}}
{dchi5lNS}{\reviewed{3.5}}
{dchi6NS}{\reviewed{2.8}}
{dchi6l}{\reviewed{1.4}}
{dchi7NS}{\reviewed{5.5}}
{min}{\reviewed{1.2}}
{max}{\reviewed{5.5}}
}}

\DeclareRobustCommand{\FirstPCAFTIjointbound}{$-0.01^{+0.04}_{-0.03}$}
\DeclareRobustCommand{\FirstPCATIGERjointbound}{$0.01^{+0.06}_{-0.06}$}
\DeclareRobustCommand{\FifthPCATIGERGRquantile}{98\%}

\DeclareRobustCommand{\TGRSIMPhenomCombinedCI}[1]{\IfEqCase{#1}{{HIER_POP}{\reviewed{\ensuremath{-19^{+28}_{-34}}}}{RESTRICTED_POP}{\reviewed{\ensuremath{-14^{+12}_{-14}}}}{HIER_POP_NEG}{\reviewed{46}}{HIER_POP_POS}{\reviewed{26}}{RESTRICTED_POP_NEG}{\reviewed{24}}{RESTRICTED_POP_POS}{\reviewed{5.7}}{HIER_MU}{\reviewed{\ensuremath{-19^{+20}_{-22}}}}{HIER_SIGMA}{\reviewed{25}}}}
\DeclareRobustCommand{\TGRSIMEOBCombinedCI}[1]{\IfEqCase{#1}{{HIER_POP}{\reviewed{\ensuremath{-49^{+95}_{-176}}}}{RESTRICTED_POP}{\reviewed{\ensuremath{-32^{+29}_{-53}}}}{HIER_POP_NEG}{\reviewed{179}}{HIER_POP_POS}{\reviewed{127}}{RESTRICTED_POP_NEG}{\reviewed{68}}{RESTRICTED_POP_POS}{\reviewed{13}}{HIER_MU}{\reviewed{\ensuremath{-51^{+53}_{-128}}}}{HIER_SIGMA}{\reviewed{119}}}}

\DeclareRobustCommand{\TGRLOSAResults}[1]{\IfEqCase{#1}{
{GW170817}{\reviewed{0.42_{-1.87}^{+2.00}} \times 10^{-6}}
{GW190425}{\reviewed{{-0.01}_{-7.40}^{+6.29}} \times 10^{-6}}
}}

\DeclareRobustCommand{\TGRMDRGravitonBound}[1]{\IfEqCase{#1}{
{gwtc3}{\reviewed{\num{2.23e-23} $\mathrm{eV}/c^2$}}
{gwtc4}{\reviewed{\num{1.92e-23} $\mathrm{eV}/c^2$}}
}}
\DeclareRobustCommand{\TGRMDRGravitonBoundNoUnits}[1]{\IfEqCase{#1}{
{gwtc3}{\reviewed{\num{2.23e-23}}}
{gwtc4}{\reviewed{\num{1.92e-23}}}
}}
\DeclareRobustCommand{\TGRMDRBoundImprovement}[1]{\IfEqCase{#1}{
{mg}{\reviewed{1.16}}
{alpha_0p0}{\reviewed{1.48}}
{alpha_0p5}{\reviewed{1.64}}
{alpha_1p5}{\reviewed{1.68}}
{alpha_2p5}{\reviewed{1.75}}
{alpha_3p0}{\reviewed{2.10}}
{alpha_3p5}{\reviewed{2.60}}
{alpha_4p0}{\reviewed{2.88}}
{alpha_m1p0}{\reviewed{1.69}}
{alpha_m2p0}{\reviewed{1.81}}
{alpha_m3p0}{\reviewed{2.00}}
{amplitude_mean}{\reviewed{1.96}}
{amplitude_min}{\reviewed{1.48}}
{amplitude_max}{\reviewed{2.88}}
}}
\DeclareRobustCommand{\TGRMDRAmplitudeBoundPeV}[1]{\IfEqCase{#1}{
{amplitude_min}{\reviewed{0.01}}
{amplitude_max}{\reviewed{351}}
}}

\DeclareRobustCommand{\TGRSSBKFiveVZeroZero}[1]{\IfEqCase{#1}{
{gwtc4}{\reviewed{\num{1.52e-14} m}}
}}
\DeclareRobustCommand{\TGRSSBKFiveVZeroZeroNoUnits}[1]{\IfEqCase{#1}{
{gwtc4}{\reviewed{\num{1.52e-14}}}
}}

\DeclareRobustCommand{\INSTRUMENTSOONEOTWO}[1]{\IfEqCase{#1}{{GW150914}{HL}{GW151012}{HL}{GW151226}{HL}{GW170104}{HL}{GW170608}{HL}{GW170729}{HLV}{GW170809}{HLV}{GW170814}{HLV}{GW170817}{HLV}{GW170818}{HLV}{GW170823}{HL}}[\textcolor{red}{???}]}

\DeclareRobustCommand{\INSTRUMENTSOTHREEA}[1]{\IfEqCase{#1}{{GW190408A}{HLV}{GW190412A}{HLV}{GW190413A}{HL}{GW190413B}{HLV}{GW190421A}{HL}{GW190424A}{L}{GW190425A}{LV}{GW190426A}{HLV}{GW190503A}{HLV}{GW190512A}{HLV}{GW190513A}{HLV}{GW190514A}{HL}{GW190517A}{HLV}{GW190519A}{HLV}{GW190521A}{HLV}{GW190521B}{HL}{GW190527A}{HL}{GW190602A}{HLV}{GW190620A}{LV}{GW190630A}{LV}{GW190701A}{HLV}{GW190706A}{HLV}{GW190707A}{HL}{GW190708A}{LV}{GW190719A}{HL}{GW190720A}{HLV}{GW190727A}{HLV}{GW190728A}{HLV}{GW190731A}{HL}{GW190803A}{HLV}{GW190814A}{HLV}{GW190828A}{HLV}{GW190828B}{HLV}{GW190909A}{HL}{GW190910A}{LV}{GW190915A}{HLV}{GW190924A}{HLV}{GW190929A}{HLV}{GW190930A}{HL}}[\textcolor{red}{???}]}

\DeclareRobustCommand{\INSTRUMENTSOTHREEB}[1]{\IfEqCase{#1}{{GW191103A}{HL}{GW191105C}{HLV}{GW191109A}{HL}{GW191113B}{HLV}{GW191118N}{LV}{GW191126C}{HL}{GW191127B}{HLV}{GW191129G}{HL}{GW191204A}{HL}{GW191204G}{HL}{GW191215G}{HLV}{GW191216G}{HV}{GW191219E}{HLV}{GW191222A}{HL}{GW191230H}{HLV}{GW200105F}{LV}{GW200112H}{LV}{GW200115A}{HLV}{GW200121A}{HV}{GW200128C}{HL}{GW200129D}{HLV}{GW200201F}{HLV}{GW200202F}{HLV}{GW200208G}{HLV}{GW200208K}{HLV}{GW200209E}{HLV}{GW200210B}{HLV}{GW200214K}{HLV}{GW200216G}{HLV}{GW200219D}{HLV}{GW200219K}{HLV}{GW200220E}{HLV}{GW200220H}{HL}{GW200224H}{HLV}{GW200225B}{HL}{GW200302A}{HV}{GW200306A}{HL}{GW200308G}{HLV}{GW200311H}{HL}{GW200311L}{HLV}{GW200316I}{HLV}{GW200322G}{HLV}}[\textcolor{red}{???}]}

\input{common__pe_macros}
\input{common__pe_macros_gwtc2point1}

\DeclareRobustCommand{\finalspingwtcthreeminus}[1]{\IfEqCase{#1}{{GW200224_222234}{0.07}{GW191129_134029}{0.05}{GW200311_115853}{0.08}{GW191230_180458}{0.12}{GW191222_033537}{0.11}{GW200225_060421}{0.13}{GW200302_015811}{0.15}{GW200128_022011}{0.10}{GW191204_171526}{0.03}{GW200112_155838}{0.06}{GW200105_162426}{0.03}{GW191105_143521}{0.05}{GW191109_010717}{0.19}{GW200209_085452}{0.11}{GW200115_042309}{0.06}{GW191127_050227}{0.29}{GW200216_220804}{0.24}{GW191215_223052}{0.07}{GW200208_130117}{0.13}{GW200219_094415}{0.13}{GW191103_012549}{0.05}{GW200316_215756}{0.04}{GW200202_154313}{0.04}{GW200129_065458}{0.05}{GW191216_213338}{0.04}}[{\red{???}}]}
\DeclareRobustCommand{\finalspingwtcthreemed}[1]{\IfEqCase{#1}{{GW200224_222234}{0.73}{GW191129_134029}{0.69}{GW200311_115853}{0.69}{GW191230_180458}{0.69}{GW191222_033537}{0.67}{GW200225_060421}{0.66}{GW200302_015811}{0.67}{GW200128_022011}{0.75}{GW191204_171526}{0.73}{GW200112_155838}{0.71}{GW200105_162426}{0.44}{GW191105_143521}{0.67}{GW191109_010717}{0.61}{GW200209_085452}{0.67}{GW200115_042309}{0.43}{GW191127_050227}{0.75}{GW200216_220804}{0.70}{GW191215_223052}{0.68}{GW200208_130117}{0.66}{GW200219_094415}{0.66}{GW191103_012549}{0.75}{GW200316_215756}{0.70}{GW200202_154313}{0.69}{GW200129_065458}{0.73}{GW191216_213338}{0.70}}[{\red{???}}]}
\DeclareRobustCommand{\finalspingwtcthreeplus}[1]{\IfEqCase{#1}{{GW200224_222234}{0.07}{GW191129_134029}{0.03}{GW200311_115853}{0.07}{GW191230_180458}{0.11}{GW191222_033537}{0.08}{GW200225_060421}{0.07}{GW200302_015811}{0.13}{GW200128_022011}{0.10}{GW191204_171526}{0.03}{GW200112_155838}{0.06}{GW200105_162426}{0.07}{GW191105_143521}{0.04}{GW191109_010717}{0.18}{GW200209_085452}{0.10}{GW200115_042309}{0.10}{GW191127_050227}{0.13}{GW200216_220804}{0.14}{GW191215_223052}{0.07}{GW200208_130117}{0.09}{GW200219_094415}{0.10}{GW191103_012549}{0.06}{GW200316_215756}{0.04}{GW200202_154313}{0.03}{GW200129_065458}{0.06}{GW191216_213338}{0.03}}[{\red{???}}]}
\DeclareRobustCommand{\finalspingwtcthreetenthpercentile}[1]{\IfEqCase{#1}{{GW200224_222234}{0.68}{GW191129_134029}{0.65}{GW200311_115853}{0.63}{GW191230_180458}{0.60}{GW191222_033537}{0.60}{GW200225_060421}{0.56}{GW200302_015811}{0.56}{GW200128_022011}{0.67}{GW191204_171526}{0.71}{GW200112_155838}{0.66}{GW200105_162426}{0.41}{GW191105_143521}{0.63}{GW191109_010717}{0.47}{GW200209_085452}{0.59}{GW200115_042309}{0.38}{GW191127_050227}{0.56}{GW200216_220804}{0.53}{GW191215_223052}{0.63}{GW200208_130117}{0.57}{GW200219_094415}{0.57}{GW191103_012549}{0.71}{GW200316_215756}{0.67}{GW200202_154313}{0.66}{GW200129_065458}{0.70}{GW191216_213338}{0.67}}[{\red{???}}]}
\DeclareRobustCommand{\finalspingwtcthreenintiethpercentile}[1]{\IfEqCase{#1}{{GW200224_222234}{0.78}{GW191129_134029}{0.71}{GW200311_115853}{0.74}{GW191230_180458}{0.77}{GW191222_033537}{0.74}{GW200225_060421}{0.71}{GW200302_015811}{0.78}{GW200128_022011}{0.83}{GW191204_171526}{0.75}{GW200112_155838}{0.75}{GW200105_162426}{0.49}{GW191105_143521}{0.70}{GW191109_010717}{0.74}{GW200209_085452}{0.74}{GW200115_042309}{0.51}{GW191127_050227}{0.86}{GW200216_220804}{0.82}{GW191215_223052}{0.73}{GW200208_130117}{0.72}{GW200219_094415}{0.74}{GW191103_012549}{0.79}{GW200316_215756}{0.73}{GW200202_154313}{0.71}{GW200129_065458}{0.78}{GW191216_213338}{0.72}}[{\red{???}}]}
\DeclareRobustCommand{\spintwoygwtcthreeminus}[1]{\IfEqCase{#1}{{GW200224_222234}{0.57}{GW191129_134029}{0.46}{GW200311_115853}{0.56}{GW191230_180458}{0.61}{GW191222_033537}{0.54}{GW200225_060421}{0.54}{GW200302_015811}{0.58}{GW200128_022011}{0.61}{GW191204_171526}{0.53}{GW200112_155838}{0.52}{GW200105_162426}{0.51}{GW191105_143521}{0.52}{GW191109_010717}{0.69}{GW200209_085452}{0.60}{GW200115_042309}{0.50}{GW191127_050227}{0.58}{GW200216_220804}{0.60}{GW191215_223052}{0.58}{GW200208_130117}{0.55}{GW200219_094415}{0.57}{GW191103_012549}{0.54}{GW200316_215756}{0.53}{GW200202_154313}{0.49}{GW200129_065458}{0.55}{GW191216_213338}{0.48}}[{\red{???}}]}
\DeclareRobustCommand{\spintwoygwtcthreemed}[1]{\IfEqCase{#1}{{GW200224_222234}{0.00}{GW191129_134029}{0.00}{GW200311_115853}{0.00}{GW191230_180458}{0.00}{GW191222_033537}{0.00}{GW200225_060421}{0.00}{GW200302_015811}{0.00}{GW200128_022011}{0.00}{GW191204_171526}{0.00}{GW200112_155838}{0.00}{GW200105_162426}{0.00}{GW191105_143521}{0.00}{GW191109_010717}{0.01}{GW200209_085452}{0.00}{GW200115_042309}{0.00}{GW191127_050227}{0.00}{GW200216_220804}{0.00}{GW191215_223052}{0.00}{GW200208_130117}{0.00}{GW200219_094415}{0.00}{GW191103_012549}{0.00}{GW200316_215756}{0.00}{GW200202_154313}{0.00}{GW200129_065458}{0.00}{GW191216_213338}{0.00}}[{\red{???}}]}
\DeclareRobustCommand{\spintwoygwtcthreeplus}[1]{\IfEqCase{#1}{{GW200224_222234}{0.55}{GW191129_134029}{0.47}{GW200311_115853}{0.58}{GW191230_180458}{0.60}{GW191222_033537}{0.56}{GW200225_060421}{0.54}{GW200302_015811}{0.57}{GW200128_022011}{0.62}{GW191204_171526}{0.50}{GW200112_155838}{0.53}{GW200105_162426}{0.51}{GW191105_143521}{0.52}{GW191109_010717}{0.69}{GW200209_085452}{0.60}{GW200115_042309}{0.51}{GW191127_050227}{0.58}{GW200216_220804}{0.57}{GW191215_223052}{0.60}{GW200208_130117}{0.55}{GW200219_094415}{0.59}{GW191103_012549}{0.55}{GW200316_215756}{0.54}{GW200202_154313}{0.50}{GW200129_065458}{0.57}{GW191216_213338}{0.47}}[{\red{???}}]}
\DeclareRobustCommand{\spintwoygwtcthreetenthpercentile}[1]{\IfEqCase{#1}{{GW200224_222234}{-0.43}{GW191129_134029}{-0.33}{GW200311_115853}{-0.40}{GW191230_180458}{-0.46}{GW191222_033537}{-0.39}{GW200225_060421}{-0.40}{GW200302_015811}{-0.42}{GW200128_022011}{-0.46}{GW191204_171526}{-0.40}{GW200112_155838}{-0.38}{GW200105_162426}{-0.35}{GW191105_143521}{-0.36}{GW191109_010717}{-0.54}{GW200209_085452}{-0.44}{GW200115_042309}{-0.35}{GW191127_050227}{-0.43}{GW200216_220804}{-0.44}{GW191215_223052}{-0.43}{GW200208_130117}{-0.41}{GW200219_094415}{-0.43}{GW191103_012549}{-0.40}{GW200316_215756}{-0.39}{GW200202_154313}{-0.35}{GW200129_065458}{-0.38}{GW191216_213338}{-0.34}}[{\red{???}}]}
\DeclareRobustCommand{\spintwoygwtcthreenintiethpercentile}[1]{\IfEqCase{#1}{{GW200224_222234}{0.40}{GW191129_134029}{0.34}{GW200311_115853}{0.43}{GW191230_180458}{0.44}{GW191222_033537}{0.41}{GW200225_060421}{0.40}{GW200302_015811}{0.42}{GW200128_022011}{0.46}{GW191204_171526}{0.37}{GW200112_155838}{0.39}{GW200105_162426}{0.36}{GW191105_143521}{0.36}{GW191109_010717}{0.56}{GW200209_085452}{0.45}{GW200115_042309}{0.37}{GW191127_050227}{0.44}{GW200216_220804}{0.43}{GW191215_223052}{0.43}{GW200208_130117}{0.40}{GW200219_094415}{0.43}{GW191103_012549}{0.40}{GW200316_215756}{0.39}{GW200202_154313}{0.35}{GW200129_065458}{0.42}{GW191216_213338}{0.33}}[{\red{???}}]}
\DeclareRobustCommand{\finalmasssourcegwtcthreeminus}[1]{\IfEqCase{#1}{{GW200224_222234}{4.5}{GW191129_134029}{1.1}{GW200311_115853}{3.9}{GW191230_180458}{10}{GW191222_033537}{9.9}{GW200225_060421}{2.8}{GW200302_015811}{6.5}{GW200128_022011}{9.9}{GW191204_171526}{0.92}{GW200112_155838}{4.3}{GW200105_162426}{1.7}{GW191105_143521}{1.2}{GW191109_010717}{15}{GW200209_085452}{8.2}{GW200115_042309}{1.6}{GW191127_050227}{21}{GW200216_220804}{13}{GW191215_223052}{3.8}{GW200208_130117}{6.4}{GW200219_094415}{7.8}{GW191103_012549}{1.7}{GW200316_215756}{1.9}{GW200202_154313}{0.66}{GW200129_065458}{3.3}{GW191216_213338}{0.93}}[{\red{???}}]}
\DeclareRobustCommand{\finalmasssourcegwtcthreemed}[1]{\IfEqCase{#1}{{GW200224_222234}{68.3}{GW191129_134029}{16.7}{GW200311_115853}{59.0}{GW191230_180458}{80}{GW191222_033537}{75.5}{GW200225_060421}{32.1}{GW200302_015811}{55.0}{GW200128_022011}{67.9}{GW191204_171526}{19.14}{GW200112_155838}{60.8}{GW200105_162426}{10.7}{GW191105_143521}{17.6}{GW191109_010717}{107}{GW200209_085452}{58.5}{GW200115_042309}{7.1}{GW191127_050227}{76}{GW200216_220804}{78}{GW191215_223052}{40.7}{GW200208_130117}{62.5}{GW200219_094415}{62.2}{GW191103_012549}{19.0}{GW200316_215756}{20.2}{GW200202_154313}{16.76}{GW200129_065458}{60.3}{GW191216_213338}{18.87}}[{\red{???}}]}
\DeclareRobustCommand{\finalmasssourcegwtcthreeplus}[1]{\IfEqCase{#1}{{GW200224_222234}{6.3}{GW191129_134029}{2.5}{GW200311_115853}{4.8}{GW191230_180458}{16}{GW191222_033537}{15.3}{GW200225_060421}{3.5}{GW200302_015811}{8.9}{GW200128_022011}{14.1}{GW191204_171526}{1.79}{GW200112_155838}{5.3}{GW200105_162426}{2.0}{GW191105_143521}{2.1}{GW191109_010717}{18}{GW200209_085452}{12.2}{GW200115_042309}{2.2}{GW191127_050227}{39}{GW200216_220804}{19}{GW191215_223052}{5.3}{GW200208_130117}{7.3}{GW200219_094415}{11.7}{GW191103_012549}{3.8}{GW200316_215756}{7.5}{GW200202_154313}{1.87}{GW200129_065458}{4.0}{GW191216_213338}{2.84}}[{\red{???}}]}
\DeclareRobustCommand{\finalmasssourcegwtcthreetenthpercentile}[1]{\IfEqCase{#1}{{GW200224_222234}{64.7}{GW191129_134029}{15.7}{GW200311_115853}{56.0}{GW191230_180458}{71}{GW191222_033537}{67.3}{GW200225_060421}{29.9}{GW200302_015811}{49.8}{GW200128_022011}{59.8}{GW191204_171526}{18.38}{GW200112_155838}{57.3}{GW200105_162426}{9.6}{GW191105_143521}{16.7}{GW191109_010717}{96}{GW200209_085452}{51.9}{GW200115_042309}{5.7}{GW191127_050227}{59}{GW200216_220804}{67}{GW191215_223052}{37.6}{GW200208_130117}{57.5}{GW200219_094415}{55.9}{GW191103_012549}{17.5}{GW200316_215756}{18.6}{GW200202_154313}{16.22}{GW200129_065458}{57.6}{GW191216_213338}{18.08}}[{\red{???}}]}
\DeclareRobustCommand{\finalmasssourcegwtcthreenintiethpercentile}[1]{\IfEqCase{#1}{{GW200224_222234}{73.1}{GW191129_134029}{18.5}{GW200311_115853}{62.7}{GW191230_180458}{91}{GW191222_033537}{87.3}{GW200225_060421}{34.7}{GW200302_015811}{61.5}{GW200128_022011}{78.6}{GW191204_171526}{20.40}{GW200112_155838}{64.6}{GW200105_162426}{12.0}{GW191105_143521}{19.1}{GW191109_010717}{121}{GW200209_085452}{67.5}{GW200115_042309}{8.7}{GW191127_050227}{105}{GW200216_220804}{92}{GW191215_223052}{44.8}{GW200208_130117}{68.2}{GW200219_094415}{71.3}{GW191103_012549}{21.4}{GW200316_215756}{24.8}{GW200202_154313}{17.99}{GW200129_065458}{63.4}{GW191216_213338}{20.65}}[{\red{???}}]}
\DeclareRobustCommand{\spinoneygwtcthreeminus}[1]{\IfEqCase{#1}{{GW200224_222234}{0.59}{GW191129_134029}{0.36}{GW200311_115853}{0.56}{GW191230_180458}{0.65}{GW191222_033537}{0.53}{GW200225_060421}{0.65}{GW200302_015811}{0.54}{GW200128_022011}{0.66}{GW191204_171526}{0.48}{GW200112_155838}{0.49}{GW200105_162426}{0.13}{GW191105_143521}{0.41}{GW191109_010717}{0.74}{GW200209_085452}{0.65}{GW200115_042309}{0.35}{GW191127_050227}{0.70}{GW200216_220804}{0.60}{GW191215_223052}{0.63}{GW200208_130117}{0.47}{GW200219_094415}{0.61}{GW191103_012549}{0.51}{GW200316_215756}{0.39}{GW200202_154313}{0.36}{GW200129_065458}{0.58}{GW191216_213338}{0.29}}[{\red{???}}]}
\DeclareRobustCommand{\spinoneygwtcthreemed}[1]{\IfEqCase{#1}{{GW200224_222234}{0.00}{GW191129_134029}{0.00}{GW200311_115853}{0.00}{GW191230_180458}{0.00}{GW191222_033537}{0.00}{GW200225_060421}{0.00}{GW200302_015811}{0.00}{GW200128_022011}{0.00}{GW191204_171526}{0.00}{GW200112_155838}{0.00}{GW200105_162426}{0.00}{GW191105_143521}{0.00}{GW191109_010717}{0.01}{GW200209_085452}{0.00}{GW200115_042309}{0.00}{GW191127_050227}{0.00}{GW200216_220804}{0.00}{GW191215_223052}{0.00}{GW200208_130117}{0.00}{GW200219_094415}{0.00}{GW191103_012549}{0.00}{GW200316_215756}{-0.01}{GW200202_154313}{0.00}{GW200129_065458}{0.00}{GW191216_213338}{0.00}}[{\red{???}}]}
\DeclareRobustCommand{\spinoneygwtcthreeplus}[1]{\IfEqCase{#1}{{GW200224_222234}{0.57}{GW191129_134029}{0.37}{GW200311_115853}{0.58}{GW191230_180458}{0.64}{GW191222_033537}{0.53}{GW200225_060421}{0.64}{GW200302_015811}{0.53}{GW200128_022011}{0.66}{GW191204_171526}{0.49}{GW200112_155838}{0.48}{GW200105_162426}{0.13}{GW191105_143521}{0.41}{GW191109_010717}{0.70}{GW200209_085452}{0.66}{GW200115_042309}{0.33}{GW191127_050227}{0.71}{GW200216_220804}{0.61}{GW191215_223052}{0.63}{GW200208_130117}{0.49}{GW200219_094415}{0.59}{GW191103_012549}{0.53}{GW200316_215756}{0.38}{GW200202_154313}{0.37}{GW200129_065458}{0.72}{GW191216_213338}{0.29}}[{\red{???}}]}
\DeclareRobustCommand{\spinoneygwtcthreetenthpercentile}[1]{\IfEqCase{#1}{{GW200224_222234}{-0.45}{GW191129_134029}{-0.25}{GW200311_115853}{-0.40}{GW191230_180458}{-0.50}{GW191222_033537}{-0.37}{GW200225_060421}{-0.52}{GW200302_015811}{-0.40}{GW200128_022011}{-0.52}{GW191204_171526}{-0.37}{GW200112_155838}{-0.35}{GW200105_162426}{-0.09}{GW191105_143521}{-0.27}{GW191109_010717}{-0.60}{GW200209_085452}{-0.49}{GW200115_042309}{-0.26}{GW191127_050227}{-0.54}{GW200216_220804}{-0.46}{GW191215_223052}{-0.47}{GW200208_130117}{-0.33}{GW200219_094415}{-0.46}{GW191103_012549}{-0.38}{GW200316_215756}{-0.29}{GW200202_154313}{-0.25}{GW200129_065458}{-0.42}{GW191216_213338}{-0.19}}[{\red{???}}]}
\DeclareRobustCommand{\spinoneygwtcthreenintiethpercentile}[1]{\IfEqCase{#1}{{GW200224_222234}{0.43}{GW191129_134029}{0.26}{GW200311_115853}{0.43}{GW191230_180458}{0.48}{GW191222_033537}{0.37}{GW200225_060421}{0.51}{GW200302_015811}{0.39}{GW200128_022011}{0.53}{GW191204_171526}{0.37}{GW200112_155838}{0.35}{GW200105_162426}{0.08}{GW191105_143521}{0.28}{GW191109_010717}{0.59}{GW200209_085452}{0.50}{GW200115_042309}{0.23}{GW191127_050227}{0.55}{GW200216_220804}{0.46}{GW191215_223052}{0.49}{GW200208_130117}{0.36}{GW200219_094415}{0.43}{GW191103_012549}{0.38}{GW200316_215756}{0.26}{GW200202_154313}{0.25}{GW200129_065458}{0.56}{GW191216_213338}{0.21}}[{\red{???}}]}
\DeclareRobustCommand{\costilttwogwtcthreeminus}[1]{\IfEqCase{#1}{{GW200224_222234}{1.01}{GW191129_134029}{1.07}{GW200311_115853}{0.84}{GW191230_180458}{0.79}{GW191222_033537}{0.81}{GW200225_060421}{0.71}{GW200302_015811}{0.98}{GW200128_022011}{0.99}{GW191204_171526}{1.06}{GW200112_155838}{1.02}{GW200105_162426}{0.89}{GW191105_143521}{0.83}{GW191109_010717}{0.66}{GW200209_085452}{0.67}{GW200115_042309}{0.64}{GW191127_050227}{1.06}{GW200216_220804}{1.03}{GW191215_223052}{0.79}{GW200208_130117}{0.76}{GW200219_094415}{0.75}{GW191103_012549}{1.19}{GW200316_215756}{1.05}{GW200202_154313}{0.97}{GW200129_065458}{1.21}{GW191216_213338}{1.16}}[{\red{???}}]}
\DeclareRobustCommand{\costilttwogwtcthreemed}[1]{\IfEqCase{#1}{{GW200224_222234}{0.19}{GW191129_134029}{0.30}{GW200311_115853}{-0.02}{GW191230_180458}{-0.12}{GW191222_033537}{-0.09}{GW200225_060421}{-0.22}{GW200302_015811}{0.12}{GW200128_022011}{0.15}{GW191204_171526}{0.45}{GW200112_155838}{0.25}{GW200105_162426}{0.02}{GW191105_143521}{-0.03}{GW191109_010717}{-0.26}{GW200209_085452}{-0.26}{GW200115_042309}{-0.31}{GW191127_050227}{0.22}{GW200216_220804}{0.18}{GW191215_223052}{-0.09}{GW200208_130117}{-0.15}{GW200219_094415}{-0.17}{GW191103_012549}{0.48}{GW200316_215756}{0.37}{GW200202_154313}{0.22}{GW200129_065458}{0.41}{GW191216_213338}{0.40}}[{\red{???}}]}
\DeclareRobustCommand{\costilttwogwtcthreeplus}[1]{\IfEqCase{#1}{{GW200224_222234}{0.71}{GW191129_134029}{0.63}{GW200311_115853}{0.87}{GW191230_180458}{0.96}{GW191222_033537}{0.95}{GW200225_060421}{1.04}{GW200302_015811}{0.78}{GW200128_022011}{0.74}{GW191204_171526}{0.50}{GW200112_155838}{0.67}{GW200105_162426}{0.84}{GW191105_143521}{0.89}{GW191109_010717}{1.04}{GW200209_085452}{1.05}{GW200115_042309}{1.13}{GW191127_050227}{0.71}{GW200216_220804}{0.75}{GW191215_223052}{0.93}{GW200208_130117}{1.00}{GW200219_094415}{0.99}{GW191103_012549}{0.48}{GW200316_215756}{0.57}{GW200202_154313}{0.68}{GW200129_065458}{0.54}{GW191216_213338}{0.54}}[{\red{???}}]}
\DeclareRobustCommand{\costilttwogwtcthreetenthpercentile}[1]{\IfEqCase{#1}{{GW200224_222234}{-0.67}{GW191129_134029}{-0.56}{GW200311_115853}{-0.75}{GW191230_180458}{-0.81}{GW191222_033537}{-0.80}{GW200225_060421}{-0.85}{GW200302_015811}{-0.72}{GW200128_022011}{-0.69}{GW191204_171526}{-0.38}{GW200112_155838}{-0.61}{GW200105_162426}{-0.74}{GW191105_143521}{-0.73}{GW191109_010717}{-0.85}{GW200209_085452}{-0.86}{GW200115_042309}{-0.90}{GW191127_050227}{-0.69}{GW200216_220804}{-0.71}{GW191215_223052}{-0.78}{GW200208_130117}{-0.82}{GW200219_094415}{-0.84}{GW191103_012549}{-0.50}{GW200316_215756}{-0.46}{GW200202_154313}{-0.55}{GW200129_065458}{-0.61}{GW191216_213338}{-0.54}}[{\red{???}}]}
\DeclareRobustCommand{\costilttwogwtcthreenintiethpercentile}[1]{\IfEqCase{#1}{{GW200224_222234}{0.81}{GW191129_134029}{0.87}{GW200311_115853}{0.72}{GW191230_180458}{0.70}{GW191222_033537}{0.71}{GW200225_060421}{0.65}{GW200302_015811}{0.82}{GW200128_022011}{0.79}{GW191204_171526}{0.89}{GW200112_155838}{0.82}{GW200105_162426}{0.74}{GW191105_143521}{0.74}{GW191109_010717}{0.62}{GW200209_085452}{0.61}{GW200115_042309}{0.65}{GW191127_050227}{0.86}{GW200216_220804}{0.85}{GW191215_223052}{0.70}{GW200208_130117}{0.70}{GW200219_094415}{0.69}{GW191103_012549}{0.91}{GW200316_215756}{0.87}{GW200202_154313}{0.82}{GW200129_065458}{0.90}{GW191216_213338}{0.89}}[{\red{???}}]}
\DeclareRobustCommand{\massonesourcegwtcthreeminus}[1]{\IfEqCase{#1}{{GW200224_222234}{4.4}{GW191129_134029}{2.1}{GW200311_115853}{3.8}{GW191230_180458}{8.6}{GW191222_033537}{8.0}{GW200225_060421}{3.0}{GW200302_015811}{8.4}{GW200128_022011}{7.6}{GW191204_171526}{1.8}{GW200112_155838}{4.5}{GW200105_162426}{1.7}{GW191105_143521}{1.6}{GW191109_010717}{11}{GW200209_085452}{6.3}{GW200115_042309}{2.5}{GW191127_050227}{20}{GW200216_220804}{13}{GW191215_223052}{4.1}{GW200208_130117}{6.2}{GW200219_094415}{6.9}{GW191103_012549}{2.2}{GW200316_215756}{2.9}{GW200202_154313}{1.4}{GW200129_065458}{3.2}{GW191216_213338}{2.2}}[{\red{???}}]}
\DeclareRobustCommand{\massonesourcegwtcthreemed}[1]{\IfEqCase{#1}{{GW200224_222234}{39.8}{GW191129_134029}{10.6}{GW200311_115853}{34.2}{GW191230_180458}{47.7}{GW191222_033537}{45.1}{GW200225_060421}{19.3}{GW200302_015811}{37.0}{GW200128_022011}{40.4}{GW191204_171526}{11.9}{GW200112_155838}{35.6}{GW200105_162426}{9.0}{GW191105_143521}{10.7}{GW191109_010717}{65}{GW200209_085452}{34.6}{GW200115_042309}{5.9}{GW191127_050227}{53}{GW200216_220804}{51}{GW191215_223052}{24.6}{GW200208_130117}{37.8}{GW200219_094415}{37.5}{GW191103_012549}{11.8}{GW200316_215756}{13.1}{GW200202_154313}{10.1}{GW200129_065458}{34.5}{GW191216_213338}{12.1}}[{\red{???}}]}
\DeclareRobustCommand{\massonesourcegwtcthreeplus}[1]{\IfEqCase{#1}{{GW200224_222234}{6.9}{GW191129_134029}{4.1}{GW200311_115853}{6.4}{GW191230_180458}{13.4}{GW191222_033537}{10.9}{GW200225_060421}{5.0}{GW200302_015811}{8.9}{GW200128_022011}{11.1}{GW191204_171526}{3.3}{GW200112_155838}{6.7}{GW200105_162426}{1.7}{GW191105_143521}{3.7}{GW191109_010717}{11}{GW200209_085452}{10.0}{GW200115_042309}{2.0}{GW191127_050227}{47}{GW200216_220804}{22}{GW191215_223052}{7.0}{GW200208_130117}{9.2}{GW200219_094415}{10.1}{GW191103_012549}{6.2}{GW200316_215756}{10.2}{GW200202_154313}{3.5}{GW200129_065458}{9.9}{GW191216_213338}{4.6}}[{\red{???}}]}
\DeclareRobustCommand{\massonesourcegwtcthreetenthpercentile}[1]{\IfEqCase{#1}{{GW200224_222234}{36.1}{GW191129_134029}{8.8}{GW200311_115853}{31.1}{GW191230_180458}{40.6}{GW191222_033537}{38.5}{GW200225_060421}{16.8}{GW200302_015811}{30.3}{GW200128_022011}{34.3}{GW191204_171526}{10.3}{GW200112_155838}{31.9}{GW200105_162426}{8.0}{GW191105_143521}{9.3}{GW191109_010717}{57}{GW200209_085452}{29.5}{GW200115_042309}{3.7}{GW191127_050227}{36}{GW200216_220804}{40}{GW191215_223052}{21.2}{GW200208_130117}{32.7}{GW200219_094415}{31.8}{GW191103_012549}{9.9}{GW200316_215756}{10.6}{GW200202_154313}{8.9}{GW200129_065458}{31.9}{GW191216_213338}{10.1}}[{\red{???}}]}
\DeclareRobustCommand{\massonesourcegwtcthreenintiethpercentile}[1]{\IfEqCase{#1}{{GW200224_222234}{45.0}{GW191129_134029}{13.7}{GW200311_115853}{38.9}{GW191230_180458}{57.6}{GW191222_033537}{53.2}{GW200225_060421}{23.0}{GW200302_015811}{43.9}{GW200128_022011}{48.7}{GW191204_171526}{14.3}{GW200112_155838}{40.8}{GW200105_162426}{10.0}{GW191105_143521}{13.3}{GW191109_010717}{73}{GW200209_085452}{41.9}{GW200115_042309}{7.4}{GW191127_050227}{88}{GW200216_220804}{68}{GW191215_223052}{29.8}{GW200208_130117}{44.9}{GW200219_094415}{45.1}{GW191103_012549}{16.1}{GW200316_215756}{20.0}{GW200202_154313}{12.7}{GW200129_065458}{42.3}{GW191216_213338}{15.2}}[{\red{???}}]}
\DeclareRobustCommand{\spintwoxgwtcthreeminus}[1]{\IfEqCase{#1}{{GW200224_222234}{0.58}{GW191129_134029}{0.47}{GW200311_115853}{0.55}{GW191230_180458}{0.60}{GW191222_033537}{0.55}{GW200225_060421}{0.55}{GW200302_015811}{0.58}{GW200128_022011}{0.61}{GW191204_171526}{0.53}{GW200112_155838}{0.51}{GW200105_162426}{0.52}{GW191105_143521}{0.51}{GW191109_010717}{0.68}{GW200209_085452}{0.58}{GW200115_042309}{0.51}{GW191127_050227}{0.60}{GW200216_220804}{0.60}{GW191215_223052}{0.58}{GW200208_130117}{0.54}{GW200219_094415}{0.60}{GW191103_012549}{0.53}{GW200316_215756}{0.55}{GW200202_154313}{0.50}{GW200129_065458}{0.57}{GW191216_213338}{0.44}}[{\red{???}}]}
\DeclareRobustCommand{\spintwoxgwtcthreemed}[1]{\IfEqCase{#1}{{GW200224_222234}{0.01}{GW191129_134029}{0.00}{GW200311_115853}{0.00}{GW191230_180458}{0.00}{GW191222_033537}{0.00}{GW200225_060421}{0.00}{GW200302_015811}{0.00}{GW200128_022011}{0.00}{GW191204_171526}{0.00}{GW200112_155838}{0.00}{GW200105_162426}{0.00}{GW191105_143521}{0.00}{GW191109_010717}{0.00}{GW200209_085452}{0.00}{GW200115_042309}{0.00}{GW191127_050227}{0.00}{GW200216_220804}{0.00}{GW191215_223052}{0.00}{GW200208_130117}{0.00}{GW200219_094415}{0.00}{GW191103_012549}{0.00}{GW200316_215756}{0.00}{GW200202_154313}{0.00}{GW200129_065458}{0.00}{GW191216_213338}{-0.01}}[{\red{???}}]}
\DeclareRobustCommand{\spintwoxgwtcthreeplus}[1]{\IfEqCase{#1}{{GW200224_222234}{0.58}{GW191129_134029}{0.46}{GW200311_115853}{0.56}{GW191230_180458}{0.60}{GW191222_033537}{0.55}{GW200225_060421}{0.55}{GW200302_015811}{0.58}{GW200128_022011}{0.61}{GW191204_171526}{0.50}{GW200112_155838}{0.53}{GW200105_162426}{0.50}{GW191105_143521}{0.51}{GW191109_010717}{0.65}{GW200209_085452}{0.60}{GW200115_042309}{0.51}{GW191127_050227}{0.60}{GW200216_220804}{0.60}{GW191215_223052}{0.58}{GW200208_130117}{0.57}{GW200219_094415}{0.58}{GW191103_012549}{0.55}{GW200316_215756}{0.53}{GW200202_154313}{0.50}{GW200129_065458}{0.55}{GW191216_213338}{0.47}}[{\red{???}}]}
\DeclareRobustCommand{\spintwoxgwtcthreetenthpercentile}[1]{\IfEqCase{#1}{{GW200224_222234}{-0.41}{GW191129_134029}{-0.33}{GW200311_115853}{-0.41}{GW191230_180458}{-0.45}{GW191222_033537}{-0.41}{GW200225_060421}{-0.40}{GW200302_015811}{-0.42}{GW200128_022011}{-0.46}{GW191204_171526}{-0.39}{GW200112_155838}{-0.37}{GW200105_162426}{-0.36}{GW191105_143521}{-0.35}{GW191109_010717}{-0.53}{GW200209_085452}{-0.44}{GW200115_042309}{-0.38}{GW191127_050227}{-0.46}{GW200216_220804}{-0.45}{GW191215_223052}{-0.42}{GW200208_130117}{-0.40}{GW200219_094415}{-0.47}{GW191103_012549}{-0.40}{GW200316_215756}{-0.41}{GW200202_154313}{-0.36}{GW200129_065458}{-0.41}{GW191216_213338}{-0.33}}[{\red{???}}]}
\DeclareRobustCommand{\spintwoxgwtcthreenintiethpercentile}[1]{\IfEqCase{#1}{{GW200224_222234}{0.44}{GW191129_134029}{0.33}{GW200311_115853}{0.40}{GW191230_180458}{0.45}{GW191222_033537}{0.39}{GW200225_060421}{0.40}{GW200302_015811}{0.43}{GW200128_022011}{0.47}{GW191204_171526}{0.38}{GW200112_155838}{0.39}{GW200105_162426}{0.35}{GW191105_143521}{0.36}{GW191109_010717}{0.51}{GW200209_085452}{0.45}{GW200115_042309}{0.36}{GW191127_050227}{0.45}{GW200216_220804}{0.45}{GW191215_223052}{0.43}{GW200208_130117}{0.42}{GW200219_094415}{0.43}{GW191103_012549}{0.40}{GW200316_215756}{0.39}{GW200202_154313}{0.35}{GW200129_065458}{0.41}{GW191216_213338}{0.31}}[{\red{???}}]}
\DeclareRobustCommand{\phitwogwtcthreeminus}[1]{\IfEqCase{#1}{{GW200224_222234}{2.9}{GW191129_134029}{2.8}{GW200311_115853}{2.8}{GW191230_180458}{2.8}{GW191222_033537}{2.9}{GW200225_060421}{2.8}{GW200302_015811}{2.9}{GW200128_022011}{2.9}{GW191204_171526}{2.8}{GW200112_155838}{2.9}{GW200105_162426}{2.8}{GW191105_143521}{2.8}{GW191109_010717}{2.7}{GW200209_085452}{2.8}{GW200115_042309}{2.8}{GW191127_050227}{2.8}{GW200216_220804}{2.9}{GW191215_223052}{2.8}{GW200208_130117}{2.8}{GW200219_094415}{2.8}{GW191103_012549}{2.8}{GW200316_215756}{2.8}{GW200202_154313}{2.8}{GW200129_065458}{2.7}{GW191216_213338}{2.8}}[{\red{???}}]}
\DeclareRobustCommand{\phitwogwtcthreemed}[1]{\IfEqCase{#1}{{GW200224_222234}{3.2}{GW191129_134029}{3.1}{GW200311_115853}{3.1}{GW191230_180458}{3.1}{GW191222_033537}{3.2}{GW200225_060421}{3.1}{GW200302_015811}{3.1}{GW200128_022011}{3.2}{GW191204_171526}{3.2}{GW200112_155838}{3.2}{GW200105_162426}{3.1}{GW191105_143521}{3.2}{GW191109_010717}{3.0}{GW200209_085452}{3.1}{GW200115_042309}{3.1}{GW191127_050227}{3.1}{GW200216_220804}{3.2}{GW191215_223052}{3.1}{GW200208_130117}{3.1}{GW200219_094415}{3.2}{GW191103_012549}{3.1}{GW200316_215756}{3.1}{GW200202_154313}{3.2}{GW200129_065458}{3.1}{GW191216_213338}{3.2}}[{\red{???}}]}
\DeclareRobustCommand{\phitwogwtcthreeplus}[1]{\IfEqCase{#1}{{GW200224_222234}{2.8}{GW191129_134029}{2.9}{GW200311_115853}{2.9}{GW191230_180458}{2.8}{GW191222_033537}{2.8}{GW200225_060421}{2.9}{GW200302_015811}{2.8}{GW200128_022011}{2.8}{GW191204_171526}{2.8}{GW200112_155838}{2.8}{GW200105_162426}{2.8}{GW191105_143521}{2.8}{GW191109_010717}{2.9}{GW200209_085452}{2.8}{GW200115_042309}{2.8}{GW191127_050227}{2.9}{GW200216_220804}{2.8}{GW191215_223052}{2.8}{GW200208_130117}{2.8}{GW200219_094415}{2.8}{GW191103_012549}{2.8}{GW200316_215756}{2.8}{GW200202_154313}{2.8}{GW200129_065458}{2.9}{GW191216_213338}{2.8}}[{\red{???}}]}
\DeclareRobustCommand{\phitwogwtcthreetenthpercentile}[1]{\IfEqCase{#1}{{GW200224_222234}{0.6}{GW191129_134029}{0.6}{GW200311_115853}{0.6}{GW191230_180458}{0.6}{GW191222_033537}{0.6}{GW200225_060421}{0.7}{GW200302_015811}{0.6}{GW200128_022011}{0.6}{GW191204_171526}{0.6}{GW200112_155838}{0.6}{GW200105_162426}{0.6}{GW191105_143521}{0.6}{GW191109_010717}{0.6}{GW200209_085452}{0.6}{GW200115_042309}{0.6}{GW191127_050227}{0.6}{GW200216_220804}{0.6}{GW191215_223052}{0.6}{GW200208_130117}{0.6}{GW200219_094415}{0.6}{GW191103_012549}{0.7}{GW200316_215756}{0.6}{GW200202_154313}{0.6}{GW200129_065458}{0.6}{GW191216_213338}{0.7}}[{\red{???}}]}
\DeclareRobustCommand{\phitwogwtcthreenintiethpercentile}[1]{\IfEqCase{#1}{{GW200224_222234}{5.7}{GW191129_134029}{5.7}{GW200311_115853}{5.6}{GW191230_180458}{5.6}{GW191222_033537}{5.7}{GW200225_060421}{5.6}{GW200302_015811}{5.7}{GW200128_022011}{5.7}{GW191204_171526}{5.6}{GW200112_155838}{5.6}{GW200105_162426}{5.6}{GW191105_143521}{5.7}{GW191109_010717}{5.6}{GW200209_085452}{5.7}{GW200115_042309}{5.6}{GW191127_050227}{5.6}{GW200216_220804}{5.7}{GW191215_223052}{5.6}{GW200208_130117}{5.7}{GW200219_094415}{5.7}{GW191103_012549}{5.7}{GW200316_215756}{5.6}{GW200202_154313}{5.7}{GW200129_065458}{5.7}{GW191216_213338}{5.6}}[{\red{???}}]}
\DeclareRobustCommand{\chipgwtcthreeminus}[1]{\IfEqCase{#1}{{GW200224_222234}{0.36}{GW191129_134029}{0.19}{GW200311_115853}{0.35}{GW191230_180458}{0.39}{GW191222_033537}{0.32}{GW200225_060421}{0.38}{GW200302_015811}{0.29}{GW200128_022011}{0.40}{GW191204_171526}{0.26}{GW200112_155838}{0.30}{GW200105_162426}{0.07}{GW191105_143521}{0.24}{GW191109_010717}{0.37}{GW200209_085452}{0.38}{GW200115_042309}{0.16}{GW191127_050227}{0.41}{GW200216_220804}{0.35}{GW191215_223052}{0.38}{GW200208_130117}{0.29}{GW200219_094415}{0.35}{GW191103_012549}{0.26}{GW200316_215756}{0.20}{GW200202_154313}{0.22}{GW200129_065458}{0.37}{GW191216_213338}{0.15}}[{\red{???}}]}
\DeclareRobustCommand{\chipgwtcthreemed}[1]{\IfEqCase{#1}{{GW200224_222234}{0.50}{GW191129_134029}{0.26}{GW200311_115853}{0.45}{GW191230_180458}{0.52}{GW191222_033537}{0.41}{GW200225_060421}{0.53}{GW200302_015811}{0.38}{GW200128_022011}{0.57}{GW191204_171526}{0.39}{GW200112_155838}{0.39}{GW200105_162426}{0.09}{GW191105_143521}{0.30}{GW191109_010717}{0.63}{GW200209_085452}{0.52}{GW200115_042309}{0.20}{GW191127_050227}{0.52}{GW200216_220804}{0.45}{GW191215_223052}{0.50}{GW200208_130117}{0.38}{GW200219_094415}{0.48}{GW191103_012549}{0.40}{GW200316_215756}{0.29}{GW200202_154313}{0.28}{GW200129_065458}{0.52}{GW191216_213338}{0.23}}[{\red{???}}]}
\DeclareRobustCommand{\chipgwtcthreeplus}[1]{\IfEqCase{#1}{{GW200224_222234}{0.37}{GW191129_134029}{0.36}{GW200311_115853}{0.40}{GW191230_180458}{0.38}{GW191222_033537}{0.41}{GW200225_060421}{0.34}{GW200302_015811}{0.44}{GW200128_022011}{0.33}{GW191204_171526}{0.35}{GW200112_155838}{0.39}{GW200105_162426}{0.17}{GW191105_143521}{0.45}{GW191109_010717}{0.29}{GW200209_085452}{0.38}{GW200115_042309}{0.34}{GW191127_050227}{0.41}{GW200216_220804}{0.42}{GW191215_223052}{0.38}{GW200208_130117}{0.41}{GW200219_094415}{0.40}{GW191103_012549}{0.41}{GW200316_215756}{0.38}{GW200202_154313}{0.40}{GW200129_065458}{0.42}{GW191216_213338}{0.35}}[{\red{???}}]}
\DeclareRobustCommand{\chipgwtcthreetenthpercentile}[1]{\IfEqCase{#1}{{GW200224_222234}{0.20}{GW191129_134029}{0.10}{GW200311_115853}{0.16}{GW191230_180458}{0.20}{GW191222_033537}{0.15}{GW200225_060421}{0.22}{GW200302_015811}{0.13}{GW200128_022011}{0.24}{GW191204_171526}{0.18}{GW200112_155838}{0.13}{GW200105_162426}{0.03}{GW191105_143521}{0.09}{GW191109_010717}{0.33}{GW200209_085452}{0.20}{GW200115_042309}{0.06}{GW191127_050227}{0.17}{GW200216_220804}{0.15}{GW191215_223052}{0.19}{GW200208_130117}{0.13}{GW200219_094415}{0.18}{GW191103_012549}{0.18}{GW200316_215756}{0.12}{GW200202_154313}{0.09}{GW200129_065458}{0.21}{GW191216_213338}{0.10}}[{\red{???}}]}
\DeclareRobustCommand{\chipgwtcthreenintiethpercentile}[1]{\IfEqCase{#1}{{GW200224_222234}{0.80}{GW191129_134029}{0.54}{GW200311_115853}{0.77}{GW191230_180458}{0.84}{GW191222_033537}{0.75}{GW200225_060421}{0.82}{GW200302_015811}{0.74}{GW200128_022011}{0.85}{GW191204_171526}{0.67}{GW200112_155838}{0.70}{GW200105_162426}{0.19}{GW191105_143521}{0.65}{GW191109_010717}{0.87}{GW200209_085452}{0.84}{GW200115_042309}{0.46}{GW191127_050227}{0.88}{GW200216_220804}{0.80}{GW191215_223052}{0.82}{GW200208_130117}{0.71}{GW200219_094415}{0.80}{GW191103_012549}{0.72}{GW200316_215756}{0.58}{GW200202_154313}{0.59}{GW200129_065458}{0.91}{GW191216_213338}{0.48}}[{\red{???}}]}
\DeclareRobustCommand{\chirpmassdetgwtcthreeminus}[1]{\IfEqCase{#1}{{GW200224_222234}{3.8}{GW191129_134029}{0.05}{GW200311_115853}{2.8}{GW191230_180458}{9.5}{GW191222_033537}{6.5}{GW200225_060421}{1.97}{GW200302_015811}{4.3}{GW200128_022011}{6.6}{GW191204_171526}{0.05}{GW200112_155838}{2.4}{GW200105_162426}{0.01}{GW191105_143521}{0.14}{GW191109_010717}{9.3}{GW200209_085452}{7.3}{GW200115_042309}{0.01}{GW191127_050227}{19}{GW200216_220804}{20}{GW191215_223052}{1.4}{GW200208_130117}{4.8}{GW200219_094415}{6.2}{GW191103_012549}{0.12}{GW200316_215756}{0.12}{GW200202_154313}{0.05}{GW200129_065458}{2.6}{GW191216_213338}{0.05}}[{\red{???}}]}
\DeclareRobustCommand{\chirpmassdetgwtcthreemed}[1]{\IfEqCase{#1}{{GW200224_222234}{41.1}{GW191129_134029}{8.49}{GW200311_115853}{32.7}{GW191230_180458}{62.8}{GW191222_033537}{51.0}{GW200225_060421}{17.65}{GW200302_015811}{30.4}{GW200128_022011}{50.5}{GW191204_171526}{9.69}{GW200112_155838}{33.9}{GW200105_162426}{3.62}{GW191105_143521}{9.58}{GW191109_010717}{60.1}{GW200209_085452}{42.9}{GW200115_042309}{2.58}{GW191127_050227}{48}{GW200216_220804}{56}{GW191215_223052}{24.9}{GW200208_130117}{38.8}{GW200219_094415}{43.7}{GW191103_012549}{9.98}{GW200316_215756}{10.68}{GW200202_154313}{8.15}{GW200129_065458}{32.1}{GW191216_213338}{8.94}}[{\red{???}}]}
\DeclareRobustCommand{\chirpmassdetgwtcthreeplus}[1]{\IfEqCase{#1}{{GW200224_222234}{3.6}{GW191129_134029}{0.06}{GW200311_115853}{2.7}{GW191230_180458}{9.4}{GW191222_033537}{7.2}{GW200225_060421}{0.98}{GW200302_015811}{7.7}{GW200128_022011}{7.5}{GW191204_171526}{0.05}{GW200112_155838}{2.9}{GW200105_162426}{0.01}{GW191105_143521}{0.12}{GW191109_010717}{9.8}{GW200209_085452}{8.7}{GW200115_042309}{0.01}{GW191127_050227}{21}{GW200216_220804}{14}{GW191215_223052}{1.5}{GW200208_130117}{5.2}{GW200219_094415}{6.3}{GW191103_012549}{0.13}{GW200316_215756}{0.12}{GW200202_154313}{0.05}{GW200129_065458}{1.8}{GW191216_213338}{0.05}}[{\red{???}}]}
\DeclareRobustCommand{\chirpmassdetgwtcthreetenthpercentile}[1]{\IfEqCase{#1}{{GW200224_222234}{38.2}{GW191129_134029}{8.45}{GW200311_115853}{30.6}{GW191230_180458}{55.5}{GW191222_033537}{46.1}{GW200225_060421}{15.96}{GW200302_015811}{27.1}{GW200128_022011}{45.4}{GW191204_171526}{9.65}{GW200112_155838}{32.1}{GW200105_162426}{3.61}{GW191105_143521}{9.48}{GW191109_010717}{52.7}{GW200209_085452}{37.3}{GW200115_042309}{2.57}{GW191127_050227}{33}{GW200216_220804}{39}{GW191215_223052}{23.8}{GW200208_130117}{35.0}{GW200219_094415}{39.0}{GW191103_012549}{9.88}{GW200316_215756}{10.59}{GW200202_154313}{8.11}{GW200129_065458}{30.2}{GW191216_213338}{8.90}}[{\red{???}}]}
\DeclareRobustCommand{\chirpmassdetgwtcthreenintiethpercentile}[1]{\IfEqCase{#1}{{GW200224_222234}{43.8}{GW191129_134029}{8.53}{GW200311_115853}{34.6}{GW191230_180458}{69.9}{GW191222_033537}{56.3}{GW200225_060421}{18.43}{GW200302_015811}{36.4}{GW200128_022011}{56.3}{GW191204_171526}{9.74}{GW200112_155838}{36.2}{GW200105_162426}{3.62}{GW191105_143521}{9.67}{GW191109_010717}{67.5}{GW200209_085452}{49.4}{GW200115_042309}{2.59}{GW191127_050227}{65}{GW200216_220804}{67}{GW191215_223052}{26.1}{GW200208_130117}{42.8}{GW200219_094415}{48.5}{GW191103_012549}{10.07}{GW200316_215756}{10.76}{GW200202_154313}{8.19}{GW200129_065458}{33.4}{GW191216_213338}{8.98}}[{\red{???}}]}
\DeclareRobustCommand{\chirpmasssourcegwtcthreeminus}[1]{\IfEqCase{#1}{{GW200224_222234}{2.5}{GW191129_134029}{0.27}{GW200311_115853}{2.0}{GW191230_180458}{4.9}{GW191222_033537}{5.0}{GW200225_060421}{1.4}{GW200302_015811}{2.9}{GW200128_022011}{4.6}{GW191204_171526}{0.25}{GW200112_155838}{2.1}{GW200105_162426}{0.08}{GW191105_143521}{0.45}{GW191109_010717}{7.5}{GW200209_085452}{3.8}{GW200115_042309}{0.07}{GW191127_050227}{9.1}{GW200216_220804}{8.5}{GW191215_223052}{1.5}{GW200208_130117}{3.1}{GW200219_094415}{3.8}{GW191103_012549}{0.57}{GW200316_215756}{0.55}{GW200202_154313}{0.20}{GW200129_065458}{2.3}{GW191216_213338}{0.19}}[{\red{???}}]}
\DeclareRobustCommand{\chirpmasssourcegwtcthreemed}[1]{\IfEqCase{#1}{{GW200224_222234}{31.0}{GW191129_134029}{7.28}{GW200311_115853}{26.6}{GW191230_180458}{35.5}{GW191222_033537}{33.8}{GW200225_060421}{14.2}{GW200302_015811}{23.3}{GW200128_022011}{30.6}{GW191204_171526}{8.53}{GW200112_155838}{27.4}{GW200105_162426}{3.42}{GW191105_143521}{7.82}{GW191109_010717}{47.5}{GW200209_085452}{26.1}{GW200115_042309}{2.43}{GW191127_050227}{29.9}{GW200216_220804}{32.9}{GW191215_223052}{18.1}{GW200208_130117}{27.7}{GW200219_094415}{27.6}{GW191103_012549}{8.34}{GW200316_215756}{8.75}{GW200202_154313}{7.49}{GW200129_065458}{27.2}{GW191216_213338}{8.33}}[{\red{???}}]}
\DeclareRobustCommand{\chirpmasssourcegwtcthreeplus}[1]{\IfEqCase{#1}{{GW200224_222234}{3.1}{GW191129_134029}{0.42}{GW200311_115853}{2.4}{GW191230_180458}{7.5}{GW191222_033537}{7.1}{GW200225_060421}{1.5}{GW200302_015811}{4.6}{GW200128_022011}{6.7}{GW191204_171526}{0.38}{GW200112_155838}{2.6}{GW200105_162426}{0.08}{GW191105_143521}{0.61}{GW191109_010717}{9.6}{GW200209_085452}{5.6}{GW200115_042309}{0.05}{GW191127_050227}{11.7}{GW200216_220804}{9.3}{GW191215_223052}{2.2}{GW200208_130117}{3.6}{GW200219_094415}{5.6}{GW191103_012549}{0.66}{GW200316_215756}{0.62}{GW200202_154313}{0.24}{GW200129_065458}{2.1}{GW191216_213338}{0.22}}[{\red{???}}]}
\DeclareRobustCommand{\chirpmasssourcegwtcthreetenthpercentile}[1]{\IfEqCase{#1}{{GW200224_222234}{29.0}{GW191129_134029}{7.07}{GW200311_115853}{25.0}{GW191230_180458}{31.6}{GW191222_033537}{29.7}{GW200225_060421}{13.1}{GW200302_015811}{21.0}{GW200128_022011}{26.8}{GW191204_171526}{8.32}{GW200112_155838}{25.7}{GW200105_162426}{3.36}{GW191105_143521}{7.46}{GW191109_010717}{41.5}{GW200209_085452}{23.1}{GW200115_042309}{2.38}{GW191127_050227}{22.8}{GW200216_220804}{26.2}{GW191215_223052}{16.9}{GW200208_130117}{25.3}{GW200219_094415}{24.6}{GW191103_012549}{7.88}{GW200316_215756}{8.31}{GW200202_154313}{7.33}{GW200129_065458}{25.4}{GW191216_213338}{8.18}}[{\red{???}}]}
\DeclareRobustCommand{\chirpmasssourcegwtcthreenintiethpercentile}[1]{\IfEqCase{#1}{{GW200224_222234}{33.3}{GW191129_134029}{7.62}{GW200311_115853}{28.4}{GW191230_180458}{41.0}{GW191222_033537}{39.4}{GW200225_060421}{15.3}{GW200302_015811}{26.7}{GW200128_022011}{35.6}{GW191204_171526}{8.83}{GW200112_155838}{29.3}{GW200105_162426}{3.48}{GW191105_143521}{8.30}{GW191109_010717}{54.5}{GW200209_085452}{30.2}{GW200115_042309}{2.47}{GW191127_050227}{38.9}{GW200216_220804}{40.0}{GW191215_223052}{19.7}{GW200208_130117}{30.5}{GW200219_094415}{32.0}{GW191103_012549}{8.88}{GW200316_215756}{9.26}{GW200202_154313}{7.68}{GW200129_065458}{28.8}{GW191216_213338}{8.51}}[{\red{???}}]}
\DeclareRobustCommand{\totalmassdetgwtcthreeminus}[1]{\IfEqCase{#1}{{GW200224_222234}{7.2}{GW191129_134029}{0.65}{GW200311_115853}{5.7}{GW191230_180458}{19}{GW191222_033537}{13}{GW200225_060421}{4.0}{GW200302_015811}{8.1}{GW200128_022011}{14}{GW191204_171526}{0.48}{GW200112_155838}{5.1}{GW200105_162426}{1.5}{GW191105_143521}{0.50}{GW191109_010717}{17}{GW200209_085452}{16}{GW200115_042309}{1.8}{GW191127_050227}{45}{GW200216_220804}{32}{GW191215_223052}{3.7}{GW200208_130117}{10}{GW200219_094415}{12}{GW191103_012549}{0.68}{GW200316_215756}{1.1}{GW200202_154313}{0.34}{GW200129_065458}{3.8}{GW191216_213338}{0.66}}[{\red{???}}]}
\DeclareRobustCommand{\totalmassdetgwtcthreemed}[1]{\IfEqCase{#1}{{GW200224_222234}{95.3}{GW191129_134029}{20.11}{GW200311_115853}{75.9}{GW191230_180458}{147}{GW191222_033537}{119}{GW200225_060421}{41.2}{GW200302_015811}{74.5}{GW200128_022011}{118}{GW191204_171526}{22.73}{GW200112_155838}{79.1}{GW200105_162426}{11.6}{GW191105_143521}{22.38}{GW191109_010717}{140}{GW200209_085452}{100}{GW200115_042309}{7.8}{GW191127_050227}{130}{GW200216_220804}{135}{GW191215_223052}{58.6}{GW200208_130117}{91}{GW200219_094415}{103}{GW191103_012549}{23.47}{GW200316_215756}{25.5}{GW200202_154313}{19.01}{GW200129_065458}{74.6}{GW191216_213338}{21.17}}[{\red{???}}]}
\DeclareRobustCommand{\totalmassdetgwtcthreeplus}[1]{\IfEqCase{#1}{{GW200224_222234}{8.4}{GW191129_134029}{2.94}{GW200311_115853}{6.2}{GW191230_180458}{21}{GW191222_033537}{16}{GW200225_060421}{3.0}{GW200302_015811}{15.7}{GW200128_022011}{18}{GW191204_171526}{1.93}{GW200112_155838}{6.5}{GW200105_162426}{1.6}{GW191105_143521}{2.35}{GW191109_010717}{21}{GW200209_085452}{20}{GW200115_042309}{1.8}{GW191127_050227}{53}{GW200216_220804}{30}{GW191215_223052}{5.1}{GW200208_130117}{11}{GW200219_094415}{14}{GW191103_012549}{4.58}{GW200316_215756}{8.8}{GW200202_154313}{1.99}{GW200129_065458}{4.5}{GW191216_213338}{2.96}}[{\red{???}}]}
\DeclareRobustCommand{\totalmassdetgwtcthreetenthpercentile}[1]{\IfEqCase{#1}{{GW200224_222234}{89.6}{GW191129_134029}{19.51}{GW200311_115853}{71.5}{GW191230_180458}{132}{GW191222_033537}{109}{GW200225_060421}{37.9}{GW200302_015811}{68.1}{GW200128_022011}{107}{GW191204_171526}{22.30}{GW200112_155838}{75.2}{GW200105_162426}{10.7}{GW191105_143521}{21.96}{GW191109_010717}{127}{GW200209_085452}{88}{GW200115_042309}{6.2}{GW191127_050227}{91}{GW200216_220804}{109}{GW191215_223052}{55.7}{GW200208_130117}{84}{GW200219_094415}{93}{GW191103_012549}{22.87}{GW200316_215756}{24.5}{GW200202_154313}{18.71}{GW200129_065458}{71.7}{GW191216_213338}{20.56}}[{\red{???}}]}
\DeclareRobustCommand{\totalmassdetgwtcthreenintiethpercentile}[1]{\IfEqCase{#1}{{GW200224_222234}{101.5}{GW191129_134029}{22.18}{GW200311_115853}{80.4}{GW191230_180458}{163}{GW191222_033537}{131}{GW200225_060421}{43.5}{GW200302_015811}{86.5}{GW200128_022011}{131}{GW191204_171526}{24.06}{GW200112_155838}{84.0}{GW200105_162426}{12.5}{GW191105_143521}{23.91}{GW191109_010717}{156}{GW200209_085452}{115}{GW200115_042309}{9.2}{GW191127_050227}{171}{GW200216_220804}{158}{GW191215_223052}{62.1}{GW200208_130117}{100}{GW200219_094415}{113}{GW191103_012549}{26.45}{GW200316_215756}{31.0}{GW200202_154313}{20.33}{GW200129_065458}{78.0}{GW191216_213338}{23.01}}[{\red{???}}]}
\DeclareRobustCommand{\redshiftgwtcthreeminus}[1]{\IfEqCase{#1}{{GW200224_222234}{0.10}{GW191129_134029}{0.06}{GW200311_115853}{0.07}{GW191230_180458}{0.27}{GW191222_033537}{0.26}{GW200225_060421}{0.10}{GW200302_015811}{0.13}{GW200128_022011}{0.29}{GW191204_171526}{0.05}{GW200112_155838}{0.08}{GW200105_162426}{0.02}{GW191105_143521}{0.09}{GW191109_010717}{0.12}{GW200209_085452}{0.25}{GW200115_042309}{0.02}{GW191127_050227}{0.29}{GW200216_220804}{0.29}{GW191215_223052}{0.15}{GW200208_130117}{0.14}{GW200219_094415}{0.22}{GW191103_012549}{0.09}{GW200316_215756}{0.08}{GW200202_154313}{0.03}{GW200129_065458}{0.07}{GW191216_213338}{0.03}}[{\red{???}}]}
\DeclareRobustCommand{\redshiftgwtcthreemed}[1]{\IfEqCase{#1}{{GW200224_222234}{0.33}{GW191129_134029}{0.17}{GW200311_115853}{0.23}{GW191230_180458}{0.76}{GW191222_033537}{0.51}{GW200225_060421}{0.22}{GW200302_015811}{0.31}{GW200128_022011}{0.66}{GW191204_171526}{0.14}{GW200112_155838}{0.24}{GW200105_162426}{0.06}{GW191105_143521}{0.23}{GW191109_010717}{0.25}{GW200209_085452}{0.64}{GW200115_042309}{0.06}{GW191127_050227}{0.57}{GW200216_220804}{0.63}{GW191215_223052}{0.38}{GW200208_130117}{0.40}{GW200219_094415}{0.57}{GW191103_012549}{0.20}{GW200316_215756}{0.22}{GW200202_154313}{0.09}{GW200129_065458}{0.18}{GW191216_213338}{0.07}}[{\red{???}}]}
\DeclareRobustCommand{\redshiftgwtcthreeplus}[1]{\IfEqCase{#1}{{GW200224_222234}{0.07}{GW191129_134029}{0.04}{GW200311_115853}{0.05}{GW191230_180458}{0.26}{GW191222_033537}{0.23}{GW200225_060421}{0.09}{GW200302_015811}{0.17}{GW200128_022011}{0.26}{GW191204_171526}{0.04}{GW200112_155838}{0.07}{GW200105_162426}{0.02}{GW191105_143521}{0.07}{GW191109_010717}{0.18}{GW200209_085452}{0.25}{GW200115_042309}{0.03}{GW191127_050227}{0.40}{GW200216_220804}{0.37}{GW191215_223052}{0.13}{GW200208_130117}{0.15}{GW200219_094415}{0.22}{GW191103_012549}{0.09}{GW200316_215756}{0.08}{GW200202_154313}{0.03}{GW200129_065458}{0.05}{GW191216_213338}{0.02}}[{\red{???}}]}
\DeclareRobustCommand{\redshiftgwtcthreetenthpercentile}[1]{\IfEqCase{#1}{{GW200224_222234}{0.25}{GW191129_134029}{0.11}{GW200311_115853}{0.17}{GW191230_180458}{0.54}{GW191222_033537}{0.31}{GW200225_060421}{0.15}{GW200302_015811}{0.20}{GW200128_022011}{0.43}{GW191204_171526}{0.10}{GW200112_155838}{0.18}{GW200105_162426}{0.04}{GW191105_143521}{0.15}{GW191109_010717}{0.15}{GW200209_085452}{0.44}{GW200115_042309}{0.05}{GW191127_050227}{0.34}{GW200216_220804}{0.39}{GW191215_223052}{0.26}{GW200208_130117}{0.29}{GW200219_094415}{0.39}{GW191103_012549}{0.12}{GW200316_215756}{0.15}{GW200202_154313}{0.06}{GW200129_065458}{0.12}{GW191216_213338}{0.05}}[{\red{???}}]}
\DeclareRobustCommand{\redshiftgwtcthreenintiethpercentile}[1]{\IfEqCase{#1}{{GW200224_222234}{0.39}{GW191129_134029}{0.20}{GW200311_115853}{0.27}{GW191230_180458}{0.96}{GW191222_033537}{0.69}{GW200225_060421}{0.29}{GW200302_015811}{0.44}{GW200128_022011}{0.86}{GW191204_171526}{0.17}{GW200112_155838}{0.30}{GW200105_162426}{0.08}{GW191105_143521}{0.28}{GW191109_010717}{0.38}{GW200209_085452}{0.83}{GW200115_042309}{0.08}{GW191127_050227}{0.87}{GW200216_220804}{0.91}{GW191215_223052}{0.48}{GW200208_130117}{0.51}{GW200219_094415}{0.75}{GW191103_012549}{0.26}{GW200316_215756}{0.28}{GW200202_154313}{0.11}{GW200129_065458}{0.22}{GW191216_213338}{0.09}}[{\red{???}}]}
\DeclareRobustCommand{\geocenttimegwtcthreeminus}[1]{\IfEqCase{#1}{{GW200224_222234}{0.0179}{GW191129_134029}{0.019}{GW200311_115853}{0.0}{GW191230_180458}{0.0}{GW191222_033537}{0.023}{GW200225_060421}{0.0049}{GW200302_015811}{0.0350}{GW200128_022011}{0.0313}{GW191204_171526}{0.0}{GW200112_155838}{0.0358}{GW200105_162426}{0.0017}{GW191105_143521}{0.023}{GW191109_010717}{0.0311}{GW200209_085452}{0.0304}{GW200115_042309}{0.0400}{GW191127_050227}{0.012}{GW200216_220804}{0.0}{GW191215_223052}{0.0439}{GW200208_130117}{0.0032}{GW200219_094415}{0.026}{GW191103_012549}{0.019}{GW200316_215756}{0.0}{GW200202_154313}{0.1}{GW200129_065458}{0.0}{GW191216_213338}{0.0045}}[{\red{???}}]}
\DeclareRobustCommand{\geocenttimegwtcthreemed}[1]{\IfEqCase{#1}{{GW200224_222234}{1266618172.3978}{GW191129_134029}{1259070047.197}{GW200311_115853}{1267963151.4}{GW191230_180458}{1261764316.4}{GW191222_033537}{1261020955.117}{GW200225_060421}{1266645879.4018}{GW200302_015811}{1267149509.5273}{GW200128_022011}{1264213229.9033}{GW191204_171526}{1259514944.1}{GW200112_155838}{1262879936.1035}{GW200105_162426}{1262276684.0349}{GW191105_143521}{1256999739.933}{GW191109_010717}{1257296855.2191}{GW200209_085452}{1265273710.1831}{GW200115_042309}{1263097407.7623}{GW191127_050227}{1258866165.541}{GW200216_220804}{1265926102.9}{GW191215_223052}{1260484270.3546}{GW200208_130117}{1265202095.9383}{GW200219_094415}{1266140673.195}{GW191103_012549}{1256779567.535}{GW200316_215756}{1268431094.2}{GW200202_154313}{1264693411.6}{GW200129_065458}{1264316116.4}{GW191216_213338}{1260567236.4871}}[{\red{???}}]}
\DeclareRobustCommand{\geocenttimegwtcthreeplus}[1]{\IfEqCase{#1}{{GW200224_222234}{0.0080}{GW191129_134029}{0.020}{GW200311_115853}{0.0}{GW191230_180458}{0.0}{GW191222_033537}{0.017}{GW200225_060421}{0.0124}{GW200302_015811}{0.0070}{GW200128_022011}{0.0100}{GW191204_171526}{0.0}{GW200112_155838}{0.0035}{GW200105_162426}{0.0448}{GW191105_143521}{0.015}{GW191109_010717}{0.0016}{GW200209_085452}{0.0067}{GW200115_042309}{0.0043}{GW191127_050227}{0.024}{GW200216_220804}{0.0}{GW191215_223052}{0.0023}{GW200208_130117}{0.0114}{GW200219_094415}{0.015}{GW191103_012549}{0.013}{GW200316_215756}{0.0}{GW200202_154313}{0.0}{GW200129_065458}{0.0}{GW191216_213338}{0.0044}}[{\red{???}}]}
\DeclareRobustCommand{\geocenttimegwtcthreetenthpercentile}[1]{\IfEqCase{#1}{{GW200224_222234}{1266618172.3803}{GW191129_134029}{1259070047.178}{GW200311_115853}{1267963151.4}{GW191230_180458}{1261764316.4}{GW191222_033537}{1261020955.095}{GW200225_060421}{1266645879.3970}{GW200302_015811}{1267149509.4933}{GW200128_022011}{1264213229.8733}{GW191204_171526}{1259514944.1}{GW200112_155838}{1262879936.0709}{GW200105_162426}{1262276684.0332}{GW191105_143521}{1256999739.910}{GW191109_010717}{1257296855.1896}{GW200209_085452}{1265273710.1818}{GW200115_042309}{1263097407.7501}{GW191127_050227}{1258866165.531}{GW200216_220804}{1265926102.9}{GW191215_223052}{1260484270.3114}{GW200208_130117}{1265202095.9356}{GW200219_094415}{1266140673.170}{GW191103_012549}{1256779567.517}{GW200316_215756}{1268431094.2}{GW200202_154313}{1264693411.6}{GW200129_065458}{1264316116.4}{GW191216_213338}{1260567236.4851}}[{\red{???}}]}
\DeclareRobustCommand{\geocenttimegwtcthreenintiethpercentile}[1]{\IfEqCase{#1}{{GW200224_222234}{1266618172.4058}{GW191129_134029}{1259070047.217}{GW200311_115853}{1267963151.4}{GW191230_180458}{1261764316.4}{GW191222_033537}{1261020955.129}{GW200225_060421}{1266645879.4139}{GW200302_015811}{1267149509.5335}{GW200128_022011}{1264213229.9121}{GW191204_171526}{1259514944.1}{GW200112_155838}{1262879936.1064}{GW200105_162426}{1262276684.0789}{GW191105_143521}{1256999739.945}{GW191109_010717}{1257296855.2207}{GW200209_085452}{1265273710.1888}{GW200115_042309}{1263097407.7666}{GW191127_050227}{1258866165.562}{GW200216_220804}{1265926102.9}{GW191215_223052}{1260484270.3569}{GW200208_130117}{1265202095.9497}{GW200219_094415}{1266140673.209}{GW191103_012549}{1256779567.546}{GW200316_215756}{1268431094.2}{GW200202_154313}{1264693411.6}{GW200129_065458}{1264316116.4}{GW191216_213338}{1260567236.4909}}[{\red{???}}]}
\DeclareRobustCommand{\luminositydistancegwtcthreeminus}[1]{\IfEqCase{#1}{{GW200224_222234}{0.63}{GW191129_134029}{0.33}{GW200311_115853}{0.40}{GW191230_180458}{2.0}{GW191222_033537}{1.7}{GW200225_060421}{0.53}{GW200302_015811}{0.77}{GW200128_022011}{2.0}{GW191204_171526}{0.25}{GW200112_155838}{0.46}{GW200105_162426}{0.11}{GW191105_143521}{0.48}{GW191109_010717}{0.65}{GW200209_085452}{1.8}{GW200115_042309}{0.10}{GW191127_050227}{1.9}{GW200216_220804}{2.0}{GW191215_223052}{0.91}{GW200208_130117}{0.85}{GW200219_094415}{1.5}{GW191103_012549}{0.47}{GW200316_215756}{0.44}{GW200202_154313}{0.16}{GW200129_065458}{0.38}{GW191216_213338}{0.13}}[{\red{???}}]}
\DeclareRobustCommand{\luminositydistancegwtcthreemed}[1]{\IfEqCase{#1}{{GW200224_222234}{1.77}{GW191129_134029}{0.82}{GW200311_115853}{1.17}{GW191230_180458}{4.8}{GW191222_033537}{3.0}{GW200225_060421}{1.15}{GW200302_015811}{1.67}{GW200128_022011}{4.0}{GW191204_171526}{0.66}{GW200112_155838}{1.25}{GW200105_162426}{0.27}{GW191105_143521}{1.15}{GW191109_010717}{1.29}{GW200209_085452}{3.9}{GW200115_042309}{0.29}{GW191127_050227}{3.4}{GW200216_220804}{3.8}{GW191215_223052}{2.10}{GW200208_130117}{2.23}{GW200219_094415}{3.4}{GW191103_012549}{0.99}{GW200316_215756}{1.12}{GW200202_154313}{0.41}{GW200129_065458}{0.90}{GW191216_213338}{0.34}}[{\red{???}}]}
\DeclareRobustCommand{\luminositydistancegwtcthreeplus}[1]{\IfEqCase{#1}{{GW200224_222234}{0.47}{GW191129_134029}{0.25}{GW200311_115853}{0.28}{GW191230_180458}{2.1}{GW191222_033537}{1.7}{GW200225_060421}{0.51}{GW200302_015811}{1.09}{GW200128_022011}{2.1}{GW191204_171526}{0.19}{GW200112_155838}{0.43}{GW200105_162426}{0.12}{GW191105_143521}{0.43}{GW191109_010717}{1.13}{GW200209_085452}{2.0}{GW200115_042309}{0.15}{GW191127_050227}{3.1}{GW200216_220804}{3.0}{GW191215_223052}{0.86}{GW200208_130117}{1.00}{GW200219_094415}{1.7}{GW191103_012549}{0.50}{GW200316_215756}{0.47}{GW200202_154313}{0.15}{GW200129_065458}{0.29}{GW191216_213338}{0.12}}[{\red{???}}]}
\DeclareRobustCommand{\luminositydistancegwtcthreetenthpercentile}[1]{\IfEqCase{#1}{{GW200224_222234}{1.29}{GW191129_134029}{0.55}{GW200311_115853}{0.87}{GW191230_180458}{3.2}{GW191222_033537}{1.6}{GW200225_060421}{0.74}{GW200302_015811}{1.03}{GW200128_022011}{2.4}{GW191204_171526}{0.47}{GW200112_155838}{0.89}{GW200105_162426}{0.18}{GW191105_143521}{0.76}{GW191109_010717}{0.76}{GW200209_085452}{2.5}{GW200115_042309}{0.21}{GW191127_050227}{1.8}{GW200216_220804}{2.2}{GW191215_223052}{1.36}{GW200208_130117}{1.54}{GW200219_094415}{2.2}{GW191103_012549}{0.60}{GW200316_215756}{0.76}{GW200202_154313}{0.28}{GW200129_065458}{0.60}{GW191216_213338}{0.23}}[{\red{???}}]}
\DeclareRobustCommand{\luminositydistancegwtcthreenintiethpercentile}[1]{\IfEqCase{#1}{{GW200224_222234}{2.15}{GW191129_134029}{1.02}{GW200311_115853}{1.39}{GW191230_180458}{6.4}{GW191222_033537}{4.3}{GW200225_060421}{1.55}{GW200302_015811}{2.49}{GW200128_022011}{5.7}{GW191204_171526}{0.82}{GW200112_155838}{1.60}{GW200105_162426}{0.36}{GW191105_143521}{1.49}{GW191109_010717}{2.13}{GW200209_085452}{5.4}{GW200115_042309}{0.40}{GW191127_050227}{5.7}{GW200216_220804}{6.0}{GW191215_223052}{2.78}{GW200208_130117}{2.98}{GW200219_094415}{4.7}{GW191103_012549}{1.38}{GW200316_215756}{1.49}{GW200202_154313}{0.53}{GW200129_065458}{1.13}{GW191216_213338}{0.44}}[{\red{???}}]}
\DeclareRobustCommand{\thetajngwtcthreeminus}[1]{\IfEqCase{#1}{{GW200224_222234}{0.42}{GW191129_134029}{1.5}{GW200311_115853}{0.40}{GW191230_180458}{1.85}{GW191222_033537}{1.3}{GW200225_060421}{1.00}{GW200302_015811}{0.98}{GW200128_022011}{1.1}{GW191204_171526}{2.04}{GW200112_155838}{0.68}{GW200105_162426}{1.2}{GW191105_143521}{0.85}{GW191109_010717}{1.18}{GW200209_085452}{1.5}{GW200115_042309}{0.44}{GW191127_050227}{1.2}{GW200216_220804}{0.69}{GW191215_223052}{0.82}{GW200208_130117}{0.58}{GW200219_094415}{0.92}{GW191103_012549}{1.1}{GW200316_215756}{1.84}{GW200202_154313}{0.59}{GW200129_065458}{0.41}{GW191216_213338}{0.81}}[{\red{???}}]}
\DeclareRobustCommand{\thetajngwtcthreemed}[1]{\IfEqCase{#1}{{GW200224_222234}{0.57}{GW191129_134029}{1.8}{GW200311_115853}{0.55}{GW191230_180458}{2.13}{GW191222_033537}{1.6}{GW200225_060421}{1.31}{GW200302_015811}{1.26}{GW200128_022011}{1.4}{GW191204_171526}{2.29}{GW200112_155838}{0.88}{GW200105_162426}{1.5}{GW191105_143521}{1.07}{GW191109_010717}{1.91}{GW200209_085452}{1.8}{GW200115_042309}{0.62}{GW191127_050227}{1.5}{GW200216_220804}{0.89}{GW191215_223052}{1.12}{GW200208_130117}{2.53}{GW200219_094415}{1.19}{GW191103_012549}{1.4}{GW200316_215756}{2.32}{GW200202_154313}{2.57}{GW200129_065458}{0.66}{GW191216_213338}{2.50}}[{\red{???}}]}
\DeclareRobustCommand{\thetajngwtcthreeplus}[1]{\IfEqCase{#1}{{GW200224_222234}{0.54}{GW191129_134029}{1.1}{GW200311_115853}{0.52}{GW191230_180458}{0.79}{GW191222_033537}{1.2}{GW200225_060421}{1.47}{GW200302_015811}{1.55}{GW200128_022011}{1.5}{GW191204_171526}{0.64}{GW200112_155838}{2.04}{GW200105_162426}{1.3}{GW191105_143521}{1.82}{GW191109_010717}{0.87}{GW200209_085452}{1.1}{GW200115_042309}{1.94}{GW191127_050227}{1.4}{GW200216_220804}{1.87}{GW191215_223052}{1.65}{GW200208_130117}{0.44}{GW200219_094415}{1.59}{GW191103_012549}{1.5}{GW200316_215756}{0.58}{GW200202_154313}{0.42}{GW200129_065458}{0.59}{GW191216_213338}{0.45}}[{\red{???}}]}
\DeclareRobustCommand{\thetajngwtcthreetenthpercentile}[1]{\IfEqCase{#1}{{GW200224_222234}{0.22}{GW191129_134029}{0.4}{GW200311_115853}{0.21}{GW191230_180458}{0.40}{GW191222_033537}{0.4}{GW200225_060421}{0.45}{GW200302_015811}{0.40}{GW200128_022011}{0.3}{GW191204_171526}{0.36}{GW200112_155838}{0.28}{GW200105_162426}{0.5}{GW191105_143521}{0.32}{GW191109_010717}{1.02}{GW200209_085452}{0.4}{GW200115_042309}{0.26}{GW191127_050227}{0.4}{GW200216_220804}{0.30}{GW191215_223052}{0.42}{GW200208_130117}{2.09}{GW200219_094415}{0.40}{GW191103_012549}{0.3}{GW200316_215756}{0.71}{GW200202_154313}{2.10}{GW200129_065458}{0.31}{GW191216_213338}{2.00}}[{\red{???}}]}
\DeclareRobustCommand{\thetajngwtcthreenintiethpercentile}[1]{\IfEqCase{#1}{{GW200224_222234}{1.00}{GW191129_134029}{2.8}{GW200311_115853}{0.97}{GW191230_180458}{2.81}{GW191222_033537}{2.7}{GW200225_060421}{2.64}{GW200302_015811}{2.68}{GW200128_022011}{2.8}{GW191204_171526}{2.84}{GW200112_155838}{2.82}{GW200105_162426}{2.7}{GW191105_143521}{2.77}{GW191109_010717}{2.63}{GW200209_085452}{2.8}{GW200115_042309}{2.12}{GW191127_050227}{2.7}{GW200216_220804}{2.57}{GW191215_223052}{2.61}{GW200208_130117}{2.90}{GW200219_094415}{2.60}{GW191103_012549}{2.8}{GW200316_215756}{2.80}{GW200202_154313}{2.92}{GW200129_065458}{1.11}{GW191216_213338}{2.88}}[{\red{???}}]}
\DeclareRobustCommand{\chieffgwtcthreeminus}[1]{\IfEqCase{#1}{{GW200224_222234}{0.15}{GW191129_134029}{0.08}{GW200311_115853}{0.20}{GW191230_180458}{0.30}{GW191222_033537}{0.25}{GW200225_060421}{0.28}{GW200302_015811}{0.25}{GW200128_022011}{0.25}{GW191204_171526}{0.05}{GW200112_155838}{0.15}{GW200105_162426}{0.18}{GW191105_143521}{0.09}{GW191109_010717}{0.31}{GW200209_085452}{0.30}{GW200115_042309}{0.41}{GW191127_050227}{0.36}{GW200216_220804}{0.36}{GW191215_223052}{0.21}{GW200208_130117}{0.27}{GW200219_094415}{0.29}{GW191103_012549}{0.10}{GW200316_215756}{0.10}{GW200202_154313}{0.06}{GW200129_065458}{0.16}{GW191216_213338}{0.06}}[{\red{???}}]}
\DeclareRobustCommand{\chieffgwtcthreemed}[1]{\IfEqCase{#1}{{GW200224_222234}{0.11}{GW191129_134029}{0.06}{GW200311_115853}{-0.02}{GW191230_180458}{-0.03}{GW191222_033537}{-0.04}{GW200225_060421}{-0.12}{GW200302_015811}{0.03}{GW200128_022011}{0.14}{GW191204_171526}{0.16}{GW200112_155838}{0.06}{GW200105_162426}{0.00}{GW191105_143521}{-0.02}{GW191109_010717}{-0.29}{GW200209_085452}{-0.10}{GW200115_042309}{-0.15}{GW191127_050227}{0.18}{GW200216_220804}{0.10}{GW191215_223052}{-0.03}{GW200208_130117}{-0.07}{GW200219_094415}{-0.08}{GW191103_012549}{0.21}{GW200316_215756}{0.13}{GW200202_154313}{0.04}{GW200129_065458}{0.11}{GW191216_213338}{0.11}}[{\red{???}}]}
\DeclareRobustCommand{\chieffgwtcthreeplus}[1]{\IfEqCase{#1}{{GW200224_222234}{0.15}{GW191129_134029}{0.16}{GW200311_115853}{0.16}{GW191230_180458}{0.26}{GW191222_033537}{0.20}{GW200225_060421}{0.17}{GW200302_015811}{0.26}{GW200128_022011}{0.24}{GW191204_171526}{0.08}{GW200112_155838}{0.15}{GW200105_162426}{0.13}{GW191105_143521}{0.13}{GW191109_010717}{0.42}{GW200209_085452}{0.24}{GW200115_042309}{0.24}{GW191127_050227}{0.34}{GW200216_220804}{0.34}{GW191215_223052}{0.17}{GW200208_130117}{0.22}{GW200219_094415}{0.23}{GW191103_012549}{0.16}{GW200316_215756}{0.27}{GW200202_154313}{0.13}{GW200129_065458}{0.11}{GW191216_213338}{0.13}}[{\red{???}}]}
\DeclareRobustCommand{\chieffgwtcthreetenthpercentile}[1]{\IfEqCase{#1}{{GW200224_222234}{-0.01}{GW191129_134029}{0.00}{GW200311_115853}{-0.17}{GW191230_180458}{-0.26}{GW191222_033537}{-0.23}{GW200225_060421}{-0.34}{GW200302_015811}{-0.15}{GW200128_022011}{-0.05}{GW191204_171526}{0.12}{GW200112_155838}{-0.05}{GW200105_162426}{-0.10}{GW191105_143521}{-0.09}{GW191109_010717}{-0.54}{GW200209_085452}{-0.33}{GW200115_042309}{-0.51}{GW191127_050227}{-0.10}{GW200216_220804}{-0.17}{GW191215_223052}{-0.19}{GW200208_130117}{-0.27}{GW200219_094415}{-0.30}{GW191103_012549}{0.13}{GW200316_215756}{0.04}{GW200202_154313}{-0.01}{GW200129_065458}{0.00}{GW191216_213338}{0.06}}[{\red{???}}]}
\DeclareRobustCommand{\chieffgwtcthreenintiethpercentile}[1]{\IfEqCase{#1}{{GW200224_222234}{0.22}{GW191129_134029}{0.19}{GW200311_115853}{0.10}{GW191230_180458}{0.18}{GW191222_033537}{0.11}{GW200225_060421}{0.02}{GW200302_015811}{0.23}{GW200128_022011}{0.33}{GW191204_171526}{0.22}{GW200112_155838}{0.18}{GW200105_162426}{0.08}{GW191105_143521}{0.07}{GW191109_010717}{0.00}{GW200209_085452}{0.09}{GW200115_042309}{0.04}{GW191127_050227}{0.45}{GW200216_220804}{0.36}{GW191215_223052}{0.10}{GW200208_130117}{0.09}{GW200219_094415}{0.10}{GW191103_012549}{0.33}{GW200316_215756}{0.32}{GW200202_154313}{0.13}{GW200129_065458}{0.20}{GW191216_213338}{0.20}}[{\red{???}}]}
\DeclareRobustCommand{\spinonegwtcthreeminus}[1]{\IfEqCase{#1}{{GW200224_222234}{0.42}{GW191129_134029}{0.22}{GW200311_115853}{0.36}{GW191230_180458}{0.46}{GW191222_033537}{0.34}{GW200225_060421}{0.51}{GW200302_015811}{0.35}{GW200128_022011}{0.53}{GW191204_171526}{0.35}{GW200112_155838}{0.31}{GW200105_162426}{0.07}{GW191105_143521}{0.21}{GW191109_010717}{0.58}{GW200209_085452}{0.46}{GW200115_042309}{0.29}{GW191127_050227}{0.58}{GW200216_220804}{0.43}{GW191215_223052}{0.42}{GW200208_130117}{0.32}{GW200219_094415}{0.43}{GW191103_012549}{0.40}{GW200316_215756}{0.28}{GW200202_154313}{0.20}{GW200129_065458}{0.47}{GW191216_213338}{0.22}}[{\red{???}}]}
\DeclareRobustCommand{\spinonegwtcthreemed}[1]{\IfEqCase{#1}{{GW200224_222234}{0.47}{GW191129_134029}{0.25}{GW200311_115853}{0.39}{GW191230_180458}{0.51}{GW191222_033537}{0.38}{GW200225_060421}{0.59}{GW200302_015811}{0.38}{GW200128_022011}{0.60}{GW191204_171526}{0.40}{GW200112_155838}{0.34}{GW200105_162426}{0.08}{GW191105_143521}{0.23}{GW191109_010717}{0.83}{GW200209_085452}{0.52}{GW200115_042309}{0.32}{GW191127_050227}{0.66}{GW200216_220804}{0.48}{GW191215_223052}{0.47}{GW200208_130117}{0.36}{GW200219_094415}{0.47}{GW191103_012549}{0.46}{GW200316_215756}{0.32}{GW200202_154313}{0.22}{GW200129_065458}{0.53}{GW191216_213338}{0.24}}[{\red{???}}]}
\DeclareRobustCommand{\spinonegwtcthreeplus}[1]{\IfEqCase{#1}{{GW200224_222234}{0.44}{GW191129_134029}{0.37}{GW200311_115853}{0.48}{GW191230_180458}{0.44}{GW191222_033537}{0.50}{GW200225_060421}{0.35}{GW200302_015811}{0.51}{GW200128_022011}{0.36}{GW191204_171526}{0.38}{GW200112_155838}{0.46}{GW200105_162426}{0.30}{GW191105_143521}{0.53}{GW191109_010717}{0.15}{GW200209_085452}{0.43}{GW200115_042309}{0.50}{GW191127_050227}{0.31}{GW200216_220804}{0.46}{GW191215_223052}{0.45}{GW200208_130117}{0.51}{GW200219_094415}{0.46}{GW191103_012549}{0.40}{GW200316_215756}{0.37}{GW200202_154313}{0.45}{GW200129_065458}{0.42}{GW191216_213338}{0.36}}[{\red{???}}]}
\DeclareRobustCommand{\spinonegwtcthreetenthpercentile}[1]{\IfEqCase{#1}{{GW200224_222234}{0.10}{GW191129_134029}{0.05}{GW200311_115853}{0.07}{GW191230_180458}{0.10}{GW191222_033537}{0.07}{GW200225_060421}{0.14}{GW200302_015811}{0.07}{GW200128_022011}{0.14}{GW191204_171526}{0.10}{GW200112_155838}{0.07}{GW200105_162426}{0.01}{GW191105_143521}{0.04}{GW191109_010717}{0.42}{GW200209_085452}{0.10}{GW200115_042309}{0.05}{GW191127_050227}{0.16}{GW200216_220804}{0.10}{GW191215_223052}{0.09}{GW200208_130117}{0.07}{GW200219_094415}{0.09}{GW191103_012549}{0.13}{GW200316_215756}{0.08}{GW200202_154313}{0.04}{GW200129_065458}{0.12}{GW191216_213338}{0.05}}[{\red{???}}]}
\DeclareRobustCommand{\spinonegwtcthreenintiethpercentile}[1]{\IfEqCase{#1}{{GW200224_222234}{0.84}{GW191129_134029}{0.54}{GW200311_115853}{0.79}{GW191230_180458}{0.90}{GW191222_033537}{0.80}{GW200225_060421}{0.89}{GW200302_015811}{0.81}{GW200128_022011}{0.92}{GW191204_171526}{0.70}{GW200112_155838}{0.71}{GW200105_162426}{0.27}{GW191105_143521}{0.64}{GW191109_010717}{0.97}{GW200209_085452}{0.90}{GW200115_042309}{0.73}{GW191127_050227}{0.94}{GW200216_220804}{0.89}{GW191215_223052}{0.85}{GW200208_130117}{0.78}{GW200219_094415}{0.88}{GW191103_012549}{0.78}{GW200316_215756}{0.61}{GW200202_154313}{0.55}{GW200129_065458}{0.93}{GW191216_213338}{0.50}}[{\red{???}}]}
\DeclareRobustCommand{\cosiotagwtcthreeminus}[1]{\IfEqCase{#1}{{GW200224_222234}{0.43}{GW191129_134029}{0.78}{GW200311_115853}{0.37}{GW191230_180458}{0.41}{GW191222_033537}{0.91}{GW200225_060421}{1.15}{GW200302_015811}{1.24}{GW200128_022011}{1.13}{GW191204_171526}{0.33}{GW200112_155838}{1.59}{GW200105_162426}{0.97}{GW191105_143521}{1.44}{GW191109_010717}{0.55}{GW200209_085452}{0.83}{GW200115_042309}{1.64}{GW191127_050227}{1.00}{GW200216_220804}{1.50}{GW191215_223052}{1.28}{GW200208_130117}{0.17}{GW200219_094415}{1.26}{GW191103_012549}{1.17}{GW200316_215756}{0.29}{GW200202_154313}{0.15}{GW200129_065458}{0.50}{GW191216_213338}{0.18}}[{\red{???}}]}
\DeclareRobustCommand{\cosiotagwtcthreemed}[1]{\IfEqCase{#1}{{GW200224_222234}{0.84}{GW191129_134029}{-0.19}{GW200311_115853}{0.86}{GW191230_180458}{-0.57}{GW191222_033537}{-0.05}{GW200225_060421}{0.21}{GW200302_015811}{0.30}{GW200128_022011}{0.16}{GW191204_171526}{-0.65}{GW200112_155838}{0.62}{GW200105_162426}{0.03}{GW191105_143521}{0.47}{GW191109_010717}{-0.40}{GW200209_085452}{-0.14}{GW200115_042309}{0.81}{GW191127_050227}{0.06}{GW200216_220804}{0.57}{GW191215_223052}{0.36}{GW200208_130117}{-0.81}{GW200219_094415}{0.33}{GW191103_012549}{0.20}{GW200316_215756}{-0.68}{GW200202_154313}{-0.84}{GW200129_065458}{0.80}{GW191216_213338}{-0.80}}[{\red{???}}]}
\DeclareRobustCommand{\cosiotagwtcthreeplus}[1]{\IfEqCase{#1}{{GW200224_222234}{0.15}{GW191129_134029}{1.16}{GW200311_115853}{0.13}{GW191230_180458}{1.53}{GW191222_033537}{1.00}{GW200225_060421}{0.74}{GW200302_015811}{0.66}{GW200128_022011}{0.80}{GW191204_171526}{1.62}{GW200112_155838}{0.36}{GW200105_162426}{0.91}{GW191105_143521}{0.50}{GW191109_010717}{1.24}{GW200209_085452}{1.11}{GW200115_042309}{0.18}{GW191127_050227}{0.88}{GW200216_220804}{0.40}{GW191215_223052}{0.60}{GW200208_130117}{0.49}{GW200219_094415}{0.63}{GW191103_012549}{0.78}{GW200316_215756}{1.57}{GW200202_154313}{0.45}{GW200129_065458}{0.19}{GW191216_213338}{0.68}}[{\red{???}}]}
\DeclareRobustCommand{\cosiotagwtcthreetenthpercentile}[1]{\IfEqCase{#1}{{GW200224_222234}{0.52}{GW191129_134029}{-0.94}{GW200311_115853}{0.57}{GW191230_180458}{-0.95}{GW191222_033537}{-0.91}{GW200225_060421}{-0.87}{GW200302_015811}{-0.89}{GW200128_022011}{-0.92}{GW191204_171526}{-0.96}{GW200112_155838}{-0.94}{GW200105_162426}{-0.89}{GW191105_143521}{-0.93}{GW191109_010717}{-0.90}{GW200209_085452}{-0.93}{GW200115_042309}{-0.52}{GW191127_050227}{-0.87}{GW200216_220804}{-0.82}{GW191215_223052}{-0.85}{GW200208_130117}{-0.97}{GW200219_094415}{-0.85}{GW191103_012549}{-0.94}{GW200316_215756}{-0.95}{GW200202_154313}{-0.98}{GW200129_065458}{0.43}{GW191216_213338}{-0.97}}[{\red{???}}]}
\DeclareRobustCommand{\cosiotagwtcthreenintiethpercentile}[1]{\IfEqCase{#1}{{GW200224_222234}{0.98}{GW191129_134029}{0.94}{GW200311_115853}{0.98}{GW191230_180458}{0.93}{GW191222_033537}{0.90}{GW200225_060421}{0.90}{GW200302_015811}{0.92}{GW200128_022011}{0.92}{GW191204_171526}{0.94}{GW200112_155838}{0.96}{GW200105_162426}{0.89}{GW191105_143521}{0.95}{GW191109_010717}{0.68}{GW200209_085452}{0.93}{GW200115_042309}{0.97}{GW191127_050227}{0.88}{GW200216_220804}{0.94}{GW191215_223052}{0.91}{GW200208_130117}{-0.47}{GW200219_094415}{0.92}{GW191103_012549}{0.94}{GW200316_215756}{0.75}{GW200202_154313}{-0.50}{GW200129_065458}{0.97}{GW191216_213338}{-0.41}}[{\red{???}}]}
\DeclareRobustCommand{\radiatedenergygwtcthreeminus}[1]{\IfEqCase{#1}{{GW200224_222234}{0.67}{GW191129_134029}{0.116}{GW200311_115853}{0.57}{GW191230_180458}{1.0}{GW191222_033537}{1.00}{GW200225_060421}{0.39}{GW200302_015811}{0.79}{GW200128_022011}{0.96}{GW191204_171526}{0.106}{GW200112_155838}{0.53}{GW200105_162426}{0.026}{GW191105_143521}{0.118}{GW191109_010717}{1.3}{GW200209_085452}{0.70}{GW200115_042309}{0.029}{GW191127_050227}{2.1}{GW200216_220804}{2.0}{GW191215_223052}{0.32}{GW200208_130117}{0.77}{GW200219_094415}{0.81}{GW191103_012549}{0.17}{GW200316_215756}{0.22}{GW200202_154313}{0.094}{GW200129_065458}{0.86}{GW191216_213338}{0.130}}[{\red{???}}]}
\DeclareRobustCommand{\radiatedenergygwtcthreemed}[1]{\IfEqCase{#1}{{GW200224_222234}{3.61}{GW191129_134029}{0.778}{GW200311_115853}{2.87}{GW191230_180458}{3.8}{GW191222_033537}{3.57}{GW200225_060421}{1.43}{GW200302_015811}{2.35}{GW200128_022011}{3.61}{GW191204_171526}{0.991}{GW200112_155838}{3.08}{GW200105_162426}{0.204}{GW191105_143521}{0.818}{GW191109_010717}{4.3}{GW200209_085452}{2.68}{GW200115_042309}{0.146}{GW191127_050227}{3.0}{GW200216_220804}{3.5}{GW191215_223052}{1.88}{GW200208_130117}{2.83}{GW200219_094415}{2.84}{GW191103_012549}{0.98}{GW200316_215756}{0.94}{GW200202_154313}{0.814}{GW200129_065458}{3.18}{GW191216_213338}{0.921}}[{\red{???}}]}
\DeclareRobustCommand{\radiatedenergygwtcthreeplus}[1]{\IfEqCase{#1}{{GW200224_222234}{0.67}{GW191129_134029}{0.072}{GW200311_115853}{0.52}{GW191230_180458}{1.2}{GW191222_033537}{1.04}{GW200225_060421}{0.27}{GW200302_015811}{1.10}{GW200128_022011}{1.26}{GW191204_171526}{0.069}{GW200112_155838}{0.59}{GW200105_162426}{0.032}{GW191105_143521}{0.088}{GW191109_010717}{2.3}{GW200209_085452}{0.83}{GW200115_042309}{0.045}{GW191127_050227}{2.6}{GW200216_220804}{2.0}{GW191215_223052}{0.37}{GW200208_130117}{0.77}{GW200219_094415}{0.91}{GW191103_012549}{0.13}{GW200316_215756}{0.11}{GW200202_154313}{0.047}{GW200129_065458}{0.42}{GW191216_213338}{0.057}}[{\red{???}}]}
\DeclareRobustCommand{\radiatedenergygwtcthreetenthpercentile}[1]{\IfEqCase{#1}{{GW200224_222234}{3.09}{GW191129_134029}{0.690}{GW200311_115853}{2.44}{GW191230_180458}{3.0}{GW191222_033537}{2.84}{GW200225_060421}{1.12}{GW200302_015811}{1.72}{GW200128_022011}{2.85}{GW191204_171526}{0.911}{GW200112_155838}{2.67}{GW200105_162426}{0.187}{GW191105_143521}{0.731}{GW191109_010717}{3.3}{GW200209_085452}{2.14}{GW200115_042309}{0.124}{GW191127_050227}{1.3}{GW200216_220804}{1.9}{GW191215_223052}{1.63}{GW200208_130117}{2.24}{GW200219_094415}{2.23}{GW191103_012549}{0.85}{GW200316_215756}{0.79}{GW200202_154313}{0.748}{GW200129_065458}{2.58}{GW191216_213338}{0.829}}[{\red{???}}]}
\DeclareRobustCommand{\radiatedenergygwtcthreenintiethpercentile}[1]{\IfEqCase{#1}{{GW200224_222234}{4.11}{GW191129_134029}{0.835}{GW200311_115853}{3.27}{GW191230_180458}{4.7}{GW191222_033537}{4.39}{GW200225_060421}{1.64}{GW200302_015811}{3.18}{GW200128_022011}{4.56}{GW191204_171526}{1.045}{GW200112_155838}{3.52}{GW200105_162426}{0.225}{GW191105_143521}{0.887}{GW191109_010717}{5.8}{GW200209_085452}{3.30}{GW200115_042309}{0.183}{GW191127_050227}{5.0}{GW200216_220804}{5.0}{GW191215_223052}{2.15}{GW200208_130117}{3.42}{GW200219_094415}{3.53}{GW191103_012549}{1.08}{GW200316_215756}{1.03}{GW200202_154313}{0.851}{GW200129_065458}{3.53}{GW191216_213338}{0.967}}[{\red{???}}]}
\DeclareRobustCommand{\costhetajngwtcthreeminus}[1]{\IfEqCase{#1}{{GW200224_222234}{0.40}{GW191129_134029}{0.78}{GW200311_115853}{0.37}{GW191230_180458}{0.45}{GW191222_033537}{0.91}{GW200225_060421}{1.19}{GW200302_015811}{1.25}{GW200128_022011}{1.19}{GW191204_171526}{0.32}{GW200112_155838}{1.61}{GW200105_162426}{0.98}{GW191105_143521}{1.45}{GW191109_010717}{0.60}{GW200209_085452}{0.75}{GW200115_042309}{1.65}{GW191127_050227}{1.07}{GW200216_220804}{1.56}{GW191215_223052}{1.37}{GW200208_130117}{0.17}{GW200219_094415}{1.30}{GW191103_012549}{1.16}{GW200316_215756}{0.29}{GW200202_154313}{0.15}{GW200129_065458}{0.47}{GW191216_213338}{0.18}}[{\red{???}}]}
\DeclareRobustCommand{\costhetajngwtcthreemed}[1]{\IfEqCase{#1}{{GW200224_222234}{0.84}{GW191129_134029}{-0.19}{GW200311_115853}{0.85}{GW191230_180458}{-0.53}{GW191222_033537}{-0.05}{GW200225_060421}{0.25}{GW200302_015811}{0.30}{GW200128_022011}{0.22}{GW191204_171526}{-0.66}{GW200112_155838}{0.64}{GW200105_162426}{0.03}{GW191105_143521}{0.48}{GW191109_010717}{-0.33}{GW200209_085452}{-0.22}{GW200115_042309}{0.81}{GW191127_050227}{0.11}{GW200216_220804}{0.63}{GW191215_223052}{0.44}{GW200208_130117}{-0.82}{GW200219_094415}{0.37}{GW191103_012549}{0.19}{GW200316_215756}{-0.68}{GW200202_154313}{-0.84}{GW200129_065458}{0.79}{GW191216_213338}{-0.80}}[{\red{???}}]}
\DeclareRobustCommand{\costhetajngwtcthreeplus}[1]{\IfEqCase{#1}{{GW200224_222234}{0.15}{GW191129_134029}{1.16}{GW200311_115853}{0.14}{GW191230_180458}{1.49}{GW191222_033537}{1.01}{GW200225_060421}{0.70}{GW200302_015811}{0.66}{GW200128_022011}{0.75}{GW191204_171526}{1.63}{GW200112_155838}{0.34}{GW200105_162426}{0.92}{GW191105_143521}{0.50}{GW191109_010717}{1.07}{GW200209_085452}{1.19}{GW200115_042309}{0.17}{GW191127_050227}{0.85}{GW200216_220804}{0.35}{GW191215_223052}{0.52}{GW200208_130117}{0.44}{GW200219_094415}{0.59}{GW191103_012549}{0.78}{GW200316_215756}{1.57}{GW200202_154313}{0.45}{GW200129_065458}{0.18}{GW191216_213338}{0.68}}[{\red{???}}]}
\DeclareRobustCommand{\costhetajngwtcthreetenthpercentile}[1]{\IfEqCase{#1}{{GW200224_222234}{0.54}{GW191129_134029}{-0.94}{GW200311_115853}{0.57}{GW191230_180458}{-0.95}{GW191222_033537}{-0.92}{GW200225_060421}{-0.88}{GW200302_015811}{-0.89}{GW200128_022011}{-0.94}{GW191204_171526}{-0.96}{GW200112_155838}{-0.95}{GW200105_162426}{-0.90}{GW191105_143521}{-0.93}{GW191109_010717}{-0.87}{GW200209_085452}{-0.94}{GW200115_042309}{-0.52}{GW191127_050227}{-0.92}{GW200216_220804}{-0.84}{GW191215_223052}{-0.86}{GW200208_130117}{-0.97}{GW200219_094415}{-0.85}{GW191103_012549}{-0.94}{GW200316_215756}{-0.94}{GW200202_154313}{-0.98}{GW200129_065458}{0.45}{GW191216_213338}{-0.97}}[{\red{???}}]}
\DeclareRobustCommand{\costhetajngwtcthreenintiethpercentile}[1]{\IfEqCase{#1}{{GW200224_222234}{0.98}{GW191129_134029}{0.94}{GW200311_115853}{0.98}{GW191230_180458}{0.92}{GW191222_033537}{0.91}{GW200225_060421}{0.90}{GW200302_015811}{0.92}{GW200128_022011}{0.94}{GW191204_171526}{0.94}{GW200112_155838}{0.96}{GW200105_162426}{0.90}{GW191105_143521}{0.95}{GW191109_010717}{0.52}{GW200209_085452}{0.93}{GW200115_042309}{0.97}{GW191127_050227}{0.91}{GW200216_220804}{0.95}{GW191215_223052}{0.91}{GW200208_130117}{-0.50}{GW200219_094415}{0.92}{GW191103_012549}{0.94}{GW200316_215756}{0.76}{GW200202_154313}{-0.50}{GW200129_065458}{0.95}{GW191216_213338}{-0.41}}[{\red{???}}]}
\DeclareRobustCommand{\totalmasssourcegwtcthreeminus}[1]{\IfEqCase{#1}{{GW200224_222234}{5.0}{GW191129_134029}{1.1}{GW200311_115853}{4.2}{GW191230_180458}{11}{GW191222_033537}{11}{GW200225_060421}{3.0}{GW200302_015811}{6.7}{GW200128_022011}{11}{GW191204_171526}{0.93}{GW200112_155838}{4.6}{GW200105_162426}{1.4}{GW191105_143521}{1.3}{GW191109_010717}{16}{GW200209_085452}{8.6}{GW200115_042309}{1.6}{GW191127_050227}{22}{GW200216_220804}{14}{GW191215_223052}{4.0}{GW200208_130117}{6.8}{GW200219_094415}{8.2}{GW191103_012549}{1.8}{GW200316_215756}{2.0}{GW200202_154313}{0.67}{GW200129_065458}{3.6}{GW191216_213338}{0.93}}[{\red{???}}]}
\DeclareRobustCommand{\totalmasssourcegwtcthreemed}[1]{\IfEqCase{#1}{{GW200224_222234}{71.9}{GW191129_134029}{17.5}{GW200311_115853}{61.9}{GW191230_180458}{83}{GW191222_033537}{79}{GW200225_060421}{33.5}{GW200302_015811}{57.3}{GW200128_022011}{72}{GW191204_171526}{20.14}{GW200112_155838}{63.9}{GW200105_162426}{11.0}{GW191105_143521}{18.5}{GW191109_010717}{112}{GW200209_085452}{61.1}{GW200115_042309}{7.3}{GW191127_050227}{80}{GW200216_220804}{81}{GW191215_223052}{42.6}{GW200208_130117}{65.4}{GW200219_094415}{65.0}{GW191103_012549}{20.0}{GW200316_215756}{21.2}{GW200202_154313}{17.58}{GW200129_065458}{63.4}{GW191216_213338}{19.80}}[{\red{???}}]}
\DeclareRobustCommand{\totalmasssourcegwtcthreeplus}[1]{\IfEqCase{#1}{{GW200224_222234}{6.8}{GW191129_134029}{2.4}{GW200311_115853}{5.3}{GW191230_180458}{17}{GW191222_033537}{16}{GW200225_060421}{3.6}{GW200302_015811}{9.6}{GW200128_022011}{15}{GW191204_171526}{1.70}{GW200112_155838}{5.7}{GW200105_162426}{1.5}{GW191105_143521}{2.1}{GW191109_010717}{20}{GW200209_085452}{12.8}{GW200115_042309}{1.7}{GW191127_050227}{39}{GW200216_220804}{20}{GW191215_223052}{5.4}{GW200208_130117}{7.8}{GW200219_094415}{12.6}{GW191103_012549}{3.7}{GW200316_215756}{7.2}{GW200202_154313}{1.78}{GW200129_065458}{4.3}{GW191216_213338}{2.72}}[{\red{???}}]}
\DeclareRobustCommand{\totalmasssourcegwtcthreetenthpercentile}[1]{\IfEqCase{#1}{{GW200224_222234}{67.9}{GW191129_134029}{16.5}{GW200311_115853}{58.5}{GW191230_180458}{75}{GW191222_033537}{70}{GW200225_060421}{31.1}{GW200302_015811}{51.9}{GW200128_022011}{63}{GW191204_171526}{19.37}{GW200112_155838}{60.1}{GW200105_162426}{10.1}{GW191105_143521}{17.4}{GW191109_010717}{99}{GW200209_085452}{54.2}{GW200115_042309}{5.8}{GW191127_050227}{61}{GW200216_220804}{70}{GW191215_223052}{39.4}{GW200208_130117}{60.0}{GW200219_094415}{58.3}{GW191103_012549}{18.5}{GW200316_215756}{19.5}{GW200202_154313}{17.04}{GW200129_065458}{60.5}{GW191216_213338}{19.02}}[{\red{???}}]}
\DeclareRobustCommand{\totalmasssourcegwtcthreenintiethpercentile}[1]{\IfEqCase{#1}{{GW200224_222234}{77.1}{GW191129_134029}{19.2}{GW200311_115853}{65.9}{GW191230_180458}{96}{GW191222_033537}{92}{GW200225_060421}{36.2}{GW200302_015811}{64.3}{GW200128_022011}{83}{GW191204_171526}{21.36}{GW200112_155838}{68.1}{GW200105_162426}{11.8}{GW191105_143521}{19.9}{GW191109_010717}{126}{GW200209_085452}{70.6}{GW200115_042309}{8.6}{GW191127_050227}{108}{GW200216_220804}{96}{GW191215_223052}{46.8}{GW200208_130117}{71.4}{GW200219_094415}{74.7}{GW191103_012549}{22.4}{GW200316_215756}{25.6}{GW200202_154313}{18.75}{GW200129_065458}{66.7}{GW191216_213338}{21.49}}[{\red{???}}]}
\DeclareRobustCommand{\phijlgwtcthreeminus}[1]{\IfEqCase{#1}{{GW200224_222234}{2.6}{GW191129_134029}{2.8}{GW200311_115853}{2.3}{GW191230_180458}{3.0}{GW191222_033537}{2.8}{GW200225_060421}{2.8}{GW200302_015811}{2.8}{GW200128_022011}{2.9}{GW191204_171526}{2.8}{GW200112_155838}{2.8}{GW200105_162426}{2.8}{GW191105_143521}{2.8}{GW191109_010717}{3.1}{GW200209_085452}{2.8}{GW200115_042309}{2.7}{GW191127_050227}{2.7}{GW200216_220804}{3.0}{GW191215_223052}{2.9}{GW200208_130117}{2.8}{GW200219_094415}{2.8}{GW191103_012549}{2.8}{GW200316_215756}{3.0}{GW200202_154313}{2.8}{GW200129_065458}{2.0}{GW191216_213338}{3.0}}[{\red{???}}]}
\DeclareRobustCommand{\phijlgwtcthreemed}[1]{\IfEqCase{#1}{{GW200224_222234}{3.0}{GW191129_134029}{3.1}{GW200311_115853}{2.7}{GW191230_180458}{3.4}{GW191222_033537}{3.1}{GW200225_060421}{3.1}{GW200302_015811}{3.1}{GW200128_022011}{3.2}{GW191204_171526}{3.2}{GW200112_155838}{3.1}{GW200105_162426}{3.2}{GW191105_143521}{3.1}{GW191109_010717}{3.5}{GW200209_085452}{3.1}{GW200115_042309}{3.0}{GW191127_050227}{3.1}{GW200216_220804}{3.4}{GW191215_223052}{3.2}{GW200208_130117}{3.2}{GW200219_094415}{3.1}{GW191103_012549}{3.1}{GW200316_215756}{3.3}{GW200202_154313}{3.1}{GW200129_065458}{2.7}{GW191216_213338}{3.3}}[{\red{???}}]}
\DeclareRobustCommand{\phijlgwtcthreeplus}[1]{\IfEqCase{#1}{{GW200224_222234}{2.8}{GW191129_134029}{2.8}{GW200311_115853}{3.3}{GW191230_180458}{2.6}{GW191222_033537}{2.8}{GW200225_060421}{2.9}{GW200302_015811}{2.8}{GW200128_022011}{2.8}{GW191204_171526}{2.8}{GW200112_155838}{2.9}{GW200105_162426}{2.8}{GW191105_143521}{2.8}{GW191109_010717}{2.4}{GW200209_085452}{2.8}{GW200115_042309}{3.0}{GW191127_050227}{2.8}{GW200216_220804}{2.5}{GW191215_223052}{2.8}{GW200208_130117}{2.7}{GW200219_094415}{2.8}{GW191103_012549}{2.9}{GW200316_215756}{2.6}{GW200202_154313}{2.9}{GW200129_065458}{3.1}{GW191216_213338}{2.7}}[{\red{???}}]}
\DeclareRobustCommand{\phijlgwtcthreetenthpercentile}[1]{\IfEqCase{#1}{{GW200224_222234}{0.8}{GW191129_134029}{0.6}{GW200311_115853}{0.6}{GW191230_180458}{0.7}{GW191222_033537}{0.7}{GW200225_060421}{0.6}{GW200302_015811}{0.6}{GW200128_022011}{0.6}{GW191204_171526}{0.6}{GW200112_155838}{0.6}{GW200105_162426}{0.7}{GW191105_143521}{0.6}{GW191109_010717}{0.8}{GW200209_085452}{0.7}{GW200115_042309}{0.7}{GW191127_050227}{0.7}{GW200216_220804}{0.8}{GW191215_223052}{0.6}{GW200208_130117}{0.7}{GW200219_094415}{0.8}{GW191103_012549}{0.7}{GW200316_215756}{0.7}{GW200202_154313}{0.6}{GW200129_065458}{1.0}{GW191216_213338}{0.6}}[{\red{???}}]}
\DeclareRobustCommand{\phijlgwtcthreenintiethpercentile}[1]{\IfEqCase{#1}{{GW200224_222234}{5.1}{GW191129_134029}{5.6}{GW200311_115853}{5.6}{GW191230_180458}{5.7}{GW191222_033537}{5.7}{GW200225_060421}{5.7}{GW200302_015811}{5.6}{GW200128_022011}{5.7}{GW191204_171526}{5.7}{GW200112_155838}{5.7}{GW200105_162426}{5.7}{GW191105_143521}{5.6}{GW191109_010717}{5.5}{GW200209_085452}{5.6}{GW200115_042309}{5.6}{GW191127_050227}{5.6}{GW200216_220804}{5.7}{GW191215_223052}{5.7}{GW200208_130117}{5.6}{GW200219_094415}{5.6}{GW191103_012549}{5.7}{GW200316_215756}{5.6}{GW200202_154313}{5.7}{GW200129_065458}{5.4}{GW191216_213338}{5.7}}[{\red{???}}]}
\DeclareRobustCommand{\masstwodetgwtcthreeminus}[1]{\IfEqCase{#1}{{GW200224_222234}{9.7}{GW191129_134029}{1.9}{GW200311_115853}{7.3}{GW191230_180458}{20}{GW191222_033537}{15}{GW200225_060421}{4.6}{GW200302_015811}{8.1}{GW200128_022011}{13}{GW191204_171526}{1.8}{GW200112_155838}{7.4}{GW200105_162426}{0.25}{GW191105_143521}{2.2}{GW191109_010717}{17}{GW200209_085452}{13}{GW200115_042309}{0.30}{GW191127_050227}{25}{GW200216_220804}{31}{GW191215_223052}{5.1}{GW200208_130117}{11.2}{GW200219_094415}{14.1}{GW191103_012549}{2.9}{GW200316_215756}{3.5}{GW200202_154313}{1.9}{GW200129_065458}{10.8}{GW191216_213338}{2.0}}[{\red{???}}]}
\DeclareRobustCommand{\masstwodetgwtcthreemed}[1]{\IfEqCase{#1}{{GW200224_222234}{43.0}{GW191129_134029}{7.8}{GW200311_115853}{34.0}{GW191230_180458}{64}{GW191222_033537}{52}{GW200225_060421}{17.2}{GW200302_015811}{26.5}{GW200128_022011}{52}{GW191204_171526}{9.3}{GW200112_155838}{35.2}{GW200105_162426}{2.02}{GW191105_143521}{9.4}{GW191109_010717}{60}{GW200209_085452}{44}{GW200115_042309}{1.53}{GW191127_050227}{38}{GW200216_220804}{51}{GW191215_223052}{24.5}{GW200208_130117}{38.5}{GW200219_094415}{44.4}{GW191103_012549}{9.4}{GW200316_215756}{9.5}{GW200202_154313}{8.0}{GW200129_065458}{34.1}{GW191216_213338}{8.2}}[{\red{???}}]}
\DeclareRobustCommand{\masstwodetgwtcthreeplus}[1]{\IfEqCase{#1}{{GW200224_222234}{5.8}{GW191129_134029}{1.7}{GW200311_115853}{4.7}{GW191230_180458}{14}{GW191222_033537}{11}{GW200225_060421}{3.0}{GW200302_015811}{12.5}{GW200128_022011}{10}{GW191204_171526}{1.5}{GW200112_155838}{5.1}{GW200105_162426}{0.35}{GW191105_143521}{1.4}{GW191109_010717}{16}{GW200209_085452}{11}{GW200115_042309}{0.91}{GW191127_050227}{31}{GW200216_220804}{23}{GW191215_223052}{4.1}{GW200208_130117}{8.6}{GW200219_094415}{9.3}{GW191103_012549}{1.8}{GW200316_215756}{2.3}{GW200202_154313}{1.2}{GW200129_065458}{3.3}{GW191216_213338}{1.7}}[{\red{???}}]}
\DeclareRobustCommand{\masstwodetgwtcthreetenthpercentile}[1]{\IfEqCase{#1}{{GW200224_222234}{35.7}{GW191129_134029}{6.2}{GW200311_115853}{28.6}{GW191230_180458}{48}{GW191222_033537}{41}{GW200225_060421}{13.7}{GW200302_015811}{19.8}{GW200128_022011}{41}{GW191204_171526}{7.8}{GW200112_155838}{29.5}{GW200105_162426}{1.87}{GW191105_143521}{7.6}{GW191109_010717}{46}{GW200209_085452}{34}{GW200115_042309}{1.29}{GW191127_050227}{17}{GW200216_220804}{24}{GW191215_223052}{20.5}{GW200208_130117}{29.6}{GW200219_094415}{33.7}{GW191103_012549}{7.2}{GW200316_215756}{6.8}{GW200202_154313}{6.5}{GW200129_065458}{25.3}{GW191216_213338}{6.7}}[{\red{???}}]}
\DeclareRobustCommand{\masstwodetgwtcthreenintiethpercentile}[1]{\IfEqCase{#1}{{GW200224_222234}{47.8}{GW191129_134029}{9.3}{GW200311_115853}{37.7}{GW191230_180458}{75}{GW191222_033537}{61}{GW200225_060421}{19.8}{GW200302_015811}{36.3}{GW200128_022011}{60}{GW191204_171526}{10.6}{GW200112_155838}{39.3}{GW200105_162426}{2.20}{GW191105_143521}{10.6}{GW191109_010717}{72}{GW200209_085452}{52}{GW200115_042309}{2.25}{GW191127_050227}{63}{GW200216_220804}{70}{GW191215_223052}{28.0}{GW200208_130117}{45.5}{GW200219_094415}{51.8}{GW191103_012549}{11.0}{GW200316_215756}{11.5}{GW200202_154313}{9.1}{GW200129_065458}{36.8}{GW191216_213338}{9.8}}[{\red{???}}]}
\DeclareRobustCommand{\ragwtcthreeminus}[1]{\IfEqCase{#1}{{GW200224_222234}{0.048}{GW191129_134029}{2.77}{GW200311_115853}{0.029}{GW191230_180458}{0.29}{GW191222_033537}{3.0}{GW200225_060421}{0.30}{GW200302_015811}{3.13}{GW200128_022011}{3.0}{GW191204_171526}{0.53}{GW200112_155838}{2.9}{GW200105_162426}{1.3}{GW191105_143521}{0.36}{GW191109_010717}{1.4}{GW200209_085452}{0.92}{GW200115_042309}{0.10}{GW191127_050227}{1.1}{GW200216_220804}{1.49}{GW191215_223052}{0.76}{GW200208_130117}{0.039}{GW200219_094415}{0.12}{GW191103_012549}{1.85}{GW200316_215756}{0.35}{GW200202_154313}{0.117}{GW200129_065458}{0.15}{GW191216_213338}{3.518}}[{\red{???}}]}
\DeclareRobustCommand{\ragwtcthreemed}[1]{\IfEqCase{#1}{{GW200224_222234}{3.050}{GW191129_134029}{5.59}{GW200311_115853}{0.038}{GW191230_180458}{1.07}{GW191222_033537}{3.6}{GW200225_060421}{1.92}{GW200302_015811}{3.83}{GW200128_022011}{3.8}{GW191204_171526}{1.27}{GW200112_155838}{3.5}{GW200105_162426}{2.0}{GW191105_143521}{0.44}{GW191109_010717}{3.7}{GW200209_085452}{2.51}{GW200115_042309}{0.74}{GW191127_050227}{1.2}{GW200216_220804}{5.31}{GW191215_223052}{2.63}{GW200208_130117}{2.438}{GW200219_094415}{0.39}{GW191103_012549}{4.35}{GW200316_215756}{1.51}{GW200202_154313}{2.523}{GW200129_065458}{5.56}{GW191216_213338}{5.557}}[{\red{???}}]}
\DeclareRobustCommand{\ragwtcthreeplus}[1]{\IfEqCase{#1}{{GW200224_222234}{0.041}{GW191129_134029}{0.42}{GW200311_115853}{0.041}{GW191230_180458}{3.73}{GW191222_033537}{1.8}{GW200225_060421}{3.25}{GW200302_015811}{0.96}{GW200128_022011}{1.1}{GW191204_171526}{1.91}{GW200112_155838}{1.7}{GW200105_162426}{3.0}{GW191105_143521}{5.68}{GW191109_010717}{1.2}{GW200209_085452}{0.63}{GW200115_042309}{4.06}{GW191127_050227}{4.9}{GW200216_220804}{0.28}{GW191215_223052}{3.28}{GW200208_130117}{0.040}{GW200219_094415}{2.85}{GW191103_012549}{0.39}{GW200316_215756}{2.01}{GW200202_154313}{0.058}{GW200129_065458}{0.47}{GW191216_213338}{0.066}}[{\red{???}}]}
\DeclareRobustCommand{\ragwtcthreetenthpercentile}[1]{\IfEqCase{#1}{{GW200224_222234}{3.015}{GW191129_134029}{3.01}{GW200311_115853}{0.014}{GW191230_180458}{0.84}{GW191222_033537}{1.1}{GW200225_060421}{1.71}{GW200302_015811}{0.89}{GW200128_022011}{0.9}{GW191204_171526}{0.82}{GW200112_155838}{0.6}{GW200105_162426}{0.8}{GW191105_143521}{0.12}{GW191109_010717}{2.4}{GW200209_085452}{2.22}{GW200115_042309}{0.66}{GW191127_050227}{0.1}{GW200216_220804}{3.96}{GW191215_223052}{1.93}{GW200208_130117}{2.409}{GW200219_094415}{0.30}{GW191103_012549}{2.56}{GW200316_215756}{1.21}{GW200202_154313}{2.425}{GW200129_065458}{5.45}{GW191216_213338}{5.260}}[{\red{???}}]}
\DeclareRobustCommand{\ragwtcthreenintiethpercentile}[1]{\IfEqCase{#1}{{GW200224_222234}{3.082}{GW191129_134029}{5.92}{GW200311_115853}{0.068}{GW191230_180458}{4.65}{GW191222_033537}{4.7}{GW200225_060421}{2.42}{GW200302_015811}{4.59}{GW200128_022011}{4.5}{GW191204_171526}{2.05}{GW200112_155838}{5.1}{GW200105_162426}{4.9}{GW191105_143521}{6.07}{GW191109_010717}{4.6}{GW200209_085452}{2.91}{GW200115_042309}{4.65}{GW191127_050227}{5.0}{GW200216_220804}{5.54}{GW191215_223052}{5.77}{GW200208_130117}{2.468}{GW200219_094415}{3.06}{GW191103_012549}{4.71}{GW200316_215756}{3.42}{GW200202_154313}{2.569}{GW200129_065458}{5.59}{GW191216_213338}{5.615}}[{\red{???}}]}
\DeclareRobustCommand{\finalmassdetgwtcthreeminus}[1]{\IfEqCase{#1}{{GW200224_222234}{6.4}{GW191129_134029}{0.67}{GW200311_115853}{5.1}{GW191230_180458}{17}{GW191222_033537}{12}{GW200225_060421}{3.6}{GW200302_015811}{7.6}{GW200128_022011}{12}{GW191204_171526}{0.50}{GW200112_155838}{4.6}{GW200105_162426}{1.8}{GW191105_143521}{0.47}{GW191109_010717}{15}{GW200209_085452}{14}{GW200115_042309}{1.7}{GW191127_050227}{43}{GW200216_220804}{30}{GW191215_223052}{3.3}{GW200208_130117}{9.1}{GW200219_094415}{11}{GW191103_012549}{0.66}{GW200316_215756}{1.1}{GW200202_154313}{0.35}{GW200129_065458}{3.4}{GW191216_213338}{0.70}}[{\red{???}}]}
\DeclareRobustCommand{\finalmassdetgwtcthreemed}[1]{\IfEqCase{#1}{{GW200224_222234}{90.5}{GW191129_134029}{19.20}{GW200311_115853}{72.4}{GW191230_180458}{140}{GW191222_033537}{114}{GW200225_060421}{39.4}{GW200302_015811}{71.6}{GW200128_022011}{112}{GW191204_171526}{21.60}{GW200112_155838}{75.3}{GW200105_162426}{11.4}{GW191105_143521}{21.36}{GW191109_010717}{135}{GW200209_085452}{96}{GW200115_042309}{7.6}{GW191127_050227}{124}{GW200216_220804}{129}{GW191215_223052}{55.9}{GW200208_130117}{87.5}{GW200219_094415}{98}{GW191103_012549}{22.27}{GW200316_215756}{24.4}{GW200202_154313}{18.12}{GW200129_065458}{70.9}{GW191216_213338}{20.18}}[{\red{???}}]}
\DeclareRobustCommand{\finalmassdetgwtcthreeplus}[1]{\IfEqCase{#1}{{GW200224_222234}{7.6}{GW191129_134029}{3.08}{GW200311_115853}{5.6}{GW191230_180458}{20}{GW191222_033537}{14}{GW200225_060421}{2.9}{GW200302_015811}{14.1}{GW200128_022011}{16}{GW191204_171526}{2.04}{GW200112_155838}{5.8}{GW200105_162426}{2.1}{GW191105_143521}{2.48}{GW191109_010717}{19}{GW200209_085452}{19}{GW200115_042309}{2.3}{GW191127_050227}{52}{GW200216_220804}{27}{GW191215_223052}{5.0}{GW200208_130117}{10.3}{GW200219_094415}{13}{GW191103_012549}{4.79}{GW200316_215756}{9.0}{GW200202_154313}{2.09}{GW200129_065458}{4.2}{GW191216_213338}{3.10}}[{\red{???}}]}
\DeclareRobustCommand{\finalmassdetgwtcthreetenthpercentile}[1]{\IfEqCase{#1}{{GW200224_222234}{85.4}{GW191129_134029}{18.57}{GW200311_115853}{68.4}{GW191230_180458}{127}{GW191222_033537}{105}{GW200225_060421}{36.5}{GW200302_015811}{65.5}{GW200128_022011}{102}{GW191204_171526}{21.14}{GW200112_155838}{71.8}{GW200105_162426}{10.2}{GW191105_143521}{20.95}{GW191109_010717}{123}{GW200209_085452}{84}{GW200115_042309}{6.1}{GW191127_050227}{88}{GW200216_220804}{106}{GW191215_223052}{53.3}{GW200208_130117}{80.3}{GW200219_094415}{89}{GW191103_012549}{21.68}{GW200316_215756}{23.3}{GW200202_154313}{17.80}{GW200129_065458}{68.3}{GW191216_213338}{19.53}}[{\red{???}}]}
\DeclareRobustCommand{\finalmassdetgwtcthreenintiethpercentile}[1]{\IfEqCase{#1}{{GW200224_222234}{96.2}{GW191129_134029}{21.37}{GW200311_115853}{76.5}{GW191230_180458}{155}{GW191222_033537}{125}{GW200225_060421}{41.6}{GW200302_015811}{82.3}{GW200128_022011}{124}{GW191204_171526}{23.02}{GW200112_155838}{79.7}{GW200105_162426}{12.7}{GW191105_143521}{22.99}{GW191109_010717}{149}{GW200209_085452}{110}{GW200115_042309}{9.3}{GW191127_050227}{164}{GW200216_220804}{150}{GW191215_223052}{59.4}{GW200208_130117}{95.3}{GW200219_094415}{108}{GW191103_012549}{25.37}{GW200316_215756}{30.0}{GW200202_154313}{19.51}{GW200129_065458}{74.2}{GW191216_213338}{22.11}}[{\red{???}}]}
\DeclareRobustCommand{\spinonexgwtcthreeminus}[1]{\IfEqCase{#1}{{GW200224_222234}{0.59}{GW191129_134029}{0.36}{GW200311_115853}{0.55}{GW191230_180458}{0.64}{GW191222_033537}{0.53}{GW200225_060421}{0.65}{GW200302_015811}{0.55}{GW200128_022011}{0.68}{GW191204_171526}{0.50}{GW200112_155838}{0.51}{GW200105_162426}{0.14}{GW191105_143521}{0.41}{GW191109_010717}{0.70}{GW200209_085452}{0.64}{GW200115_042309}{0.33}{GW191127_050227}{0.70}{GW200216_220804}{0.60}{GW191215_223052}{0.64}{GW200208_130117}{0.47}{GW200219_094415}{0.61}{GW191103_012549}{0.53}{GW200316_215756}{0.40}{GW200202_154313}{0.37}{GW200129_065458}{0.81}{GW191216_213338}{0.29}}[{\red{???}}]}
\DeclareRobustCommand{\spinonexgwtcthreemed}[1]{\IfEqCase{#1}{{GW200224_222234}{0.01}{GW191129_134029}{0.00}{GW200311_115853}{0.00}{GW191230_180458}{0.00}{GW191222_033537}{0.00}{GW200225_060421}{0.00}{GW200302_015811}{0.00}{GW200128_022011}{0.00}{GW191204_171526}{0.00}{GW200112_155838}{0.00}{GW200105_162426}{0.00}{GW191105_143521}{0.00}{GW191109_010717}{0.00}{GW200209_085452}{0.00}{GW200115_042309}{0.00}{GW191127_050227}{0.00}{GW200216_220804}{0.00}{GW191215_223052}{0.00}{GW200208_130117}{0.00}{GW200219_094415}{0.00}{GW191103_012549}{0.00}{GW200316_215756}{0.00}{GW200202_154313}{0.00}{GW200129_065458}{-0.02}{GW191216_213338}{0.00}}[{\red{???}}]}
\DeclareRobustCommand{\spinonexgwtcthreeplus}[1]{\IfEqCase{#1}{{GW200224_222234}{0.61}{GW191129_134029}{0.37}{GW200311_115853}{0.54}{GW191230_180458}{0.65}{GW191222_033537}{0.54}{GW200225_060421}{0.65}{GW200302_015811}{0.55}{GW200128_022011}{0.67}{GW191204_171526}{0.48}{GW200112_155838}{0.48}{GW200105_162426}{0.13}{GW191105_143521}{0.42}{GW191109_010717}{0.68}{GW200209_085452}{0.65}{GW200115_042309}{0.33}{GW191127_050227}{0.67}{GW200216_220804}{0.59}{GW191215_223052}{0.63}{GW200208_130117}{0.52}{GW200219_094415}{0.61}{GW191103_012549}{0.51}{GW200316_215756}{0.38}{GW200202_154313}{0.37}{GW200129_065458}{0.65}{GW191216_213338}{0.29}}[{\red{???}}]}
\DeclareRobustCommand{\spinonexgwtcthreetenthpercentile}[1]{\IfEqCase{#1}{{GW200224_222234}{-0.43}{GW191129_134029}{-0.25}{GW200311_115853}{-0.42}{GW191230_180458}{-0.49}{GW191222_033537}{-0.39}{GW200225_060421}{-0.52}{GW200302_015811}{-0.41}{GW200128_022011}{-0.53}{GW191204_171526}{-0.39}{GW200112_155838}{-0.37}{GW200105_162426}{-0.09}{GW191105_143521}{-0.27}{GW191109_010717}{-0.56}{GW200209_085452}{-0.48}{GW200115_042309}{-0.24}{GW191127_050227}{-0.54}{GW200216_220804}{-0.46}{GW191215_223052}{-0.50}{GW200208_130117}{-0.34}{GW200219_094415}{-0.45}{GW191103_012549}{-0.39}{GW200316_215756}{-0.27}{GW200202_154313}{-0.25}{GW200129_065458}{-0.72}{GW191216_213338}{-0.21}}[{\red{???}}]}
\DeclareRobustCommand{\spinonexgwtcthreenintiethpercentile}[1]{\IfEqCase{#1}{{GW200224_222234}{0.49}{GW191129_134029}{0.26}{GW200311_115853}{0.38}{GW191230_180458}{0.49}{GW191222_033537}{0.38}{GW200225_060421}{0.52}{GW200302_015811}{0.40}{GW200128_022011}{0.52}{GW191204_171526}{0.36}{GW200112_155838}{0.35}{GW200105_162426}{0.08}{GW191105_143521}{0.29}{GW191109_010717}{0.57}{GW200209_085452}{0.49}{GW200115_042309}{0.25}{GW191127_050227}{0.52}{GW200216_220804}{0.44}{GW191215_223052}{0.48}{GW200208_130117}{0.37}{GW200219_094415}{0.45}{GW191103_012549}{0.38}{GW200316_215756}{0.27}{GW200202_154313}{0.24}{GW200129_065458}{0.46}{GW191216_213338}{0.20}}[{\red{???}}]}
\DeclareRobustCommand{\loglikelihoodgwtcthreeminus}[1]{\IfEqCase{#1}{{GW200224_222234}{5.0}{GW191129_134029}{5.3}{GW200311_115853}{4.7}{GW191230_180458}{4.4}{GW191222_033537}{4.3}{GW200225_060421}{5.0}{GW200302_015811}{4.6}{GW200128_022011}{4.5}{GW191204_171526}{5.1}{GW200112_155838}{4.9}{GW200105_162426}{5.5}{GW191105_143521}{5.9}{GW191109_010717}{6.7}{GW200209_085452}{4.7}{GW200115_042309}{6.4}{GW191127_050227}{5.3}{GW200216_220804}{4.2}{GW191215_223052}{5.0}{GW200208_130117}{5.2}{GW200219_094415}{5.1}{GW191103_012549}{5.1}{GW200316_215756}{6.6}{GW200202_154313}{5.7}{GW200129_065458}{21.5}{GW191216_213338}{6.7}}[{\red{???}}]}
\DeclareRobustCommand{\loglikelihoodgwtcthreemed}[1]{\IfEqCase{#1}{{GW200224_222234}{188.2}{GW191129_134029}{73.6}{GW200311_115853}{148.0}{GW191230_180458}{46.2}{GW191222_033537}{68.9}{GW200225_060421}{66.4}{GW200302_015811}{47.0}{GW200128_022011}{48.7}{GW191204_171526}{138.2}{GW200112_155838}{183.8}{GW200105_162426}{82.7}{GW191105_143521}{34.7}{GW191109_010717}{138.1}{GW200209_085452}{37.1}{GW200115_042309}{47.8}{GW191127_050227}{32.1}{GW200216_220804}{25.0}{GW191215_223052}{51.9}{GW200208_130117}{49.4}{GW200219_094415}{48.6}{GW191103_012549}{27.5}{GW200316_215756}{40.3}{GW200202_154313}{44.9}{GW200129_065458}{343.3}{GW191216_213338}{156.6}}[{\red{???}}]}
\DeclareRobustCommand{\loglikelihoodgwtcthreeplus}[1]{\IfEqCase{#1}{{GW200224_222234}{3.3}{GW191129_134029}{5.1}{GW200311_115853}{4.6}{GW191230_180458}{2.8}{GW191222_033537}{2.8}{GW200225_060421}{3.8}{GW200302_015811}{4.0}{GW200128_022011}{3.5}{GW191204_171526}{4.4}{GW200112_155838}{3.4}{GW200105_162426}{4.0}{GW191105_143521}{5.4}{GW191109_010717}{7.4}{GW200209_085452}{3.2}{GW200115_042309}{8.1}{GW191127_050227}{4.9}{GW200216_220804}{3.1}{GW191215_223052}{4.2}{GW200208_130117}{2.9}{GW200219_094415}{3.4}{GW191103_012549}{4.8}{GW200316_215756}{4.8}{GW200202_154313}{6.9}{GW200129_065458}{6.3}{GW191216_213338}{8.0}}[{\red{???}}]}
\DeclareRobustCommand{\loglikelihoodgwtcthreetenthpercentile}[1]{\IfEqCase{#1}{{GW200224_222234}{184.6}{GW191129_134029}{69.5}{GW200311_115853}{144.4}{GW191230_180458}{43.0}{GW191222_033537}{65.7}{GW200225_060421}{62.7}{GW200302_015811}{43.7}{GW200128_022011}{45.3}{GW191204_171526}{134.4}{GW200112_155838}{180.1}{GW200105_162426}{78.6}{GW191105_143521}{30.2}{GW191109_010717}{132.9}{GW200209_085452}{33.6}{GW200115_042309}{42.9}{GW191127_050227}{28.0}{GW200216_220804}{21.9}{GW191215_223052}{48.2}{GW200208_130117}{45.7}{GW200219_094415}{44.8}{GW191103_012549}{23.6}{GW200316_215756}{35.4}{GW200202_154313}{40.4}{GW200129_065458}{330.2}{GW191216_213338}{151.5}}[{\red{???}}]}
\DeclareRobustCommand{\loglikelihoodgwtcthreenintiethpercentile}[1]{\IfEqCase{#1}{{GW200224_222234}{190.9}{GW191129_134029}{77.9}{GW200311_115853}{151.8}{GW191230_180458}{48.5}{GW191222_033537}{71.1}{GW200225_060421}{69.4}{GW200302_015811}{50.1}{GW200128_022011}{51.4}{GW191204_171526}{141.8}{GW200112_155838}{186.6}{GW200105_162426}{85.8}{GW191105_143521}{39.1}{GW191109_010717}{144.1}{GW200209_085452}{39.7}{GW200115_042309}{54.8}{GW191127_050227}{35.9}{GW200216_220804}{27.5}{GW191215_223052}{55.2}{GW200208_130117}{51.8}{GW200219_094415}{51.4}{GW191103_012549}{31.6}{GW200316_215756}{44.2}{GW200202_154313}{51.1}{GW200129_065458}{348.4}{GW191216_213338}{163.5}}[{\red{???}}]}
\DeclareRobustCommand{\tilttwogwtcthreeminus}[1]{\IfEqCase{#1}{{GW200224_222234}{0.92}{GW191129_134029}{0.89}{GW200311_115853}{1.0}{GW191230_180458}{1.1}{GW191222_033537}{1.1}{GW200225_060421}{1.18}{GW200302_015811}{1.0}{GW200128_022011}{0.95}{GW191204_171526}{0.77}{GW200112_155838}{0.91}{GW200105_162426}{1.0}{GW191105_143521}{1.07}{GW191109_010717}{1.16}{GW200209_085452}{1.17}{GW200115_042309}{1.27}{GW191127_050227}{0.98}{GW200216_220804}{1.0}{GW191215_223052}{1.09}{GW200208_130117}{1.17}{GW200219_094415}{1.1}{GW191103_012549}{0.77}{GW200316_215756}{0.83}{GW200202_154313}{0.90}{GW200129_065458}{0.84}{GW191216_213338}{0.83}}[{\red{???}}]}
\DeclareRobustCommand{\tilttwogwtcthreemed}[1]{\IfEqCase{#1}{{GW200224_222234}{1.38}{GW191129_134029}{1.27}{GW200311_115853}{1.6}{GW191230_180458}{1.7}{GW191222_033537}{1.7}{GW200225_060421}{1.79}{GW200302_015811}{1.4}{GW200128_022011}{1.42}{GW191204_171526}{1.11}{GW200112_155838}{1.32}{GW200105_162426}{1.5}{GW191105_143521}{1.60}{GW191109_010717}{1.84}{GW200209_085452}{1.84}{GW200115_042309}{1.89}{GW191127_050227}{1.35}{GW200216_220804}{1.4}{GW191215_223052}{1.66}{GW200208_130117}{1.72}{GW200219_094415}{1.7}{GW191103_012549}{1.07}{GW200316_215756}{1.19}{GW200202_154313}{1.34}{GW200129_065458}{1.15}{GW191216_213338}{1.16}}[{\red{???}}]}
\DeclareRobustCommand{\tilttwogwtcthreeplus}[1]{\IfEqCase{#1}{{GW200224_222234}{1.16}{GW191129_134029}{1.19}{GW200311_115853}{1.0}{GW191230_180458}{1.0}{GW191222_033537}{1.0}{GW200225_060421}{0.97}{GW200302_015811}{1.1}{GW200128_022011}{1.15}{GW191204_171526}{1.12}{GW200112_155838}{1.14}{GW200105_162426}{1.1}{GW191105_143521}{0.99}{GW191109_010717}{0.92}{GW200209_085452}{0.93}{GW200115_042309}{0.93}{GW191127_050227}{1.22}{GW200216_220804}{1.2}{GW191215_223052}{0.99}{GW200208_130117}{0.99}{GW200219_094415}{1.0}{GW191103_012549}{1.29}{GW200316_215756}{1.13}{GW200202_154313}{1.07}{GW200129_065458}{1.36}{GW191216_213338}{1.28}}[{\red{???}}]}
\DeclareRobustCommand{\tilttwogwtcthreetenthpercentile}[1]{\IfEqCase{#1}{{GW200224_222234}{0.63}{GW191129_134029}{0.52}{GW200311_115853}{0.8}{GW191230_180458}{0.8}{GW191222_033537}{0.8}{GW200225_060421}{0.87}{GW200302_015811}{0.6}{GW200128_022011}{0.66}{GW191204_171526}{0.47}{GW200112_155838}{0.60}{GW200105_162426}{0.7}{GW191105_143521}{0.74}{GW191109_010717}{0.91}{GW200209_085452}{0.92}{GW200115_042309}{0.86}{GW191127_050227}{0.53}{GW200216_220804}{0.6}{GW191215_223052}{0.79}{GW200208_130117}{0.79}{GW200219_094415}{0.8}{GW191103_012549}{0.43}{GW200316_215756}{0.51}{GW200202_154313}{0.62}{GW200129_065458}{0.45}{GW191216_213338}{0.47}}[{\red{???}}]}
\DeclareRobustCommand{\tilttwogwtcthreenintiethpercentile}[1]{\IfEqCase{#1}{{GW200224_222234}{2.30}{GW191129_134029}{2.17}{GW200311_115853}{2.4}{GW191230_180458}{2.5}{GW191222_033537}{2.5}{GW200225_060421}{2.59}{GW200302_015811}{2.4}{GW200128_022011}{2.34}{GW191204_171526}{1.96}{GW200112_155838}{2.23}{GW200105_162426}{2.4}{GW191105_143521}{2.39}{GW191109_010717}{2.59}{GW200209_085452}{2.60}{GW200115_042309}{2.69}{GW191127_050227}{2.34}{GW200216_220804}{2.4}{GW191215_223052}{2.46}{GW200208_130117}{2.54}{GW200219_094415}{2.6}{GW191103_012549}{2.09}{GW200316_215756}{2.05}{GW200202_154313}{2.16}{GW200129_065458}{2.23}{GW191216_213338}{2.15}}[{\red{???}}]}
\DeclareRobustCommand{\tiltonegwtcthreeminus}[1]{\IfEqCase{#1}{{GW200224_222234}{0.84}{GW191129_134029}{0.87}{GW200311_115853}{1.02}{GW191230_180458}{1.0}{GW191222_033537}{1.10}{GW200225_060421}{0.90}{GW200302_015811}{0.97}{GW200128_022011}{0.79}{GW191204_171526}{0.72}{GW200112_155838}{0.95}{GW200105_162426}{1.1}{GW191105_143521}{1.05}{GW191109_010717}{0.90}{GW200209_085452}{1.01}{GW200115_042309}{1.37}{GW191127_050227}{0.84}{GW200216_220804}{0.92}{GW191215_223052}{0.96}{GW200208_130117}{1.15}{GW200219_094415}{1.06}{GW191103_012549}{0.68}{GW200316_215756}{0.80}{GW200202_154313}{0.94}{GW200129_065458}{0.92}{GW191216_213338}{0.73}}[{\red{???}}]}
\DeclareRobustCommand{\tiltonegwtcthreemed}[1]{\IfEqCase{#1}{{GW200224_222234}{1.27}{GW191129_134029}{1.29}{GW200311_115853}{1.66}{GW191230_180458}{1.6}{GW191222_033537}{1.73}{GW200225_060421}{1.88}{GW200302_015811}{1.51}{GW200128_022011}{1.20}{GW191204_171526}{1.13}{GW200112_155838}{1.43}{GW200105_162426}{1.6}{GW191105_143521}{1.64}{GW191109_010717}{2.22}{GW200209_085452}{1.76}{GW200115_042309}{2.24}{GW191127_050227}{1.19}{GW200216_220804}{1.33}{GW191215_223052}{1.63}{GW200208_130117}{1.82}{GW200219_094415}{1.80}{GW191103_012549}{1.00}{GW200316_215756}{1.08}{GW200202_154313}{1.39}{GW200129_065458}{1.45}{GW191216_213338}{1.01}}[{\red{???}}]}
\DeclareRobustCommand{\tiltonegwtcthreeplus}[1]{\IfEqCase{#1}{{GW200224_222234}{1.03}{GW191129_134029}{1.01}{GW200311_115853}{0.97}{GW191230_180458}{1.0}{GW191222_033537}{0.94}{GW200225_060421}{0.80}{GW200302_015811}{1.06}{GW200128_022011}{1.03}{GW191204_171526}{0.91}{GW200112_155838}{1.02}{GW200105_162426}{1.2}{GW191105_143521}{0.95}{GW191109_010717}{0.67}{GW200209_085452}{0.91}{GW200115_042309}{0.66}{GW191127_050227}{1.06}{GW200216_220804}{1.14}{GW191215_223052}{0.83}{GW200208_130117}{0.93}{GW200219_094415}{0.92}{GW191103_012549}{0.98}{GW200316_215756}{1.07}{GW200202_154313}{1.06}{GW200129_065458}{1.04}{GW191216_213338}{1.20}}[{\red{???}}]}
\DeclareRobustCommand{\tiltonegwtcthreetenthpercentile}[1]{\IfEqCase{#1}{{GW200224_222234}{0.60}{GW191129_134029}{0.59}{GW200311_115853}{0.89}{GW191230_180458}{0.8}{GW191222_033537}{0.88}{GW200225_060421}{1.24}{GW200302_015811}{0.75}{GW200128_022011}{0.59}{GW191204_171526}{0.56}{GW200112_155838}{0.67}{GW200105_162426}{0.6}{GW191105_143521}{0.82}{GW191109_010717}{1.55}{GW200209_085452}{0.98}{GW200115_042309}{1.15}{GW191127_050227}{0.50}{GW200216_220804}{0.57}{GW191215_223052}{0.92}{GW200208_130117}{0.91}{GW200219_094415}{0.98}{GW191103_012549}{0.44}{GW200316_215756}{0.40}{GW200202_154313}{0.64}{GW200129_065458}{0.70}{GW191216_213338}{0.39}}[{\red{???}}]}
\DeclareRobustCommand{\tiltonegwtcthreenintiethpercentile}[1]{\IfEqCase{#1}{{GW200224_222234}{2.03}{GW191129_134029}{2.03}{GW200311_115853}{2.42}{GW191230_180458}{2.4}{GW191222_033537}{2.49}{GW200225_060421}{2.51}{GW200302_015811}{2.35}{GW200128_022011}{1.96}{GW191204_171526}{1.76}{GW200112_155838}{2.22}{GW200105_162426}{2.5}{GW191105_143521}{2.38}{GW191109_010717}{2.77}{GW200209_085452}{2.49}{GW200115_042309}{2.79}{GW191127_050227}{1.96}{GW200216_220804}{2.23}{GW191215_223052}{2.27}{GW200208_130117}{2.58}{GW200219_094415}{2.56}{GW191103_012549}{1.69}{GW200316_215756}{1.86}{GW200202_154313}{2.20}{GW200129_065458}{2.18}{GW191216_213338}{1.89}}[{\red{???}}]}
\DeclareRobustCommand{\psigwtcthreeminus}[1]{\IfEqCase{#1}{{GW200224_222234}{1.6}{GW191129_134029}{1.9}{GW200311_115853}{1.5}{GW191230_180458}{1.7}{GW191222_033537}{0.99}{GW200225_060421}{1.5}{GW200302_015811}{1.3}{GW200128_022011}{1.4}{GW191204_171526}{1.5}{GW200112_155838}{1.5}{GW200105_162426}{2.3}{GW191105_143521}{1.4}{GW191109_010717}{1.9}{GW200209_085452}{1.1}{GW200115_042309}{2.1}{GW191127_050227}{1.4}{GW200216_220804}{1.4}{GW191215_223052}{1.4}{GW200208_130117}{1.1}{GW200219_094415}{1.4}{GW191103_012549}{1.5}{GW200316_215756}{1.5}{GW200202_154313}{1.3}{GW200129_065458}{0.99}{GW191216_213338}{1.2}}[{\red{???}}]}
\DeclareRobustCommand{\psigwtcthreemed}[1]{\IfEqCase{#1}{{GW200224_222234}{1.7}{GW191129_134029}{2.1}{GW200311_115853}{1.8}{GW191230_180458}{1.9}{GW191222_033537}{1.11}{GW200225_060421}{1.7}{GW200302_015811}{1.4}{GW200128_022011}{1.6}{GW191204_171526}{1.6}{GW200112_155838}{1.6}{GW200105_162426}{2.4}{GW191105_143521}{1.5}{GW191109_010717}{2.0}{GW200209_085452}{1.3}{GW200115_042309}{2.3}{GW191127_050227}{1.6}{GW200216_220804}{1.6}{GW191215_223052}{1.6}{GW200208_130117}{1.3}{GW200219_094415}{1.7}{GW191103_012549}{1.6}{GW200316_215756}{1.6}{GW200202_154313}{1.5}{GW200129_065458}{1.18}{GW191216_213338}{1.7}}[{\red{???}}]}
\DeclareRobustCommand{\psigwtcthreeplus}[1]{\IfEqCase{#1}{{GW200224_222234}{1.4}{GW191129_134029}{3.6}{GW200311_115853}{1.1}{GW191230_180458}{3.6}{GW191222_033537}{1.88}{GW200225_060421}{1.3}{GW200302_015811}{1.5}{GW200128_022011}{1.4}{GW191204_171526}{1.3}{GW200112_155838}{1.4}{GW200105_162426}{3.5}{GW191105_143521}{1.5}{GW191109_010717}{1.1}{GW200209_085452}{1.6}{GW200115_042309}{3.5}{GW191127_050227}{1.4}{GW200216_220804}{1.4}{GW191215_223052}{1.4}{GW200208_130117}{1.6}{GW200219_094415}{1.2}{GW191103_012549}{1.4}{GW200316_215756}{1.4}{GW200202_154313}{1.5}{GW200129_065458}{1.65}{GW191216_213338}{3.7}}[{\red{???}}]}
\DeclareRobustCommand{\psigwtcthreetenthpercentile}[1]{\IfEqCase{#1}{{GW200224_222234}{0.2}{GW191129_134029}{0.4}{GW200311_115853}{0.5}{GW191230_180458}{0.4}{GW191222_033537}{0.24}{GW200225_060421}{0.3}{GW200302_015811}{0.3}{GW200128_022011}{0.3}{GW191204_171526}{0.4}{GW200112_155838}{0.3}{GW200105_162426}{0.3}{GW191105_143521}{0.2}{GW191109_010717}{0.2}{GW200209_085452}{0.3}{GW200115_042309}{0.4}{GW191127_050227}{0.3}{GW200216_220804}{0.3}{GW191215_223052}{0.3}{GW200208_130117}{0.4}{GW200219_094415}{0.5}{GW191103_012549}{0.3}{GW200316_215756}{0.2}{GW200202_154313}{0.3}{GW200129_065458}{0.40}{GW191216_213338}{0.7}}[{\red{???}}]}
\DeclareRobustCommand{\psigwtcthreenintiethpercentile}[1]{\IfEqCase{#1}{{GW200224_222234}{3.0}{GW191129_134029}{5.0}{GW200311_115853}{2.7}{GW191230_180458}{4.9}{GW191222_033537}{2.83}{GW200225_060421}{2.8}{GW200302_015811}{2.8}{GW200128_022011}{2.8}{GW191204_171526}{2.8}{GW200112_155838}{2.8}{GW200105_162426}{5.2}{GW191105_143521}{2.9}{GW191109_010717}{3.0}{GW200209_085452}{2.8}{GW200115_042309}{5.2}{GW191127_050227}{2.8}{GW200216_220804}{2.8}{GW191215_223052}{2.9}{GW200208_130117}{2.7}{GW200219_094415}{2.7}{GW191103_012549}{2.8}{GW200316_215756}{3.0}{GW200202_154313}{2.8}{GW200129_065458}{2.46}{GW191216_213338}{4.8}}[{\red{???}}]}
\DeclareRobustCommand{\decgwtcthreeminus}[1]{\IfEqCase{#1}{{GW200224_222234}{0.076}{GW191129_134029}{0.62}{GW200311_115853}{0.093}{GW191230_180458}{0.52}{GW191222_033537}{0.54}{GW200225_060421}{0.73}{GW200302_015811}{0.55}{GW200128_022011}{0.73}{GW191204_171526}{0.21}{GW200112_155838}{0.57}{GW200105_162426}{0.82}{GW191105_143521}{0.14}{GW191109_010717}{0.20}{GW200209_085452}{1.22}{GW200115_042309}{0.62}{GW191127_050227}{1.86}{GW200216_220804}{0.59}{GW191215_223052}{0.87}{GW200208_130117}{0.063}{GW200219_094415}{0.19}{GW191103_012549}{1.06}{GW200316_215756}{1.433}{GW200202_154313}{0.11}{GW200129_065458}{0.53}{GW191216_213338}{1.09}}[{\red{???}}]}
\DeclareRobustCommand{\decgwtcthreemed}[1]{\IfEqCase{#1}{{GW200224_222234}{-0.168}{GW191129_134029}{-0.58}{GW200311_115853}{-0.133}{GW191230_180458}{-0.62}{GW191222_033537}{-0.68}{GW200225_060421}{0.94}{GW200302_015811}{-0.51}{GW200128_022011}{-0.38}{GW191204_171526}{-0.54}{GW200112_155838}{-0.21}{GW200105_162426}{-0.05}{GW191105_143521}{-0.61}{GW191109_010717}{-0.59}{GW200209_085452}{0.88}{GW200115_042309}{-0.03}{GW191127_050227}{1.00}{GW200216_220804}{0.75}{GW191215_223052}{-0.24}{GW200208_130117}{-0.597}{GW200219_094415}{-0.46}{GW191103_012549}{0.65}{GW200316_215756}{0.819}{GW200202_154313}{0.38}{GW200129_065458}{0.09}{GW191216_213338}{0.44}}[{\red{???}}]}
\DeclareRobustCommand{\decgwtcthreeplus}[1]{\IfEqCase{#1}{{GW200224_222234}{0.095}{GW191129_134029}{1.30}{GW200311_115853}{0.079}{GW191230_180458}{0.98}{GW191222_033537}{1.56}{GW200225_060421}{0.53}{GW200302_015811}{1.65}{GW200128_022011}{1.35}{GW191204_171526}{0.90}{GW200112_155838}{1.12}{GW200105_162426}{0.86}{GW191105_143521}{1.86}{GW191109_010717}{0.99}{GW200209_085452}{0.48}{GW200115_042309}{0.52}{GW191127_050227}{0.50}{GW200216_220804}{0.37}{GW191215_223052}{0.93}{GW200208_130117}{0.074}{GW200219_094415}{1.39}{GW191103_012549}{0.73}{GW200316_215756}{0.060}{GW200202_154313}{0.13}{GW200129_065458}{0.39}{GW191216_213338}{0.47}}[{\red{???}}]}
\DeclareRobustCommand{\decgwtcthreetenthpercentile}[1]{\IfEqCase{#1}{{GW200224_222234}{-0.227}{GW191129_134029}{-1.06}{GW200311_115853}{-0.205}{GW191230_180458}{-1.07}{GW191222_033537}{-1.15}{GW200225_060421}{0.40}{GW200302_015811}{-1.03}{GW200128_022011}{-1.01}{GW191204_171526}{-0.74}{GW200112_155838}{-0.66}{GW200105_162426}{-0.73}{GW191105_143521}{-0.73}{GW191109_010717}{-0.77}{GW200209_085452}{-0.19}{GW200115_042309}{-0.53}{GW191127_050227}{-0.79}{GW200216_220804}{0.25}{GW191215_223052}{-0.96}{GW200208_130117}{-0.645}{GW200219_094415}{-0.60}{GW191103_012549}{-0.29}{GW200316_215756}{-0.432}{GW200202_154313}{0.30}{GW200129_065458}{0.00}{GW191216_213338}{-0.20}}[{\red{???}}]}
\DeclareRobustCommand{\decgwtcthreenintiethpercentile}[1]{\IfEqCase{#1}{{GW200224_222234}{-0.099}{GW191129_134029}{0.63}{GW200311_115853}{-0.071}{GW191230_180458}{0.07}{GW191222_033537}{0.72}{GW200225_060421}{1.39}{GW200302_015811}{0.96}{GW200128_022011}{0.86}{GW191204_171526}{0.08}{GW200112_155838}{0.80}{GW200105_162426}{0.71}{GW191105_143521}{0.92}{GW191109_010717}{0.14}{GW200209_085452}{1.33}{GW200115_042309}{0.15}{GW191127_050227}{1.46}{GW200216_220804}{1.05}{GW191215_223052}{0.49}{GW200208_130117}{-0.544}{GW200219_094415}{0.72}{GW191103_012549}{1.32}{GW200316_215756}{0.866}{GW200202_154313}{0.48}{GW200129_065458}{0.40}{GW191216_213338}{0.79}}[{\red{???}}]}
\DeclareRobustCommand{\symmetricmassratiogwtcthreeminus}[1]{\IfEqCase{#1}{{GW200224_222234}{0.017}{GW191129_134029}{0.048}{GW200311_115853}{0.019}{GW191230_180458}{0.030}{GW191222_033537}{0.028}{GW200225_060421}{0.030}{GW200302_015811}{0.041}{GW200128_022011}{0.024}{GW191204_171526}{0.031}{GW200112_155838}{0.019}{GW200105_162426}{0.028}{GW191105_143521}{0.037}{GW191109_010717}{0.023}{GW200209_085452}{0.025}{GW200115_042309}{0.046}{GW191127_050227}{0.126}{GW200216_220804}{0.093}{GW191215_223052}{0.029}{GW200208_130117}{0.032}{GW200219_094415}{0.032}{GW191103_012549}{0.061}{GW200316_215756}{0.089}{GW200202_154313}{0.037}{GW200129_065458}{0.035}{GW191216_213338}{0.046}}[{\red{???}}]}
\DeclareRobustCommand{\symmetricmassratiogwtcthreemed}[1]{\IfEqCase{#1}{{GW200224_222234}{0.248}{GW191129_134029}{0.237}{GW200311_115853}{0.248}{GW191230_180458}{0.246}{GW191222_033537}{0.247}{GW200225_060421}{0.244}{GW200302_015811}{0.229}{GW200128_022011}{0.247}{GW191204_171526}{0.242}{GW200112_155838}{0.247}{GW200105_162426}{0.144}{GW191105_143521}{0.243}{GW191109_010717}{0.244}{GW200209_085452}{0.247}{GW200115_042309}{0.157}{GW191127_050227}{0.217}{GW200216_220804}{0.235}{GW191215_223052}{0.244}{GW200208_130117}{0.244}{GW200219_094415}{0.246}{GW191103_012549}{0.240}{GW200316_215756}{0.234}{GW200202_154313}{0.244}{GW200129_065458}{0.248}{GW191216_213338}{0.238}}[{\red{???}}]}
\DeclareRobustCommand{\symmetricmassratiogwtcthreeplus}[1]{\IfEqCase{#1}{{GW200224_222234}{0.002}{GW191129_134029}{0.012}{GW200311_115853}{0.002}{GW191230_180458}{0.004}{GW191222_033537}{0.003}{GW200225_060421}{0.006}{GW200302_015811}{0.021}{GW200128_022011}{0.003}{GW191204_171526}{0.008}{GW200112_155838}{0.003}{GW200105_162426}{0.036}{GW191105_143521}{0.006}{GW191109_010717}{0.006}{GW200209_085452}{0.003}{GW200115_042309}{0.083}{GW191127_050227}{0.033}{GW200216_220804}{0.015}{GW191215_223052}{0.006}{GW200208_130117}{0.006}{GW200219_094415}{0.004}{GW191103_012549}{0.009}{GW200316_215756}{0.016}{GW200202_154313}{0.006}{GW200129_065458}{0.002}{GW191216_213338}{0.012}}[{\red{???}}]}
\DeclareRobustCommand{\symmetricmassratiogwtcthreetenthpercentile}[1]{\IfEqCase{#1}{{GW200224_222234}{0.236}{GW191129_134029}{0.202}{GW200311_115853}{0.235}{GW191230_180458}{0.226}{GW191222_033537}{0.228}{GW200225_060421}{0.223}{GW200302_015811}{0.197}{GW200128_022011}{0.230}{GW191204_171526}{0.220}{GW200112_155838}{0.234}{GW200105_162426}{0.128}{GW191105_143521}{0.217}{GW191109_010717}{0.227}{GW200209_085452}{0.230}{GW200115_042309}{0.121}{GW191127_050227}{0.115}{GW200216_220804}{0.161}{GW191215_223052}{0.224}{GW200208_130117}{0.221}{GW200219_094415}{0.224}{GW191103_012549}{0.198}{GW200316_215756}{0.171}{GW200202_154313}{0.218}{GW200129_065458}{0.222}{GW191216_213338}{0.207}}[{\red{???}}]}
\DeclareRobustCommand{\symmetricmassratiogwtcthreenintiethpercentile}[1]{\IfEqCase{#1}{{GW200224_222234}{0.250}{GW191129_134029}{0.249}{GW200311_115853}{0.250}{GW191230_180458}{0.250}{GW191222_033537}{0.250}{GW200225_060421}{0.250}{GW200302_015811}{0.248}{GW200128_022011}{0.250}{GW191204_171526}{0.249}{GW200112_155838}{0.250}{GW200105_162426}{0.163}{GW191105_143521}{0.250}{GW191109_010717}{0.249}{GW200209_085452}{0.250}{GW200115_042309}{0.231}{GW191127_050227}{0.249}{GW200216_220804}{0.250}{GW191215_223052}{0.250}{GW200208_130117}{0.250}{GW200219_094415}{0.250}{GW191103_012549}{0.250}{GW200316_215756}{0.249}{GW200202_154313}{0.250}{GW200129_065458}{0.250}{GW191216_213338}{0.249}}[{\red{???}}]}
\DeclareRobustCommand{\masstwosourcegwtcthreeminus}[1]{\IfEqCase{#1}{{GW200224_222234}{7.0}{GW191129_134029}{1.7}{GW200311_115853}{5.9}{GW191230_180458}{10.2}{GW191222_033537}{10.5}{GW200225_060421}{3.5}{GW200302_015811}{5.8}{GW200128_022011}{8.0}{GW191204_171526}{1.6}{GW200112_155838}{5.9}{GW200105_162426}{0.24}{GW191105_143521}{1.9}{GW191109_010717}{13}{GW200209_085452}{7.0}{GW200115_042309}{0.28}{GW191127_050227}{14}{GW200216_220804}{16}{GW191215_223052}{3.8}{GW200208_130117}{7.4}{GW200219_094415}{8.4}{GW191103_012549}{2.4}{GW200316_215756}{2.9}{GW200202_154313}{1.7}{GW200129_065458}{9.3}{GW191216_213338}{1.9}}[{\red{???}}]}
\DeclareRobustCommand{\masstwosourcegwtcthreemed}[1]{\IfEqCase{#1}{{GW200224_222234}{32.4}{GW191129_134029}{6.7}{GW200311_115853}{27.7}{GW191230_180458}{36.0}{GW191222_033537}{34.7}{GW200225_060421}{14.0}{GW200302_015811}{20.3}{GW200128_022011}{31.1}{GW191204_171526}{8.2}{GW200112_155838}{28.3}{GW200105_162426}{1.91}{GW191105_143521}{7.7}{GW191109_010717}{47}{GW200209_085452}{26.6}{GW200115_042309}{1.44}{GW191127_050227}{24}{GW200216_220804}{30}{GW191215_223052}{17.8}{GW200208_130117}{27.4}{GW200219_094415}{27.9}{GW191103_012549}{7.9}{GW200316_215756}{7.8}{GW200202_154313}{7.3}{GW200129_065458}{28.9}{GW191216_213338}{7.7}}[{\red{???}}]}
\DeclareRobustCommand{\masstwosourcegwtcthreeplus}[1]{\IfEqCase{#1}{{GW200224_222234}{4.7}{GW191129_134029}{1.5}{GW200311_115853}{4.1}{GW191230_180458}{9.7}{GW191222_033537}{9.3}{GW200225_060421}{2.8}{GW200302_015811}{8.0}{GW200128_022011}{8.7}{GW191204_171526}{1.4}{GW200112_155838}{4.4}{GW200105_162426}{0.33}{GW191105_143521}{1.4}{GW191109_010717}{15}{GW200209_085452}{7.0}{GW200115_042309}{0.85}{GW191127_050227}{17}{GW200216_220804}{14}{GW191215_223052}{3.7}{GW200208_130117}{6.1}{GW200219_094415}{7.4}{GW191103_012549}{1.7}{GW200316_215756}{1.9}{GW200202_154313}{1.1}{GW200129_065458}{3.4}{GW191216_213338}{1.6}}[{\red{???}}]}
\DeclareRobustCommand{\masstwosourcegwtcthreetenthpercentile}[1]{\IfEqCase{#1}{{GW200224_222234}{27.1}{GW191129_134029}{5.3}{GW200311_115853}{23.2}{GW191230_180458}{28.0}{GW191222_033537}{26.6}{GW200225_060421}{11.2}{GW200302_015811}{15.5}{GW200128_022011}{24.8}{GW191204_171526}{6.9}{GW200112_155838}{23.7}{GW200105_162426}{1.77}{GW191105_143521}{6.2}{GW191109_010717}{36}{GW200209_085452}{21.3}{GW200115_042309}{1.21}{GW191127_050227}{12}{GW200216_220804}{16}{GW191215_223052}{14.7}{GW200208_130117}{21.5}{GW200219_094415}{21.6}{GW191103_012549}{6.0}{GW200316_215756}{5.6}{GW200202_154313}{6.0}{GW200129_065458}{21.3}{GW191216_213338}{6.3}}[{\red{???}}]}
\DeclareRobustCommand{\masstwosourcegwtcthreenintiethpercentile}[1]{\IfEqCase{#1}{{GW200224_222234}{36.1}{GW191129_134029}{8.0}{GW200311_115853}{31.0}{GW191230_180458}{43.4}{GW191222_033537}{42.1}{GW200225_060421}{16.3}{GW200302_015811}{26.6}{GW200128_022011}{37.8}{GW191204_171526}{9.4}{GW200112_155838}{31.8}{GW200105_162426}{2.08}{GW191105_143521}{8.8}{GW191109_010717}{58}{GW200209_085452}{32.0}{GW200115_042309}{2.12}{GW191127_050227}{37}{GW200216_220804}{41}{GW191215_223052}{20.7}{GW200208_130117}{32.3}{GW200219_094415}{33.8}{GW191103_012549}{9.3}{GW200316_215756}{9.4}{GW200202_154313}{8.3}{GW200129_065458}{31.7}{GW191216_213338}{9.1}}[{\red{???}}]}
\DeclareRobustCommand{\iotagwtcthreeminus}[1]{\IfEqCase{#1}{{GW200224_222234}{0.42}{GW191129_134029}{1.5}{GW200311_115853}{0.39}{GW191230_180458}{1.92}{GW191222_033537}{1.3}{GW200225_060421}{1.0}{GW200302_015811}{0.99}{GW200128_022011}{1.1}{GW191204_171526}{2.03}{GW200112_155838}{0.69}{GW200105_162426}{1.2}{GW191105_143521}{0.86}{GW191109_010717}{1.41}{GW200209_085452}{1.5}{GW200115_042309}{0.46}{GW191127_050227}{1.2}{GW200216_220804}{0.73}{GW191215_223052}{0.91}{GW200208_130117}{0.62}{GW200219_094415}{0.96}{GW191103_012549}{1.1}{GW200316_215756}{1.84}{GW200202_154313}{0.59}{GW200129_065458}{0.50}{GW191216_213338}{0.80}}[{\red{???}}]}
\DeclareRobustCommand{\iotagwtcthreemed}[1]{\IfEqCase{#1}{{GW200224_222234}{0.58}{GW191129_134029}{1.8}{GW200311_115853}{0.54}{GW191230_180458}{2.17}{GW191222_033537}{1.6}{GW200225_060421}{1.4}{GW200302_015811}{1.27}{GW200128_022011}{1.4}{GW191204_171526}{2.28}{GW200112_155838}{0.90}{GW200105_162426}{1.5}{GW191105_143521}{1.08}{GW191109_010717}{1.98}{GW200209_085452}{1.7}{GW200115_042309}{0.63}{GW191127_050227}{1.5}{GW200216_220804}{0.96}{GW191215_223052}{1.21}{GW200208_130117}{2.52}{GW200219_094415}{1.23}{GW191103_012549}{1.4}{GW200316_215756}{2.32}{GW200202_154313}{2.57}{GW200129_065458}{0.65}{GW191216_213338}{2.50}}[{\red{???}}]}
\DeclareRobustCommand{\iotagwtcthreeplus}[1]{\IfEqCase{#1}{{GW200224_222234}{0.57}{GW191129_134029}{1.1}{GW200311_115853}{0.51}{GW191230_180458}{0.76}{GW191222_033537}{1.2}{GW200225_060421}{1.4}{GW200302_015811}{1.55}{GW200128_022011}{1.5}{GW191204_171526}{0.66}{GW200112_155838}{2.00}{GW200105_162426}{1.3}{GW191105_143521}{1.80}{GW191109_010717}{0.85}{GW200209_085452}{1.2}{GW200115_042309}{1.92}{GW191127_050227}{1.3}{GW200216_220804}{1.79}{GW191215_223052}{1.55}{GW200208_130117}{0.45}{GW200219_094415}{1.53}{GW191103_012549}{1.5}{GW200316_215756}{0.59}{GW200202_154313}{0.42}{GW200129_065458}{0.62}{GW191216_213338}{0.46}}[{\red{???}}]}
\DeclareRobustCommand{\iotagwtcthreetenthpercentile}[1]{\IfEqCase{#1}{{GW200224_222234}{0.22}{GW191129_134029}{0.4}{GW200311_115853}{0.21}{GW191230_180458}{0.38}{GW191222_033537}{0.4}{GW200225_060421}{0.5}{GW200302_015811}{0.40}{GW200128_022011}{0.4}{GW191204_171526}{0.36}{GW200112_155838}{0.29}{GW200105_162426}{0.5}{GW191105_143521}{0.32}{GW191109_010717}{0.83}{GW200209_085452}{0.4}{GW200115_042309}{0.25}{GW191127_050227}{0.5}{GW200216_220804}{0.34}{GW191215_223052}{0.43}{GW200208_130117}{2.05}{GW200219_094415}{0.40}{GW191103_012549}{0.3}{GW200316_215756}{0.72}{GW200202_154313}{2.09}{GW200129_065458}{0.23}{GW191216_213338}{2.00}}[{\red{???}}]}
\DeclareRobustCommand{\iotagwtcthreenintiethpercentile}[1]{\IfEqCase{#1}{{GW200224_222234}{1.03}{GW191129_134029}{2.8}{GW200311_115853}{0.96}{GW191230_180458}{2.83}{GW191222_033537}{2.7}{GW200225_060421}{2.6}{GW200302_015811}{2.67}{GW200128_022011}{2.7}{GW191204_171526}{2.85}{GW200112_155838}{2.81}{GW200105_162426}{2.7}{GW191105_143521}{2.76}{GW191109_010717}{2.69}{GW200209_085452}{2.8}{GW200115_042309}{2.12}{GW191127_050227}{2.6}{GW200216_220804}{2.53}{GW191215_223052}{2.59}{GW200208_130117}{2.89}{GW200219_094415}{2.58}{GW191103_012549}{2.8}{GW200316_215756}{2.81}{GW200202_154313}{2.92}{GW200129_065458}{1.12}{GW191216_213338}{2.88}}[{\red{???}}]}
\DeclareRobustCommand{\costiltonegwtcthreeminus}[1]{\IfEqCase{#1}{{GW200224_222234}{0.96}{GW191129_134029}{0.95}{GW200311_115853}{0.78}{GW191230_180458}{0.83}{GW191222_033537}{0.73}{GW200225_060421}{0.59}{GW200302_015811}{0.90}{GW200128_022011}{0.98}{GW191204_171526}{0.88}{GW200112_155838}{0.91}{GW200105_162426}{0.93}{GW191105_143521}{0.78}{GW191109_010717}{0.36}{GW200209_085452}{0.70}{GW200115_042309}{0.35}{GW191127_050227}{1.00}{GW200216_220804}{1.02}{GW191215_223052}{0.72}{GW200208_130117}{0.68}{GW200219_094415}{0.69}{GW191103_012549}{0.94}{GW200316_215756}{1.02}{GW200202_154313}{0.95}{GW200129_065458}{0.92}{GW191216_213338}{1.13}}[{\red{???}}]}
\DeclareRobustCommand{\costiltonegwtcthreemed}[1]{\IfEqCase{#1}{{GW200224_222234}{0.30}{GW191129_134029}{0.27}{GW200311_115853}{-0.09}{GW191230_180458}{-0.04}{GW191222_033537}{-0.16}{GW200225_060421}{-0.30}{GW200302_015811}{0.06}{GW200128_022011}{0.36}{GW191204_171526}{0.43}{GW200112_155838}{0.14}{GW200105_162426}{0.01}{GW191105_143521}{-0.07}{GW191109_010717}{-0.61}{GW200209_085452}{-0.19}{GW200115_042309}{-0.62}{GW191127_050227}{0.37}{GW200216_220804}{0.24}{GW191215_223052}{-0.06}{GW200208_130117}{-0.24}{GW200219_094415}{-0.23}{GW191103_012549}{0.54}{GW200316_215756}{0.47}{GW200202_154313}{0.18}{GW200129_065458}{0.13}{GW191216_213338}{0.54}}[{\red{???}}]}
\DeclareRobustCommand{\costiltonegwtcthreeplus}[1]{\IfEqCase{#1}{{GW200224_222234}{0.61}{GW191129_134029}{0.64}{GW200311_115853}{0.89}{GW191230_180458}{0.86}{GW191222_033537}{0.97}{GW200225_060421}{0.86}{GW200302_015811}{0.80}{GW200128_022011}{0.56}{GW191204_171526}{0.49}{GW200112_155838}{0.75}{GW200105_162426}{0.90}{GW191105_143521}{0.90}{GW191109_010717}{0.85}{GW200209_085452}{0.92}{GW200115_042309}{1.26}{GW191127_050227}{0.56}{GW200216_220804}{0.68}{GW191215_223052}{0.85}{GW200208_130117}{1.03}{GW200219_094415}{0.96}{GW191103_012549}{0.41}{GW200316_215756}{0.49}{GW200202_154313}{0.72}{GW200129_065458}{0.74}{GW191216_213338}{0.43}}[{\red{???}}]}
\DeclareRobustCommand{\costiltonegwtcthreetenthpercentile}[1]{\IfEqCase{#1}{{GW200224_222234}{-0.44}{GW191129_134029}{-0.45}{GW200311_115853}{-0.75}{GW191230_180458}{-0.73}{GW191222_033537}{-0.80}{GW200225_060421}{-0.81}{GW200302_015811}{-0.70}{GW200128_022011}{-0.38}{GW191204_171526}{-0.19}{GW200112_155838}{-0.60}{GW200105_162426}{-0.82}{GW191105_143521}{-0.72}{GW191109_010717}{-0.93}{GW200209_085452}{-0.79}{GW200115_042309}{-0.94}{GW191127_050227}{-0.38}{GW200216_220804}{-0.62}{GW191215_223052}{-0.64}{GW200208_130117}{-0.85}{GW200219_094415}{-0.83}{GW191103_012549}{-0.12}{GW200316_215756}{-0.29}{GW200202_154313}{-0.59}{GW200129_065458}{-0.57}{GW191216_213338}{-0.32}}[{\red{???}}]}
\DeclareRobustCommand{\costiltonegwtcthreenintiethpercentile}[1]{\IfEqCase{#1}{{GW200224_222234}{0.82}{GW191129_134029}{0.83}{GW200311_115853}{0.63}{GW191230_180458}{0.69}{GW191222_033537}{0.64}{GW200225_060421}{0.32}{GW200302_015811}{0.73}{GW200128_022011}{0.83}{GW191204_171526}{0.85}{GW200112_155838}{0.78}{GW200105_162426}{0.82}{GW191105_143521}{0.68}{GW191109_010717}{0.02}{GW200209_085452}{0.55}{GW200115_042309}{0.41}{GW191127_050227}{0.88}{GW200216_220804}{0.84}{GW191215_223052}{0.61}{GW200208_130117}{0.61}{GW200219_094415}{0.56}{GW191103_012549}{0.90}{GW200316_215756}{0.92}{GW200202_154313}{0.80}{GW200129_065458}{0.76}{GW191216_213338}{0.92}}[{\red{???}}]}
\DeclareRobustCommand{\phionetwogwtcthreeminus}[1]{\IfEqCase{#1}{{GW200224_222234}{2.7}{GW191129_134029}{2.8}{GW200311_115853}{2.8}{GW191230_180458}{2.9}{GW191222_033537}{2.8}{GW200225_060421}{2.8}{GW200302_015811}{2.8}{GW200128_022011}{3.0}{GW191204_171526}{2.8}{GW200112_155838}{2.7}{GW200105_162426}{2.8}{GW191105_143521}{2.8}{GW191109_010717}{3.5}{GW200209_085452}{2.8}{GW200115_042309}{2.8}{GW191127_050227}{2.8}{GW200216_220804}{2.9}{GW191215_223052}{2.9}{GW200208_130117}{2.7}{GW200219_094415}{2.9}{GW191103_012549}{2.8}{GW200316_215756}{2.8}{GW200202_154313}{2.9}{GW200129_065458}{2.6}{GW191216_213338}{2.8}}[{\red{???}}]}
\DeclareRobustCommand{\phionetwogwtcthreemed}[1]{\IfEqCase{#1}{{GW200224_222234}{3.1}{GW191129_134029}{3.1}{GW200311_115853}{3.1}{GW191230_180458}{3.1}{GW191222_033537}{3.2}{GW200225_060421}{3.2}{GW200302_015811}{3.1}{GW200128_022011}{3.3}{GW191204_171526}{3.1}{GW200112_155838}{3.1}{GW200105_162426}{3.2}{GW191105_143521}{3.2}{GW191109_010717}{3.7}{GW200209_085452}{3.1}{GW200115_042309}{3.2}{GW191127_050227}{3.2}{GW200216_220804}{3.2}{GW191215_223052}{3.2}{GW200208_130117}{3.0}{GW200219_094415}{3.2}{GW191103_012549}{3.1}{GW200316_215756}{3.1}{GW200202_154313}{3.2}{GW200129_065458}{3.0}{GW191216_213338}{3.2}}[{\red{???}}]}
\DeclareRobustCommand{\phionetwogwtcthreeplus}[1]{\IfEqCase{#1}{{GW200224_222234}{2.8}{GW191129_134029}{2.8}{GW200311_115853}{2.8}{GW191230_180458}{2.9}{GW191222_033537}{2.8}{GW200225_060421}{2.8}{GW200302_015811}{2.8}{GW200128_022011}{2.7}{GW191204_171526}{2.8}{GW200112_155838}{2.9}{GW200105_162426}{2.8}{GW191105_143521}{2.8}{GW191109_010717}{2.3}{GW200209_085452}{2.9}{GW200115_042309}{2.8}{GW191127_050227}{2.8}{GW200216_220804}{2.8}{GW191215_223052}{2.8}{GW200208_130117}{2.9}{GW200219_094415}{2.8}{GW191103_012549}{2.9}{GW200316_215756}{2.8}{GW200202_154313}{2.8}{GW200129_065458}{2.9}{GW191216_213338}{2.8}}[{\red{???}}]}
\DeclareRobustCommand{\phionetwogwtcthreetenthpercentile}[1]{\IfEqCase{#1}{{GW200224_222234}{0.7}{GW191129_134029}{0.7}{GW200311_115853}{0.7}{GW191230_180458}{0.6}{GW191222_033537}{0.7}{GW200225_060421}{0.7}{GW200302_015811}{0.7}{GW200128_022011}{0.6}{GW191204_171526}{0.7}{GW200112_155838}{0.7}{GW200105_162426}{0.7}{GW191105_143521}{0.7}{GW191109_010717}{0.5}{GW200209_085452}{0.5}{GW200115_042309}{0.7}{GW191127_050227}{0.6}{GW200216_220804}{0.6}{GW191215_223052}{0.6}{GW200208_130117}{0.6}{GW200219_094415}{0.6}{GW191103_012549}{0.6}{GW200316_215756}{0.6}{GW200202_154313}{0.7}{GW200129_065458}{0.6}{GW191216_213338}{0.7}}[{\red{???}}]}
\DeclareRobustCommand{\phionetwogwtcthreenintiethpercentile}[1]{\IfEqCase{#1}{{GW200224_222234}{5.6}{GW191129_134029}{5.6}{GW200311_115853}{5.6}{GW191230_180458}{5.7}{GW191222_033537}{5.6}{GW200225_060421}{5.6}{GW200302_015811}{5.6}{GW200128_022011}{5.7}{GW191204_171526}{5.6}{GW200112_155838}{5.6}{GW200105_162426}{5.6}{GW191105_143521}{5.6}{GW191109_010717}{5.8}{GW200209_085452}{5.7}{GW200115_042309}{5.6}{GW191127_050227}{5.7}{GW200216_220804}{5.6}{GW191215_223052}{5.6}{GW200208_130117}{5.6}{GW200219_094415}{5.7}{GW191103_012549}{5.6}{GW200316_215756}{5.6}{GW200202_154313}{5.6}{GW200129_065458}{5.6}{GW191216_213338}{5.6}}[{\red{???}}]}
\DeclareRobustCommand{\massratiogwtcthreeminus}[1]{\IfEqCase{#1}{{GW200224_222234}{0.26}{GW191129_134029}{0.29}{GW200311_115853}{0.27}{GW191230_180458}{0.32}{GW191222_033537}{0.32}{GW200225_060421}{0.28}{GW200302_015811}{0.21}{GW200128_022011}{0.29}{GW191204_171526}{0.26}{GW200112_155838}{0.26}{GW200105_162426}{0.056}{GW191105_143521}{0.31}{GW191109_010717}{0.24}{GW200209_085452}{0.30}{GW200115_042309}{0.097}{GW191127_050227}{0.35}{GW200216_220804}{0.40}{GW191215_223052}{0.27}{GW200208_130117}{0.29}{GW200219_094415}{0.32}{GW191103_012549}{0.37}{GW200316_215756}{0.38}{GW200202_154313}{0.31}{GW200129_065458}{0.41}{GW191216_213338}{0.29}}[{\red{???}}]}
\DeclareRobustCommand{\massratiogwtcthreemed}[1]{\IfEqCase{#1}{{GW200224_222234}{0.82}{GW191129_134029}{0.63}{GW200311_115853}{0.82}{GW191230_180458}{0.78}{GW191222_033537}{0.79}{GW200225_060421}{0.73}{GW200302_015811}{0.55}{GW200128_022011}{0.79}{GW191204_171526}{0.69}{GW200112_155838}{0.81}{GW200105_162426}{0.211}{GW191105_143521}{0.72}{GW191109_010717}{0.73}{GW200209_085452}{0.79}{GW200115_042309}{0.243}{GW191127_050227}{0.47}{GW200216_220804}{0.61}{GW191215_223052}{0.73}{GW200208_130117}{0.73}{GW200219_094415}{0.77}{GW191103_012549}{0.67}{GW200316_215756}{0.59}{GW200202_154313}{0.72}{GW200129_065458}{0.85}{GW191216_213338}{0.63}}[{\red{???}}]}
\DeclareRobustCommand{\massratiogwtcthreeplus}[1]{\IfEqCase{#1}{{GW200224_222234}{0.16}{GW191129_134029}{0.31}{GW200311_115853}{0.16}{GW191230_180458}{0.20}{GW191222_033537}{0.18}{GW200225_060421}{0.23}{GW200302_015811}{0.36}{GW200128_022011}{0.18}{GW191204_171526}{0.25}{GW200112_155838}{0.17}{GW200105_162426}{0.095}{GW191105_143521}{0.24}{GW191109_010717}{0.21}{GW200209_085452}{0.19}{GW200115_042309}{0.432}{GW191127_050227}{0.47}{GW200216_220804}{0.35}{GW191215_223052}{0.24}{GW200208_130117}{0.23}{GW200219_094415}{0.21}{GW191103_012549}{0.29}{GW200316_215756}{0.34}{GW200202_154313}{0.24}{GW200129_065458}{0.12}{GW191216_213338}{0.31}}[{\red{???}}]}
\DeclareRobustCommand{\massratiogwtcthreetenthpercentile}[1]{\IfEqCase{#1}{{GW200224_222234}{0.62}{GW191129_134029}{0.39}{GW200311_115853}{0.61}{GW191230_180458}{0.52}{GW191222_033537}{0.54}{GW200225_060421}{0.50}{GW200302_015811}{0.37}{GW200128_022011}{0.56}{GW191204_171526}{0.48}{GW200112_155838}{0.60}{GW200105_162426}{0.177}{GW191105_143521}{0.47}{GW191109_010717}{0.54}{GW200209_085452}{0.56}{GW200115_042309}{0.164}{GW191127_050227}{0.15}{GW200216_220804}{0.25}{GW191215_223052}{0.51}{GW200208_130117}{0.49}{GW200219_094415}{0.51}{GW191103_012549}{0.37}{GW200316_215756}{0.28}{GW200202_154313}{0.48}{GW200129_065458}{0.50}{GW191216_213338}{0.41}}[{\red{???}}]}
\DeclareRobustCommand{\massratiogwtcthreenintiethpercentile}[1]{\IfEqCase{#1}{{GW200224_222234}{0.96}{GW191129_134029}{0.91}{GW200311_115853}{0.96}{GW191230_180458}{0.95}{GW191222_033537}{0.96}{GW200225_060421}{0.93}{GW200302_015811}{0.84}{GW200128_022011}{0.96}{GW191204_171526}{0.91}{GW200112_155838}{0.96}{GW200105_162426}{0.259}{GW191105_143521}{0.93}{GW191109_010717}{0.91}{GW200209_085452}{0.96}{GW200115_042309}{0.571}{GW191127_050227}{0.88}{GW200216_220804}{0.92}{GW191215_223052}{0.94}{GW200208_130117}{0.94}{GW200219_094415}{0.95}{GW191103_012549}{0.92}{GW200316_215756}{0.89}{GW200202_154313}{0.94}{GW200129_065458}{0.96}{GW191216_213338}{0.90}}[{\red{???}}]}
\DeclareRobustCommand{\comovingdistgwtcthreeminus}[1]{\IfEqCase{#1}{{GW200224_222234}{400}{GW191129_134029}{260}{GW200311_115853}{280}{GW191230_180458}{850}{GW191222_033537}{920}{GW200225_060421}{390}{GW200302_015811}{510}{GW200128_022011}{970}{GW191204_171526}{200}{GW200112_155838}{330}{GW200105_162426}{100}{GW191105_143521}{350}{GW191109_010717}{470}{GW200209_085452}{840}{GW200115_042309}{92}{GW191127_050227}{1000}{GW200216_220804}{1000}{GW191215_223052}{560}{GW200208_130117}{500}{GW200219_094415}{750}{GW191103_012549}{360}{GW200316_215756}{320}{GW200202_154313}{140}{GW200129_065458}{290}{GW191216_213338}{120}}[{\red{???}}]}
\DeclareRobustCommand{\comovingdistgwtcthreemed}[1]{\IfEqCase{#1}{{GW200224_222234}{1340}{GW191129_134029}{700}{GW200311_115853}{950}{GW191230_180458}{2740}{GW191222_033537}{1980}{GW200225_060421}{940}{GW200302_015811}{1270}{GW200128_022011}{2440}{GW191204_171526}{580}{GW200112_155838}{1010}{GW200105_162426}{260}{GW191105_143521}{940}{GW191109_010717}{1040}{GW200209_085452}{2390}{GW200115_042309}{274}{GW191127_050227}{2200}{GW200216_220804}{2350}{GW191215_223052}{1520}{GW200208_130117}{1590}{GW200219_094415}{2180}{GW191103_012549}{830}{GW200316_215756}{920}{GW200202_154313}{380}{GW200129_065458}{760}{GW191216_213338}{320}}[{\red{???}}]}
\DeclareRobustCommand{\comovingdistgwtcthreeplus}[1]{\IfEqCase{#1}{{GW200224_222234}{270}{GW191129_134029}{180}{GW200311_115853}{180}{GW191230_180458}{690}{GW191222_033537}{710}{GW200225_060421}{330}{GW200302_015811}{590}{GW200128_022011}{750}{GW191204_171526}{140}{GW200112_155838}{270}{GW200105_162426}{100}{GW191105_143521}{280}{GW191109_010717}{660}{GW200209_085452}{720}{GW200115_042309}{126}{GW191127_050227}{1100}{GW200216_220804}{1040}{GW191215_223052}{440}{GW200208_130117}{500}{GW200219_094415}{680}{GW191103_012549}{340}{GW200316_215756}{310}{GW200202_154313}{120}{GW200129_065458}{200}{GW191216_213338}{100}}[{\red{???}}]}
\DeclareRobustCommand{\comovingdistgwtcthreetenthpercentile}[1]{\IfEqCase{#1}{{GW200224_222234}{1030}{GW191129_134029}{490}{GW200311_115853}{740}{GW191230_180458}{2080}{GW191222_033537}{1260}{GW200225_060421}{640}{GW200302_015811}{860}{GW200128_022011}{1690}{GW191204_171526}{420}{GW200112_155838}{750}{GW200105_162426}{170}{GW191105_143521}{660}{GW191109_010717}{660}{GW200209_085452}{1750}{GW200115_042309}{202}{GW191127_050227}{1400}{GW200216_220804}{1560}{GW191215_223052}{1080}{GW200208_130117}{1190}{GW200219_094415}{1550}{GW191103_012549}{530}{GW200316_215756}{660}{GW200202_154313}{270}{GW200129_065458}{540}{GW191216_213338}{220}}[{\red{???}}]}
\DeclareRobustCommand{\comovingdistgwtcthreenintiethpercentile}[1]{\IfEqCase{#1}{{GW200224_222234}{1550}{GW191129_134029}{850}{GW200311_115853}{1100}{GW191230_180458}{3280}{GW191222_033537}{2550}{GW200225_060421}{1200}{GW200302_015811}{1730}{GW200128_022011}{3040}{GW191204_171526}{700}{GW200112_155838}{1230}{GW200105_162426}{340}{GW191105_143521}{1160}{GW191109_010717}{1540}{GW200209_085452}{2950}{GW200115_042309}{365}{GW191127_050227}{3100}{GW200216_220804}{3160}{GW191215_223052}{1880}{GW200208_130117}{1970}{GW200219_094415}{2710}{GW191103_012549}{1090}{GW200316_215756}{1160}{GW200202_154313}{480}{GW200129_065458}{930}{GW191216_213338}{400}}[{\red{???}}]}
\DeclareRobustCommand{\phasegwtcthreeminus}[1]{\IfEqCase{#1}{{GW200224_222234}{2.8}{GW191129_134029}{2.9}{GW200311_115853}{2.7}{GW191230_180458}{2.8}{GW191222_033537}{2.8}{GW200225_060421}{2.4}{GW200302_015811}{2.2}{GW200128_022011}{3.1}{GW191204_171526}{2.7}{GW200112_155838}{2.9}{GW200105_162426}{3.1}{GW191105_143521}{2.8}{GW191109_010717}{1.4}{GW200209_085452}{2.9}{GW200115_042309}{2.8}{GW191127_050227}{2.8}{GW200216_220804}{2.6}{GW191215_223052}{2.4}{GW200208_130117}{3.5}{GW200219_094415}{3.5}{GW191103_012549}{2.6}{GW200316_215756}{2.4}{GW200202_154313}{2.6}{GW200129_065458}{3.0}{GW191216_213338}{2.1}}[{\red{???}}]}
\DeclareRobustCommand{\phasegwtcthreemed}[1]{\IfEqCase{#1}{{GW200224_222234}{3.0}{GW191129_134029}{3.4}{GW200311_115853}{3.1}{GW191230_180458}{3.1}{GW191222_033537}{3.1}{GW200225_060421}{2.8}{GW200302_015811}{2.5}{GW200128_022011}{3.4}{GW191204_171526}{3.1}{GW200112_155838}{3.4}{GW200105_162426}{4.2}{GW191105_143521}{3.1}{GW191109_010717}{1.6}{GW200209_085452}{3.2}{GW200115_042309}{3.2}{GW191127_050227}{3.0}{GW200216_220804}{3.0}{GW191215_223052}{2.6}{GW200208_130117}{3.8}{GW200219_094415}{3.8}{GW191103_012549}{3.0}{GW200316_215756}{2.9}{GW200202_154313}{3.0}{GW200129_065458}{3.5}{GW191216_213338}{2.2}}[{\red{???}}]}
\DeclareRobustCommand{\phasegwtcthreeplus}[1]{\IfEqCase{#1}{{GW200224_222234}{3.0}{GW191129_134029}{2.5}{GW200311_115853}{2.8}{GW191230_180458}{2.9}{GW191222_033537}{3.0}{GW200225_060421}{3.2}{GW200302_015811}{3.4}{GW200128_022011}{2.7}{GW191204_171526}{2.8}{GW200112_155838}{2.5}{GW200105_162426}{1.4}{GW191105_143521}{2.9}{GW191109_010717}{4.3}{GW200209_085452}{2.8}{GW200115_042309}{2.8}{GW191127_050227}{2.9}{GW200216_220804}{3.0}{GW191215_223052}{3.4}{GW200208_130117}{2.2}{GW200219_094415}{2.2}{GW191103_012549}{3.0}{GW200316_215756}{2.8}{GW200202_154313}{2.9}{GW200129_065458}{2.4}{GW191216_213338}{3.9}}[{\red{???}}]}
\DeclareRobustCommand{\phasegwtcthreetenthpercentile}[1]{\IfEqCase{#1}{{GW200224_222234}{0.4}{GW191129_134029}{0.9}{GW200311_115853}{0.8}{GW191230_180458}{0.6}{GW191222_033537}{0.5}{GW200225_060421}{0.7}{GW200302_015811}{0.7}{GW200128_022011}{0.6}{GW191204_171526}{0.8}{GW200112_155838}{1.0}{GW200105_162426}{1.8}{GW191105_143521}{0.6}{GW191109_010717}{0.3}{GW200209_085452}{0.5}{GW200115_042309}{0.6}{GW191127_050227}{0.5}{GW200216_220804}{0.7}{GW191215_223052}{0.4}{GW200208_130117}{0.7}{GW200219_094415}{0.8}{GW191103_012549}{0.7}{GW200316_215756}{1.0}{GW200202_154313}{0.7}{GW200129_065458}{0.9}{GW191216_213338}{0.3}}[{\red{???}}]}
\DeclareRobustCommand{\phasegwtcthreenintiethpercentile}[1]{\IfEqCase{#1}{{GW200224_222234}{5.8}{GW191129_134029}{5.5}{GW200311_115853}{5.5}{GW191230_180458}{5.6}{GW191222_033537}{5.8}{GW200225_060421}{5.5}{GW200302_015811}{5.5}{GW200128_022011}{5.8}{GW191204_171526}{5.4}{GW200112_155838}{5.5}{GW200105_162426}{5.3}{GW191105_143521}{5.7}{GW191109_010717}{5.0}{GW200209_085452}{5.7}{GW200115_042309}{5.7}{GW191127_050227}{5.7}{GW200216_220804}{5.6}{GW191215_223052}{5.7}{GW200208_130117}{5.7}{GW200219_094415}{5.7}{GW191103_012549}{5.6}{GW200316_215756}{5.2}{GW200202_154313}{5.5}{GW200129_065458}{5.6}{GW191216_213338}{6.0}}[{\red{???}}]}
\DeclareRobustCommand{\phionegwtcthreeminus}[1]{\IfEqCase{#1}{{GW200224_222234}{2.9}{GW191129_134029}{2.9}{GW200311_115853}{2.7}{GW191230_180458}{2.8}{GW191222_033537}{2.8}{GW200225_060421}{2.9}{GW200302_015811}{2.8}{GW200128_022011}{2.8}{GW191204_171526}{2.8}{GW200112_155838}{2.8}{GW200105_162426}{2.9}{GW191105_143521}{2.8}{GW191109_010717}{2.7}{GW200209_085452}{2.8}{GW200115_042309}{3.0}{GW191127_050227}{2.8}{GW200216_220804}{2.9}{GW191215_223052}{2.8}{GW200208_130117}{2.8}{GW200219_094415}{2.9}{GW191103_012549}{2.8}{GW200316_215756}{3.0}{GW200202_154313}{2.9}{GW200129_065458}{2.8}{GW191216_213338}{2.7}}[{\red{???}}]}
\DeclareRobustCommand{\phionegwtcthreemed}[1]{\IfEqCase{#1}{{GW200224_222234}{3.1}{GW191129_134029}{3.2}{GW200311_115853}{3.1}{GW191230_180458}{3.2}{GW191222_033537}{3.1}{GW200225_060421}{3.2}{GW200302_015811}{3.2}{GW200128_022011}{3.1}{GW191204_171526}{3.1}{GW200112_155838}{3.1}{GW200105_162426}{3.2}{GW191105_143521}{3.1}{GW191109_010717}{3.1}{GW200209_085452}{3.2}{GW200115_042309}{3.3}{GW191127_050227}{3.1}{GW200216_220804}{3.2}{GW191215_223052}{3.1}{GW200208_130117}{3.0}{GW200219_094415}{3.2}{GW191103_012549}{3.1}{GW200316_215756}{3.3}{GW200202_154313}{3.2}{GW200129_065458}{3.1}{GW191216_213338}{3.0}}[{\red{???}}]}
\DeclareRobustCommand{\phionegwtcthreeplus}[1]{\IfEqCase{#1}{{GW200224_222234}{2.9}{GW191129_134029}{2.8}{GW200311_115853}{2.9}{GW191230_180458}{2.8}{GW191222_033537}{2.9}{GW200225_060421}{2.8}{GW200302_015811}{2.8}{GW200128_022011}{2.8}{GW191204_171526}{2.9}{GW200112_155838}{2.9}{GW200105_162426}{2.8}{GW191105_143521}{2.8}{GW191109_010717}{2.9}{GW200209_085452}{2.8}{GW200115_042309}{2.7}{GW191127_050227}{2.9}{GW200216_220804}{2.8}{GW191215_223052}{2.9}{GW200208_130117}{2.9}{GW200219_094415}{2.8}{GW191103_012549}{2.9}{GW200316_215756}{2.7}{GW200202_154313}{2.8}{GW200129_065458}{2.9}{GW191216_213338}{3.0}}[{\red{???}}]}
\DeclareRobustCommand{\phionegwtcthreetenthpercentile}[1]{\IfEqCase{#1}{{GW200224_222234}{0.6}{GW191129_134029}{0.6}{GW200311_115853}{0.7}{GW191230_180458}{0.6}{GW191222_033537}{0.6}{GW200225_060421}{0.6}{GW200302_015811}{0.6}{GW200128_022011}{0.7}{GW191204_171526}{0.7}{GW200112_155838}{0.6}{GW200105_162426}{0.6}{GW191105_143521}{0.6}{GW191109_010717}{0.6}{GW200209_085452}{0.6}{GW200115_042309}{0.6}{GW191127_050227}{0.7}{GW200216_220804}{0.7}{GW191215_223052}{0.6}{GW200208_130117}{0.6}{GW200219_094415}{0.6}{GW191103_012549}{0.7}{GW200316_215756}{0.6}{GW200202_154313}{0.6}{GW200129_065458}{0.6}{GW191216_213338}{0.6}}[{\red{???}}]}
\DeclareRobustCommand{\phionegwtcthreenintiethpercentile}[1]{\IfEqCase{#1}{{GW200224_222234}{5.7}{GW191129_134029}{5.7}{GW200311_115853}{5.6}{GW191230_180458}{5.6}{GW191222_033537}{5.6}{GW200225_060421}{5.7}{GW200302_015811}{5.7}{GW200128_022011}{5.6}{GW191204_171526}{5.6}{GW200112_155838}{5.6}{GW200105_162426}{5.6}{GW191105_143521}{5.7}{GW191109_010717}{5.6}{GW200209_085452}{5.7}{GW200115_042309}{5.7}{GW191127_050227}{5.7}{GW200216_220804}{5.6}{GW191215_223052}{5.7}{GW200208_130117}{5.7}{GW200219_094415}{5.7}{GW191103_012549}{5.6}{GW200316_215756}{5.7}{GW200202_154313}{5.7}{GW200129_065458}{5.7}{GW191216_213338}{5.6}}[{\red{???}}]}
\DeclareRobustCommand{\spintwogwtcthreeminus}[1]{\IfEqCase{#1}{{GW200224_222234}{0.39}{GW191129_134029}{0.31}{GW200311_115853}{0.37}{GW191230_180458}{0.44}{GW191222_033537}{0.38}{GW200225_060421}{0.39}{GW200302_015811}{0.41}{GW200128_022011}{0.45}{GW191204_171526}{0.40}{GW200112_155838}{0.34}{GW200105_162426}{0.30}{GW191105_143521}{0.31}{GW191109_010717}{0.58}{GW200209_085452}{0.44}{GW200115_042309}{0.39}{GW191127_050227}{0.49}{GW200216_220804}{0.46}{GW191215_223052}{0.40}{GW200208_130117}{0.39}{GW200219_094415}{0.43}{GW191103_012549}{0.44}{GW200316_215756}{0.39}{GW200202_154313}{0.30}{GW200129_065458}{0.42}{GW191216_213338}{0.32}}[{\red{???}}]}
\DeclareRobustCommand{\spintwogwtcthreemed}[1]{\IfEqCase{#1}{{GW200224_222234}{0.44}{GW191129_134029}{0.35}{GW200311_115853}{0.41}{GW191230_180458}{0.49}{GW191222_033537}{0.41}{GW200225_060421}{0.42}{GW200302_015811}{0.45}{GW200128_022011}{0.50}{GW191204_171526}{0.46}{GW200112_155838}{0.39}{GW200105_162426}{0.33}{GW191105_143521}{0.34}{GW191109_010717}{0.65}{GW200209_085452}{0.49}{GW200115_042309}{0.44}{GW191127_050227}{0.54}{GW200216_220804}{0.51}{GW191215_223052}{0.44}{GW200208_130117}{0.43}{GW200219_094415}{0.48}{GW191103_012549}{0.50}{GW200316_215756}{0.44}{GW200202_154313}{0.33}{GW200129_065458}{0.49}{GW191216_213338}{0.36}}[{\red{???}}]}
\DeclareRobustCommand{\spintwogwtcthreeplus}[1]{\IfEqCase{#1}{{GW200224_222234}{0.48}{GW191129_134029}{0.51}{GW200311_115853}{0.51}{GW191230_180458}{0.45}{GW191222_033537}{0.50}{GW200225_060421}{0.49}{GW200302_015811}{0.48}{GW200128_022011}{0.44}{GW191204_171526}{0.41}{GW200112_155838}{0.50}{GW200105_162426}{0.56}{GW191105_143521}{0.54}{GW191109_010717}{0.32}{GW200209_085452}{0.45}{GW200115_042309}{0.48}{GW191127_050227}{0.41}{GW200216_220804}{0.44}{GW191215_223052}{0.49}{GW200208_130117}{0.49}{GW200219_094415}{0.46}{GW191103_012549}{0.44}{GW200316_215756}{0.47}{GW200202_154313}{0.53}{GW200129_065458}{0.44}{GW191216_213338}{0.50}}[{\red{???}}]}
\DeclareRobustCommand{\spintwogwtcthreetenthpercentile}[1]{\IfEqCase{#1}{{GW200224_222234}{0.09}{GW191129_134029}{0.07}{GW200311_115853}{0.08}{GW191230_180458}{0.10}{GW191222_033537}{0.07}{GW200225_060421}{0.08}{GW200302_015811}{0.09}{GW200128_022011}{0.10}{GW191204_171526}{0.11}{GW200112_155838}{0.08}{GW200105_162426}{0.06}{GW191105_143521}{0.06}{GW191109_010717}{0.14}{GW200209_085452}{0.10}{GW200115_042309}{0.09}{GW191127_050227}{0.11}{GW200216_220804}{0.10}{GW191215_223052}{0.09}{GW200208_130117}{0.08}{GW200219_094415}{0.10}{GW191103_012549}{0.10}{GW200316_215756}{0.10}{GW200202_154313}{0.06}{GW200129_065458}{0.13}{GW191216_213338}{0.08}}[{\red{???}}]}
\DeclareRobustCommand{\spintwogwtcthreenintiethpercentile}[1]{\IfEqCase{#1}{{GW200224_222234}{0.85}{GW191129_134029}{0.76}{GW200311_115853}{0.84}{GW191230_180458}{0.89}{GW191222_033537}{0.84}{GW200225_060421}{0.85}{GW200302_015811}{0.87}{GW200128_022011}{0.89}{GW191204_171526}{0.79}{GW200112_155838}{0.81}{GW200105_162426}{0.79}{GW191105_143521}{0.78}{GW191109_010717}{0.94}{GW200209_085452}{0.90}{GW200115_042309}{0.84}{GW191127_050227}{0.91}{GW200216_220804}{0.90}{GW191215_223052}{0.86}{GW200208_130117}{0.85}{GW200219_094415}{0.88}{GW191103_012549}{0.88}{GW200316_215756}{0.83}{GW200202_154313}{0.77}{GW200129_065458}{0.86}{GW191216_213338}{0.76}}[{\red{???}}]}
\DeclareRobustCommand{\spinonezgwtcthreeminus}[1]{\IfEqCase{#1}{{GW200224_222234}{0.31}{GW191129_134029}{0.20}{GW200311_115853}{0.41}{GW191230_180458}{0.49}{GW191222_033537}{0.45}{GW200225_060421}{0.48}{GW200302_015811}{0.36}{GW200128_022011}{0.38}{GW191204_171526}{0.27}{GW200112_155838}{0.31}{GW200105_162426}{0.21}{GW191105_143521}{0.30}{GW191109_010717}{0.41}{GW200209_085452}{0.52}{GW200115_042309}{0.55}{GW191127_050227}{0.45}{GW200216_220804}{0.44}{GW191215_223052}{0.39}{GW200208_130117}{0.48}{GW200219_094415}{0.50}{GW191103_012549}{0.32}{GW200316_215756}{0.24}{GW200202_154313}{0.23}{GW200129_065458}{0.35}{GW191216_213338}{0.20}}[{\red{???}}]}
\DeclareRobustCommand{\spinonezgwtcthreemed}[1]{\IfEqCase{#1}{{GW200224_222234}{0.11}{GW191129_134029}{0.05}{GW200311_115853}{-0.02}{GW191230_180458}{-0.01}{GW191222_033537}{-0.03}{GW200225_060421}{-0.13}{GW200302_015811}{0.01}{GW200128_022011}{0.17}{GW191204_171526}{0.16}{GW200112_155838}{0.03}{GW200105_162426}{0.00}{GW191105_143521}{-0.01}{GW191109_010717}{-0.44}{GW200209_085452}{-0.06}{GW200115_042309}{-0.15}{GW191127_050227}{0.18}{GW200216_220804}{0.07}{GW191215_223052}{-0.02}{GW200208_130117}{-0.05}{GW200219_094415}{-0.07}{GW191103_012549}{0.24}{GW200316_215756}{0.12}{GW200202_154313}{0.02}{GW200129_065458}{0.05}{GW191216_213338}{0.11}}[{\red{???}}]}
\DeclareRobustCommand{\spinonezgwtcthreeplus}[1]{\IfEqCase{#1}{{GW200224_222234}{0.42}{GW191129_134029}{0.24}{GW200311_115853}{0.31}{GW191230_180458}{0.44}{GW191222_033537}{0.33}{GW200225_060421}{0.32}{GW200302_015811}{0.43}{GW200128_022011}{0.47}{GW191204_171526}{0.21}{GW200112_155838}{0.36}{GW200105_162426}{0.16}{GW191105_143521}{0.21}{GW191109_010717}{0.60}{GW200209_085452}{0.37}{GW200115_042309}{0.25}{GW191127_050227}{0.51}{GW200216_220804}{0.53}{GW191215_223052}{0.29}{GW200208_130117}{0.31}{GW200219_094415}{0.38}{GW191103_012549}{0.29}{GW200316_215756}{0.34}{GW200202_154313}{0.23}{GW200129_065458}{0.41}{GW191216_213338}{0.26}}[{\red{???}}]}
\DeclareRobustCommand{\spinonezgwtcthreetenthpercentile}[1]{\IfEqCase{#1}{{GW200224_222234}{-0.12}{GW191129_134029}{-0.08}{GW200311_115853}{-0.32}{GW191230_180458}{-0.37}{GW191222_033537}{-0.37}{GW200225_060421}{-0.52}{GW200302_015811}{-0.25}{GW200128_022011}{-0.12}{GW191204_171526}{-0.04}{GW200112_155838}{-0.18}{GW200105_162426}{-0.11}{GW191105_143521}{-0.22}{GW191109_010717}{-0.79}{GW200209_085452}{-0.46}{GW200115_042309}{-0.62}{GW191127_050227}{-0.15}{GW200216_220804}{-0.24}{GW191215_223052}{-0.30}{GW200208_130117}{-0.42}{GW200219_094415}{-0.45}{GW191103_012549}{-0.02}{GW200316_215756}{-0.05}{GW200202_154313}{-0.12}{GW200129_065458}{-0.22}{GW191216_213338}{-0.03}}[{\red{???}}]}
\DeclareRobustCommand{\spinonezgwtcthreenintiethpercentile}[1]{\IfEqCase{#1}{{GW200224_222234}{0.43}{GW191129_134029}{0.23}{GW200311_115853}{0.20}{GW191230_180458}{0.31}{GW191222_033537}{0.20}{GW200225_060421}{0.11}{GW200302_015811}{0.33}{GW200128_022011}{0.54}{GW191204_171526}{0.33}{GW200112_155838}{0.29}{GW200105_162426}{0.09}{GW191105_143521}{0.14}{GW191109_010717}{0.01}{GW200209_085452}{0.21}{GW200115_042309}{0.05}{GW191127_050227}{0.60}{GW200216_220804}{0.49}{GW191215_223052}{0.20}{GW200208_130117}{0.16}{GW200219_094415}{0.20}{GW191103_012549}{0.46}{GW200316_215756}{0.39}{GW200202_154313}{0.20}{GW200129_065458}{0.36}{GW191216_213338}{0.31}}[{\red{???}}]}
\DeclareRobustCommand{\spintwozgwtcthreeminus}[1]{\IfEqCase{#1}{{GW200224_222234}{0.44}{GW191129_134029}{0.30}{GW200311_115853}{0.45}{GW191230_180458}{0.56}{GW191222_033537}{0.50}{GW200225_060421}{0.55}{GW200302_015811}{0.47}{GW200128_022011}{0.47}{GW191204_171526}{0.35}{GW200112_155838}{0.35}{GW200105_162426}{0.42}{GW191105_143521}{0.34}{GW191109_010717}{0.58}{GW200209_085452}{0.58}{GW200115_042309}{0.59}{GW191127_050227}{0.52}{GW200216_220804}{0.52}{GW191215_223052}{0.50}{GW200208_130117}{0.52}{GW200219_094415}{0.57}{GW191103_012549}{0.42}{GW200316_215756}{0.35}{GW200202_154313}{0.28}{GW200129_065458}{0.50}{GW191216_213338}{0.37}}[{\red{???}}]}
\DeclareRobustCommand{\spintwozgwtcthreemed}[1]{\IfEqCase{#1}{{GW200224_222234}{0.05}{GW191129_134029}{0.07}{GW200311_115853}{0.00}{GW191230_180458}{-0.03}{GW191222_033537}{-0.02}{GW200225_060421}{-0.05}{GW200302_015811}{0.03}{GW200128_022011}{0.04}{GW191204_171526}{0.17}{GW200112_155838}{0.06}{GW200105_162426}{0.00}{GW191105_143521}{0.00}{GW191109_010717}{-0.10}{GW200209_085452}{-0.08}{GW200115_042309}{-0.08}{GW191127_050227}{0.06}{GW200216_220804}{0.05}{GW191215_223052}{-0.02}{GW200208_130117}{-0.03}{GW200219_094415}{-0.04}{GW191103_012549}{0.18}{GW200316_215756}{0.12}{GW200202_154313}{0.05}{GW200129_065458}{0.15}{GW191216_213338}{0.11}}[{\red{???}}]}
\DeclareRobustCommand{\spintwozgwtcthreeplus}[1]{\IfEqCase{#1}{{GW200224_222234}{0.45}{GW191129_134029}{0.50}{GW200311_115853}{0.41}{GW191230_180458}{0.46}{GW191222_033537}{0.42}{GW200225_060421}{0.40}{GW200302_015811}{0.54}{GW200128_022011}{0.52}{GW191204_171526}{0.43}{GW200112_155838}{0.46}{GW200105_162426}{0.38}{GW191105_143521}{0.42}{GW191109_010717}{0.62}{GW200209_085452}{0.41}{GW200115_042309}{0.42}{GW191127_050227}{0.64}{GW200216_220804}{0.61}{GW191215_223052}{0.41}{GW200208_130117}{0.42}{GW200219_094415}{0.47}{GW191103_012549}{0.52}{GW200316_215756}{0.49}{GW200202_154313}{0.42}{GW200129_065458}{0.49}{GW191216_213338}{0.41}}[{\red{???}}]}
\DeclareRobustCommand{\spintwozgwtcthreetenthpercentile}[1]{\IfEqCase{#1}{{GW200224_222234}{-0.25}{GW191129_134029}{-0.13}{GW200311_115853}{-0.33}{GW191230_180458}{-0.46}{GW191222_033537}{-0.39}{GW200225_060421}{-0.47}{GW200302_015811}{-0.31}{GW200128_022011}{-0.30}{GW191204_171526}{-0.09}{GW200112_155838}{-0.19}{GW200105_162426}{-0.26}{GW191105_143521}{-0.24}{GW191109_010717}{-0.56}{GW200209_085452}{-0.53}{GW200115_042309}{-0.57}{GW191127_050227}{-0.31}{GW200216_220804}{-0.32}{GW191215_223052}{-0.39}{GW200208_130117}{-0.44}{GW200219_094415}{-0.48}{GW191103_012549}{-0.13}{GW200316_215756}{-0.12}{GW200202_154313}{-0.13}{GW200129_065458}{-0.21}{GW191216_213338}{-0.14}}[{\red{???}}]}
\DeclareRobustCommand{\spintwozgwtcthreenintiethpercentile}[1]{\IfEqCase{#1}{{GW200224_222234}{0.40}{GW191129_134029}{0.44}{GW200311_115853}{0.29}{GW191230_180458}{0.31}{GW191222_033537}{0.28}{GW200225_060421}{0.22}{GW200302_015811}{0.44}{GW200128_022011}{0.45}{GW191204_171526}{0.51}{GW200112_155838}{0.40}{GW200105_162426}{0.26}{GW191105_143521}{0.29}{GW191109_010717}{0.34}{GW200209_085452}{0.21}{GW200115_042309}{0.22}{GW191127_050227}{0.58}{GW200216_220804}{0.53}{GW191215_223052}{0.27}{GW200208_130117}{0.27}{GW200219_094415}{0.29}{GW191103_012549}{0.60}{GW200316_215756}{0.50}{GW200202_154313}{0.35}{GW200129_065458}{0.53}{GW191216_213338}{0.43}}[{\red{???}}]}
\DeclareRobustCommand{\massonedetgwtcthreeminus}[1]{\IfEqCase{#1}{{GW200224_222234}{5.4}{GW191129_134029}{2.3}{GW200311_115853}{4.6}{GW191230_180458}{13}{GW191222_033537}{9.5}{GW200225_060421}{3.2}{GW200302_015811}{9.7}{GW200128_022011}{9.9}{GW191204_171526}{2.0}{GW200112_155838}{5.2}{GW200105_162426}{1.8}{GW191105_143521}{1.8}{GW191109_010717}{8.9}{GW200209_085452}{9.9}{GW200115_042309}{2.7}{GW191127_050227}{37}{GW200216_220804}{21}{GW191215_223052}{4.8}{GW200208_130117}{8.4}{GW200219_094415}{8.9}{GW191103_012549}{2.3}{GW200316_215756}{3.4}{GW200202_154313}{1.5}{GW200129_065458}{3.3}{GW191216_213338}{2.4}}[{\red{???}}]}
\DeclareRobustCommand{\massonedetgwtcthreemed}[1]{\IfEqCase{#1}{{GW200224_222234}{52.5}{GW191129_134029}{12.3}{GW200311_115853}{41.8}{GW191230_180458}{83}{GW191222_033537}{67.2}{GW200225_060421}{23.6}{GW200302_015811}{48.5}{GW200128_022011}{66.1}{GW191204_171526}{13.4}{GW200112_155838}{44.0}{GW200105_162426}{9.6}{GW191105_143521}{13.0}{GW191109_010717}{81.2}{GW200209_085452}{56.5}{GW200115_042309}{6.3}{GW191127_050227}{86}{GW200216_220804}{84}{GW191215_223052}{33.7}{GW200208_130117}{53.0}{GW200219_094415}{58.6}{GW191103_012549}{14.0}{GW200316_215756}{16.0}{GW200202_154313}{11.0}{GW200129_065458}{40.2}{GW191216_213338}{13.0}}[{\red{???}}]}
\DeclareRobustCommand{\massonedetgwtcthreeplus}[1]{\IfEqCase{#1}{{GW200224_222234}{9.1}{GW191129_134029}{4.9}{GW200311_115853}{8.2}{GW191230_180458}{19}{GW191222_033537}{14.7}{GW200225_060421}{5.6}{GW200302_015811}{10.8}{GW200128_022011}{16.6}{GW191204_171526}{3.8}{GW200112_155838}{8.2}{GW200105_162426}{1.9}{GW191105_143521}{4.5}{GW191109_010717}{12.9}{GW200209_085452}{15.4}{GW200115_042309}{2.1}{GW191127_050227}{60}{GW200216_220804}{28}{GW191215_223052}{9.4}{GW200208_130117}{12.2}{GW200219_094415}{13.5}{GW191103_012549}{7.4}{GW200316_215756}{12.3}{GW200202_154313}{3.8}{GW200129_065458}{12.2}{GW191216_213338}{5.0}}[{\red{???}}]}
\DeclareRobustCommand{\massonedetgwtcthreetenthpercentile}[1]{\IfEqCase{#1}{{GW200224_222234}{48.1}{GW191129_134029}{10.2}{GW200311_115853}{38.1}{GW191230_180458}{73}{GW191222_033537}{59.4}{GW200225_060421}{20.8}{GW200302_015811}{40.7}{GW200128_022011}{58.0}{GW191204_171526}{11.7}{GW200112_155838}{39.8}{GW200105_162426}{8.5}{GW191105_143521}{11.4}{GW191109_010717}{74.0}{GW200209_085452}{48.5}{GW200115_042309}{3.9}{GW191127_050227}{55}{GW200216_220804}{68}{GW191215_223052}{29.6}{GW200208_130117}{46.1}{GW200219_094415}{51.1}{GW191103_012549}{11.9}{GW200316_215756}{13.0}{GW200202_154313}{9.7}{GW200129_065458}{37.5}{GW191216_213338}{10.8}}[{\red{???}}]}
\DeclareRobustCommand{\massonedetgwtcthreenintiethpercentile}[1]{\IfEqCase{#1}{{GW200224_222234}{59.3}{GW191129_134029}{15.9}{GW200311_115853}{47.8}{GW191230_180458}{98}{GW191222_033537}{77.9}{GW200225_060421}{27.9}{GW200302_015811}{56.6}{GW200128_022011}{78.3}{GW191204_171526}{16.2}{GW200112_155838}{50.3}{GW200105_162426}{10.6}{GW191105_143521}{16.3}{GW191109_010717}{90.3}{GW200209_085452}{67.7}{GW200115_042309}{7.9}{GW191127_050227}{133}{GW200216_220804}{104}{GW191215_223052}{40.5}{GW200208_130117}{62.2}{GW200219_094415}{69.0}{GW191103_012549}{19.3}{GW200316_215756}{24.2}{GW200202_154313}{13.8}{GW200129_065458}{50.4}{GW191216_213338}{16.3}}[{\red{???}}]}
\DeclareRobustCommand{\chieffinfinityonlyprecavggwtcthreeminus}[1]{\IfEqCase{#1}{{GW200224_222234}{0.15}{GW191129_134029}{0.08}{GW200311_115853}{0.20}{GW191230_180458}{0.30}{GW191222_033537}{0.25}{GW200225_060421}{0.28}{GW200302_015811}{0.25}{GW200128_022011}{0.25}{GW191204_171526}{0.05}{GW200112_155838}{0.15}{GW200105_162426}{0.18}{GW191105_143521}{0.09}{GW191109_010717}{0.31}{GW200209_085452}{0.30}{GW200115_042309}{0.41}{GW191127_050227}{0.36}{GW200216_220804}{0.36}{GW191215_223052}{0.21}{GW200208_130117}{0.27}{GW200219_094415}{0.29}{GW191103_012549}{0.10}{GW200316_215756}{0.10}{GW200202_154313}{0.06}{GW200129_065458}{0.16}{GW191216_213338}{0.06}}[{\red{???}}]}
\DeclareRobustCommand{\chieffinfinityonlyprecavggwtcthreemed}[1]{\IfEqCase{#1}{{GW200224_222234}{0.11}{GW191129_134029}{0.06}{GW200311_115853}{-0.02}{GW191230_180458}{-0.03}{GW191222_033537}{-0.04}{GW200225_060421}{-0.12}{GW200302_015811}{0.03}{GW200128_022011}{0.14}{GW191204_171526}{0.16}{GW200112_155838}{0.06}{GW200105_162426}{0.00}{GW191105_143521}{-0.02}{GW191109_010717}{-0.29}{GW200209_085452}{-0.10}{GW200115_042309}{-0.15}{GW191127_050227}{0.18}{GW200216_220804}{0.10}{GW191215_223052}{-0.03}{GW200208_130117}{-0.07}{GW200219_094415}{-0.08}{GW191103_012549}{0.21}{GW200316_215756}{0.13}{GW200202_154313}{0.04}{GW200129_065458}{0.11}{GW191216_213338}{0.11}}[{\red{???}}]}
\DeclareRobustCommand{\chieffinfinityonlyprecavggwtcthreeplus}[1]{\IfEqCase{#1}{{GW200224_222234}{0.15}{GW191129_134029}{0.16}{GW200311_115853}{0.16}{GW191230_180458}{0.26}{GW191222_033537}{0.20}{GW200225_060421}{0.17}{GW200302_015811}{0.26}{GW200128_022011}{0.24}{GW191204_171526}{0.08}{GW200112_155838}{0.15}{GW200105_162426}{0.13}{GW191105_143521}{0.13}{GW191109_010717}{0.42}{GW200209_085452}{0.24}{GW200115_042309}{0.24}{GW191127_050227}{0.34}{GW200216_220804}{0.34}{GW191215_223052}{0.17}{GW200208_130117}{0.22}{GW200219_094415}{0.23}{GW191103_012549}{0.16}{GW200316_215756}{0.27}{GW200202_154313}{0.13}{GW200129_065458}{0.11}{GW191216_213338}{0.13}}[{\red{???}}]}
\DeclareRobustCommand{\chieffinfinityonlyprecavggwtcthreetenthpercentile}[1]{\IfEqCase{#1}{{GW200224_222234}{-0.01}{GW191129_134029}{0.00}{GW200311_115853}{-0.17}{GW191230_180458}{-0.26}{GW191222_033537}{-0.23}{GW200225_060421}{-0.34}{GW200302_015811}{-0.15}{GW200128_022011}{-0.05}{GW191204_171526}{0.12}{GW200112_155838}{-0.05}{GW200105_162426}{-0.10}{GW191105_143521}{-0.09}{GW191109_010717}{-0.54}{GW200209_085452}{-0.33}{GW200115_042309}{-0.51}{GW191127_050227}{-0.10}{GW200216_220804}{-0.17}{GW191215_223052}{-0.19}{GW200208_130117}{-0.27}{GW200219_094415}{-0.30}{GW191103_012549}{0.13}{GW200316_215756}{0.04}{GW200202_154313}{-0.01}{GW200129_065458}{0.00}{GW191216_213338}{0.06}}[{\red{???}}]}
\DeclareRobustCommand{\chieffinfinityonlyprecavggwtcthreenintiethpercentile}[1]{\IfEqCase{#1}{{GW200224_222234}{0.22}{GW191129_134029}{0.19}{GW200311_115853}{0.10}{GW191230_180458}{0.18}{GW191222_033537}{0.11}{GW200225_060421}{0.02}{GW200302_015811}{0.23}{GW200128_022011}{0.33}{GW191204_171526}{0.22}{GW200112_155838}{0.18}{GW200105_162426}{0.08}{GW191105_143521}{0.07}{GW191109_010717}{0.00}{GW200209_085452}{0.09}{GW200115_042309}{0.04}{GW191127_050227}{0.45}{GW200216_220804}{0.36}{GW191215_223052}{0.10}{GW200208_130117}{0.09}{GW200219_094415}{0.10}{GW191103_012549}{0.33}{GW200316_215756}{0.32}{GW200202_154313}{0.13}{GW200129_065458}{0.20}{GW191216_213338}{0.20}}[{\red{???}}]}
\DeclareRobustCommand{\chipinfinityonlyprecavggwtcthreeminus}[1]{\IfEqCase{#1}{{GW200224_222234}{0.37}{GW191129_134029}{0.19}{GW200311_115853}{0.35}{GW191230_180458}{0.39}{GW191222_033537}{0.32}{GW200225_060421}{0.38}{GW200302_015811}{0.29}{GW200128_022011}{0.40}{GW191204_171526}{0.26}{GW200112_155838}{0.30}{GW200105_162426}{0.07}{GW191105_143521}{0.24}{GW191109_010717}{0.38}{GW200209_085452}{0.38}{GW200115_042309}{0.16}{GW191127_050227}{0.41}{GW200216_220804}{0.35}{GW191215_223052}{0.38}{GW200208_130117}{0.29}{GW200219_094415}{0.35}{GW191103_012549}{0.27}{GW200316_215756}{0.20}{GW200202_154313}{0.22}{GW200129_065458}{0.39}{GW191216_213338}{0.15}}[{\red{???}}]}
\DeclareRobustCommand{\chipinfinityonlyprecavggwtcthreemed}[1]{\IfEqCase{#1}{{GW200224_222234}{0.50}{GW191129_134029}{0.26}{GW200311_115853}{0.45}{GW191230_180458}{0.52}{GW191222_033537}{0.41}{GW200225_060421}{0.53}{GW200302_015811}{0.38}{GW200128_022011}{0.56}{GW191204_171526}{0.39}{GW200112_155838}{0.40}{GW200105_162426}{0.09}{GW191105_143521}{0.30}{GW191109_010717}{0.63}{GW200209_085452}{0.52}{GW200115_042309}{0.20}{GW191127_050227}{0.52}{GW200216_220804}{0.45}{GW191215_223052}{0.50}{GW200208_130117}{0.38}{GW200219_094415}{0.47}{GW191103_012549}{0.40}{GW200316_215756}{0.29}{GW200202_154313}{0.28}{GW200129_065458}{0.54}{GW191216_213338}{0.23}}[{\red{???}}]}
\DeclareRobustCommand{\chipinfinityonlyprecavggwtcthreeplus}[1]{\IfEqCase{#1}{{GW200224_222234}{0.37}{GW191129_134029}{0.36}{GW200311_115853}{0.39}{GW191230_180458}{0.38}{GW191222_033537}{0.42}{GW200225_060421}{0.35}{GW200302_015811}{0.44}{GW200128_022011}{0.34}{GW191204_171526}{0.35}{GW200112_155838}{0.38}{GW200105_162426}{0.17}{GW191105_143521}{0.45}{GW191109_010717}{0.28}{GW200209_085452}{0.38}{GW200115_042309}{0.34}{GW191127_050227}{0.40}{GW200216_220804}{0.42}{GW191215_223052}{0.38}{GW200208_130117}{0.42}{GW200219_094415}{0.40}{GW191103_012549}{0.41}{GW200316_215756}{0.38}{GW200202_154313}{0.40}{GW200129_065458}{0.39}{GW191216_213338}{0.35}}[{\red{???}}]}
\DeclareRobustCommand{\chipinfinityonlyprecavggwtcthreetenthpercentile}[1]{\IfEqCase{#1}{{GW200224_222234}{0.20}{GW191129_134029}{0.10}{GW200311_115853}{0.16}{GW191230_180458}{0.20}{GW191222_033537}{0.14}{GW200225_060421}{0.22}{GW200302_015811}{0.13}{GW200128_022011}{0.23}{GW191204_171526}{0.18}{GW200112_155838}{0.14}{GW200105_162426}{0.03}{GW191105_143521}{0.09}{GW191109_010717}{0.33}{GW200209_085452}{0.20}{GW200115_042309}{0.06}{GW191127_050227}{0.17}{GW200216_220804}{0.15}{GW191215_223052}{0.19}{GW200208_130117}{0.13}{GW200219_094415}{0.18}{GW191103_012549}{0.18}{GW200316_215756}{0.12}{GW200202_154313}{0.09}{GW200129_065458}{0.21}{GW191216_213338}{0.10}}[{\red{???}}]}
\DeclareRobustCommand{\chipinfinityonlyprecavggwtcthreenintiethpercentile}[1]{\IfEqCase{#1}{{GW200224_222234}{0.80}{GW191129_134029}{0.54}{GW200311_115853}{0.77}{GW191230_180458}{0.84}{GW191222_033537}{0.75}{GW200225_060421}{0.82}{GW200302_015811}{0.74}{GW200128_022011}{0.84}{GW191204_171526}{0.67}{GW200112_155838}{0.70}{GW200105_162426}{0.19}{GW191105_143521}{0.65}{GW191109_010717}{0.87}{GW200209_085452}{0.84}{GW200115_042309}{0.46}{GW191127_050227}{0.88}{GW200216_220804}{0.80}{GW191215_223052}{0.82}{GW200208_130117}{0.71}{GW200219_094415}{0.80}{GW191103_012549}{0.72}{GW200316_215756}{0.58}{GW200202_154313}{0.59}{GW200129_065458}{0.91}{GW191216_213338}{0.48}}[{\red{???}}]}
\DeclareRobustCommand{\logpriorgwtcthreeminus}[1]{\IfEqCase{#1}{{GW200224_222234}{10.9}{GW200311_115853}{10.8}{GW191230_180458}{10.8}{GW191222_033537}{9.0}{GW200225_060421}{9.3}{GW200302_015811}{9.0}{GW200128_022011}{9.2}{GW200112_155838}{9.0}{GW191105_143521}{11.1}{GW191109_010717}{9.4}{GW200209_085452}{11.0}{GW191127_050227}{11.0}{GW200216_220804}{10.5}{GW191215_223052}{10.9}{GW200208_130117}{10.9}{GW200219_094415}{10.8}{GW191103_012549}{9.0}{GW200202_154313}{10.9}{GW200129_065458}{11.1}}[{\red{???}}]}
\DeclareRobustCommand{\logpriorgwtcthreemed}[1]{\IfEqCase{#1}{{GW200224_222234}{127.6}{GW200311_115853}{132.7}{GW191230_180458}{129.7}{GW191222_033537}{90.7}{GW200225_060421}{100.3}{GW200302_015811}{82.6}{GW200128_022011}{93.6}{GW200112_155838}{72.1}{GW191105_143521}{129.4}{GW191109_010717}{90.9}{GW200209_085452}{132.4}{GW191127_050227}{129.5}{GW200216_220804}{131.3}{GW191215_223052}{131.5}{GW200208_130117}{128.4}{GW200219_094415}{132.2}{GW191103_012549}{100.7}{GW200202_154313}{135.3}{GW200129_065458}{132.8}}[{\red{???}}]}
\DeclareRobustCommand{\logpriorgwtcthreeplus}[1]{\IfEqCase{#1}{{GW200224_222234}{8.7}{GW200311_115853}{8.6}{GW191230_180458}{8.8}{GW191222_033537}{6.9}{GW200225_060421}{6.9}{GW200302_015811}{6.9}{GW200128_022011}{6.9}{GW200112_155838}{6.8}{GW191105_143521}{8.8}{GW191109_010717}{7.2}{GW200209_085452}{8.7}{GW191127_050227}{8.9}{GW200216_220804}{8.7}{GW191215_223052}{8.8}{GW200208_130117}{8.7}{GW200219_094415}{8.8}{GW191103_012549}{6.9}{GW200202_154313}{8.7}{GW200129_065458}{8.8}}[{\red{???}}]}
\DeclareRobustCommand{\logpriorgwtcthreetenthpercentile}[1]{\IfEqCase{#1}{{GW200224_222234}{119.3}{GW200311_115853}{124.6}{GW191230_180458}{121.4}{GW191222_033537}{83.9}{GW200225_060421}{93.4}{GW200302_015811}{75.8}{GW200128_022011}{86.7}{GW200112_155838}{65.4}{GW191105_143521}{121.1}{GW191109_010717}{83.8}{GW200209_085452}{124.1}{GW191127_050227}{121.2}{GW200216_220804}{123.3}{GW191215_223052}{123.1}{GW200208_130117}{120.2}{GW200219_094415}{124.0}{GW191103_012549}{93.9}{GW200202_154313}{127.1}{GW200129_065458}{124.4}}[{\red{???}}]}
\DeclareRobustCommand{\logpriorgwtcthreenintiethpercentile}[1]{\IfEqCase{#1}{{GW200224_222234}{134.5}{GW200311_115853}{139.6}{GW191230_180458}{136.7}{GW191222_033537}{96.3}{GW200225_060421}{105.9}{GW200302_015811}{88.2}{GW200128_022011}{99.2}{GW200112_155838}{77.6}{GW191105_143521}{136.5}{GW191109_010717}{96.6}{GW200209_085452}{139.4}{GW191127_050227}{136.7}{GW200216_220804}{138.2}{GW191215_223052}{138.6}{GW200208_130117}{135.4}{GW200219_094415}{139.2}{GW191103_012549}{106.3}{GW200202_154313}{142.2}{GW200129_065458}{139.9}}[{\red{???}}]}
\DeclareRobustCommand{\networkmatchedfiltersnrgwtcthreeminus}[1]{\IfEqCase{#1}{{GW200224_222234}{0.2}{GW191129_134029}{0.3}{GW200311_115853}{0.2}{GW191230_180458}{0.4}{GW191222_033537}{0.3}{GW200225_060421}{0.4}{GW200302_015811}{0.4}{GW200128_022011}{0.4}{GW191204_171526}{0.2}{GW200112_155838}{0.2}{GW200105_162426}{0.4}{GW191105_143521}{0.5}{GW191109_010717}{0.5}{GW200209_085452}{0.5}{GW200115_042309}{0.5}{GW191127_050227}{0.6}{GW200216_220804}{0.6}{GW191215_223052}{0.4}{GW200208_130117}{0.5}{GW200219_094415}{0.5}{GW191103_012549}{0.5}{GW200316_215756}{0.7}{GW200202_154313}{0.4}{GW200129_065458}{0.2}{GW191216_213338}{0.2}}[{\red{???}}]}
\DeclareRobustCommand{\networkmatchedfiltersnrgwtcthreemed}[1]{\IfEqCase{#1}{{GW200224_222234}{20.0}{GW191129_134029}{13.1}{GW200311_115853}{17.8}{GW191230_180458}{10.4}{GW191222_033537}{12.5}{GW200225_060421}{12.5}{GW200302_015811}{10.8}{GW200128_022011}{10.6}{GW191204_171526}{17.5}{GW200112_155838}{19.8}{GW200105_162426}{13.7}{GW191105_143521}{9.7}{GW191109_010717}{17.2}{GW200209_085452}{9.6}{GW200115_042309}{11.3}{GW191127_050227}{9.1}{GW200216_220804}{8.1}{GW191215_223052}{11.2}{GW200208_130117}{10.8}{GW200219_094415}{10.7}{GW191103_012549}{8.9}{GW200316_215756}{10.3}{GW200202_154313}{10.8}{GW200129_065458}{26.8}{GW191216_213338}{18.6}}[{\red{???}}]}
\DeclareRobustCommand{\networkmatchedfiltersnrgwtcthreeplus}[1]{\IfEqCase{#1}{{GW200224_222234}{0.2}{GW191129_134029}{0.2}{GW200311_115853}{0.2}{GW191230_180458}{0.3}{GW191222_033537}{0.2}{GW200225_060421}{0.3}{GW200302_015811}{0.3}{GW200128_022011}{0.3}{GW191204_171526}{0.2}{GW200112_155838}{0.1}{GW200105_162426}{0.2}{GW191105_143521}{0.3}{GW191109_010717}{0.5}{GW200209_085452}{0.4}{GW200115_042309}{0.3}{GW191127_050227}{0.5}{GW200216_220804}{0.3}{GW191215_223052}{0.3}{GW200208_130117}{0.3}{GW200219_094415}{0.3}{GW191103_012549}{0.3}{GW200316_215756}{0.4}{GW200202_154313}{0.2}{GW200129_065458}{0.2}{GW191216_213338}{0.2}}[{\red{???}}]}
\DeclareRobustCommand{\networkmatchedfiltersnrgwtcthreetenthpercentile}[1]{\IfEqCase{#1}{{GW200224_222234}{19.8}{GW191129_134029}{12.9}{GW200311_115853}{17.7}{GW191230_180458}{10.1}{GW191222_033537}{12.2}{GW200225_060421}{12.2}{GW200302_015811}{10.5}{GW200128_022011}{10.3}{GW191204_171526}{17.3}{GW200112_155838}{19.6}{GW200105_162426}{13.4}{GW191105_143521}{9.3}{GW191109_010717}{16.9}{GW200209_085452}{9.2}{GW200115_042309}{10.9}{GW191127_050227}{8.7}{GW200216_220804}{7.7}{GW191215_223052}{10.9}{GW200208_130117}{10.5}{GW200219_094415}{10.3}{GW191103_012549}{8.5}{GW200316_215756}{9.8}{GW200202_154313}{10.5}{GW200129_065458}{26.6}{GW191216_213338}{18.4}}[{\red{???}}]}
\DeclareRobustCommand{\networkmatchedfiltersnrgwtcthreenintiethpercentile}[1]{\IfEqCase{#1}{{GW200224_222234}{20.1}{GW191129_134029}{13.3}{GW200311_115853}{18.0}{GW191230_180458}{10.7}{GW191222_033537}{12.6}{GW200225_060421}{12.7}{GW200302_015811}{11.1}{GW200128_022011}{10.9}{GW191204_171526}{17.6}{GW200112_155838}{19.9}{GW200105_162426}{13.9}{GW191105_143521}{9.9}{GW191109_010717}{17.7}{GW200209_085452}{9.9}{GW200115_042309}{11.5}{GW191127_050227}{9.5}{GW200216_220804}{8.4}{GW191215_223052}{11.5}{GW200208_130117}{11.0}{GW200219_094415}{10.9}{GW191103_012549}{9.1}{GW200316_215756}{10.6}{GW200202_154313}{11.0}{GW200129_065458}{26.9}{GW191216_213338}{18.7}}[{\red{???}}]}
\DeclareRobustCommand{\networkoptimalsnrgwtcthreeminus}[1]{\IfEqCase{#1}{{GW200224_222234}{1.7}{GW191129_134029}{1.7}{GW200311_115853}{1.7}{GW191230_180458}{1.7}{GW191222_033537}{1.7}{GW200225_060421}{1.7}{GW200302_015811}{1.7}{GW200128_022011}{1.7}{GW191204_171526}{1.7}{GW200112_155838}{1.7}{GW200105_162426}{1.7}{GW191105_143521}{1.8}{GW191109_010717}{1.7}{GW200209_085452}{1.7}{GW200115_042309}{1.7}{GW191127_050227}{1.8}{GW200216_220804}{1.7}{GW191215_223052}{1.7}{GW200208_130117}{1.7}{GW200219_094415}{1.8}{GW191103_012549}{1.8}{GW200316_215756}{1.8}{GW200202_154313}{1.7}{GW200129_065458}{1.6}{GW191216_213338}{1.7}}[{\red{???}}]}
\DeclareRobustCommand{\networkoptimalsnrgwtcthreemed}[1]{\IfEqCase{#1}{{GW200224_222234}{19.8}{GW191129_134029}{12.9}{GW200311_115853}{17.6}{GW191230_180458}{10.2}{GW191222_033537}{12.1}{GW200225_060421}{12.2}{GW200302_015811}{10.5}{GW200128_022011}{10.4}{GW191204_171526}{17.2}{GW200112_155838}{19.6}{GW200105_162426}{13.3}{GW191105_143521}{9.2}{GW191109_010717}{17.0}{GW200209_085452}{9.3}{GW200115_042309}{10.9}{GW191127_050227}{8.6}{GW200216_220804}{7.6}{GW191215_223052}{10.9}{GW200208_130117}{10.4}{GW200219_094415}{10.3}{GW191103_012549}{8.4}{GW200316_215756}{9.8}{GW200202_154313}{10.4}{GW200129_065458}{26.6}{GW191216_213338}{18.3}}[{\red{???}}]}
\DeclareRobustCommand{\networkoptimalsnrgwtcthreeplus}[1]{\IfEqCase{#1}{{GW200224_222234}{1.7}{GW191129_134029}{1.7}{GW200311_115853}{1.7}{GW191230_180458}{1.7}{GW191222_033537}{1.7}{GW200225_060421}{1.7}{GW200302_015811}{1.7}{GW200128_022011}{1.6}{GW191204_171526}{1.7}{GW200112_155838}{1.7}{GW200105_162426}{1.7}{GW191105_143521}{1.7}{GW191109_010717}{1.7}{GW200209_085452}{1.7}{GW200115_042309}{1.7}{GW191127_050227}{1.8}{GW200216_220804}{1.8}{GW191215_223052}{1.7}{GW200208_130117}{1.7}{GW200219_094415}{1.7}{GW191103_012549}{1.7}{GW200316_215756}{1.7}{GW200202_154313}{1.7}{GW200129_065458}{1.7}{GW191216_213338}{1.7}}[{\red{???}}]}
\DeclareRobustCommand{\networkoptimalsnrgwtcthreetenthpercentile}[1]{\IfEqCase{#1}{{GW200224_222234}{18.5}{GW191129_134029}{11.6}{GW200311_115853}{16.3}{GW191230_180458}{8.8}{GW191222_033537}{10.8}{GW200225_060421}{10.8}{GW200302_015811}{9.2}{GW200128_022011}{9.0}{GW191204_171526}{15.9}{GW200112_155838}{18.3}{GW200105_162426}{12.0}{GW191105_143521}{7.8}{GW191109_010717}{15.7}{GW200209_085452}{8.0}{GW200115_042309}{9.5}{GW191127_050227}{7.2}{GW200216_220804}{6.2}{GW191215_223052}{9.6}{GW200208_130117}{9.0}{GW200219_094415}{8.9}{GW191103_012549}{7.0}{GW200316_215756}{8.5}{GW200202_154313}{9.1}{GW200129_065458}{25.3}{GW191216_213338}{17.0}}[{\red{???}}]}
\DeclareRobustCommand{\networkoptimalsnrgwtcthreenintiethpercentile}[1]{\IfEqCase{#1}{{GW200224_222234}{21.1}{GW191129_134029}{14.2}{GW200311_115853}{18.9}{GW191230_180458}{11.5}{GW191222_033537}{13.4}{GW200225_060421}{13.5}{GW200302_015811}{11.8}{GW200128_022011}{11.7}{GW191204_171526}{18.5}{GW200112_155838}{20.9}{GW200105_162426}{14.6}{GW191105_143521}{10.5}{GW191109_010717}{18.4}{GW200209_085452}{10.7}{GW200115_042309}{12.2}{GW191127_050227}{10.0}{GW200216_220804}{8.9}{GW191215_223052}{12.2}{GW200208_130117}{11.7}{GW200219_094415}{11.6}{GW191103_012549}{9.7}{GW200316_215756}{11.2}{GW200202_154313}{11.8}{GW200129_065458}{27.9}{GW191216_213338}{19.6}}[{\red{???}}]}
\DeclareRobustCommand{\PEpercentBNSgwtcthree}[1]{\IfEqCase{#1}{{GW200224_222234}{0}{GW191129_134029}{0}{GW200311_115853}{0}{GW191230_180458}{0}{GW191222_033537}{0}{GW200225_060421}{0}{GW200302_015811}{0}{GW200128_022011}{0}{GW191204_171526}{0}{GW200112_155838}{0}{GW200105_162426}{0}{GW191105_143521}{0}{GW191109_010717}{0}{GW200209_085452}{0}{GW200115_042309}{1}{GW191127_050227}{0}{GW200216_220804}{0}{GW191215_223052}{0}{GW200208_130117}{0}{GW200219_094415}{0}{GW191103_012549}{0}{GW200316_215756}{0}{GW200202_154313}{0}{GW200129_065458}{0}{GW191216_213338}{0}}[{\red{???}}]}
\DeclareRobustCommand{\PEpercentNSBHgwtcthree}[1]{\IfEqCase{#1}{{GW200224_222234}{0}{GW191129_134029}{0}{GW200311_115853}{0}{GW191230_180458}{0}{GW191222_033537}{0}{GW200225_060421}{0}{GW200302_015811}{0}{GW200128_022011}{0}{GW191204_171526}{0}{GW200112_155838}{0}{GW200105_162426}{99}{GW191105_143521}{0}{GW191109_010717}{0}{GW200209_085452}{0}{GW200115_042309}{99}{GW191127_050227}{0}{GW200216_220804}{0}{GW191215_223052}{0}{GW200208_130117}{0}{GW200219_094415}{0}{GW191103_012549}{0}{GW200316_215756}{0}{GW200202_154313}{0}{GW200129_065458}{0}{GW191216_213338}{0}}[{\red{???}}]}
\DeclareRobustCommand{\PEpercentBBHgwtcthree}[1]{\IfEqCase{#1}{{GW200224_222234}{100}{GW191129_134029}{100}{GW200311_115853}{100}{GW191230_180458}{100}{GW191222_033537}{100}{GW200225_060421}{100}{GW200302_015811}{100}{GW200128_022011}{100}{GW191204_171526}{100}{GW200112_155838}{100}{GW200105_162426}{1}{GW191105_143521}{100}{GW191109_010717}{100}{GW200209_085452}{100}{GW200115_042309}{0}{GW191127_050227}{100}{GW200216_220804}{100}{GW191215_223052}{100}{GW200208_130117}{100}{GW200219_094415}{100}{GW191103_012549}{100}{GW200316_215756}{100}{GW200202_154313}{100}{GW200129_065458}{100}{GW191216_213338}{100}}[{\red{???}}]}
\DeclareRobustCommand{\PEpercentMassGapgwtcthree}[1]{\IfEqCase{#1}{{GW200224_222234}{0}{GW191129_134029}{0}{GW200311_115853}{0}{GW191230_180458}{0}{GW191222_033537}{0}{GW200225_060421}{0}{GW200302_015811}{0}{GW200128_022011}{0}{GW191204_171526}{0}{GW200112_155838}{0}{GW200105_162426}{0}{GW191105_143521}{0}{GW191109_010717}{0}{GW200209_085452}{0}{GW200115_042309}{0}{GW191127_050227}{0}{GW200216_220804}{0}{GW191215_223052}{0}{GW200208_130117}{0}{GW200219_094415}{0}{GW191103_012549}{0}{GW200316_215756}{0}{GW200202_154313}{0}{GW200129_065458}{0}{GW191216_213338}{0}}[{\red{???}}]}
\DeclareRobustCommand{\percentmassonelessthanthreegwtcthree}[1]{\IfEqCase{#1}{{GW200224_222234}{0}{GW191129_134029}{0}{GW200311_115853}{0}{GW191230_180458}{0}{GW191222_033537}{0}{GW200225_060421}{0}{GW200302_015811}{0}{GW200128_022011}{0}{GW191204_171526}{0}{GW200112_155838}{0}{GW200105_162426}{0}{GW191105_143521}{0}{GW191109_010717}{0}{GW200209_085452}{0}{GW200115_042309}{1}{GW191127_050227}{0}{GW200216_220804}{0}{GW191215_223052}{0}{GW200208_130117}{0}{GW200219_094415}{0}{GW191103_012549}{0}{GW200316_215756}{0}{GW200202_154313}{0}{GW200129_065458}{0}{GW191216_213338}{0}}[{\red{???}}]}
\DeclareRobustCommand{\percentmasstwolessthanthreegwtcthree}[1]{\IfEqCase{#1}{{GW200224_222234}{0}{GW191129_134029}{0}{GW200311_115853}{0}{GW191230_180458}{0}{GW191222_033537}{0}{GW200225_060421}{0}{GW200302_015811}{0}{GW200128_022011}{0}{GW191204_171526}{0}{GW200112_155838}{0}{GW200105_162426}{99}{GW191105_143521}{0}{GW191109_010717}{0}{GW200209_085452}{0}{GW200115_042309}{100}{GW191127_050227}{0}{GW200216_220804}{0}{GW191215_223052}{0}{GW200208_130117}{0}{GW200219_094415}{0}{GW191103_012549}{0}{GW200316_215756}{0}{GW200202_154313}{0}{GW200129_065458}{0}{GW191216_213338}{0}}[{\red{???}}]}
\DeclareRobustCommand{\percentmassonelessthanfivegwtcthree}[1]{\IfEqCase{#1}{{GW200224_222234}{0}{GW191129_134029}{0}{GW200311_115853}{0}{GW191230_180458}{0}{GW191222_033537}{0}{GW200225_060421}{0}{GW200302_015811}{0}{GW200128_022011}{0}{GW191204_171526}{0}{GW200112_155838}{0}{GW200105_162426}{1}{GW191105_143521}{0}{GW191109_010717}{0}{GW200209_085452}{0}{GW200115_042309}{29}{GW191127_050227}{0}{GW200216_220804}{0}{GW191215_223052}{0}{GW200208_130117}{0}{GW200219_094415}{0}{GW191103_012549}{0}{GW200316_215756}{0}{GW200202_154313}{0}{GW200129_065458}{0}{GW191216_213338}{0}}[{\red{???}}]}
\DeclareRobustCommand{\percentmasstwolessthanfivegwtcthree}[1]{\IfEqCase{#1}{{GW200224_222234}{0}{GW191129_134029}{5}{GW200311_115853}{0}{GW191230_180458}{0}{GW191222_033537}{0}{GW200225_060421}{0}{GW200302_015811}{0}{GW200128_022011}{0}{GW191204_171526}{0}{GW200112_155838}{0}{GW200105_162426}{100}{GW191105_143521}{1}{GW191109_010717}{0}{GW200209_085452}{0}{GW200115_042309}{100}{GW191127_050227}{0}{GW200216_220804}{0}{GW191215_223052}{0}{GW200208_130117}{0}{GW200219_094415}{0}{GW191103_012549}{2}{GW200316_215756}{5}{GW200202_154313}{1}{GW200129_065458}{0}{GW191216_213338}{2}}[{\red{???}}]}
\DeclareRobustCommand{\percentmassonemorethansixtyfivegwtcthree}[1]{\IfEqCase{#1}{{GW200224_222234}{0}{GW191129_134029}{0}{GW200311_115853}{0}{GW191230_180458}{2}{GW191222_033537}{0}{GW200225_060421}{0}{GW200302_015811}{0}{GW200128_022011}{0}{GW191204_171526}{0}{GW200112_155838}{0}{GW200105_162426}{0}{GW191105_143521}{0}{GW191109_010717}{51}{GW200209_085452}{0}{GW200115_042309}{0}{GW191127_050227}{30}{GW200216_220804}{13}{GW191215_223052}{0}{GW200208_130117}{0}{GW200219_094415}{0}{GW191103_012549}{0}{GW200316_215756}{0}{GW200202_154313}{0}{GW200129_065458}{0}{GW191216_213338}{0}}[{\red{???}}]}
\DeclareRobustCommand{\percentmasstwomorethansixtyfivegwtcthree}[1]{\IfEqCase{#1}{{GW200224_222234}{0}{GW191129_134029}{0}{GW200311_115853}{0}{GW191230_180458}{0}{GW191222_033537}{0}{GW200225_060421}{0}{GW200302_015811}{0}{GW200128_022011}{0}{GW191204_171526}{0}{GW200112_155838}{0}{GW200105_162426}{0}{GW191105_143521}{0}{GW191109_010717}{2}{GW200209_085452}{0}{GW200115_042309}{0}{GW191127_050227}{0}{GW200216_220804}{0}{GW191215_223052}{0}{GW200208_130117}{0}{GW200219_094415}{0}{GW191103_012549}{0}{GW200316_215756}{0}{GW200202_154313}{0}{GW200129_065458}{0}{GW191216_213338}{0}}[{\red{???}}]}
\DeclareRobustCommand{\percentmassonemorethanonetwentygwtcthree}[1]{\IfEqCase{#1}{{GW200224_222234}{0}{GW191129_134029}{0}{GW200311_115853}{0}{GW191230_180458}{0}{GW191222_033537}{0}{GW200225_060421}{0}{GW200302_015811}{0}{GW200128_022011}{0}{GW191204_171526}{0}{GW200112_155838}{0}{GW200105_162426}{0}{GW191105_143521}{0}{GW191109_010717}{0}{GW200209_085452}{0}{GW200115_042309}{0}{GW191127_050227}{1}{GW200216_220804}{0}{GW191215_223052}{0}{GW200208_130117}{0}{GW200219_094415}{0}{GW191103_012549}{0}{GW200316_215756}{0}{GW200202_154313}{0}{GW200129_065458}{0}{GW191216_213338}{0}}[{\red{???}}]}
\DeclareRobustCommand{\percentmasstwomorethanonetwentygwtcthree}[1]{\IfEqCase{#1}{{GW200224_222234}{0}{GW191129_134029}{0}{GW200311_115853}{0}{GW191230_180458}{0}{GW191222_033537}{0}{GW200225_060421}{0}{GW200302_015811}{0}{GW200128_022011}{0}{GW191204_171526}{0}{GW200112_155838}{0}{GW200105_162426}{0}{GW191105_143521}{0}{GW191109_010717}{0}{GW200209_085452}{0}{GW200115_042309}{0}{GW191127_050227}{0}{GW200216_220804}{0}{GW191215_223052}{0}{GW200208_130117}{0}{GW200219_094415}{0}{GW191103_012549}{0}{GW200316_215756}{0}{GW200202_154313}{0}{GW200129_065458}{0}{GW191216_213338}{0}}[{\red{???}}]}
\DeclareRobustCommand{\percentmassfinalmorethanonehundredgwtcthree}[1]{\IfEqCase{#1}{{GW200224_222234}{0}{GW191129_134029}{0}{GW200311_115853}{0}{GW191230_180458}{2}{GW191222_033537}{0}{GW200225_060421}{0}{GW200302_015811}{0}{GW200128_022011}{0}{GW191204_171526}{0}{GW200112_155838}{0}{GW200105_162426}{0}{GW191105_143521}{0}{GW191109_010717}{78}{GW200209_085452}{0}{GW200115_042309}{0}{GW191127_050227}{14}{GW200216_220804}{4}{GW191215_223052}{0}{GW200208_130117}{0}{GW200219_094415}{0}{GW191103_012549}{0}{GW200316_215756}{0}{GW200202_154313}{0}{GW200129_065458}{0}{GW191216_213338}{0}}[{\red{???}}]}
\DeclareRobustCommand{\percentchieffmorethanzerogwtcthree}[1]{\IfEqCase{#1}{{GW200224_222234}{87}{GW191129_134029}{91}{GW200311_115853}{42}{GW191230_180458}{43}{GW191222_033537}{37}{GW200225_060421}{15}{GW200302_015811}{60}{GW200128_022011}{83}{GW191204_171526}{100}{GW200112_155838}{77}{GW200105_162426}{53}{GW191105_143521}{37}{GW191109_010717}{10}{GW200209_085452}{26}{GW200115_042309}{18}{GW191127_050227}{79}{GW200216_220804}{69}{GW191215_223052}{38}{GW200208_130117}{30}{GW200219_094415}{29}{GW191103_012549}{100}{GW200316_215756}{98}{GW200202_154313}{86}{GW200129_065458}{89}{GW191216_213338}{100}}[{\red{???}}]}
\DeclareRobustCommand{\percentchiefflessthanzerogwtcthree}[1]{\IfEqCase{#1}{{GW200224_222234}{13}{GW191129_134029}{9}{GW200311_115853}{58}{GW191230_180458}{57}{GW191222_033537}{63}{GW200225_060421}{85}{GW200302_015811}{40}{GW200128_022011}{17}{GW191204_171526}{0}{GW200112_155838}{23}{GW200105_162426}{47}{GW191105_143521}{63}{GW191109_010717}{90}{GW200209_085452}{74}{GW200115_042309}{82}{GW191127_050227}{21}{GW200216_220804}{31}{GW191215_223052}{62}{GW200208_130117}{70}{GW200219_094415}{71}{GW191103_012549}{0}{GW200316_215756}{2}{GW200202_154313}{14}{GW200129_065458}{11}{GW191216_213338}{0}}[{\red{???}}]}
\DeclareRobustCommand{\percentchionemorethanpointeightgwtcthree}[1]{\IfEqCase{#1}{{GW200224_222234}{14}{GW191129_134029}{1}{GW200311_115853}{10}{GW191230_180458}{21}{GW191222_033537}{10}{GW200225_060421}{21}{GW200302_015811}{11}{GW200128_022011}{26}{GW191204_171526}{4}{GW200112_155838}{5}{GW200105_162426}{0}{GW191105_143521}{4}{GW191109_010717}{55}{GW200209_085452}{21}{GW200115_042309}{6}{GW191127_050227}{33}{GW200216_220804}{19}{GW191215_223052}{15}{GW200208_130117}{9}{GW200219_094415}{17}{GW191103_012549}{9}{GW200316_215756}{2}{GW200202_154313}{2}{GW200129_065458}{26}{GW191216_213338}{1}}[{\red{???}}]}
\DeclareRobustCommand{\percentchitwomorethanpointeightgwtcthree}[1]{\IfEqCase{#1}{{GW200224_222234}{14}{GW191129_134029}{8}{GW200311_115853}{13}{GW191230_180458}{19}{GW191222_033537}{13}{GW200225_060421}{14}{GW200302_015811}{16}{GW200128_022011}{19}{GW191204_171526}{9}{GW200112_155838}{11}{GW200105_162426}{10}{GW191105_143521}{9}{GW191109_010717}{33}{GW200209_085452}{19}{GW200115_042309}{13}{GW191127_050227}{22}{GW200216_220804}{21}{GW191215_223052}{16}{GW200208_130117}{14}{GW200219_094415}{18}{GW191103_012549}{17}{GW200316_215756}{13}{GW200202_154313}{8}{GW200129_065458}{15}{GW191216_213338}{8}}[{\red{???}}]}
\DeclareRobustCommand{\percentanychimorethanpointeightgwtcthree}[1]{\IfEqCase{#1}{{GW200224_222234}{26}{GW191129_134029}{9}{GW200311_115853}{21}{GW191230_180458}{36}{GW191222_033537}{21}{GW200225_060421}{33}{GW200302_015811}{25}{GW200128_022011}{41}{GW191204_171526}{13}{GW200112_155838}{15}{GW200105_162426}{10}{GW191105_143521}{12}{GW191109_010717}{72}{GW200209_085452}{37}{GW200115_042309}{18}{GW191127_050227}{49}{GW200216_220804}{36}{GW191215_223052}{29}{GW200208_130117}{22}{GW200219_094415}{32}{GW191103_012549}{24}{GW200316_215756}{14}{GW200202_154313}{10}{GW200129_065458}{36}{GW191216_213338}{8}}[{\red{???}}]}
\newcommand{\chirpmassdetgwtcthreeuncert}[1]{\ensuremath{ \chirpmassdetgwtcthreemed{#1}_{-\chirpmassdetgwtcthreeminus{#1}}^{+\chirpmassdetgwtcthreeplus{#1}}  } }

\newcommand{\massratiogwtcthreeuncert}[1]{\ensuremath{ \massratiogwtcthreemed{#1}_{-\massratiogwtcthreeminus{#1}}^{+\massratiogwtcthreeplus{#1}}  } }

\newcommand{\chieffgwtcthreeuncert}[1]{\ensuremath{ \chieffgwtcthreemed{#1}_{-\chieffgwtcthreeminus{#1}}^{+\chieffgwtcthreeplus{#1}}  } }

\newcommand{\networkmatchedfiltersnrgwtcthreeuncert}[1]{\ensuremath{ \networkmatchedfiltersnrgwtcthreemed{#1}_{-\networkmatchedfiltersnrgwtcthreeminus{#1}}^{+\networkmatchedfiltersnrgwtcthreeplus{#1}}  } }

\newcommand{\imrppnrtMCobsCOMPACTOneSevenZeroEightOneSevenHighSpinCat}{\macro{\ensuremath{1.1976_{-0.0002}^{+0.0004}}}}

\newcommand{\imrppnrtCHIEFFCOMPACTOneSevenZeroEightOneSevenHighSpinCat}{\macro{\ensuremath{0.02_{-0.02}^{+0.06}}}}

\newcommand{\imrppnrtMASSRATIOCOMPACTOneSevenZeroEightOneSevenHighSpinCat}{\macro{\ensuremath{0.72_{-0.21}^{+0.24}}}}

\newcommand{\imrppnrtPEMATCHSNRCOMPACTOneSevenZeroEightOneSevenHighSpinCat}{\macro{\ensuremath{32.7_{-0.1}^{+0.1}}}}

\newcommand{\loglikelihoodfourtwofiveminus}[1]{\IfEqCase{#1}{{AlignedSpinInspiralTidalHS}{5.5}{AlignedSpinInspiralTidalLS}{5.4}{AlignedSpinTidalHS}{6.5}{AlignedSpinTidalLS}{6.3}{IMRPhenomDNRTidal-HS}{5.4}{IMRPhenomDNRTidal-LS}{5.5}{IMRPhenomPv2NRTidal-HS}{5.7}{IMRPhenomPv2NRTidal-LS}{5.5}{SEOBNRv4TsurrogateHS}{5.2}{SEOBNRv4TsurrogateLS}{5.4}{SEOBNRv4TsurrogatehighspinRIFT}{18.6}{SEOBNRv4TsurrogatelowspinRIFT}{5.0}{TEOBResumS-HS}{15.8}{TEOBResumS-LS}{5.1}{TaylorF2-HS}{5.5}{TaylorF2-LS}{5.4}{PrecessingSpinIMRTidalHS}{5.7}{PrecessingSpinIMRTidalLS}{5.5}{PublicationSamples}{5.6}}}
\newcommand{\loglikelihoodfourtwofivemed}[1]{\IfEqCase{#1}{{AlignedSpinInspiralTidalHS}{-240965.7}{AlignedSpinInspiralTidalLS}{-240964.4}{AlignedSpinTidalHS}{-1045826.3}{AlignedSpinTidalLS}{-1045825.4}{IMRPhenomDNRTidal-HS}{-1045828.5}{IMRPhenomDNRTidal-LS}{-1045827.1}{IMRPhenomPv2NRTidal-HS}{-500483.9}{IMRPhenomPv2NRTidal-LS}{-500482.8}{SEOBNRv4TsurrogateHS}{-1045828.0}{SEOBNRv4TsurrogateLS}{-1045827.1}{SEOBNRv4TsurrogatehighspinRIFT}{66.9}{SEOBNRv4TsurrogatelowspinRIFT}{68.5}{TEOBResumS-HS}{67.1}{TEOBResumS-LS}{68.8}{TaylorF2-HS}{-240965.7}{TaylorF2-LS}{-240964.4}{PrecessingSpinIMRTidalHS}{-500483.9}{PrecessingSpinIMRTidalLS}{-500482.8}{PublicationSamples}{-500483.9}}}
\newcommand{\loglikelihoodfourtwofiveplus}[1]{\IfEqCase{#1}{{AlignedSpinInspiralTidalHS}{3.8}{AlignedSpinInspiralTidalLS}{3.2}{AlignedSpinTidalHS}{1045896.4}{AlignedSpinTidalLS}{1045896.5}{IMRPhenomDNRTidal-HS}{4.5}{IMRPhenomDNRTidal-LS}{3.7}{IMRPhenomPv2NRTidal-HS}{4.5}{IMRPhenomPv2NRTidal-LS}{3.8}{SEOBNRv4TsurrogateHS}{3.9}{SEOBNRv4TsurrogateLS}{3.5}{SEOBNRv4TsurrogatehighspinRIFT}{4.3}{SEOBNRv4TsurrogatelowspinRIFT}{3.4}{TEOBResumS-HS}{4.4}{TEOBResumS-LS}{3.3}{TaylorF2-HS}{3.8}{TaylorF2-LS}{3.2}{PrecessingSpinIMRTidalHS}{4.5}{PrecessingSpinIMRTidalLS}{3.8}{PublicationSamples}{4.5}}}
\newcommand{\chiefffourtwofiveminus}[1]{\IfEqCase{#1}{{AlignedSpinInspiralTidalHS}{0.04}{AlignedSpinInspiralTidalLS}{0.01}{AlignedSpinTidalHS}{0.03}{AlignedSpinTidalLS}{0.01}{IMRPhenomDNRTidal-HS}{0.03}{IMRPhenomDNRTidal-LS}{0.01}{IMRPhenomPv2NRTidal-HS}{0.05}{IMRPhenomPv2NRTidal-LS}{0.01}{SEOBNRv4TsurrogateHS}{0.03}{SEOBNRv4TsurrogateLS}{0.01}{SEOBNRv4TsurrogatehighspinRIFT}{0.03}{SEOBNRv4TsurrogatelowspinRIFT}{0.01}{TEOBResumS-HS}{0.03}{TEOBResumS-LS}{0.01}{TaylorF2-HS}{0.04}{TaylorF2-LS}{0.01}{PrecessingSpinIMRTidalHS}{0.05}{PrecessingSpinIMRTidalLS}{0.01}{PublicationSamples}{0.05}}}
\newcommand{\chiefffourtwofivemed}[1]{\IfEqCase{#1}{{AlignedSpinInspiralTidalHS}{0.05}{AlignedSpinInspiralTidalLS}{0.01}{AlignedSpinTidalHS}{0.04}{AlignedSpinTidalLS}{0.01}{IMRPhenomDNRTidal-HS}{0.04}{IMRPhenomDNRTidal-LS}{0.01}{IMRPhenomPv2NRTidal-HS}{0.06}{IMRPhenomPv2NRTidal-LS}{0.01}{SEOBNRv4TsurrogateHS}{0.04}{SEOBNRv4TsurrogateLS}{0.01}{SEOBNRv4TsurrogatehighspinRIFT}{0.03}{SEOBNRv4TsurrogatelowspinRIFT}{0.01}{TEOBResumS-HS}{0.04}{TEOBResumS-LS}{0.01}{TaylorF2-HS}{0.05}{TaylorF2-LS}{0.01}{PrecessingSpinIMRTidalHS}{0.06}{PrecessingSpinIMRTidalLS}{0.01}{PublicationSamples}{0.06}}}
\newcommand{\chiefffourtwofiveplus}[1]{\IfEqCase{#1}{{AlignedSpinInspiralTidalHS}{0.09}{AlignedSpinInspiralTidalLS}{0.01}{AlignedSpinTidalHS}{0.08}{AlignedSpinTidalLS}{0.01}{IMRPhenomDNRTidal-HS}{0.10}{IMRPhenomDNRTidal-LS}{0.01}{IMRPhenomPv2NRTidal-HS}{0.11}{IMRPhenomPv2NRTidal-LS}{0.01}{SEOBNRv4TsurrogateHS}{0.07}{SEOBNRv4TsurrogateLS}{0.01}{SEOBNRv4TsurrogatehighspinRIFT}{0.07}{SEOBNRv4TsurrogatelowspinRIFT}{0.02}{TEOBResumS-HS}{0.06}{TEOBResumS-LS}{0.02}{TaylorF2-HS}{0.09}{TaylorF2-LS}{0.01}{PrecessingSpinIMRTidalHS}{0.11}{PrecessingSpinIMRTidalLS}{0.01}{PublicationSamples}{0.11}}}
\newcommand{\totalmasssourcefourtwofiveminus}[1]{\IfEqCase{#1}{{AlignedSpinInspiralTidalHS}{0.09}{AlignedSpinInspiralTidalLS}{0.05}{AlignedSpinTidalHS}{0.08}{AlignedSpinTidalLS}{0.05}{IMRPhenomDNRTidal-HS}{0.09}{IMRPhenomDNRTidal-LS}{0.05}{IMRPhenomPv2NRTidal-HS}{0.1}{IMRPhenomPv2NRTidal-LS}{0.05}{SEOBNRv4TsurrogateHS}{0.07}{SEOBNRv4TsurrogateLS}{0.05}{SEOBNRv4TsurrogatehighspinRIFT}{0.07}{SEOBNRv4TsurrogatelowspinRIFT}{0.05}{TEOBResumS-HS}{0.08}{TEOBResumS-LS}{0.05}{TaylorF2-HS}{0.09}{TaylorF2-LS}{0.05}{PrecessingSpinIMRTidalHS}{0.1}{PrecessingSpinIMRTidalLS}{0.05}{PublicationSamples}{0.1}}}
\newcommand{\totalmasssourcefourtwofivemed}[1]{\IfEqCase{#1}{{AlignedSpinInspiralTidalHS}{3.37}{AlignedSpinInspiralTidalLS}{3.31}{AlignedSpinTidalHS}{3.35}{AlignedSpinTidalLS}{3.31}{IMRPhenomDNRTidal-HS}{3.36}{IMRPhenomDNRTidal-LS}{3.31}{IMRPhenomPv2NRTidal-HS}{3.4}{IMRPhenomPv2NRTidal-LS}{3.31}{SEOBNRv4TsurrogateHS}{3.35}{SEOBNRv4TsurrogateLS}{3.31}{SEOBNRv4TsurrogatehighspinRIFT}{3.34}{SEOBNRv4TsurrogatelowspinRIFT}{3.31}{TEOBResumS-HS}{3.35}{TEOBResumS-LS}{3.31}{TaylorF2-HS}{3.37}{TaylorF2-LS}{3.31}{PrecessingSpinIMRTidalHS}{3.4}{PrecessingSpinIMRTidalLS}{3.31}{PublicationSamples}{3.4}}}
\newcommand{\totalmasssourcefourtwofiveplus}[1]{\IfEqCase{#1}{{AlignedSpinInspiralTidalHS}{0.2}{AlignedSpinInspiralTidalLS}{0.06}{AlignedSpinTidalHS}{0.3}{AlignedSpinTidalLS}{0.06}{IMRPhenomDNRTidal-HS}{0.4}{IMRPhenomDNRTidal-LS}{0.06}{IMRPhenomPv2NRTidal-HS}{0.3}{IMRPhenomPv2NRTidal-LS}{0.06}{SEOBNRv4TsurrogateHS}{0.2}{SEOBNRv4TsurrogateLS}{0.06}{SEOBNRv4TsurrogatehighspinRIFT}{0.2}{SEOBNRv4TsurrogatelowspinRIFT}{0.06}{TEOBResumS-HS}{0.2}{TEOBResumS-LS}{0.06}{TaylorF2-HS}{0.2}{TaylorF2-LS}{0.06}{PrecessingSpinIMRTidalHS}{0.3}{PrecessingSpinIMRTidalLS}{0.06}{PublicationSamples}{0.3}}}
\newcommand{\chipfourtwofiveminus}[1]{\IfEqCase{#1}{{AlignedSpinInspiralTidalHS}{0.00}{AlignedSpinInspiralTidalLS}{0.00}{AlignedSpinTidalHS}{0.00}{AlignedSpinTidalLS}{0.00}{IMRPhenomDNRTidal-HS}{0.00}{IMRPhenomDNRTidal-LS}{0.00}{IMRPhenomPv2NRTidal-HS}{0.27}{IMRPhenomPv2NRTidal-LS}{0.02}{SEOBNRv4TsurrogateHS}{0.00}{SEOBNRv4TsurrogateLS}{0.00}{SEOBNRv4TsurrogatehighspinRIFT}{0.00}{SEOBNRv4TsurrogatelowspinRIFT}{0.00}{TEOBResumS-HS}{0.00}{TEOBResumS-LS}{0.00}{TaylorF2-HS}{0.00}{TaylorF2-LS}{0.00}{PrecessingSpinIMRTidalHS}{0.27}{PrecessingSpinIMRTidalLS}{0.02}{PublicationSamples}{0.27}}}
\newcommand{\chipfourtwofivemed}[1]{\IfEqCase{#1}{{AlignedSpinInspiralTidalHS}{0.00}{AlignedSpinInspiralTidalLS}{0.00}{AlignedSpinTidalHS}{0.00}{AlignedSpinTidalLS}{0.00}{IMRPhenomDNRTidal-HS}{0.00}{IMRPhenomDNRTidal-LS}{0.00}{IMRPhenomPv2NRTidal-HS}{0.34}{IMRPhenomPv2NRTidal-LS}{0.03}{SEOBNRv4TsurrogateHS}{0.00}{SEOBNRv4TsurrogateLS}{0.00}{SEOBNRv4TsurrogatehighspinRIFT}{0.00}{SEOBNRv4TsurrogatelowspinRIFT}{0.00}{TEOBResumS-HS}{0.00}{TEOBResumS-LS}{0.00}{TaylorF2-HS}{0.00}{TaylorF2-LS}{0.00}{PrecessingSpinIMRTidalHS}{0.34}{PrecessingSpinIMRTidalLS}{0.03}{PublicationSamples}{0.34}}}
\newcommand{\chipfourtwofiveplus}[1]{\IfEqCase{#1}{{AlignedSpinInspiralTidalHS}{0.00}{AlignedSpinInspiralTidalLS}{0.00}{AlignedSpinTidalHS}{0.00}{AlignedSpinTidalLS}{0.00}{IMRPhenomDNRTidal-HS}{0.00}{IMRPhenomDNRTidal-LS}{0.00}{IMRPhenomPv2NRTidal-HS}{0.43}{IMRPhenomPv2NRTidal-LS}{0.02}{SEOBNRv4TsurrogateHS}{0.00}{SEOBNRv4TsurrogateLS}{0.00}{SEOBNRv4TsurrogatehighspinRIFT}{0.00}{SEOBNRv4TsurrogatelowspinRIFT}{0.00}{TEOBResumS-HS}{0.00}{TEOBResumS-LS}{0.00}{TaylorF2-HS}{0.00}{TaylorF2-LS}{0.00}{PrecessingSpinIMRTidalHS}{0.43}{PrecessingSpinIMRTidalLS}{0.02}{PublicationSamples}{0.43}}}
\newcommand{\spinoneyfourtwofiveminus}[1]{\IfEqCase{#1}{{AlignedSpinInspiralTidalHS}{0.00}{AlignedSpinInspiralTidalLS}{0.00}{AlignedSpinTidalHS}{0.00}{AlignedSpinTidalLS}{0.00}{IMRPhenomDNRTidal-HS}{0.00}{IMRPhenomDNRTidal-LS}{0.00}{IMRPhenomPv2NRTidal-HS}{0.49}{IMRPhenomPv2NRTidal-LS}{0.03}{SEOBNRv4TsurrogateHS}{0.00}{SEOBNRv4TsurrogateLS}{0.00}{SEOBNRv4TsurrogatehighspinRIFT}{0.00}{SEOBNRv4TsurrogatelowspinRIFT}{0.00}{TEOBResumS-HS}{0.00}{TEOBResumS-LS}{0.00}{TaylorF2-HS}{0.00}{TaylorF2-LS}{0.00}{PrecessingSpinIMRTidalHS}{0.48}{PrecessingSpinIMRTidalLS}{0.03}{PublicationSamples}{0.48}}}
\newcommand{\spinoneyfourtwofivemed}[1]{\IfEqCase{#1}{{AlignedSpinInspiralTidalHS}{0.00}{AlignedSpinInspiralTidalLS}{0.00}{AlignedSpinTidalHS}{0.00}{AlignedSpinTidalLS}{0.00}{IMRPhenomDNRTidal-HS}{0.00}{IMRPhenomDNRTidal-LS}{0.00}{IMRPhenomPv2NRTidal-HS}{0.003}{IMRPhenomPv2NRTidal-LS}{0.00}{SEOBNRv4TsurrogateHS}{0.00}{SEOBNRv4TsurrogateLS}{0.00}{SEOBNRv4TsurrogatehighspinRIFT}{0.00}{SEOBNRv4TsurrogatelowspinRIFT}{0.00}{TEOBResumS-HS}{0.00}{TEOBResumS-LS}{0.00}{TaylorF2-HS}{0.00}{TaylorF2-LS}{0.00}{PrecessingSpinIMRTidalHS}{0.003}{PrecessingSpinIMRTidalLS}{0.00}{PublicationSamples}{0.003}}}
\newcommand{\spinoneyfourtwofiveplus}[1]{\IfEqCase{#1}{{AlignedSpinInspiralTidalHS}{0.00}{AlignedSpinInspiralTidalLS}{0.00}{AlignedSpinTidalHS}{0.00}{AlignedSpinTidalLS}{0.00}{IMRPhenomDNRTidal-HS}{0.00}{IMRPhenomDNRTidal-LS}{0.00}{IMRPhenomPv2NRTidal-HS}{0.48}{IMRPhenomPv2NRTidal-LS}{0.03}{SEOBNRv4TsurrogateHS}{0.00}{SEOBNRv4TsurrogateLS}{0.00}{SEOBNRv4TsurrogatehighspinRIFT}{0.00}{SEOBNRv4TsurrogatelowspinRIFT}{0.00}{TEOBResumS-HS}{0.00}{TEOBResumS-LS}{0.00}{TaylorF2-HS}{0.00}{TaylorF2-LS}{0.00}{PrecessingSpinIMRTidalHS}{0.48}{PrecessingSpinIMRTidalLS}{0.03}{PublicationSamples}{0.48}}}
\newcommand{\phitwofourtwofiveminus}[1]{\IfEqCase{#1}{{AlignedSpinInspiralTidalHS}{0.00}{AlignedSpinInspiralTidalLS}{0.00}{AlignedSpinTidalHS}{0.00}{AlignedSpinTidalLS}{0.00}{IMRPhenomDNRTidal-HS}{0.00}{IMRPhenomDNRTidal-LS}{0.00}{IMRPhenomPv2NRTidal-HS}{2.83}{IMRPhenomPv2NRTidal-LS}{2.82}{SEOBNRv4TsurrogateHS}{0.00}{SEOBNRv4TsurrogateLS}{0.00}{SEOBNRv4TsurrogatehighspinRIFT}{0.00}{SEOBNRv4TsurrogatelowspinRIFT}{0.00}{TEOBResumS-HS}{0.00}{TEOBResumS-LS}{0.00}{TaylorF2-HS}{0.00}{TaylorF2-LS}{0.00}{PrecessingSpinIMRTidalHS}{2.84}{PrecessingSpinIMRTidalLS}{2.82}{PublicationSamples}{2.84}}}
\newcommand{\phitwofourtwofivemed}[1]{\IfEqCase{#1}{{AlignedSpinInspiralTidalHS}{0.00}{AlignedSpinInspiralTidalLS}{0.00}{AlignedSpinTidalHS}{0.00}{AlignedSpinTidalLS}{0.00}{IMRPhenomDNRTidal-HS}{0.00}{IMRPhenomDNRTidal-LS}{0.00}{IMRPhenomPv2NRTidal-HS}{3.14}{IMRPhenomPv2NRTidal-LS}{3.13}{SEOBNRv4TsurrogateHS}{0.00}{SEOBNRv4TsurrogateLS}{0.00}{SEOBNRv4TsurrogatehighspinRIFT}{0.00}{SEOBNRv4TsurrogatelowspinRIFT}{0.00}{TEOBResumS-HS}{0.00}{TEOBResumS-LS}{0.00}{TaylorF2-HS}{0.00}{TaylorF2-LS}{0.00}{PrecessingSpinIMRTidalHS}{3.15}{PrecessingSpinIMRTidalLS}{3.13}{PublicationSamples}{3.15}}}
\newcommand{\phitwofourtwofiveplus}[1]{\IfEqCase{#1}{{AlignedSpinInspiralTidalHS}{0.00}{AlignedSpinInspiralTidalLS}{0.00}{AlignedSpinTidalHS}{0.00}{AlignedSpinTidalLS}{0.00}{IMRPhenomDNRTidal-HS}{0.00}{IMRPhenomDNRTidal-LS}{0.00}{IMRPhenomPv2NRTidal-HS}{2.84}{IMRPhenomPv2NRTidal-LS}{2.83}{SEOBNRv4TsurrogateHS}{0.00}{SEOBNRv4TsurrogateLS}{0.00}{SEOBNRv4TsurrogatehighspinRIFT}{0.00}{SEOBNRv4TsurrogatelowspinRIFT}{0.00}{TEOBResumS-HS}{0.00}{TEOBResumS-LS}{0.00}{TaylorF2-HS}{0.00}{TaylorF2-LS}{0.00}{PrecessingSpinIMRTidalHS}{2.83}{PrecessingSpinIMRTidalLS}{2.83}{PublicationSamples}{2.83}}}
\newcommand{\phionetwofourtwofiveminus}[1]{\IfEqCase{#1}{{AlignedSpinInspiralTidalHS}{0.00}{AlignedSpinInspiralTidalLS}{0.00}{AlignedSpinTidalHS}{0.00}{AlignedSpinTidalLS}{0.00}{IMRPhenomDNRTidal-HS}{0.00}{IMRPhenomDNRTidal-LS}{0.00}{IMRPhenomPv2NRTidal-HS}{2.88}{IMRPhenomPv2NRTidal-LS}{2.74}{SEOBNRv4TsurrogateHS}{0.00}{SEOBNRv4TsurrogateLS}{0.00}{SEOBNRv4TsurrogatehighspinRIFT}{0.00}{SEOBNRv4TsurrogatelowspinRIFT}{0.00}{TEOBResumS-HS}{0.00}{TEOBResumS-LS}{0.00}{TaylorF2-HS}{0.00}{TaylorF2-LS}{0.00}{PrecessingSpinIMRTidalHS}{2.87}{PrecessingSpinIMRTidalLS}{2.75}{PublicationSamples}{2.88}}}
\newcommand{\phionetwofourtwofivemed}[1]{\IfEqCase{#1}{{AlignedSpinInspiralTidalHS}{0.00}{AlignedSpinInspiralTidalLS}{0.00}{AlignedSpinTidalHS}{0.00}{AlignedSpinTidalLS}{0.00}{IMRPhenomDNRTidal-HS}{0.00}{IMRPhenomDNRTidal-LS}{0.00}{IMRPhenomPv2NRTidal-HS}{3.19}{IMRPhenomPv2NRTidal-LS}{3.05}{SEOBNRv4TsurrogateHS}{0.00}{SEOBNRv4TsurrogateLS}{0.00}{SEOBNRv4TsurrogatehighspinRIFT}{0.00}{SEOBNRv4TsurrogatelowspinRIFT}{0.00}{TEOBResumS-HS}{0.00}{TEOBResumS-LS}{0.00}{TaylorF2-HS}{0.00}{TaylorF2-LS}{0.00}{PrecessingSpinIMRTidalHS}{3.18}{PrecessingSpinIMRTidalLS}{3.05}{PublicationSamples}{3.18}}}
\newcommand{\phionetwofourtwofiveplus}[1]{\IfEqCase{#1}{{AlignedSpinInspiralTidalHS}{0.00}{AlignedSpinInspiralTidalLS}{0.00}{AlignedSpinTidalHS}{0.00}{AlignedSpinTidalLS}{0.00}{IMRPhenomDNRTidal-HS}{0.00}{IMRPhenomDNRTidal-LS}{0.00}{IMRPhenomPv2NRTidal-HS}{2.75}{IMRPhenomPv2NRTidal-LS}{2.88}{SEOBNRv4TsurrogateHS}{0.00}{SEOBNRv4TsurrogateLS}{0.00}{SEOBNRv4TsurrogatehighspinRIFT}{0.00}{SEOBNRv4TsurrogatelowspinRIFT}{0.00}{TEOBResumS-HS}{0.00}{TEOBResumS-LS}{0.00}{TaylorF2-HS}{0.00}{TaylorF2-LS}{0.00}{PrecessingSpinIMRTidalHS}{2.76}{PrecessingSpinIMRTidalLS}{2.88}{PublicationSamples}{2.76}}}
\newcommand{\rafourtwofiveminus}[1]{\IfEqCase{#1}{{AlignedSpinInspiralTidalHS}{1.04004}{AlignedSpinInspiralTidalLS}{1.10804}{AlignedSpinTidalHS}{1.04466}{AlignedSpinTidalLS}{1.05437}{IMRPhenomDNRTidal-HS}{1.04415}{IMRPhenomDNRTidal-LS}{1.34502}{IMRPhenomPv2NRTidal-HS}{1.14841}{IMRPhenomPv2NRTidal-LS}{1.49247}{SEOBNRv4TsurrogateHS}{1.04813}{SEOBNRv4TsurrogateLS}{0.99881}{SEOBNRv4TsurrogatehighspinRIFT}{1.03845}{SEOBNRv4TsurrogatelowspinRIFT}{1.07158}{TEOBResumS-HS}{1.04161}{TEOBResumS-LS}{1.06268}{TaylorF2-HS}{1.03744}{TaylorF2-LS}{1.11687}{PrecessingSpinIMRTidalHS}{1.14713}{PrecessingSpinIMRTidalLS}{1.36500}{PublicationSamples}{1.14187}}}
\newcommand{\rafourtwofivemed}[1]{\IfEqCase{#1}{{AlignedSpinInspiralTidalHS}{1.47807}{AlignedSpinInspiralTidalLS}{1.57610}{AlignedSpinTidalHS}{1.52327}{AlignedSpinTidalLS}{1.53218}{IMRPhenomDNRTidal-HS}{1.52282}{IMRPhenomDNRTidal-LS}{1.86790}{IMRPhenomPv2NRTidal-HS}{1.62902}{IMRPhenomPv2NRTidal-LS}{1.99502}{SEOBNRv4TsurrogateHS}{1.52665}{SEOBNRv4TsurrogateLS}{1.46873}{SEOBNRv4TsurrogatehighspinRIFT}{1.50830}{SEOBNRv4TsurrogatelowspinRIFT}{1.53881}{TEOBResumS-HS}{1.51000}{TEOBResumS-LS}{1.50223}{TaylorF2-HS}{1.47697}{TaylorF2-LS}{1.57735}{PrecessingSpinIMRTidalHS}{1.62833}{PrecessingSpinIMRTidalLS}{1.86958}{PublicationSamples}{1.62336}}}
\newcommand{\rafourtwofiveplus}[1]{\IfEqCase{#1}{{AlignedSpinInspiralTidalHS}{3.24632}{AlignedSpinInspiralTidalLS}{3.13801}{AlignedSpinTidalHS}{3.25020}{AlignedSpinTidalLS}{3.18422}{IMRPhenomDNRTidal-HS}{3.29588}{IMRPhenomDNRTidal-LS}{2.86054}{IMRPhenomPv2NRTidal-HS}{3.13029}{IMRPhenomPv2NRTidal-LS}{2.74917}{SEOBNRv4TsurrogateHS}{3.18219}{SEOBNRv4TsurrogateLS}{3.22628}{SEOBNRv4TsurrogatehighspinRIFT}{3.24313}{SEOBNRv4TsurrogatelowspinRIFT}{3.18050}{TEOBResumS-HS}{3.24215}{TEOBResumS-LS}{3.23570}{TaylorF2-HS}{3.25577}{TaylorF2-LS}{3.13908}{PrecessingSpinIMRTidalHS}{3.12911}{PrecessingSpinIMRTidalLS}{2.87696}{PublicationSamples}{3.13407}}}
\newcommand{\phijlfourtwofiveminus}[1]{\IfEqCase{#1}{{AlignedSpinInspiralTidalHS}{0.97}{AlignedSpinInspiralTidalLS}{0.88}{AlignedSpinTidalHS}{0.78}{AlignedSpinTidalLS}{0.77}{IMRPhenomDNRTidal-HS}{0.74}{IMRPhenomDNRTidal-LS}{0.78}{IMRPhenomPv2NRTidal-HS}{2.89}{IMRPhenomPv2NRTidal-LS}{2.58}{SEOBNRv4TsurrogateHS}{0.89}{SEOBNRv4TsurrogateLS}{0.85}{SEOBNRv4TsurrogatehighspinRIFT}{0.00}{SEOBNRv4TsurrogatelowspinRIFT}{0.00}{TEOBResumS-HS}{0.00}{TEOBResumS-LS}{0.00}{TaylorF2-HS}{0.97}{TaylorF2-LS}{0.88}{PrecessingSpinIMRTidalHS}{2.88}{PrecessingSpinIMRTidalLS}{2.57}{PublicationSamples}{2.89}}}
\newcommand{\phijlfourtwofivemed}[1]{\IfEqCase{#1}{{AlignedSpinInspiralTidalHS}{1.22}{AlignedSpinInspiralTidalLS}{1.11}{AlignedSpinTidalHS}{0.78}{AlignedSpinTidalLS}{0.77}{IMRPhenomDNRTidal-HS}{0.96}{IMRPhenomDNRTidal-LS}{1.02}{IMRPhenomPv2NRTidal-HS}{3.23}{IMRPhenomPv2NRTidal-LS}{2.87}{SEOBNRv4TsurrogateHS}{1.13}{SEOBNRv4TsurrogateLS}{1.09}{SEOBNRv4TsurrogatehighspinRIFT}{0.00}{SEOBNRv4TsurrogatelowspinRIFT}{0.00}{TEOBResumS-HS}{0.00}{TEOBResumS-LS}{0.00}{TaylorF2-HS}{1.22}{TaylorF2-LS}{1.11}{PrecessingSpinIMRTidalHS}{3.23}{PrecessingSpinIMRTidalLS}{2.86}{PublicationSamples}{3.23}}}
\newcommand{\phijlfourtwofiveplus}[1]{\IfEqCase{#1}{{AlignedSpinInspiralTidalHS}{1.64}{AlignedSpinInspiralTidalLS}{1.75}{AlignedSpinTidalHS}{2.37}{AlignedSpinTidalLS}{2.37}{IMRPhenomDNRTidal-HS}{1.88}{IMRPhenomDNRTidal-LS}{1.83}{IMRPhenomPv2NRTidal-HS}{2.75}{IMRPhenomPv2NRTidal-LS}{3.05}{SEOBNRv4TsurrogateHS}{1.72}{SEOBNRv4TsurrogateLS}{1.76}{SEOBNRv4TsurrogatehighspinRIFT}{3.14}{SEOBNRv4TsurrogatelowspinRIFT}{3.14}{TEOBResumS-HS}{3.14}{TEOBResumS-LS}{3.14}{TaylorF2-HS}{1.63}{TaylorF2-LS}{1.75}{PrecessingSpinIMRTidalHS}{2.76}{PrecessingSpinIMRTidalLS}{3.07}{PublicationSamples}{2.75}}}
\newcommand{\tilttwofourtwofiveminus}[1]{\IfEqCase{#1}{{AlignedSpinInspiralTidalHS}{0.00}{AlignedSpinInspiralTidalLS}{0.00}{AlignedSpinTidalHS}{0.00}{AlignedSpinTidalLS}{0.00}{IMRPhenomDNRTidal-HS}{0.00}{IMRPhenomDNRTidal-LS}{0.00}{IMRPhenomPv2NRTidal-HS}{0.87}{IMRPhenomPv2NRTidal-LS}{0.78}{SEOBNRv4TsurrogateHS}{0.00}{SEOBNRv4TsurrogateLS}{0.00}{SEOBNRv4TsurrogatehighspinRIFT}{0.00}{SEOBNRv4TsurrogatelowspinRIFT}{0.00}{TEOBResumS-HS}{0.00}{TEOBResumS-LS}{0.00}{TaylorF2-HS}{0.00}{TaylorF2-LS}{0.00}{PrecessingSpinIMRTidalHS}{0.87}{PrecessingSpinIMRTidalLS}{0.79}{PublicationSamples}{0.87}}}
\newcommand{\tilttwofourtwofivemed}[1]{\IfEqCase{#1}{{AlignedSpinInspiralTidalHS}{0.00}{AlignedSpinInspiralTidalLS}{0.00}{AlignedSpinTidalHS}{0.00}{AlignedSpinTidalLS}{0.00}{IMRPhenomDNRTidal-HS}{0.00}{IMRPhenomDNRTidal-LS}{0.00}{IMRPhenomPv2NRTidal-HS}{1.41}{IMRPhenomPv2NRTidal-LS}{1.09}{SEOBNRv4TsurrogateHS}{0.00}{SEOBNRv4TsurrogateLS}{0.00}{SEOBNRv4TsurrogatehighspinRIFT}{0.00}{SEOBNRv4TsurrogatelowspinRIFT}{0.00}{TEOBResumS-HS}{0.00}{TEOBResumS-LS}{0.00}{TaylorF2-HS}{0.00}{TaylorF2-LS}{0.00}{PrecessingSpinIMRTidalHS}{1.41}{PrecessingSpinIMRTidalLS}{1.09}{PublicationSamples}{1.41}}}
\newcommand{\tilttwofourtwofiveplus}[1]{\IfEqCase{#1}{{AlignedSpinInspiralTidalHS}{3.14}{AlignedSpinInspiralTidalLS}{3.14}{AlignedSpinTidalHS}{3.14}{AlignedSpinTidalLS}{3.14}{IMRPhenomDNRTidal-HS}{3.14}{IMRPhenomDNRTidal-LS}{3.14}{IMRPhenomPv2NRTidal-HS}{0.94}{IMRPhenomPv2NRTidal-LS}{1.21}{SEOBNRv4TsurrogateHS}{3.14}{SEOBNRv4TsurrogateLS}{3.14}{SEOBNRv4TsurrogatehighspinRIFT}{3.14}{SEOBNRv4TsurrogatelowspinRIFT}{3.14}{TEOBResumS-HS}{3.14}{TEOBResumS-LS}{3.14}{TaylorF2-HS}{3.14}{TaylorF2-LS}{3.14}{PrecessingSpinIMRTidalHS}{0.94}{PrecessingSpinIMRTidalLS}{1.20}{PublicationSamples}{0.94}}}
\newcommand{\costhetajnfourtwofiveminus}[1]{\IfEqCase{#1}{{AlignedSpinInspiralTidalHS}{1.30}{AlignedSpinInspiralTidalLS}{1.40}{AlignedSpinTidalHS}{1.44}{AlignedSpinTidalLS}{1.44}{IMRPhenomDNRTidal-HS}{1.53}{IMRPhenomDNRTidal-LS}{1.48}{IMRPhenomPv2NRTidal-HS}{1.43}{IMRPhenomPv2NRTidal-LS}{1.44}{SEOBNRv4TsurrogateHS}{1.38}{SEOBNRv4TsurrogateLS}{1.42}{SEOBNRv4TsurrogatehighspinRIFT}{1.41}{SEOBNRv4TsurrogatelowspinRIFT}{1.42}{TEOBResumS-HS}{1.42}{TEOBResumS-LS}{1.40}{TaylorF2-HS}{1.30}{TaylorF2-LS}{1.40}{PrecessingSpinIMRTidalHS}{1.43}{PrecessingSpinIMRTidalLS}{1.44}{PublicationSamples}{1.43}}}
\newcommand{\costhetajnfourtwofivemed}[1]{\IfEqCase{#1}{{AlignedSpinInspiralTidalHS}{0.34}{AlignedSpinInspiralTidalLS}{0.44}{AlignedSpinTidalHS}{0.49}{AlignedSpinTidalLS}{0.49}{IMRPhenomDNRTidal-HS}{0.58}{IMRPhenomDNRTidal-LS}{0.53}{IMRPhenomPv2NRTidal-HS}{0.47}{IMRPhenomPv2NRTidal-LS}{0.48}{SEOBNRv4TsurrogateHS}{0.43}{SEOBNRv4TsurrogateLS}{0.46}{SEOBNRv4TsurrogatehighspinRIFT}{0.45}{SEOBNRv4TsurrogatelowspinRIFT}{0.46}{TEOBResumS-HS}{0.46}{TEOBResumS-LS}{0.44}{TaylorF2-HS}{0.34}{TaylorF2-LS}{0.44}{PrecessingSpinIMRTidalHS}{0.47}{PrecessingSpinIMRTidalLS}{0.48}{PublicationSamples}{0.47}}}
\newcommand{\costhetajnfourtwofiveplus}[1]{\IfEqCase{#1}{{AlignedSpinInspiralTidalHS}{0.62}{AlignedSpinInspiralTidalLS}{0.53}{AlignedSpinTidalHS}{0.49}{AlignedSpinTidalLS}{0.49}{IMRPhenomDNRTidal-HS}{0.40}{IMRPhenomDNRTidal-LS}{0.45}{IMRPhenomPv2NRTidal-HS}{0.50}{IMRPhenomPv2NRTidal-LS}{0.49}{SEOBNRv4TsurrogateHS}{0.54}{SEOBNRv4TsurrogateLS}{0.51}{SEOBNRv4TsurrogatehighspinRIFT}{0.52}{SEOBNRv4TsurrogatelowspinRIFT}{0.51}{TEOBResumS-HS}{0.51}{TEOBResumS-LS}{0.54}{TaylorF2-HS}{0.63}{TaylorF2-LS}{0.53}{PrecessingSpinIMRTidalHS}{0.50}{PrecessingSpinIMRTidalLS}{0.49}{PublicationSamples}{0.50}}}
\newcommand{\spintwofourtwofiveminus}[1]{\IfEqCase{#1}{{AlignedSpinInspiralTidalHS}{0.07}{AlignedSpinInspiralTidalLS}{0.01}{AlignedSpinTidalHS}{0.07}{AlignedSpinTidalLS}{0.01}{IMRPhenomDNRTidal-HS}{0.10}{IMRPhenomDNRTidal-LS}{0.01}{IMRPhenomPv2NRTidal-HS}{0.25}{IMRPhenomPv2NRTidal-LS}{0.02}{SEOBNRv4TsurrogateHS}{0.06}{SEOBNRv4TsurrogateLS}{0.01}{SEOBNRv4TsurrogatehighspinRIFT}{0.06}{SEOBNRv4TsurrogatelowspinRIFT}{0.01}{TEOBResumS-HS}{0.06}{TEOBResumS-LS}{0.01}{TaylorF2-HS}{0.07}{TaylorF2-LS}{0.01}{PrecessingSpinIMRTidalHS}{0.25}{PrecessingSpinIMRTidalLS}{0.02}{PublicationSamples}{0.25}}}
\newcommand{\spintwofourtwofivemed}[1]{\IfEqCase{#1}{{AlignedSpinInspiralTidalHS}{0.08}{AlignedSpinInspiralTidalLS}{0.01}{AlignedSpinTidalHS}{0.07}{AlignedSpinTidalLS}{0.01}{IMRPhenomDNRTidal-HS}{0.11}{IMRPhenomDNRTidal-LS}{0.01}{IMRPhenomPv2NRTidal-HS}{0.28}{IMRPhenomPv2NRTidal-LS}{0.03}{SEOBNRv4TsurrogateHS}{0.06}{SEOBNRv4TsurrogateLS}{0.01}{SEOBNRv4TsurrogatehighspinRIFT}{0.06}{SEOBNRv4TsurrogatelowspinRIFT}{0.01}{TEOBResumS-HS}{0.06}{TEOBResumS-LS}{0.01}{TaylorF2-HS}{0.08}{TaylorF2-LS}{0.01}{PrecessingSpinIMRTidalHS}{0.28}{PrecessingSpinIMRTidalLS}{0.03}{PublicationSamples}{0.28}}}
\newcommand{\spintwofourtwofiveplus}[1]{\IfEqCase{#1}{{AlignedSpinInspiralTidalHS}{0.26}{AlignedSpinInspiralTidalLS}{0.03}{AlignedSpinTidalHS}{0.27}{AlignedSpinTidalLS}{0.03}{IMRPhenomDNRTidal-HS}{0.38}{IMRPhenomDNRTidal-LS}{0.03}{IMRPhenomPv2NRTidal-HS}{0.51}{IMRPhenomPv2NRTidal-LS}{0.02}{SEOBNRv4TsurrogateHS}{0.19}{SEOBNRv4TsurrogateLS}{0.03}{SEOBNRv4TsurrogatehighspinRIFT}{0.19}{SEOBNRv4TsurrogatelowspinRIFT}{0.03}{TEOBResumS-HS}{0.19}{TEOBResumS-LS}{0.03}{TaylorF2-HS}{0.26}{TaylorF2-LS}{0.03}{PrecessingSpinIMRTidalHS}{0.51}{PrecessingSpinIMRTidalLS}{0.02}{PublicationSamples}{0.51}}}
\newcommand{\massonedetfourtwofiveminus}[1]{\IfEqCase{#1}{{AlignedSpinInspiralTidalHS}{0.3}{AlignedSpinInspiralTidalLS}{0.09}{AlignedSpinTidalHS}{0.2}{AlignedSpinTidalLS}{0.09}{IMRPhenomDNRTidal-HS}{0.3}{IMRPhenomDNRTidal-LS}{0.09}{IMRPhenomPv2NRTidal-HS}{0.4}{IMRPhenomPv2NRTidal-LS}{0.09}{SEOBNRv4TsurrogateHS}{0.2}{SEOBNRv4TsurrogateLS}{0.09}{SEOBNRv4TsurrogatehighspinRIFT}{0.2}{SEOBNRv4TsurrogatelowspinRIFT}{0.09}{TEOBResumS-HS}{0.2}{TEOBResumS-LS}{0.09}{TaylorF2-HS}{0.3}{TaylorF2-LS}{0.09}{PrecessingSpinIMRTidalHS}{0.4}{PrecessingSpinIMRTidalLS}{0.09}{PublicationSamples}{0.4}}}
\newcommand{\massonedetfourtwofivemed}[1]{\IfEqCase{#1}{{AlignedSpinInspiralTidalHS}{2.0}{AlignedSpinInspiralTidalLS}{1.81}{AlignedSpinTidalHS}{2.0}{AlignedSpinTidalLS}{1.81}{IMRPhenomDNRTidal-HS}{2.0}{IMRPhenomDNRTidal-LS}{1.81}{IMRPhenomPv2NRTidal-HS}{2.1}{IMRPhenomPv2NRTidal-LS}{1.80}{SEOBNRv4TsurrogateHS}{2.0}{SEOBNRv4TsurrogateLS}{1.80}{SEOBNRv4TsurrogatehighspinRIFT}{1.9}{SEOBNRv4TsurrogatelowspinRIFT}{1.81}{TEOBResumS-HS}{2.0}{TEOBResumS-LS}{1.81}{TaylorF2-HS}{2.0}{TaylorF2-LS}{1.81}{PrecessingSpinIMRTidalHS}{2.1}{PrecessingSpinIMRTidalLS}{1.80}{PublicationSamples}{2.1}}}
\newcommand{\massonedetfourtwofiveplus}[1]{\IfEqCase{#1}{{AlignedSpinInspiralTidalHS}{0.5}{AlignedSpinInspiralTidalLS}{0.2}{AlignedSpinTidalHS}{0.6}{AlignedSpinTidalLS}{0.2}{IMRPhenomDNRTidal-HS}{0.7}{IMRPhenomDNRTidal-LS}{0.2}{IMRPhenomPv2NRTidal-HS}{0.6}{IMRPhenomPv2NRTidal-LS}{0.2}{SEOBNRv4TsurrogateHS}{0.5}{SEOBNRv4TsurrogateLS}{0.2}{SEOBNRv4TsurrogatehighspinRIFT}{0.5}{SEOBNRv4TsurrogatelowspinRIFT}{0.2}{TEOBResumS-HS}{0.5}{TEOBResumS-LS}{0.2}{TaylorF2-HS}{0.5}{TaylorF2-LS}{0.2}{PrecessingSpinIMRTidalHS}{0.6}{PrecessingSpinIMRTidalLS}{0.2}{PublicationSamples}{0.6}}}
\newcommand{\spintwoxfourtwofiveminus}[1]{\IfEqCase{#1}{{AlignedSpinInspiralTidalHS}{0.00}{AlignedSpinInspiralTidalLS}{0.00}{AlignedSpinTidalHS}{0.00}{AlignedSpinTidalLS}{0.00}{IMRPhenomDNRTidal-HS}{0.00}{IMRPhenomDNRTidal-LS}{0.00}{IMRPhenomPv2NRTidal-HS}{0.47}{IMRPhenomPv2NRTidal-LS}{0.03}{SEOBNRv4TsurrogateHS}{0.00}{SEOBNRv4TsurrogateLS}{0.00}{SEOBNRv4TsurrogatehighspinRIFT}{0.00}{SEOBNRv4TsurrogatelowspinRIFT}{0.00}{TEOBResumS-HS}{0.00}{TEOBResumS-LS}{0.00}{TaylorF2-HS}{0.00}{TaylorF2-LS}{0.00}{PrecessingSpinIMRTidalHS}{0.47}{PrecessingSpinIMRTidalLS}{0.03}{PublicationSamples}{0.47}}}
\newcommand{\spintwoxfourtwofivemed}[1]{\IfEqCase{#1}{{AlignedSpinInspiralTidalHS}{0.00}{AlignedSpinInspiralTidalLS}{0.00}{AlignedSpinTidalHS}{0.00}{AlignedSpinTidalLS}{0.00}{IMRPhenomDNRTidal-HS}{0.00}{IMRPhenomDNRTidal-LS}{0.00}{IMRPhenomPv2NRTidal-HS}{0.0007}{IMRPhenomPv2NRTidal-LS}{0.00}{SEOBNRv4TsurrogateHS}{0.00}{SEOBNRv4TsurrogateLS}{0.00}{SEOBNRv4TsurrogatehighspinRIFT}{0.00}{SEOBNRv4TsurrogatelowspinRIFT}{0.00}{TEOBResumS-HS}{0.00}{TEOBResumS-LS}{0.00}{TaylorF2-HS}{0.00}{TaylorF2-LS}{0.00}{PrecessingSpinIMRTidalHS}{0.0006}{PrecessingSpinIMRTidalLS}{0.00}{PublicationSamples}{0.0007}}}
\newcommand{\spintwoxfourtwofiveplus}[1]{\IfEqCase{#1}{{AlignedSpinInspiralTidalHS}{0.00}{AlignedSpinInspiralTidalLS}{0.00}{AlignedSpinTidalHS}{0.00}{AlignedSpinTidalLS}{0.00}{IMRPhenomDNRTidal-HS}{0.00}{IMRPhenomDNRTidal-LS}{0.00}{IMRPhenomPv2NRTidal-HS}{0.48}{IMRPhenomPv2NRTidal-LS}{0.03}{SEOBNRv4TsurrogateHS}{0.00}{SEOBNRv4TsurrogateLS}{0.00}{SEOBNRv4TsurrogatehighspinRIFT}{0.00}{SEOBNRv4TsurrogatelowspinRIFT}{0.00}{TEOBResumS-HS}{0.00}{TEOBResumS-LS}{0.00}{TaylorF2-HS}{0.00}{TaylorF2-LS}{0.00}{PrecessingSpinIMRTidalHS}{0.47}{PrecessingSpinIMRTidalLS}{0.03}{PublicationSamples}{0.47}}}
\newcommand{\massratiofourtwofiveminus}[1]{\IfEqCase{#1}{{AlignedSpinInspiralTidalHS}{0.24}{AlignedSpinInspiralTidalLS}{0.15}{AlignedSpinTidalHS}{0.30}{AlignedSpinTidalLS}{0.15}{IMRPhenomDNRTidal-HS}{0.31}{IMRPhenomDNRTidal-LS}{0.15}{IMRPhenomPv2NRTidal-HS}{0.25}{IMRPhenomPv2NRTidal-LS}{0.15}{SEOBNRv4TsurrogateHS}{0.29}{SEOBNRv4TsurrogateLS}{0.15}{SEOBNRv4TsurrogatehighspinRIFT}{0.27}{SEOBNRv4TsurrogatelowspinRIFT}{0.15}{TEOBResumS-HS}{0.27}{TEOBResumS-LS}{0.15}{TaylorF2-HS}{0.24}{TaylorF2-LS}{0.15}{PrecessingSpinIMRTidalHS}{0.25}{PrecessingSpinIMRTidalLS}{0.15}{PublicationSamples}{0.25}}}
\newcommand{\massratiofourtwofivemed}[1]{\IfEqCase{#1}{{AlignedSpinInspiralTidalHS}{0.70}{AlignedSpinInspiralTidalLS}{0.89}{AlignedSpinTidalHS}{0.74}{AlignedSpinTidalLS}{0.89}{IMRPhenomDNRTidal-HS}{0.72}{IMRPhenomDNRTidal-LS}{0.89}{IMRPhenomPv2NRTidal-HS}{0.67}{IMRPhenomPv2NRTidal-LS}{0.90}{SEOBNRv4TsurrogateHS}{0.77}{SEOBNRv4TsurrogateLS}{0.90}{SEOBNRv4TsurrogatehighspinRIFT}{0.78}{SEOBNRv4TsurrogatelowspinRIFT}{0.89}{TEOBResumS-HS}{0.75}{TEOBResumS-LS}{0.89}{TaylorF2-HS}{0.70}{TaylorF2-LS}{0.89}{PrecessingSpinIMRTidalHS}{0.67}{PrecessingSpinIMRTidalLS}{0.90}{PublicationSamples}{0.67}}}
\newcommand{\massratiofourtwofiveplus}[1]{\IfEqCase{#1}{{AlignedSpinInspiralTidalHS}{0.26}{AlignedSpinInspiralTidalLS}{0.10}{AlignedSpinTidalHS}{0.22}{AlignedSpinTidalLS}{0.09}{IMRPhenomDNRTidal-HS}{0.25}{IMRPhenomDNRTidal-LS}{0.09}{IMRPhenomPv2NRTidal-HS}{0.29}{IMRPhenomPv2NRTidal-LS}{0.09}{SEOBNRv4TsurrogateHS}{0.20}{SEOBNRv4TsurrogateLS}{0.09}{SEOBNRv4TsurrogatehighspinRIFT}{0.19}{SEOBNRv4TsurrogatelowspinRIFT}{0.10}{TEOBResumS-HS}{0.22}{TEOBResumS-LS}{0.10}{TaylorF2-HS}{0.26}{TaylorF2-LS}{0.10}{PrecessingSpinIMRTidalHS}{0.29}{PrecessingSpinIMRTidalLS}{0.09}{PublicationSamples}{0.29}}}
\newcommand{\spinonefourtwofiveminus}[1]{\IfEqCase{#1}{{AlignedSpinInspiralTidalHS}{0.06}{AlignedSpinInspiralTidalLS}{0.01}{AlignedSpinTidalHS}{0.06}{AlignedSpinTidalLS}{0.01}{IMRPhenomDNRTidal-HS}{0.08}{IMRPhenomDNRTidal-LS}{0.01}{IMRPhenomPv2NRTidal-HS}{0.25}{IMRPhenomPv2NRTidal-LS}{0.03}{SEOBNRv4TsurrogateHS}{0.05}{SEOBNRv4TsurrogateLS}{0.01}{SEOBNRv4TsurrogatehighspinRIFT}{0.05}{SEOBNRv4TsurrogatelowspinRIFT}{0.01}{TEOBResumS-HS}{0.05}{TEOBResumS-LS}{0.01}{TaylorF2-HS}{0.06}{TaylorF2-LS}{0.01}{PrecessingSpinIMRTidalHS}{0.25}{PrecessingSpinIMRTidalLS}{0.03}{PublicationSamples}{0.25}}}
\newcommand{\spinonefourtwofivemed}[1]{\IfEqCase{#1}{{AlignedSpinInspiralTidalHS}{0.06}{AlignedSpinInspiralTidalLS}{0.01}{AlignedSpinTidalHS}{0.06}{AlignedSpinTidalLS}{0.01}{IMRPhenomDNRTidal-HS}{0.09}{IMRPhenomDNRTidal-LS}{0.01}{IMRPhenomPv2NRTidal-HS}{0.27}{IMRPhenomPv2NRTidal-LS}{0.03}{SEOBNRv4TsurrogateHS}{0.06}{SEOBNRv4TsurrogateLS}{0.01}{SEOBNRv4TsurrogatehighspinRIFT}{0.06}{SEOBNRv4TsurrogatelowspinRIFT}{0.01}{TEOBResumS-HS}{0.06}{TEOBResumS-LS}{0.01}{TaylorF2-HS}{0.06}{TaylorF2-LS}{0.01}{PrecessingSpinIMRTidalHS}{0.27}{PrecessingSpinIMRTidalLS}{0.03}{PublicationSamples}{0.27}}}
\newcommand{\spinonefourtwofiveplus}[1]{\IfEqCase{#1}{{AlignedSpinInspiralTidalHS}{0.17}{AlignedSpinInspiralTidalLS}{0.03}{AlignedSpinTidalHS}{0.19}{AlignedSpinTidalLS}{0.03}{IMRPhenomDNRTidal-HS}{0.25}{IMRPhenomDNRTidal-LS}{0.03}{IMRPhenomPv2NRTidal-HS}{0.51}{IMRPhenomPv2NRTidal-LS}{0.02}{SEOBNRv4TsurrogateHS}{0.15}{SEOBNRv4TsurrogateLS}{0.03}{SEOBNRv4TsurrogatehighspinRIFT}{0.15}{SEOBNRv4TsurrogatelowspinRIFT}{0.03}{TEOBResumS-HS}{0.14}{TEOBResumS-LS}{0.03}{TaylorF2-HS}{0.17}{TaylorF2-LS}{0.03}{PrecessingSpinIMRTidalHS}{0.51}{PrecessingSpinIMRTidalLS}{0.02}{PublicationSamples}{0.51}}}
\newcommand{\costiltonefourtwofiveminus}[1]{\IfEqCase{#1}{{AlignedSpinInspiralTidalHS}{2.00}{AlignedSpinInspiralTidalLS}{2.00}{AlignedSpinTidalHS}{2.00}{AlignedSpinTidalLS}{2.00}{IMRPhenomDNRTidal-HS}{2.00}{IMRPhenomDNRTidal-LS}{2.00}{IMRPhenomPv2NRTidal-HS}{0.65}{IMRPhenomPv2NRTidal-LS}{1.10}{SEOBNRv4TsurrogateHS}{2.00}{SEOBNRv4TsurrogateLS}{2.00}{SEOBNRv4TsurrogatehighspinRIFT}{2.00}{SEOBNRv4TsurrogatelowspinRIFT}{2.00}{TEOBResumS-HS}{2.00}{TEOBResumS-LS}{2.00}{TaylorF2-HS}{2.00}{TaylorF2-LS}{2.00}{PrecessingSpinIMRTidalHS}{0.65}{PrecessingSpinIMRTidalLS}{1.10}{PublicationSamples}{0.65}}}
\newcommand{\costiltonefourtwofivemed}[1]{\IfEqCase{#1}{{AlignedSpinInspiralTidalHS}{1.00}{AlignedSpinInspiralTidalLS}{1.00}{AlignedSpinTidalHS}{1.00}{AlignedSpinTidalLS}{1.00}{IMRPhenomDNRTidal-HS}{1.00}{IMRPhenomDNRTidal-LS}{1.00}{IMRPhenomPv2NRTidal-HS}{0.26}{IMRPhenomPv2NRTidal-LS}{0.51}{SEOBNRv4TsurrogateHS}{1.00}{SEOBNRv4TsurrogateLS}{1.00}{SEOBNRv4TsurrogatehighspinRIFT}{1.00}{SEOBNRv4TsurrogatelowspinRIFT}{1.00}{TEOBResumS-HS}{1.00}{TEOBResumS-LS}{1.00}{TaylorF2-HS}{1.00}{TaylorF2-LS}{1.00}{PrecessingSpinIMRTidalHS}{0.26}{PrecessingSpinIMRTidalLS}{0.51}{PublicationSamples}{0.26}}}
\newcommand{\costiltonefourtwofiveplus}[1]{\IfEqCase{#1}{{AlignedSpinInspiralTidalHS}{0.00}{AlignedSpinInspiralTidalLS}{0.00}{AlignedSpinTidalHS}{0.00}{AlignedSpinTidalLS}{0.00}{IMRPhenomDNRTidal-HS}{0.00}{IMRPhenomDNRTidal-LS}{0.00}{IMRPhenomPv2NRTidal-HS}{0.61}{IMRPhenomPv2NRTidal-LS}{0.45}{SEOBNRv4TsurrogateHS}{0.00}{SEOBNRv4TsurrogateLS}{0.00}{SEOBNRv4TsurrogatehighspinRIFT}{0.00}{SEOBNRv4TsurrogatelowspinRIFT}{0.00}{TEOBResumS-HS}{0.00}{TEOBResumS-LS}{0.00}{TaylorF2-HS}{0.00}{TaylorF2-LS}{0.00}{PrecessingSpinIMRTidalHS}{0.61}{PrecessingSpinIMRTidalLS}{0.44}{PublicationSamples}{0.61}}}
\newcommand{\phasefourtwofiveminus}[1]{\IfEqCase{#1}{{AlignedSpinInspiralTidalHS}{3.10}{AlignedSpinInspiralTidalLS}{2.74}{AlignedSpinTidalHS}{2.81}{AlignedSpinTidalLS}{2.76}{IMRPhenomDNRTidal-HS}{2.83}{IMRPhenomDNRTidal-LS}{2.62}{IMRPhenomPv2NRTidal-HS}{2.82}{IMRPhenomPv2NRTidal-LS}{2.85}{SEOBNRv4TsurrogateHS}{2.74}{SEOBNRv4TsurrogateLS}{2.81}{SEOBNRv4TsurrogatehighspinRIFT}{2.78}{SEOBNRv4TsurrogatelowspinRIFT}{2.79}{TEOBResumS-HS}{2.78}{TEOBResumS-LS}{2.83}{TaylorF2-HS}{3.13}{TaylorF2-LS}{2.74}{PrecessingSpinIMRTidalHS}{2.82}{PrecessingSpinIMRTidalLS}{2.86}{PublicationSamples}{2.82}}}
\newcommand{\phasefourtwofivemed}[1]{\IfEqCase{#1}{{AlignedSpinInspiralTidalHS}{3.44}{AlignedSpinInspiralTidalLS}{3.01}{AlignedSpinTidalHS}{3.13}{AlignedSpinTidalLS}{3.07}{IMRPhenomDNRTidal-HS}{3.13}{IMRPhenomDNRTidal-LS}{2.88}{IMRPhenomPv2NRTidal-HS}{3.13}{IMRPhenomPv2NRTidal-LS}{3.20}{SEOBNRv4TsurrogateHS}{3.11}{SEOBNRv4TsurrogateLS}{3.16}{SEOBNRv4TsurrogatehighspinRIFT}{3.12}{SEOBNRv4TsurrogatelowspinRIFT}{3.12}{TEOBResumS-HS}{3.09}{TEOBResumS-LS}{3.14}{TaylorF2-HS}{3.46}{TaylorF2-LS}{3.00}{PrecessingSpinIMRTidalHS}{3.12}{PrecessingSpinIMRTidalLS}{3.20}{PublicationSamples}{3.13}}}
\newcommand{\phasefourtwofiveplus}[1]{\IfEqCase{#1}{{AlignedSpinInspiralTidalHS}{2.49}{AlignedSpinInspiralTidalLS}{2.93}{AlignedSpinTidalHS}{2.82}{AlignedSpinTidalLS}{2.88}{IMRPhenomDNRTidal-HS}{2.82}{IMRPhenomDNRTidal-LS}{3.04}{IMRPhenomPv2NRTidal-HS}{2.86}{IMRPhenomPv2NRTidal-LS}{2.74}{SEOBNRv4TsurrogateHS}{2.81}{SEOBNRv4TsurrogateLS}{2.74}{SEOBNRv4TsurrogatehighspinRIFT}{2.81}{SEOBNRv4TsurrogatelowspinRIFT}{2.79}{TEOBResumS-HS}{2.87}{TEOBResumS-LS}{2.86}{TaylorF2-HS}{2.47}{TaylorF2-LS}{2.94}{PrecessingSpinIMRTidalHS}{2.87}{PrecessingSpinIMRTidalLS}{2.73}{PublicationSamples}{2.86}}}
\newcommand{\masstwodetfourtwofiveminus}[1]{\IfEqCase{#1}{{AlignedSpinInspiralTidalHS}{0.3}{AlignedSpinInspiralTidalLS}{0.1}{AlignedSpinTidalHS}{0.3}{AlignedSpinTidalLS}{0.1}{IMRPhenomDNRTidal-HS}{0.3}{IMRPhenomDNRTidal-LS}{0.1}{IMRPhenomPv2NRTidal-HS}{0.3}{IMRPhenomPv2NRTidal-LS}{0.1}{SEOBNRv4TsurrogateHS}{0.3}{SEOBNRv4TsurrogateLS}{0.1}{SEOBNRv4TsurrogatehighspinRIFT}{0.3}{SEOBNRv4TsurrogatelowspinRIFT}{0.1}{TEOBResumS-HS}{0.3}{TEOBResumS-LS}{0.1}{TaylorF2-HS}{0.3}{TaylorF2-LS}{0.1}{PrecessingSpinIMRTidalHS}{0.3}{PrecessingSpinIMRTidalLS}{0.1}{PublicationSamples}{0.3}}}
\newcommand{\masstwodetfourtwofivemed}[1]{\IfEqCase{#1}{{AlignedSpinInspiralTidalHS}{1.4}{AlignedSpinInspiralTidalLS}{1.61}{AlignedSpinTidalHS}{1.5}{AlignedSpinTidalLS}{1.61}{IMRPhenomDNRTidal-HS}{1.5}{IMRPhenomDNRTidal-LS}{1.62}{IMRPhenomPv2NRTidal-HS}{1.4}{IMRPhenomPv2NRTidal-LS}{1.62}{SEOBNRv4TsurrogateHS}{1.5}{SEOBNRv4TsurrogateLS}{1.62}{SEOBNRv4TsurrogatehighspinRIFT}{1.5}{SEOBNRv4TsurrogatelowspinRIFT}{1.61}{TEOBResumS-HS}{1.5}{TEOBResumS-LS}{1.61}{TaylorF2-HS}{1.4}{TaylorF2-LS}{1.61}{PrecessingSpinIMRTidalHS}{1.4}{PrecessingSpinIMRTidalLS}{1.62}{PublicationSamples}{1.4}}}
\newcommand{\masstwodetfourtwofiveplus}[1]{\IfEqCase{#1}{{AlignedSpinInspiralTidalHS}{0.2}{AlignedSpinInspiralTidalLS}{0.09}{AlignedSpinTidalHS}{0.2}{AlignedSpinTidalLS}{0.08}{IMRPhenomDNRTidal-HS}{0.2}{IMRPhenomDNRTidal-LS}{0.08}{IMRPhenomPv2NRTidal-HS}{0.3}{IMRPhenomPv2NRTidal-LS}{0.08}{SEOBNRv4TsurrogateHS}{0.2}{SEOBNRv4TsurrogateLS}{0.08}{SEOBNRv4TsurrogatehighspinRIFT}{0.2}{SEOBNRv4TsurrogatelowspinRIFT}{0.08}{TEOBResumS-HS}{0.2}{TEOBResumS-LS}{0.09}{TaylorF2-HS}{0.2}{TaylorF2-LS}{0.09}{PrecessingSpinIMRTidalHS}{0.3}{PrecessingSpinIMRTidalLS}{0.08}{PublicationSamples}{0.3}}}
\newcommand{\masstwosourcefourtwofiveminus}[1]{\IfEqCase{#1}{{AlignedSpinInspiralTidalHS}{0.3}{AlignedSpinInspiralTidalLS}{0.1}{AlignedSpinTidalHS}{0.3}{AlignedSpinTidalLS}{0.1}{IMRPhenomDNRTidal-HS}{0.3}{IMRPhenomDNRTidal-LS}{0.1}{IMRPhenomPv2NRTidal-HS}{0.3}{IMRPhenomPv2NRTidal-LS}{0.1}{SEOBNRv4TsurrogateHS}{0.3}{SEOBNRv4TsurrogateLS}{0.1}{SEOBNRv4TsurrogatehighspinRIFT}{0.3}{SEOBNRv4TsurrogatelowspinRIFT}{0.1}{TEOBResumS-HS}{0.3}{TEOBResumS-LS}{0.1}{TaylorF2-HS}{0.3}{TaylorF2-LS}{0.1}{PrecessingSpinIMRTidalHS}{0.3}{PrecessingSpinIMRTidalLS}{0.1}{PublicationSamples}{0.3}}}
\newcommand{\masstwosourcefourtwofivemed}[1]{\IfEqCase{#1}{{AlignedSpinInspiralTidalHS}{1.4}{AlignedSpinInspiralTidalLS}{1.56}{AlignedSpinTidalHS}{1.4}{AlignedSpinTidalLS}{1.56}{IMRPhenomDNRTidal-HS}{1.4}{IMRPhenomDNRTidal-LS}{1.56}{IMRPhenomPv2NRTidal-HS}{1.4}{IMRPhenomPv2NRTidal-LS}{1.57}{SEOBNRv4TsurrogateHS}{1.4}{SEOBNRv4TsurrogateLS}{1.56}{SEOBNRv4TsurrogatehighspinRIFT}{1.5}{SEOBNRv4TsurrogatelowspinRIFT}{1.56}{TEOBResumS-HS}{1.4}{TEOBResumS-LS}{1.56}{TaylorF2-HS}{1.4}{TaylorF2-LS}{1.56}{PrecessingSpinIMRTidalHS}{1.4}{PrecessingSpinIMRTidalLS}{1.57}{PublicationSamples}{1.4}}}
\newcommand{\masstwosourcefourtwofiveplus}[1]{\IfEqCase{#1}{{AlignedSpinInspiralTidalHS}{0.2}{AlignedSpinInspiralTidalLS}{0.09}{AlignedSpinTidalHS}{0.2}{AlignedSpinTidalLS}{0.08}{IMRPhenomDNRTidal-HS}{0.2}{IMRPhenomDNRTidal-LS}{0.08}{IMRPhenomPv2NRTidal-HS}{0.3}{IMRPhenomPv2NRTidal-LS}{0.08}{SEOBNRv4TsurrogateHS}{0.2}{SEOBNRv4TsurrogateLS}{0.08}{SEOBNRv4TsurrogatehighspinRIFT}{0.2}{SEOBNRv4TsurrogatelowspinRIFT}{0.08}{TEOBResumS-HS}{0.2}{TEOBResumS-LS}{0.09}{TaylorF2-HS}{0.2}{TaylorF2-LS}{0.09}{PrecessingSpinIMRTidalHS}{0.3}{PrecessingSpinIMRTidalLS}{0.08}{PublicationSamples}{0.3}}}
\newcommand{\decfourtwofiveminus}[1]{\IfEqCase{#1}{{AlignedSpinInspiralTidalHS}{0.92098}{AlignedSpinInspiralTidalLS}{0.94910}{AlignedSpinTidalHS}{0.88455}{AlignedSpinTidalLS}{0.88762}{IMRPhenomDNRTidal-HS}{0.92222}{IMRPhenomDNRTidal-LS}{0.90469}{IMRPhenomPv2NRTidal-HS}{0.90104}{IMRPhenomPv2NRTidal-LS}{0.97042}{SEOBNRv4TsurrogateHS}{0.87704}{SEOBNRv4TsurrogateLS}{0.89335}{SEOBNRv4TsurrogatehighspinRIFT}{0.89796}{SEOBNRv4TsurrogatelowspinRIFT}{0.90269}{TEOBResumS-HS}{0.88789}{TEOBResumS-LS}{0.89751}{TaylorF2-HS}{0.92183}{TaylorF2-LS}{0.95570}{PrecessingSpinIMRTidalHS}{0.89977}{PrecessingSpinIMRTidalLS}{0.96778}{PublicationSamples}{0.89807}}}
\newcommand{\decfourtwofivemed}[1]{\IfEqCase{#1}{{AlignedSpinInspiralTidalHS}{-0.14949}{AlignedSpinInspiralTidalLS}{-0.10865}{AlignedSpinTidalHS}{-0.15883}{AlignedSpinTidalLS}{-0.13405}{IMRPhenomDNRTidal-HS}{-0.18418}{IMRPhenomDNRTidal-LS}{-0.06597}{IMRPhenomPv2NRTidal-HS}{-0.12984}{IMRPhenomPv2NRTidal-LS}{-0.05685}{SEOBNRv4TsurrogateHS}{-0.14893}{SEOBNRv4TsurrogateLS}{-0.17562}{SEOBNRv4TsurrogatehighspinRIFT}{-0.15053}{SEOBNRv4TsurrogatelowspinRIFT}{-0.12883}{TEOBResumS-HS}{-0.15241}{TEOBResumS-LS}{-0.16207}{TaylorF2-HS}{-0.14824}{TaylorF2-LS}{-0.10463}{PrecessingSpinIMRTidalHS}{-0.13006}{PrecessingSpinIMRTidalLS}{-0.06120}{PublicationSamples}{-0.13133}}}
\newcommand{\decfourtwofiveplus}[1]{\IfEqCase{#1}{{AlignedSpinInspiralTidalHS}{0.99909}{AlignedSpinInspiralTidalLS}{0.93177}{AlignedSpinTidalHS}{0.97810}{AlignedSpinTidalLS}{0.98373}{IMRPhenomDNRTidal-HS}{0.95052}{IMRPhenomDNRTidal-LS}{0.93229}{IMRPhenomPv2NRTidal-HS}{0.96946}{IMRPhenomPv2NRTidal-LS}{0.91574}{SEOBNRv4TsurrogateHS}{1.00436}{SEOBNRv4TsurrogateLS}{1.00055}{SEOBNRv4TsurrogatehighspinRIFT}{0.99086}{SEOBNRv4TsurrogatelowspinRIFT}{0.97101}{TEOBResumS-HS}{0.98792}{TEOBResumS-LS}{1.00135}{TaylorF2-HS}{0.99948}{TaylorF2-LS}{0.92354}{PrecessingSpinIMRTidalHS}{0.96811}{PrecessingSpinIMRTidalLS}{0.91709}{PublicationSamples}{0.96897}}}
\newcommand{\psifourtwofiveminus}[1]{\IfEqCase{#1}{{AlignedSpinInspiralTidalHS}{1.40}{AlignedSpinInspiralTidalLS}{1.41}{AlignedSpinTidalHS}{1.63}{AlignedSpinTidalLS}{1.69}{IMRPhenomDNRTidal-HS}{1.36}{IMRPhenomDNRTidal-LS}{1.41}{IMRPhenomPv2NRTidal-HS}{1.46}{IMRPhenomPv2NRTidal-LS}{1.40}{SEOBNRv4TsurrogateHS}{1.36}{SEOBNRv4TsurrogateLS}{1.43}{SEOBNRv4TsurrogatehighspinRIFT}{2.83}{SEOBNRv4TsurrogatelowspinRIFT}{2.86}{TEOBResumS-HS}{2.86}{TEOBResumS-LS}{2.86}{TaylorF2-HS}{1.41}{TaylorF2-LS}{1.41}{PrecessingSpinIMRTidalHS}{1.46}{PrecessingSpinIMRTidalLS}{1.39}{PublicationSamples}{1.46}}}
\newcommand{\psifourtwofivemed}[1]{\IfEqCase{#1}{{AlignedSpinInspiralTidalHS}{1.55}{AlignedSpinInspiralTidalLS}{1.56}{AlignedSpinTidalHS}{1.80}{AlignedSpinTidalLS}{1.86}{IMRPhenomDNRTidal-HS}{1.50}{IMRPhenomDNRTidal-LS}{1.56}{IMRPhenomPv2NRTidal-HS}{1.61}{IMRPhenomPv2NRTidal-LS}{1.54}{SEOBNRv4TsurrogateHS}{1.50}{SEOBNRv4TsurrogateLS}{1.56}{SEOBNRv4TsurrogatehighspinRIFT}{3.13}{SEOBNRv4TsurrogatelowspinRIFT}{3.14}{TEOBResumS-HS}{3.16}{TEOBResumS-LS}{3.16}{TaylorF2-HS}{1.56}{TaylorF2-LS}{1.55}{PrecessingSpinIMRTidalHS}{1.62}{PrecessingSpinIMRTidalLS}{1.54}{PublicationSamples}{1.61}}}
\newcommand{\psifourtwofiveplus}[1]{\IfEqCase{#1}{{AlignedSpinInspiralTidalHS}{1.42}{AlignedSpinInspiralTidalLS}{1.44}{AlignedSpinTidalHS}{3.54}{AlignedSpinTidalLS}{3.48}{IMRPhenomDNRTidal-HS}{1.50}{IMRPhenomDNRTidal-LS}{1.43}{IMRPhenomPv2NRTidal-HS}{1.38}{IMRPhenomPv2NRTidal-LS}{1.43}{SEOBNRv4TsurrogateHS}{1.47}{SEOBNRv4TsurrogateLS}{1.42}{SEOBNRv4TsurrogatehighspinRIFT}{2.85}{SEOBNRv4TsurrogatelowspinRIFT}{2.83}{TEOBResumS-HS}{2.82}{TEOBResumS-LS}{2.84}{TaylorF2-HS}{1.41}{TaylorF2-LS}{1.44}{PrecessingSpinIMRTidalHS}{1.38}{PrecessingSpinIMRTidalLS}{1.43}{PublicationSamples}{1.38}}}
\newcommand{\networkoptimalsnrfourtwofiveminus}[1]{\IfEqCase{#1}{{AlignedSpinInspiralTidalHS}{1.7}{AlignedSpinInspiralTidalLS}{1.7}{IMRPhenomDNRTidal-HS}{1.7}{IMRPhenomDNRTidal-LS}{1.7}{IMRPhenomPv2NRTidal-HS}{1.7}{IMRPhenomPv2NRTidal-LS}{1.7}{SEOBNRv4TsurrogateHS}{1.7}{SEOBNRv4TsurrogateLS}{1.7}{TaylorF2-HS}{1.7}{TaylorF2-LS}{1.7}{PrecessingSpinIMRTidalHS}{1.7}{PrecessingSpinIMRTidalLS}{1.7}{PublicationSamples}{1.7}}}
\newcommand{\networkoptimalsnrfourtwofivemed}[1]{\IfEqCase{#1}{{AlignedSpinInspiralTidalHS}{12.1}{AlignedSpinInspiralTidalLS}{12.2}{IMRPhenomDNRTidal-HS}{12.0}{IMRPhenomDNRTidal-LS}{12.1}{IMRPhenomPv2NRTidal-HS}{12.0}{IMRPhenomPv2NRTidal-LS}{12.1}{SEOBNRv4TsurrogateHS}{12.0}{SEOBNRv4TsurrogateLS}{12.1}{TaylorF2-HS}{12.1}{TaylorF2-LS}{12.2}{PrecessingSpinIMRTidalHS}{12.0}{PrecessingSpinIMRTidalLS}{12.1}{PublicationSamples}{12.0}}}
\newcommand{\networkoptimalsnrfourtwofiveplus}[1]{\IfEqCase{#1}{{AlignedSpinInspiralTidalHS}{1.7}{AlignedSpinInspiralTidalLS}{1.7}{IMRPhenomDNRTidal-HS}{1.7}{IMRPhenomDNRTidal-LS}{1.7}{IMRPhenomPv2NRTidal-HS}{1.7}{IMRPhenomPv2NRTidal-LS}{1.7}{SEOBNRv4TsurrogateHS}{1.7}{SEOBNRv4TsurrogateLS}{1.7}{TaylorF2-HS}{1.7}{TaylorF2-LS}{1.7}{PrecessingSpinIMRTidalHS}{1.7}{PrecessingSpinIMRTidalLS}{1.7}{PublicationSamples}{1.7}}}
\newcommand{\thetajnfourtwofiveminus}[1]{\IfEqCase{#1}{{AlignedSpinInspiralTidalHS}{0.97}{AlignedSpinInspiralTidalLS}{0.88}{AlignedSpinTidalHS}{0.83}{AlignedSpinTidalLS}{0.83}{IMRPhenomDNRTidal-HS}{0.74}{IMRPhenomDNRTidal-LS}{0.78}{IMRPhenomPv2NRTidal-HS}{0.85}{IMRPhenomPv2NRTidal-LS}{0.84}{SEOBNRv4TsurrogateHS}{0.89}{SEOBNRv4TsurrogateLS}{0.85}{SEOBNRv4TsurrogatehighspinRIFT}{0.86}{SEOBNRv4TsurrogatelowspinRIFT}{0.86}{TEOBResumS-HS}{0.86}{TEOBResumS-LS}{0.88}{TaylorF2-HS}{0.97}{TaylorF2-LS}{0.88}{PrecessingSpinIMRTidalHS}{0.85}{PrecessingSpinIMRTidalLS}{0.84}{PublicationSamples}{0.85}}}
\newcommand{\thetajnfourtwofivemed}[1]{\IfEqCase{#1}{{AlignedSpinInspiralTidalHS}{1.22}{AlignedSpinInspiralTidalLS}{1.11}{AlignedSpinTidalHS}{1.06}{AlignedSpinTidalLS}{1.06}{IMRPhenomDNRTidal-HS}{0.96}{IMRPhenomDNRTidal-LS}{1.02}{IMRPhenomPv2NRTidal-HS}{1.08}{IMRPhenomPv2NRTidal-LS}{1.07}{SEOBNRv4TsurrogateHS}{1.13}{SEOBNRv4TsurrogateLS}{1.09}{SEOBNRv4TsurrogatehighspinRIFT}{1.10}{SEOBNRv4TsurrogatelowspinRIFT}{1.09}{TEOBResumS-HS}{1.09}{TEOBResumS-LS}{1.12}{TaylorF2-HS}{1.22}{TaylorF2-LS}{1.11}{PrecessingSpinIMRTidalHS}{1.08}{PrecessingSpinIMRTidalLS}{1.07}{PublicationSamples}{1.08}}}
\newcommand{\thetajnfourtwofiveplus}[1]{\IfEqCase{#1}{{AlignedSpinInspiralTidalHS}{1.64}{AlignedSpinInspiralTidalLS}{1.75}{AlignedSpinTidalHS}{1.79}{AlignedSpinTidalLS}{1.78}{IMRPhenomDNRTidal-HS}{1.88}{IMRPhenomDNRTidal-LS}{1.83}{IMRPhenomPv2NRTidal-HS}{1.77}{IMRPhenomPv2NRTidal-LS}{1.78}{SEOBNRv4TsurrogateHS}{1.72}{SEOBNRv4TsurrogateLS}{1.76}{SEOBNRv4TsurrogatehighspinRIFT}{1.76}{SEOBNRv4TsurrogatelowspinRIFT}{1.75}{TEOBResumS-HS}{1.77}{TEOBResumS-LS}{1.74}{TaylorF2-HS}{1.63}{TaylorF2-LS}{1.75}{PrecessingSpinIMRTidalHS}{1.77}{PrecessingSpinIMRTidalLS}{1.78}{PublicationSamples}{1.77}}}
\newcommand{\totalmassdetfourtwofiveminus}[1]{\IfEqCase{#1}{{AlignedSpinInspiralTidalHS}{0.06}{AlignedSpinInspiralTidalLS}{0.007}{AlignedSpinTidalHS}{0.04}{AlignedSpinTidalLS}{0.007}{IMRPhenomDNRTidal-HS}{0.06}{IMRPhenomDNRTidal-LS}{0.006}{IMRPhenomPv2NRTidal-HS}{0.08}{IMRPhenomPv2NRTidal-LS}{0.006}{SEOBNRv4TsurrogateHS}{0.03}{SEOBNRv4TsurrogateLS}{0.006}{SEOBNRv4TsurrogatehighspinRIFT}{0.03}{SEOBNRv4TsurrogatelowspinRIFT}{0.007}{TEOBResumS-HS}{0.04}{TEOBResumS-LS}{0.007}{TaylorF2-HS}{0.06}{TaylorF2-LS}{0.007}{PrecessingSpinIMRTidalHS}{0.08}{PrecessingSpinIMRTidalLS}{0.006}{PublicationSamples}{0.08}}}
\newcommand{\totalmassdetfourtwofivemed}[1]{\IfEqCase{#1}{{AlignedSpinInspiralTidalHS}{3.48}{AlignedSpinInspiralTidalLS}{3.42}{AlignedSpinTidalHS}{3.46}{AlignedSpinTidalLS}{3.42}{IMRPhenomDNRTidal-HS}{3.47}{IMRPhenomDNRTidal-LS}{3.42}{IMRPhenomPv2NRTidal-HS}{3.50}{IMRPhenomPv2NRTidal-LS}{3.42}{SEOBNRv4TsurrogateHS}{3.45}{SEOBNRv4TsurrogateLS}{3.42}{SEOBNRv4TsurrogatehighspinRIFT}{3.45}{SEOBNRv4TsurrogatelowspinRIFT}{3.42}{TEOBResumS-HS}{3.46}{TEOBResumS-LS}{3.42}{TaylorF2-HS}{3.48}{TaylorF2-LS}{3.42}{PrecessingSpinIMRTidalHS}{3.50}{PrecessingSpinIMRTidalLS}{3.42}{PublicationSamples}{3.50}}}
\newcommand{\totalmassdetfourtwofiveplus}[1]{\IfEqCase{#1}{{AlignedSpinInspiralTidalHS}{0.3}{AlignedSpinInspiralTidalLS}{0.04}{AlignedSpinTidalHS}{0.3}{AlignedSpinTidalLS}{0.04}{IMRPhenomDNRTidal-HS}{0.4}{IMRPhenomDNRTidal-LS}{0.04}{IMRPhenomPv2NRTidal-HS}{0.3}{IMRPhenomPv2NRTidal-LS}{0.04}{SEOBNRv4TsurrogateHS}{0.2}{SEOBNRv4TsurrogateLS}{0.04}{SEOBNRv4TsurrogatehighspinRIFT}{0.2}{SEOBNRv4TsurrogatelowspinRIFT}{0.04}{TEOBResumS-HS}{0.2}{TEOBResumS-LS}{0.04}{TaylorF2-HS}{0.3}{TaylorF2-LS}{0.04}{PrecessingSpinIMRTidalHS}{0.3}{PrecessingSpinIMRTidalLS}{0.04}{PublicationSamples}{0.3}}}
\newcommand{\redshiftfourtwofiveminus}[1]{\IfEqCase{#1}{{AlignedSpinInspiralTidalHS}{0.02}{AlignedSpinInspiralTidalLS}{0.02}{AlignedSpinTidalHS}{0.02}{AlignedSpinTidalLS}{0.02}{IMRPhenomDNRTidal-HS}{0.02}{IMRPhenomDNRTidal-LS}{0.02}{IMRPhenomPv2NRTidal-HS}{0.02}{IMRPhenomPv2NRTidal-LS}{0.02}{SEOBNRv4TsurrogateHS}{0.02}{SEOBNRv4TsurrogateLS}{0.02}{SEOBNRv4TsurrogatehighspinRIFT}{0.02}{SEOBNRv4TsurrogatelowspinRIFT}{0.02}{TEOBResumS-HS}{0.02}{TEOBResumS-LS}{0.02}{TaylorF2-HS}{0.02}{TaylorF2-LS}{0.02}{PrecessingSpinIMRTidalHS}{0.02}{PrecessingSpinIMRTidalLS}{0.02}{PublicationSamples}{0.02}}}
\newcommand{\redshiftfourtwofivemed}[1]{\IfEqCase{#1}{{AlignedSpinInspiralTidalHS}{0.04}{AlignedSpinInspiralTidalLS}{0.04}{AlignedSpinTidalHS}{0.04}{AlignedSpinTidalLS}{0.03}{IMRPhenomDNRTidal-HS}{0.04}{IMRPhenomDNRTidal-LS}{0.03}{IMRPhenomPv2NRTidal-HS}{0.03}{IMRPhenomPv2NRTidal-LS}{0.03}{SEOBNRv4TsurrogateHS}{0.03}{SEOBNRv4TsurrogateLS}{0.03}{SEOBNRv4TsurrogatehighspinRIFT}{0.04}{SEOBNRv4TsurrogatelowspinRIFT}{0.03}{TEOBResumS-HS}{0.04}{TEOBResumS-LS}{0.03}{TaylorF2-HS}{0.04}{TaylorF2-LS}{0.04}{PrecessingSpinIMRTidalHS}{0.03}{PrecessingSpinIMRTidalLS}{0.03}{PublicationSamples}{0.03}}}
\newcommand{\redshiftfourtwofiveplus}[1]{\IfEqCase{#1}{{AlignedSpinInspiralTidalHS}{0.02}{AlignedSpinInspiralTidalLS}{0.01}{AlignedSpinTidalHS}{0.02}{AlignedSpinTidalLS}{0.01}{IMRPhenomDNRTidal-HS}{0.01}{IMRPhenomDNRTidal-LS}{0.01}{IMRPhenomPv2NRTidal-HS}{0.01}{IMRPhenomPv2NRTidal-LS}{0.01}{SEOBNRv4TsurrogateHS}{0.01}{SEOBNRv4TsurrogateLS}{0.01}{SEOBNRv4TsurrogatehighspinRIFT}{0.02}{SEOBNRv4TsurrogatelowspinRIFT}{0.02}{TEOBResumS-HS}{0.02}{TEOBResumS-LS}{0.01}{TaylorF2-HS}{0.02}{TaylorF2-LS}{0.02}{PrecessingSpinIMRTidalHS}{0.01}{PrecessingSpinIMRTidalLS}{0.01}{PublicationSamples}{0.01}}}
\newcommand{\iotafourtwofiveminus}[1]{\IfEqCase{#1}{{AlignedSpinInspiralTidalHS}{0.97}{AlignedSpinInspiralTidalLS}{0.88}{AlignedSpinTidalHS}{0.83}{AlignedSpinTidalLS}{0.83}{IMRPhenomDNRTidal-HS}{0.74}{IMRPhenomDNRTidal-LS}{0.78}{IMRPhenomPv2NRTidal-HS}{0.85}{IMRPhenomPv2NRTidal-LS}{0.84}{SEOBNRv4TsurrogateHS}{0.89}{SEOBNRv4TsurrogateLS}{0.85}{SEOBNRv4TsurrogatehighspinRIFT}{0.86}{SEOBNRv4TsurrogatelowspinRIFT}{0.86}{TEOBResumS-HS}{0.86}{TEOBResumS-LS}{0.88}{TaylorF2-HS}{0.97}{TaylorF2-LS}{0.88}{PrecessingSpinIMRTidalHS}{0.85}{PrecessingSpinIMRTidalLS}{0.84}{PublicationSamples}{0.85}}}
\newcommand{\iotafourtwofivemed}[1]{\IfEqCase{#1}{{AlignedSpinInspiralTidalHS}{1.22}{AlignedSpinInspiralTidalLS}{1.11}{AlignedSpinTidalHS}{1.06}{AlignedSpinTidalLS}{1.06}{IMRPhenomDNRTidal-HS}{0.96}{IMRPhenomDNRTidal-LS}{1.02}{IMRPhenomPv2NRTidal-HS}{1.09}{IMRPhenomPv2NRTidal-LS}{1.07}{SEOBNRv4TsurrogateHS}{1.13}{SEOBNRv4TsurrogateLS}{1.09}{SEOBNRv4TsurrogatehighspinRIFT}{1.10}{SEOBNRv4TsurrogatelowspinRIFT}{1.09}{TEOBResumS-HS}{1.09}{TEOBResumS-LS}{1.12}{TaylorF2-HS}{1.22}{TaylorF2-LS}{1.11}{PrecessingSpinIMRTidalHS}{1.09}{PrecessingSpinIMRTidalLS}{1.07}{PublicationSamples}{1.09}}}
\newcommand{\iotafourtwofiveplus}[1]{\IfEqCase{#1}{{AlignedSpinInspiralTidalHS}{1.64}{AlignedSpinInspiralTidalLS}{1.75}{AlignedSpinTidalHS}{1.79}{AlignedSpinTidalLS}{1.78}{IMRPhenomDNRTidal-HS}{1.88}{IMRPhenomDNRTidal-LS}{1.83}{IMRPhenomPv2NRTidal-HS}{1.77}{IMRPhenomPv2NRTidal-LS}{1.78}{SEOBNRv4TsurrogateHS}{1.72}{SEOBNRv4TsurrogateLS}{1.76}{SEOBNRv4TsurrogatehighspinRIFT}{1.76}{SEOBNRv4TsurrogatelowspinRIFT}{1.75}{TEOBResumS-HS}{1.77}{TEOBResumS-LS}{1.74}{TaylorF2-HS}{1.63}{TaylorF2-LS}{1.75}{PrecessingSpinIMRTidalHS}{1.77}{PrecessingSpinIMRTidalLS}{1.78}{PublicationSamples}{1.77}}}
\newcommand{\spinonexfourtwofiveminus}[1]{\IfEqCase{#1}{{AlignedSpinInspiralTidalHS}{0.00}{AlignedSpinInspiralTidalLS}{0.00}{AlignedSpinTidalHS}{0.00}{AlignedSpinTidalLS}{0.00}{IMRPhenomDNRTidal-HS}{0.00}{IMRPhenomDNRTidal-LS}{0.00}{IMRPhenomPv2NRTidal-HS}{0.50}{IMRPhenomPv2NRTidal-LS}{0.03}{SEOBNRv4TsurrogateHS}{0.00}{SEOBNRv4TsurrogateLS}{0.00}{SEOBNRv4TsurrogatehighspinRIFT}{0.00}{SEOBNRv4TsurrogatelowspinRIFT}{0.00}{TEOBResumS-HS}{0.00}{TEOBResumS-LS}{0.00}{TaylorF2-HS}{0.00}{TaylorF2-LS}{0.00}{PrecessingSpinIMRTidalHS}{0.50}{PrecessingSpinIMRTidalLS}{0.03}{PublicationSamples}{0.50}}}
\newcommand{\spinonexfourtwofivemed}[1]{\IfEqCase{#1}{{AlignedSpinInspiralTidalHS}{0.00}{AlignedSpinInspiralTidalLS}{0.00}{AlignedSpinTidalHS}{0.00}{AlignedSpinTidalLS}{0.00}{IMRPhenomDNRTidal-HS}{0.00}{IMRPhenomDNRTidal-LS}{0.00}{IMRPhenomPv2NRTidal-HS}{0.00}{IMRPhenomPv2NRTidal-LS}{0.00009}{SEOBNRv4TsurrogateHS}{0.00}{SEOBNRv4TsurrogateLS}{0.00}{SEOBNRv4TsurrogatehighspinRIFT}{0.00}{SEOBNRv4TsurrogatelowspinRIFT}{0.00}{TEOBResumS-HS}{0.00}{TEOBResumS-LS}{0.00}{TaylorF2-HS}{0.00}{TaylorF2-LS}{0.00}{PrecessingSpinIMRTidalHS}{0.00}{PrecessingSpinIMRTidalLS}{0.0001}{PublicationSamples}{0.00}}}
\newcommand{\spinonexfourtwofiveplus}[1]{\IfEqCase{#1}{{AlignedSpinInspiralTidalHS}{0.00}{AlignedSpinInspiralTidalLS}{0.00}{AlignedSpinTidalHS}{0.00}{AlignedSpinTidalLS}{0.00}{IMRPhenomDNRTidal-HS}{0.00}{IMRPhenomDNRTidal-LS}{0.00}{IMRPhenomPv2NRTidal-HS}{0.47}{IMRPhenomPv2NRTidal-LS}{0.03}{SEOBNRv4TsurrogateHS}{0.00}{SEOBNRv4TsurrogateLS}{0.00}{SEOBNRv4TsurrogatehighspinRIFT}{0.00}{SEOBNRv4TsurrogatelowspinRIFT}{0.00}{TEOBResumS-HS}{0.00}{TEOBResumS-LS}{0.00}{TaylorF2-HS}{0.00}{TaylorF2-LS}{0.00}{PrecessingSpinIMRTidalHS}{0.47}{PrecessingSpinIMRTidalLS}{0.03}{PublicationSamples}{0.47}}}
\newcommand{\chirpmassdetfourtwofiveminus}[1]{\IfEqCase{#1}{{AlignedSpinInspiralTidalHS}{0.0005}{AlignedSpinInspiralTidalLS}{0.0003}{AlignedSpinTidalHS}{0.0005}{AlignedSpinTidalLS}{0.0003}{IMRPhenomDNRTidal-HS}{0.0005}{IMRPhenomDNRTidal-LS}{0.0003}{IMRPhenomPv2NRTidal-HS}{0.0006}{IMRPhenomPv2NRTidal-LS}{0.0003}{SEOBNRv4TsurrogateHS}{0.0005}{SEOBNRv4TsurrogateLS}{0.0003}{SEOBNRv4TsurrogatehighspinRIFT}{0.0005}{SEOBNRv4TsurrogatelowspinRIFT}{0.0004}{TEOBResumS-HS}{0.0005}{TEOBResumS-LS}{0.0003}{TaylorF2-HS}{0.0005}{TaylorF2-LS}{0.0003}{PrecessingSpinIMRTidalHS}{0.0006}{PrecessingSpinIMRTidalLS}{0.0003}{PublicationSamples}{0.0006}}}
\newcommand{\chirpmassdetfourtwofivemed}[1]{\IfEqCase{#1}{{AlignedSpinInspiralTidalHS}{1.49}{AlignedSpinInspiralTidalLS}{1.49}{AlignedSpinTidalHS}{1.49}{AlignedSpinTidalLS}{1.49}{IMRPhenomDNRTidal-HS}{1.49}{IMRPhenomDNRTidal-LS}{1.49}{IMRPhenomPv2NRTidal-HS}{1.4873}{IMRPhenomPv2NRTidal-LS}{1.49}{SEOBNRv4TsurrogateHS}{1.49}{SEOBNRv4TsurrogateLS}{1.49}{SEOBNRv4TsurrogatehighspinRIFT}{1.49}{SEOBNRv4TsurrogatelowspinRIFT}{1.49}{TEOBResumS-HS}{1.49}{TEOBResumS-LS}{1.49}{TaylorF2-HS}{1.49}{TaylorF2-LS}{1.49}{PrecessingSpinIMRTidalHS}{1.49}{PrecessingSpinIMRTidalLS}{1.49}{PublicationSamples}{1.49}}}
\newcommand{\chirpmassdetfourtwofiveplus}[1]{\IfEqCase{#1}{{AlignedSpinInspiralTidalHS}{0.0007}{AlignedSpinInspiralTidalLS}{0.0003}{AlignedSpinTidalHS}{0.0007}{AlignedSpinTidalLS}{0.0003}{IMRPhenomDNRTidal-HS}{0.0008}{IMRPhenomDNRTidal-LS}{0.0004}{IMRPhenomPv2NRTidal-HS}{0.0008}{IMRPhenomPv2NRTidal-LS}{0.0003}{SEOBNRv4TsurrogateHS}{0.0006}{SEOBNRv4TsurrogateLS}{0.0003}{SEOBNRv4TsurrogatehighspinRIFT}{0.0006}{SEOBNRv4TsurrogatelowspinRIFT}{0.0003}{TEOBResumS-HS}{0.0005}{TEOBResumS-LS}{0.0003}{TaylorF2-HS}{0.0007}{TaylorF2-LS}{0.0003}{PrecessingSpinIMRTidalHS}{0.0008}{PrecessingSpinIMRTidalLS}{0.0003}{PublicationSamples}{0.0008}}}
\newcommand{\cosiotafourtwofiveminus}[1]{\IfEqCase{#1}{{AlignedSpinInspiralTidalHS}{1.30}{AlignedSpinInspiralTidalLS}{1.40}{AlignedSpinTidalHS}{1.44}{AlignedSpinTidalLS}{1.44}{IMRPhenomDNRTidal-HS}{1.53}{IMRPhenomDNRTidal-LS}{1.48}{IMRPhenomPv2NRTidal-HS}{1.42}{IMRPhenomPv2NRTidal-LS}{1.44}{SEOBNRv4TsurrogateHS}{1.38}{SEOBNRv4TsurrogateLS}{1.42}{SEOBNRv4TsurrogatehighspinRIFT}{1.41}{SEOBNRv4TsurrogatelowspinRIFT}{1.42}{TEOBResumS-HS}{1.42}{TEOBResumS-LS}{1.40}{TaylorF2-HS}{1.30}{TaylorF2-LS}{1.40}{PrecessingSpinIMRTidalHS}{1.42}{PrecessingSpinIMRTidalLS}{1.44}{PublicationSamples}{1.42}}}
\newcommand{\cosiotafourtwofivemed}[1]{\IfEqCase{#1}{{AlignedSpinInspiralTidalHS}{0.34}{AlignedSpinInspiralTidalLS}{0.44}{AlignedSpinTidalHS}{0.49}{AlignedSpinTidalLS}{0.49}{IMRPhenomDNRTidal-HS}{0.58}{IMRPhenomDNRTidal-LS}{0.53}{IMRPhenomPv2NRTidal-HS}{0.46}{IMRPhenomPv2NRTidal-LS}{0.48}{SEOBNRv4TsurrogateHS}{0.43}{SEOBNRv4TsurrogateLS}{0.46}{SEOBNRv4TsurrogatehighspinRIFT}{0.45}{SEOBNRv4TsurrogatelowspinRIFT}{0.46}{TEOBResumS-HS}{0.46}{TEOBResumS-LS}{0.44}{TaylorF2-HS}{0.34}{TaylorF2-LS}{0.44}{PrecessingSpinIMRTidalHS}{0.46}{PrecessingSpinIMRTidalLS}{0.48}{PublicationSamples}{0.46}}}
\newcommand{\cosiotafourtwofiveplus}[1]{\IfEqCase{#1}{{AlignedSpinInspiralTidalHS}{0.62}{AlignedSpinInspiralTidalLS}{0.53}{AlignedSpinTidalHS}{0.49}{AlignedSpinTidalLS}{0.49}{IMRPhenomDNRTidal-HS}{0.40}{IMRPhenomDNRTidal-LS}{0.45}{IMRPhenomPv2NRTidal-HS}{0.51}{IMRPhenomPv2NRTidal-LS}{0.49}{SEOBNRv4TsurrogateHS}{0.54}{SEOBNRv4TsurrogateLS}{0.51}{SEOBNRv4TsurrogatehighspinRIFT}{0.52}{SEOBNRv4TsurrogatelowspinRIFT}{0.51}{TEOBResumS-HS}{0.51}{TEOBResumS-LS}{0.54}{TaylorF2-HS}{0.63}{TaylorF2-LS}{0.53}{PrecessingSpinIMRTidalHS}{0.51}{PrecessingSpinIMRTidalLS}{0.49}{PublicationSamples}{0.51}}}
\newcommand{\comovingdistfourtwofiveminus}[1]{\IfEqCase{#1}{{AlignedSpinInspiralTidalHS}{72}{AlignedSpinInspiralTidalLS}{71}{AlignedSpinTidalHS}{69}{AlignedSpinTidalLS}{70}{IMRPhenomDNRTidal-HS}{69}{IMRPhenomDNRTidal-LS}{69}{IMRPhenomPv2NRTidal-HS}{67}{IMRPhenomPv2NRTidal-LS}{68}{SEOBNRv4TsurrogateHS}{68}{SEOBNRv4TsurrogateLS}{69}{SEOBNRv4TsurrogatehighspinRIFT}{70}{SEOBNRv4TsurrogatelowspinRIFT}{68}{TEOBResumS-HS}{70}{TEOBResumS-LS}{69}{TaylorF2-HS}{72}{TaylorF2-LS}{71}{PrecessingSpinIMRTidalHS}{67}{PrecessingSpinIMRTidalLS}{68}{PublicationSamples}{67}}}
\newcommand{\comovingdistfourtwofivemed}[1]{\IfEqCase{#1}{{AlignedSpinInspiralTidalHS}{156}{AlignedSpinInspiralTidalLS}{157}{AlignedSpinTidalHS}{155}{AlignedSpinTidalLS}{153}{IMRPhenomDNRTidal-HS}{155}{IMRPhenomDNRTidal-LS}{153}{IMRPhenomPv2NRTidal-HS}{151}{IMRPhenomPv2NRTidal-LS}{151}{SEOBNRv4TsurrogateHS}{153}{SEOBNRv4TsurrogateLS}{153}{SEOBNRv4TsurrogatehighspinRIFT}{157}{SEOBNRv4TsurrogatelowspinRIFT}{152}{TEOBResumS-HS}{157}{TEOBResumS-LS}{153}{TaylorF2-HS}{156}{TaylorF2-LS}{157}{PrecessingSpinIMRTidalHS}{151}{PrecessingSpinIMRTidalLS}{151}{PublicationSamples}{151}}}
\newcommand{\comovingdistfourtwofiveplus}[1]{\IfEqCase{#1}{{AlignedSpinInspiralTidalHS}{67}{AlignedSpinInspiralTidalLS}{65}{AlignedSpinTidalHS}{65}{AlignedSpinTidalLS}{63}{IMRPhenomDNRTidal-HS}{63}{IMRPhenomDNRTidal-LS}{64}{IMRPhenomPv2NRTidal-HS}{64}{IMRPhenomPv2NRTidal-LS}{64}{SEOBNRv4TsurrogateHS}{64}{SEOBNRv4TsurrogateLS}{64}{SEOBNRv4TsurrogatehighspinRIFT}{70}{SEOBNRv4TsurrogatelowspinRIFT}{65}{TEOBResumS-HS}{68}{TEOBResumS-LS}{64}{TaylorF2-HS}{67}{TaylorF2-LS}{65}{PrecessingSpinIMRTidalHS}{64}{PrecessingSpinIMRTidalLS}{64}{PublicationSamples}{64}}}
\newcommand{\logpriorfourtwofiveminus}[1]{\IfEqCase{#1}{{AlignedSpinInspiralTidalHS}{8.6}{AlignedSpinInspiralTidalLS}{8.5}{IMRPhenomDNRTidal-HS}{8.6}{IMRPhenomDNRTidal-LS}{8.6}{IMRPhenomPv2NRTidal-HS}{8.6}{IMRPhenomPv2NRTidal-LS}{8.4}{SEOBNRv4TsurrogateHS}{8.4}{SEOBNRv4TsurrogateLS}{8.8}{TaylorF2-HS}{8.6}{TaylorF2-LS}{8.5}{PrecessingSpinIMRTidalHS}{8.6}{PrecessingSpinIMRTidalLS}{8.4}{PublicationSamples}{8.6}}}
\newcommand{\logpriorfourtwofivemed}[1]{\IfEqCase{#1}{{AlignedSpinInspiralTidalHS}{102.5}{AlignedSpinInspiralTidalLS}{106.7}{IMRPhenomDNRTidal-HS}{94.5}{IMRPhenomDNRTidal-LS}{99.2}{IMRPhenomPv2NRTidal-HS}{98.4}{IMRPhenomPv2NRTidal-LS}{97.8}{SEOBNRv4TsurrogateHS}{95.6}{SEOBNRv4TsurrogateLS}{99.0}{TaylorF2-HS}{102.5}{TaylorF2-LS}{106.7}{PrecessingSpinIMRTidalHS}{98.4}{PrecessingSpinIMRTidalLS}{97.8}{PublicationSamples}{98.4}}}
\newcommand{\logpriorfourtwofiveplus}[1]{\IfEqCase{#1}{{AlignedSpinInspiralTidalHS}{6.8}{AlignedSpinInspiralTidalLS}{6.9}{IMRPhenomDNRTidal-HS}{6.8}{IMRPhenomDNRTidal-LS}{6.9}{IMRPhenomPv2NRTidal-HS}{6.7}{IMRPhenomPv2NRTidal-LS}{6.7}{SEOBNRv4TsurrogateHS}{6.9}{SEOBNRv4TsurrogateLS}{6.9}{TaylorF2-HS}{6.8}{TaylorF2-LS}{6.9}{PrecessingSpinIMRTidalHS}{6.7}{PrecessingSpinIMRTidalLS}{6.7}{PublicationSamples}{6.7}}}
\newcommand{\tiltonefourtwofiveminus}[1]{\IfEqCase{#1}{{AlignedSpinInspiralTidalHS}{0.00}{AlignedSpinInspiralTidalLS}{0.00}{AlignedSpinTidalHS}{0.00}{AlignedSpinTidalLS}{0.00}{IMRPhenomDNRTidal-HS}{0.00}{IMRPhenomDNRTidal-LS}{0.00}{IMRPhenomPv2NRTidal-HS}{0.80}{IMRPhenomPv2NRTidal-LS}{0.74}{SEOBNRv4TsurrogateHS}{0.00}{SEOBNRv4TsurrogateLS}{0.00}{SEOBNRv4TsurrogatehighspinRIFT}{0.00}{SEOBNRv4TsurrogatelowspinRIFT}{0.00}{TEOBResumS-HS}{0.00}{TEOBResumS-LS}{0.00}{TaylorF2-HS}{0.00}{TaylorF2-LS}{0.00}{PrecessingSpinIMRTidalHS}{0.80}{PrecessingSpinIMRTidalLS}{0.74}{PublicationSamples}{0.79}}}
\newcommand{\tiltonefourtwofivemed}[1]{\IfEqCase{#1}{{AlignedSpinInspiralTidalHS}{0.00}{AlignedSpinInspiralTidalLS}{0.00}{AlignedSpinTidalHS}{0.00}{AlignedSpinTidalLS}{0.00}{IMRPhenomDNRTidal-HS}{0.00}{IMRPhenomDNRTidal-LS}{0.00}{IMRPhenomPv2NRTidal-HS}{1.31}{IMRPhenomPv2NRTidal-LS}{1.03}{SEOBNRv4TsurrogateHS}{0.00}{SEOBNRv4TsurrogateLS}{0.00}{SEOBNRv4TsurrogatehighspinRIFT}{0.00}{SEOBNRv4TsurrogatelowspinRIFT}{0.00}{TEOBResumS-HS}{0.00}{TEOBResumS-LS}{0.00}{TaylorF2-HS}{0.00}{TaylorF2-LS}{0.00}{PrecessingSpinIMRTidalHS}{1.31}{PrecessingSpinIMRTidalLS}{1.03}{PublicationSamples}{1.31}}}
\newcommand{\tiltonefourtwofiveplus}[1]{\IfEqCase{#1}{{AlignedSpinInspiralTidalHS}{3.14}{AlignedSpinInspiralTidalLS}{3.14}{AlignedSpinTidalHS}{3.14}{AlignedSpinTidalLS}{3.14}{IMRPhenomDNRTidal-HS}{3.14}{IMRPhenomDNRTidal-LS}{3.14}{IMRPhenomPv2NRTidal-HS}{0.66}{IMRPhenomPv2NRTidal-LS}{1.16}{SEOBNRv4TsurrogateHS}{3.14}{SEOBNRv4TsurrogateLS}{3.14}{SEOBNRv4TsurrogatehighspinRIFT}{3.14}{SEOBNRv4TsurrogatelowspinRIFT}{3.14}{TEOBResumS-HS}{3.14}{TEOBResumS-LS}{3.14}{TaylorF2-HS}{3.14}{TaylorF2-LS}{3.14}{PrecessingSpinIMRTidalHS}{0.66}{PrecessingSpinIMRTidalLS}{1.17}{PublicationSamples}{0.66}}}
\newcommand{\spintwoyfourtwofiveminus}[1]{\IfEqCase{#1}{{AlignedSpinInspiralTidalHS}{0.00}{AlignedSpinInspiralTidalLS}{0.00}{AlignedSpinTidalHS}{0.00}{AlignedSpinTidalLS}{0.00}{IMRPhenomDNRTidal-HS}{0.00}{IMRPhenomDNRTidal-LS}{0.00}{IMRPhenomPv2NRTidal-HS}{0.48}{IMRPhenomPv2NRTidal-LS}{0.03}{SEOBNRv4TsurrogateHS}{0.00}{SEOBNRv4TsurrogateLS}{0.00}{SEOBNRv4TsurrogatehighspinRIFT}{0.00}{SEOBNRv4TsurrogatelowspinRIFT}{0.00}{TEOBResumS-HS}{0.00}{TEOBResumS-LS}{0.00}{TaylorF2-HS}{0.00}{TaylorF2-LS}{0.00}{PrecessingSpinIMRTidalHS}{0.48}{PrecessingSpinIMRTidalLS}{0.03}{PublicationSamples}{0.48}}}
\newcommand{\spintwoyfourtwofivemed}[1]{\IfEqCase{#1}{{AlignedSpinInspiralTidalHS}{0.00}{AlignedSpinInspiralTidalLS}{0.00}{AlignedSpinTidalHS}{0.00}{AlignedSpinTidalLS}{0.00}{IMRPhenomDNRTidal-HS}{0.00}{IMRPhenomDNRTidal-LS}{0.00}{IMRPhenomPv2NRTidal-HS}{0.00002}{IMRPhenomPv2NRTidal-LS}{0.00003}{SEOBNRv4TsurrogateHS}{0.00}{SEOBNRv4TsurrogateLS}{0.00}{SEOBNRv4TsurrogatehighspinRIFT}{0.00}{SEOBNRv4TsurrogatelowspinRIFT}{0.00}{TEOBResumS-HS}{0.00}{TEOBResumS-LS}{0.00}{TaylorF2-HS}{0.00}{TaylorF2-LS}{0.00}{PrecessingSpinIMRTidalHS}{0.00}{PrecessingSpinIMRTidalLS}{0.00002}{PublicationSamples}{0.00}}}
\newcommand{\spintwoyfourtwofiveplus}[1]{\IfEqCase{#1}{{AlignedSpinInspiralTidalHS}{0.00}{AlignedSpinInspiralTidalLS}{0.00}{AlignedSpinTidalHS}{0.00}{AlignedSpinTidalLS}{0.00}{IMRPhenomDNRTidal-HS}{0.00}{IMRPhenomDNRTidal-LS}{0.00}{IMRPhenomPv2NRTidal-HS}{0.48}{IMRPhenomPv2NRTidal-LS}{0.03}{SEOBNRv4TsurrogateHS}{0.00}{SEOBNRv4TsurrogateLS}{0.00}{SEOBNRv4TsurrogatehighspinRIFT}{0.00}{SEOBNRv4TsurrogatelowspinRIFT}{0.00}{TEOBResumS-HS}{0.00}{TEOBResumS-LS}{0.00}{TaylorF2-HS}{0.00}{TaylorF2-LS}{0.00}{PrecessingSpinIMRTidalHS}{0.48}{PrecessingSpinIMRTidalLS}{0.03}{PublicationSamples}{0.48}}}
\newcommand{\spintwozfourtwofiveminus}[1]{\IfEqCase{#1}{{AlignedSpinInspiralTidalHS}{0.18}{AlignedSpinInspiralTidalLS}{0.02}{AlignedSpinTidalHS}{0.24}{AlignedSpinTidalLS}{0.02}{IMRPhenomDNRTidal-HS}{0.39}{IMRPhenomDNRTidal-LS}{0.02}{IMRPhenomPv2NRTidal-HS}{0.18}{IMRPhenomPv2NRTidal-LS}{0.02}{SEOBNRv4TsurrogateHS}{0.16}{SEOBNRv4TsurrogateLS}{0.02}{SEOBNRv4TsurrogatehighspinRIFT}{0.18}{SEOBNRv4TsurrogatelowspinRIFT}{0.02}{TEOBResumS-HS}{0.18}{TEOBResumS-LS}{0.02}{TaylorF2-HS}{0.18}{TaylorF2-LS}{0.02}{PrecessingSpinIMRTidalHS}{0.18}{PrecessingSpinIMRTidalLS}{0.02}{PublicationSamples}{0.18}}}
\newcommand{\spintwozfourtwofivemed}[1]{\IfEqCase{#1}{{AlignedSpinInspiralTidalHS}{0.04}{AlignedSpinInspiralTidalLS}{0.008}{AlignedSpinTidalHS}{0.03}{AlignedSpinTidalLS}{0.009}{IMRPhenomDNRTidal-HS}{0.03}{IMRPhenomDNRTidal-LS}{0.009}{IMRPhenomPv2NRTidal-HS}{0.03}{IMRPhenomPv2NRTidal-LS}{0.009}{SEOBNRv4TsurrogateHS}{0.02}{SEOBNRv4TsurrogateLS}{0.009}{SEOBNRv4TsurrogatehighspinRIFT}{0.03}{SEOBNRv4TsurrogatelowspinRIFT}{0.01}{TEOBResumS-HS}{0.03}{TEOBResumS-LS}{0.009}{TaylorF2-HS}{0.04}{TaylorF2-LS}{0.008}{PrecessingSpinIMRTidalHS}{0.03}{PrecessingSpinIMRTidalLS}{0.009}{PublicationSamples}{0.03}}}
\newcommand{\spintwozfourtwofiveplus}[1]{\IfEqCase{#1}{{AlignedSpinInspiralTidalHS}{0.30}{AlignedSpinInspiralTidalLS}{0.03}{AlignedSpinTidalHS}{0.26}{AlignedSpinTidalLS}{0.03}{IMRPhenomDNRTidal-HS}{0.37}{IMRPhenomDNRTidal-LS}{0.03}{IMRPhenomPv2NRTidal-HS}{0.30}{IMRPhenomPv2NRTidal-LS}{0.03}{SEOBNRv4TsurrogateHS}{0.20}{SEOBNRv4TsurrogateLS}{0.03}{SEOBNRv4TsurrogatehighspinRIFT}{0.21}{SEOBNRv4TsurrogatelowspinRIFT}{0.03}{TEOBResumS-HS}{0.21}{TEOBResumS-LS}{0.03}{TaylorF2-HS}{0.30}{TaylorF2-LS}{0.03}{PrecessingSpinIMRTidalHS}{0.30}{PrecessingSpinIMRTidalLS}{0.03}{PublicationSamples}{0.30}}}
\newcommand{\massonesourcefourtwofiveminus}[1]{\IfEqCase{#1}{{AlignedSpinInspiralTidalHS}{0.3}{AlignedSpinInspiralTidalLS}{0.10}{AlignedSpinTidalHS}{0.2}{AlignedSpinTidalLS}{0.09}{IMRPhenomDNRTidal-HS}{0.3}{IMRPhenomDNRTidal-LS}{0.09}{IMRPhenomPv2NRTidal-HS}{0.3}{IMRPhenomPv2NRTidal-LS}{0.09}{SEOBNRv4TsurrogateHS}{0.2}{SEOBNRv4TsurrogateLS}{0.09}{SEOBNRv4TsurrogatehighspinRIFT}{0.2}{SEOBNRv4TsurrogatelowspinRIFT}{0.09}{TEOBResumS-HS}{0.2}{TEOBResumS-LS}{0.09}{TaylorF2-HS}{0.3}{TaylorF2-LS}{0.10}{PrecessingSpinIMRTidalHS}{0.3}{PrecessingSpinIMRTidalLS}{0.09}{PublicationSamples}{0.3}}}
\newcommand{\massonesourcefourtwofivemed}[1]{\IfEqCase{#1}{{AlignedSpinInspiralTidalHS}{2.0}{AlignedSpinInspiralTidalLS}{1.75}{AlignedSpinTidalHS}{1.9}{AlignedSpinTidalLS}{1.75}{IMRPhenomDNRTidal-HS}{2.0}{IMRPhenomDNRTidal-LS}{1.75}{IMRPhenomPv2NRTidal-HS}{2.0}{IMRPhenomPv2NRTidal-LS}{1.74}{SEOBNRv4TsurrogateHS}{1.9}{SEOBNRv4TsurrogateLS}{1.74}{SEOBNRv4TsurrogatehighspinRIFT}{1.9}{SEOBNRv4TsurrogatelowspinRIFT}{1.75}{TEOBResumS-HS}{1.9}{TEOBResumS-LS}{1.75}{TaylorF2-HS}{2.0}{TaylorF2-LS}{1.75}{PrecessingSpinIMRTidalHS}{2.0}{PrecessingSpinIMRTidalLS}{1.74}{PublicationSamples}{2.0}}}
\newcommand{\massonesourcefourtwofiveplus}[1]{\IfEqCase{#1}{{AlignedSpinInspiralTidalHS}{0.5}{AlignedSpinInspiralTidalLS}{0.2}{AlignedSpinTidalHS}{0.6}{AlignedSpinTidalLS}{0.2}{IMRPhenomDNRTidal-HS}{0.7}{IMRPhenomDNRTidal-LS}{0.2}{IMRPhenomPv2NRTidal-HS}{0.6}{IMRPhenomPv2NRTidal-LS}{0.2}{SEOBNRv4TsurrogateHS}{0.5}{SEOBNRv4TsurrogateLS}{0.2}{SEOBNRv4TsurrogatehighspinRIFT}{0.5}{SEOBNRv4TsurrogatelowspinRIFT}{0.2}{TEOBResumS-HS}{0.5}{TEOBResumS-LS}{0.2}{TaylorF2-HS}{0.5}{TaylorF2-LS}{0.2}{PrecessingSpinIMRTidalHS}{0.6}{PrecessingSpinIMRTidalLS}{0.2}{PublicationSamples}{0.6}}}
\newcommand{\geocenttimefourtwofiveminus}[1]{\IfEqCase{#1}{{AlignedSpinInspiralTidalHS}{0.007}{AlignedSpinInspiralTidalLS}{0.008}{AlignedSpinTidalHS}{0.03}{AlignedSpinTidalLS}{0.02}{IMRPhenomDNRTidal-HS}{0.008}{IMRPhenomDNRTidal-LS}{0.01}{IMRPhenomPv2NRTidal-HS}{0.009}{IMRPhenomPv2NRTidal-LS}{0.01}{SEOBNRv4TsurrogateHS}{0.01}{SEOBNRv4TsurrogateLS}{0.008}{SEOBNRv4TsurrogatehighspinRIFT}{0.0}{SEOBNRv4TsurrogatelowspinRIFT}{0.0}{TEOBResumS-HS}{0.0}{TEOBResumS-LS}{0.0}{TaylorF2-HS}{0.007}{TaylorF2-LS}{0.008}{PrecessingSpinIMRTidalHS}{0.009}{PrecessingSpinIMRTidalLS}{0.01}{PublicationSamples}{0.009}}}
\newcommand{\geocenttimefourtwofivemed}[1]{\IfEqCase{#1}{{AlignedSpinInspiralTidalHS}{1240215503.0}{AlignedSpinInspiralTidalLS}{1240215503.0}{AlignedSpinTidalHS}{1240215503.0}{AlignedSpinTidalLS}{1240215503.0}{IMRPhenomDNRTidal-HS}{1240215503.0}{IMRPhenomDNRTidal-LS}{1240215503.0}{IMRPhenomPv2NRTidal-HS}{1240215503.0}{IMRPhenomPv2NRTidal-LS}{1240215503.0}{SEOBNRv4TsurrogateHS}{1240215503.0}{SEOBNRv4TsurrogateLS}{1240215503.0}{SEOBNRv4TsurrogatehighspinRIFT}{1240215503.0}{SEOBNRv4TsurrogatelowspinRIFT}{1240215503.0}{TEOBResumS-HS}{1240215503.0}{TEOBResumS-LS}{1240215503.0}{TaylorF2-HS}{1240215503.0}{TaylorF2-LS}{1240215503.0}{PrecessingSpinIMRTidalHS}{1240215503.0}{PrecessingSpinIMRTidalLS}{1240215503.0}{PublicationSamples}{1240215503.0}}}
\newcommand{\geocenttimefourtwofiveplus}[1]{\IfEqCase{#1}{{AlignedSpinInspiralTidalHS}{0.04}{AlignedSpinInspiralTidalLS}{0.03}{AlignedSpinTidalHS}{0.02}{AlignedSpinTidalLS}{0.02}{IMRPhenomDNRTidal-HS}{0.04}{IMRPhenomDNRTidal-LS}{0.03}{IMRPhenomPv2NRTidal-HS}{0.03}{IMRPhenomPv2NRTidal-LS}{0.03}{SEOBNRv4TsurrogateHS}{0.04}{SEOBNRv4TsurrogateLS}{0.04}{SEOBNRv4TsurrogatehighspinRIFT}{0.0}{SEOBNRv4TsurrogatelowspinRIFT}{0.0}{TEOBResumS-HS}{0.0}{TEOBResumS-LS}{0.0}{TaylorF2-HS}{0.04}{TaylorF2-LS}{0.03}{PrecessingSpinIMRTidalHS}{0.03}{PrecessingSpinIMRTidalLS}{0.03}{PublicationSamples}{0.03}}}
\newcommand{\costilttwofourtwofiveminus}[1]{\IfEqCase{#1}{{AlignedSpinInspiralTidalHS}{2.00}{AlignedSpinInspiralTidalLS}{2.00}{AlignedSpinTidalHS}{2.00}{AlignedSpinTidalLS}{2.00}{IMRPhenomDNRTidal-HS}{2.00}{IMRPhenomDNRTidal-LS}{2.00}{IMRPhenomPv2NRTidal-HS}{0.87}{IMRPhenomPv2NRTidal-LS}{1.13}{SEOBNRv4TsurrogateHS}{2.00}{SEOBNRv4TsurrogateLS}{2.00}{SEOBNRv4TsurrogatehighspinRIFT}{2.00}{SEOBNRv4TsurrogatelowspinRIFT}{2.00}{TEOBResumS-HS}{2.00}{TEOBResumS-LS}{2.00}{TaylorF2-HS}{2.00}{TaylorF2-LS}{2.00}{PrecessingSpinIMRTidalHS}{0.87}{PrecessingSpinIMRTidalLS}{1.12}{PublicationSamples}{0.86}}}
\newcommand{\costilttwofourtwofivemed}[1]{\IfEqCase{#1}{{AlignedSpinInspiralTidalHS}{1.00}{AlignedSpinInspiralTidalLS}{1.00}{AlignedSpinTidalHS}{1.00}{AlignedSpinTidalLS}{1.00}{IMRPhenomDNRTidal-HS}{1.00}{IMRPhenomDNRTidal-LS}{1.00}{IMRPhenomPv2NRTidal-HS}{0.16}{IMRPhenomPv2NRTidal-LS}{0.46}{SEOBNRv4TsurrogateHS}{1.00}{SEOBNRv4TsurrogateLS}{1.00}{SEOBNRv4TsurrogatehighspinRIFT}{1.00}{SEOBNRv4TsurrogatelowspinRIFT}{1.00}{TEOBResumS-HS}{1.00}{TEOBResumS-LS}{1.00}{TaylorF2-HS}{1.00}{TaylorF2-LS}{1.00}{PrecessingSpinIMRTidalHS}{0.16}{PrecessingSpinIMRTidalLS}{0.46}{PublicationSamples}{0.16}}}
\newcommand{\costilttwofourtwofiveplus}[1]{\IfEqCase{#1}{{AlignedSpinInspiralTidalHS}{0.00}{AlignedSpinInspiralTidalLS}{0.00}{AlignedSpinTidalHS}{0.00}{AlignedSpinTidalLS}{0.00}{IMRPhenomDNRTidal-HS}{0.00}{IMRPhenomDNRTidal-LS}{0.00}{IMRPhenomPv2NRTidal-HS}{0.70}{IMRPhenomPv2NRTidal-LS}{0.49}{SEOBNRv4TsurrogateHS}{0.00}{SEOBNRv4TsurrogateLS}{0.00}{SEOBNRv4TsurrogatehighspinRIFT}{0.00}{SEOBNRv4TsurrogatelowspinRIFT}{0.00}{TEOBResumS-HS}{0.00}{TEOBResumS-LS}{0.00}{TaylorF2-HS}{0.00}{TaylorF2-LS}{0.00}{PrecessingSpinIMRTidalHS}{0.70}{PrecessingSpinIMRTidalLS}{0.49}{PublicationSamples}{0.70}}}
\newcommand{\luminositydistancefourtwofiveminus}[1]{\IfEqCase{#1}{{AlignedSpinInspiralTidalHS}{0.08}{AlignedSpinInspiralTidalLS}{0.07}{AlignedSpinTidalHS}{0.07}{AlignedSpinTidalLS}{0.07}{IMRPhenomDNRTidal-HS}{0.07}{IMRPhenomDNRTidal-LS}{0.07}{IMRPhenomPv2NRTidal-HS}{0.07}{IMRPhenomPv2NRTidal-LS}{0.07}{SEOBNRv4TsurrogateHS}{0.07}{SEOBNRv4TsurrogateLS}{0.07}{SEOBNRv4TsurrogatehighspinRIFT}{0.07}{SEOBNRv4TsurrogatelowspinRIFT}{0.07}{TEOBResumS-HS}{0.07}{TEOBResumS-LS}{0.07}{TaylorF2-HS}{0.08}{TaylorF2-LS}{0.08}{PrecessingSpinIMRTidalHS}{0.07}{PrecessingSpinIMRTidalLS}{0.07}{PublicationSamples}{0.07}}}
\newcommand{\luminositydistancefourtwofivemed}[1]{\IfEqCase{#1}{{AlignedSpinInspiralTidalHS}{0.16}{AlignedSpinInspiralTidalLS}{0.16}{AlignedSpinTidalHS}{0.16}{AlignedSpinTidalLS}{0.16}{IMRPhenomDNRTidal-HS}{0.16}{IMRPhenomDNRTidal-LS}{0.16}{IMRPhenomPv2NRTidal-HS}{0.16}{IMRPhenomPv2NRTidal-LS}{0.16}{SEOBNRv4TsurrogateHS}{0.16}{SEOBNRv4TsurrogateLS}{0.16}{SEOBNRv4TsurrogatehighspinRIFT}{0.16}{SEOBNRv4TsurrogatelowspinRIFT}{0.16}{TEOBResumS-HS}{0.16}{TEOBResumS-LS}{0.16}{TaylorF2-HS}{0.16}{TaylorF2-LS}{0.16}{PrecessingSpinIMRTidalHS}{0.16}{PrecessingSpinIMRTidalLS}{0.16}{PublicationSamples}{0.16}}}
\newcommand{\luminositydistancefourtwofiveplus}[1]{\IfEqCase{#1}{{AlignedSpinInspiralTidalHS}{0.07}{AlignedSpinInspiralTidalLS}{0.07}{AlignedSpinTidalHS}{0.07}{AlignedSpinTidalLS}{0.07}{IMRPhenomDNRTidal-HS}{0.07}{IMRPhenomDNRTidal-LS}{0.07}{IMRPhenomPv2NRTidal-HS}{0.07}{IMRPhenomPv2NRTidal-LS}{0.07}{SEOBNRv4TsurrogateHS}{0.07}{SEOBNRv4TsurrogateLS}{0.07}{SEOBNRv4TsurrogatehighspinRIFT}{0.08}{SEOBNRv4TsurrogatelowspinRIFT}{0.07}{TEOBResumS-HS}{0.07}{TEOBResumS-LS}{0.07}{TaylorF2-HS}{0.07}{TaylorF2-LS}{0.07}{PrecessingSpinIMRTidalHS}{0.07}{PrecessingSpinIMRTidalLS}{0.07}{PublicationSamples}{0.07}}}
\newcommand{\spinonezfourtwofiveminus}[1]{\IfEqCase{#1}{{AlignedSpinInspiralTidalHS}{0.14}{AlignedSpinInspiralTidalLS}{0.02}{AlignedSpinTidalHS}{0.15}{AlignedSpinTidalLS}{0.02}{IMRPhenomDNRTidal-HS}{0.22}{IMRPhenomDNRTidal-LS}{0.02}{IMRPhenomPv2NRTidal-HS}{0.12}{IMRPhenomPv2NRTidal-LS}{0.02}{SEOBNRv4TsurrogateHS}{0.11}{SEOBNRv4TsurrogateLS}{0.02}{SEOBNRv4TsurrogatehighspinRIFT}{0.14}{SEOBNRv4TsurrogatelowspinRIFT}{0.02}{TEOBResumS-HS}{0.13}{TEOBResumS-LS}{0.02}{TaylorF2-HS}{0.14}{TaylorF2-LS}{0.02}{PrecessingSpinIMRTidalHS}{0.12}{PrecessingSpinIMRTidalLS}{0.02}{PublicationSamples}{0.12}}}
\newcommand{\spinonezfourtwofivemed}[1]{\IfEqCase{#1}{{AlignedSpinInspiralTidalHS}{0.04}{AlignedSpinInspiralTidalLS}{0.01}{AlignedSpinTidalHS}{0.04}{AlignedSpinTidalLS}{0.01}{IMRPhenomDNRTidal-HS}{0.06}{IMRPhenomDNRTidal-LS}{0.01}{IMRPhenomPv2NRTidal-HS}{0.06}{IMRPhenomPv2NRTidal-LS}{0.01}{SEOBNRv4TsurrogateHS}{0.04}{SEOBNRv4TsurrogateLS}{0.01}{SEOBNRv4TsurrogatehighspinRIFT}{0.04}{SEOBNRv4TsurrogatelowspinRIFT}{0.01}{TEOBResumS-HS}{0.04}{TEOBResumS-LS}{0.01}{TaylorF2-HS}{0.04}{TaylorF2-LS}{0.01}{PrecessingSpinIMRTidalHS}{0.06}{PrecessingSpinIMRTidalLS}{0.01}{PublicationSamples}{0.06}}}
\newcommand{\spinonezfourtwofiveplus}[1]{\IfEqCase{#1}{{AlignedSpinInspiralTidalHS}{0.19}{AlignedSpinInspiralTidalLS}{0.03}{AlignedSpinTidalHS}{0.20}{AlignedSpinTidalLS}{0.03}{IMRPhenomDNRTidal-HS}{0.26}{IMRPhenomDNRTidal-LS}{0.03}{IMRPhenomPv2NRTidal-HS}{0.18}{IMRPhenomPv2NRTidal-LS}{0.03}{SEOBNRv4TsurrogateHS}{0.16}{SEOBNRv4TsurrogateLS}{0.03}{SEOBNRv4TsurrogatehighspinRIFT}{0.16}{SEOBNRv4TsurrogatelowspinRIFT}{0.03}{TEOBResumS-HS}{0.16}{TEOBResumS-LS}{0.03}{TaylorF2-HS}{0.19}{TaylorF2-LS}{0.03}{PrecessingSpinIMRTidalHS}{0.18}{PrecessingSpinIMRTidalLS}{0.03}{PublicationSamples}{0.18}}}
\newcommand{\networkmatchedfiltersnrfourtwofiveminus}[1]{\IfEqCase{#1}{{AlignedSpinInspiralTidalHS}{0.4}{AlignedSpinInspiralTidalLS}{0.4}{IMRPhenomDNRTidal-HS}{0.4}{IMRPhenomDNRTidal-LS}{0.4}{IMRPhenomPv2NRTidal-HS}{0.4}{IMRPhenomPv2NRTidal-LS}{0.4}{SEOBNRv4TsurrogateHS}{0.4}{SEOBNRv4TsurrogateLS}{0.4}{TaylorF2-HS}{0.4}{TaylorF2-LS}{0.4}{PrecessingSpinIMRTidalHS}{0.4}{PrecessingSpinIMRTidalLS}{0.4}{PublicationSamples}{0.4}}}
\newcommand{\networkmatchedfiltersnrfourtwofivemed}[1]{\IfEqCase{#1}{{AlignedSpinInspiralTidalHS}{12.4}{AlignedSpinInspiralTidalLS}{12.5}{IMRPhenomDNRTidal-HS}{12.3}{IMRPhenomDNRTidal-LS}{12.4}{IMRPhenomPv2NRTidal-HS}{12.4}{IMRPhenomPv2NRTidal-LS}{12.5}{SEOBNRv4TsurrogateHS}{12.4}{SEOBNRv4TsurrogateLS}{12.4}{TaylorF2-HS}{12.4}{TaylorF2-LS}{12.5}{PrecessingSpinIMRTidalHS}{12.4}{PrecessingSpinIMRTidalLS}{12.5}{PublicationSamples}{12.4}}}
\newcommand{\networkmatchedfiltersnrfourtwofiveplus}[1]{\IfEqCase{#1}{{AlignedSpinInspiralTidalHS}{0.3}{AlignedSpinInspiralTidalLS}{0.2}{IMRPhenomDNRTidal-HS}{0.3}{IMRPhenomDNRTidal-LS}{0.3}{IMRPhenomPv2NRTidal-HS}{0.3}{IMRPhenomPv2NRTidal-LS}{0.3}{SEOBNRv4TsurrogateHS}{0.3}{SEOBNRv4TsurrogateLS}{0.3}{TaylorF2-HS}{0.3}{TaylorF2-LS}{0.2}{PrecessingSpinIMRTidalHS}{0.3}{PrecessingSpinIMRTidalLS}{0.3}{PublicationSamples}{0.3}}}
\newcommand{\chirpmasssourcefourtwofiveminus}[1]{\IfEqCase{#1}{{AlignedSpinInspiralTidalHS}{0.02}{AlignedSpinInspiralTidalLS}{0.02}{AlignedSpinTidalHS}{0.02}{AlignedSpinTidalLS}{0.02}{IMRPhenomDNRTidal-HS}{0.02}{IMRPhenomDNRTidal-LS}{0.02}{IMRPhenomPv2NRTidal-HS}{0.02}{IMRPhenomPv2NRTidal-LS}{0.02}{SEOBNRv4TsurrogateHS}{0.02}{SEOBNRv4TsurrogateLS}{0.02}{SEOBNRv4TsurrogatehighspinRIFT}{0.02}{SEOBNRv4TsurrogatelowspinRIFT}{0.02}{TEOBResumS-HS}{0.02}{TEOBResumS-LS}{0.02}{TaylorF2-HS}{0.02}{TaylorF2-LS}{0.02}{PrecessingSpinIMRTidalHS}{0.02}{PrecessingSpinIMRTidalLS}{0.02}{PublicationSamples}{0.02}}}
\newcommand{\chirpmasssourcefourtwofivemed}[1]{\IfEqCase{#1}{{AlignedSpinInspiralTidalHS}{1.44}{AlignedSpinInspiralTidalLS}{1.44}{AlignedSpinTidalHS}{1.44}{AlignedSpinTidalLS}{1.44}{IMRPhenomDNRTidal-HS}{1.44}{IMRPhenomDNRTidal-LS}{1.44}{IMRPhenomPv2NRTidal-HS}{1.44}{IMRPhenomPv2NRTidal-LS}{1.44}{SEOBNRv4TsurrogateHS}{1.44}{SEOBNRv4TsurrogateLS}{1.44}{SEOBNRv4TsurrogatehighspinRIFT}{1.44}{SEOBNRv4TsurrogatelowspinRIFT}{1.44}{TEOBResumS-HS}{1.44}{TEOBResumS-LS}{1.44}{TaylorF2-HS}{1.44}{TaylorF2-LS}{1.44}{PrecessingSpinIMRTidalHS}{1.44}{PrecessingSpinIMRTidalLS}{1.44}{PublicationSamples}{1.44}}}
\newcommand{\chirpmasssourcefourtwofiveplus}[1]{\IfEqCase{#1}{{AlignedSpinInspiralTidalHS}{0.02}{AlignedSpinInspiralTidalLS}{0.02}{AlignedSpinTidalHS}{0.02}{AlignedSpinTidalLS}{0.02}{IMRPhenomDNRTidal-HS}{0.02}{IMRPhenomDNRTidal-LS}{0.02}{IMRPhenomPv2NRTidal-HS}{0.02}{IMRPhenomPv2NRTidal-LS}{0.02}{SEOBNRv4TsurrogateHS}{0.02}{SEOBNRv4TsurrogateLS}{0.02}{SEOBNRv4TsurrogatehighspinRIFT}{0.02}{SEOBNRv4TsurrogatelowspinRIFT}{0.02}{TEOBResumS-HS}{0.02}{TEOBResumS-LS}{0.02}{TaylorF2-HS}{0.02}{TaylorF2-LS}{0.02}{PrecessingSpinIMRTidalHS}{0.02}{PrecessingSpinIMRTidalLS}{0.02}{PublicationSamples}{0.02}}}
\newcommand{\phionefourtwofiveminus}[1]{\IfEqCase{#1}{{AlignedSpinInspiralTidalHS}{0.00}{AlignedSpinInspiralTidalLS}{0.00}{AlignedSpinTidalHS}{0.00}{AlignedSpinTidalLS}{0.00}{IMRPhenomDNRTidal-HS}{0.00}{IMRPhenomDNRTidal-LS}{0.00}{IMRPhenomPv2NRTidal-HS}{2.73}{IMRPhenomPv2NRTidal-LS}{2.85}{SEOBNRv4TsurrogateHS}{0.00}{SEOBNRv4TsurrogateLS}{0.00}{SEOBNRv4TsurrogatehighspinRIFT}{0.00}{SEOBNRv4TsurrogatelowspinRIFT}{0.00}{TEOBResumS-HS}{0.00}{TEOBResumS-LS}{0.00}{TaylorF2-HS}{0.00}{TaylorF2-LS}{0.00}{PrecessingSpinIMRTidalHS}{2.73}{PrecessingSpinIMRTidalLS}{2.85}{PublicationSamples}{2.73}}}
\newcommand{\phionefourtwofivemed}[1]{\IfEqCase{#1}{{AlignedSpinInspiralTidalHS}{0.00}{AlignedSpinInspiralTidalLS}{0.00}{AlignedSpinTidalHS}{0.00}{AlignedSpinTidalLS}{0.00}{IMRPhenomDNRTidal-HS}{0.00}{IMRPhenomDNRTidal-LS}{0.00}{IMRPhenomPv2NRTidal-HS}{3.05}{IMRPhenomPv2NRTidal-LS}{3.15}{SEOBNRv4TsurrogateHS}{0.00}{SEOBNRv4TsurrogateLS}{0.00}{SEOBNRv4TsurrogatehighspinRIFT}{0.00}{SEOBNRv4TsurrogatelowspinRIFT}{0.00}{TEOBResumS-HS}{0.00}{TEOBResumS-LS}{0.00}{TaylorF2-HS}{0.00}{TaylorF2-LS}{0.00}{PrecessingSpinIMRTidalHS}{3.05}{PrecessingSpinIMRTidalLS}{3.15}{PublicationSamples}{3.06}}}
\newcommand{\phionefourtwofiveplus}[1]{\IfEqCase{#1}{{AlignedSpinInspiralTidalHS}{0.00}{AlignedSpinInspiralTidalLS}{0.00}{AlignedSpinTidalHS}{0.00}{AlignedSpinTidalLS}{0.00}{IMRPhenomDNRTidal-HS}{0.00}{IMRPhenomDNRTidal-LS}{0.00}{IMRPhenomPv2NRTidal-HS}{2.90}{IMRPhenomPv2NRTidal-LS}{2.83}{SEOBNRv4TsurrogateHS}{0.00}{SEOBNRv4TsurrogateLS}{0.00}{SEOBNRv4TsurrogatehighspinRIFT}{0.00}{SEOBNRv4TsurrogatelowspinRIFT}{0.00}{TEOBResumS-HS}{0.00}{TEOBResumS-LS}{0.00}{TaylorF2-HS}{0.00}{TaylorF2-LS}{0.00}{PrecessingSpinIMRTidalHS}{2.90}{PrecessingSpinIMRTidalLS}{2.83}{PublicationSamples}{2.90}}}
\newcommand{\symmetricmassratiofourtwofiveminus}[1]{\IfEqCase{#1}{{AlignedSpinInspiralTidalHS}{0.03}{AlignedSpinInspiralTidalLS}{0.005}{AlignedSpinTidalHS}{0.03}{AlignedSpinTidalLS}{0.005}{IMRPhenomDNRTidal-HS}{0.04}{IMRPhenomDNRTidal-LS}{0.005}{IMRPhenomPv2NRTidal-HS}{0.03}{IMRPhenomPv2NRTidal-LS}{0.005}{SEOBNRv4TsurrogateHS}{0.03}{SEOBNRv4TsurrogateLS}{0.004}{SEOBNRv4TsurrogatehighspinRIFT}{0.02}{SEOBNRv4TsurrogatelowspinRIFT}{0.005}{TEOBResumS-HS}{0.03}{TEOBResumS-LS}{0.005}{TaylorF2-HS}{0.03}{TaylorF2-LS}{0.005}{PrecessingSpinIMRTidalHS}{0.03}{PrecessingSpinIMRTidalLS}{0.005}{PublicationSamples}{0.03}}}
\newcommand{\symmetricmassratiofourtwofivemed}[1]{\IfEqCase{#1}{{AlignedSpinInspiralTidalHS}{0.242}{AlignedSpinInspiralTidalLS}{0.249}{AlignedSpinTidalHS}{0.245}{AlignedSpinTidalLS}{0.249}{IMRPhenomDNRTidal-HS}{0.243}{IMRPhenomDNRTidal-LS}{0.249}{IMRPhenomPv2NRTidal-HS}{0.240}{IMRPhenomPv2NRTidal-LS}{0.249}{SEOBNRv4TsurrogateHS}{0.246}{SEOBNRv4TsurrogateLS}{0.249}{SEOBNRv4TsurrogatehighspinRIFT}{0.246}{SEOBNRv4TsurrogatelowspinRIFT}{0.249}{TEOBResumS-HS}{0.245}{TEOBResumS-LS}{0.249}{TaylorF2-HS}{0.242}{TaylorF2-LS}{0.249}{PrecessingSpinIMRTidalHS}{0.240}{PrecessingSpinIMRTidalLS}{0.249}{PublicationSamples}{0.240}}}
\newcommand{\symmetricmassratiofourtwofiveplus}[1]{\IfEqCase{#1}{{AlignedSpinInspiralTidalHS}{0.007}{AlignedSpinInspiralTidalLS}{0.0008}{AlignedSpinTidalHS}{0.005}{AlignedSpinTidalLS}{0.0008}{IMRPhenomDNRTidal-HS}{0.007}{IMRPhenomDNRTidal-LS}{0.0008}{IMRPhenomPv2NRTidal-HS}{0.010}{IMRPhenomPv2NRTidal-LS}{0.0007}{SEOBNRv4TsurrogateHS}{0.004}{SEOBNRv4TsurrogateLS}{0.0007}{SEOBNRv4TsurrogatehighspinRIFT}{0.004}{SEOBNRv4TsurrogatelowspinRIFT}{0.0008}{TEOBResumS-HS}{0.005}{TEOBResumS-LS}{0.0009}{TaylorF2-HS}{0.007}{TaylorF2-LS}{0.0008}{PrecessingSpinIMRTidalHS}{0.010}{PrecessingSpinIMRTidalLS}{0.0007}{PublicationSamples}{0.010}}}

\acrodef{MDC}[MDC]{mock data challenge}
\acrodef{MECO}[MECO]{minimum energy circular orbit}
\acrodef{QNM}[QNM]{quasi-normal mode}

\begin{document}

\newcommand{\TGRIINUMTESTS}{\reviewed{eight}\xspace}
\newcommand{\TGRIINUMEVENTS}{\reviewed{42}\xspace}

\DeclareRobustCommand{\TGRTIGERMinimumSNR}{6}

\DeclareRobustCommand{\TGRTIGERMaxFracDiffEndInspiralvsMECO}{\reviewed{3\%}}

\DeclareRobustCommand{\TGRFTIEndInspiraltoTIGEREndInspiralRatioMin}{\reviewed{1.4}}
\DeclareRobustCommand{\TGRFTIEndInspiraltoTIGEREndInspiralRatioMax}{\reviewed{3.6}}

\DeclareRobustCommand{\TGRTIGERCredibleLevel}[1]{\IfEqCase{#1}{
{GW230628_231200}{\reviewed{\num{94.3}}}
{GW231110_040320}{\reviewed{\num{98.9}}}
}}

\DeclareRobustCommand{\TGRFTIMinimumSNR}{10}
\DeclareRobustCommand{\TGRFTIMinimumCycles}{5}
\DeclareRobustCommand{\TGRFTITaperingFreqOld}{0.35}
\DeclareRobustCommand{\TGRFTITaperingFreqNew}{1.0}
\DeclareRobustCommand{\TGRFTITaperingCycles}{1}

\DeclareRobustCommand{\TGRFTIPulsarBound}[1]{\IfEqCase{#1}{
{dchiMinus2}{\reviewed{\num{2.6e-10}}}
{dchi0}{\reviewed{\num{1.0e-4}}}
{dchi1}{\reviewed{\num{7.0e-2}}}
{dchi2}{\reviewed{8.2}}
{dchi3}{\reviewed{\num{9.0e2}}}
{dchi4}{\reviewed{\num{1.2e6}}}
{dchi6}{\reviewed{\num{2.3e10}}}
{dchi7}{\reviewed{\num{1.1e13}}}
}}

\DeclareRobustCommand{\TGRFTICredibleLevel}[1]{\IfEqCase{#1}{
{GW230628_231200}{\reviewed{\num{92.7}}}
{GW231110_040320}{\reviewed{\num{98.2}}}
}}

\DeclareRobustCommand{\TGRPCAFTITIGERInspiralSNR}{14.5}
\DeclareRobustCommand{\TGRPCAFTITIGERChirpMass}{20.5M_{\odot}}
\DeclareRobustCommand{\TGRPCAFTITIGEREvents}{two}

\DeclareRobustCommand{\TGRSIMPhenomMinimumSNR}{6}
\DeclareRobustCommand{\TGRSIMPhenomNumberOFouraEvents}{\reviewed{seven}}
\DeclareRobustCommand{\TGRSIMPhenomNumberGWTCThreeEvents}{\reviewed{eight}}
\DeclareRobustCommand{\TGRSIMEOBNumberOFouraEvents}{\reviewed{six}}

\DeclareRobustCommand{\TGRLOSAmaxMtdet}{10\,\mathrm{\Msun}}
\DeclareRobustCommand{\TGRLOSAminq}{0.25}
\DeclareRobustCommand{\TGRLOSAmaxchip}{0.4}
\DeclareRobustCommand{\TGRLOSAtsdprodmax}{0.01}

\DeclareRobustCommand{\TGRMDRQuantileGRnew}[1]{\IfEqCase{#1}{
{alpha_0p0}{\reviewed{71.67}}
{alpha_0p5}{\reviewed{84.80}}
{alpha_1p5}{\reviewed{10.46}}
{alpha_2p5}{\reviewed{90.02}}
{alpha_3p0}{\reviewed{79.77}}
{alpha_3p5}{\reviewed{84.23}}
{alpha_4p0}{\reviewed{82.49}}
{alpha_m1p0}{\reviewed{10.32}}
{alpha_m2p0}{\reviewed{3.40}}
{alpha_m3p0}{\reviewed{1.13}}
}}
\DeclareRobustCommand{\TGRMDRQuantileGRold}[1]{\IfEqCase{#1}{
{alpha_0p0}{\reviewed{84.11}}
{alpha_0p5}{\reviewed{92.18}}
{alpha_1p5}{\reviewed{1.29}}
{alpha_2p5}{\reviewed{96.36}}
{alpha_3p0}{\reviewed{90.45}}
{alpha_3p5}{\reviewed{85.01}}
{alpha_4p0}{\reviewed{75.34}}
{alpha_m1p0}{\reviewed{45.54}}
{alpha_m2p0}{\reviewed{22.34}}
{alpha_m3p0}{\reviewed{15.69}}
}}
\DeclareRobustCommand{\TGRMDREventNumber}[1]{\IfEqCase{#1}{
{gwtc3}{\reviewed{43}}
{gwtc4new}{\reviewed{40}}
{gwtc4cumulative}{\reviewed{83}}
}}
\DeclareRobustCommand{\TGRMDRExpectedBoundImprovement}{\reviewed{\num{1.39}}}
\DeclareRobustCommand{\TGRMDRGravitonOrbitScaleOoM}{\reviewed{\num{11}}}
\DeclareRobustCommand{\TGRMDRGravitonOrbitEoMCorrectionScaleOoM}{\reviewed{\num{-22}}}
\DeclareRobustCommand{\TGRMDRInjectionRecoveryMismatch}{\reviewed{\num{0.0054}}}
\DeclareRobustCommand{\TGRMDRInjectionRecoveryPriorFactor}{\reviewed{\num{3.31e+05}}}

\title{\paperscommonname II. Parameterized Tests}

\iftoggle{endauthorlist}{
 \let\mymaketitle\maketitle
 \let\myauthor\author
 \let\myaffiliation\affiliation
 \author{\LVKCollabAuthors}
 {\def\thefootnote{}\footnotetext{\LVKCorrespondence}}
 \email{~~~~~~lvc.publications@ligo.org}
}{
 \iftoggle{fullauthorlist}{

\author[0000-0003-4786-2698]{A.~G.~Abac}
\affiliation{Max Planck Institute for Gravitational Physics (Albert Einstein Institute), D-14476 Potsdam, Germany}
\author{I.~Abouelfettouh}
\affiliation{LIGO Hanford Observatory, Richland, WA 99352, USA}
\author{F.~Acernese}
\affiliation{Dipartimento di Farmacia, Universit\`a di Salerno, I-84084 Fisciano, Salerno, Italy}
\affiliation{INFN, Sezione di Napoli, I-80126 Napoli, Italy}
\author[0000-0002-8648-0767]{K.~Ackley}
\affiliation{University of Warwick, Coventry CV4 7AL, United Kingdom}
\author[0000-0001-5525-6255]{C.~Adamcewicz}
\affiliation{OzGrav, School of Physics \& Astronomy, Monash University, Clayton 3800, Victoria, Australia}
\author[0009-0004-2101-5428]{S.~Adhicary}
\affiliation{The Pennsylvania State University, University Park, PA 16802, USA}
\author{D.~Adhikari}
\affiliation{Max Planck Institute for Gravitational Physics (Albert Einstein Institute), D-30167 Hannover, Germany}
\affiliation{Leibniz Universit\"{a}t Hannover, D-30167 Hannover, Germany}
\author[0000-0002-4559-8427]{N.~Adhikari}
\affiliation{University of Wisconsin-Milwaukee, Milwaukee, WI 53201, USA}
\author[0000-0002-5731-5076]{R.~X.~Adhikari}
\affiliation{LIGO Laboratory, California Institute of Technology, Pasadena, CA 91125, USA}
\author{V.~K.~Adkins}
\affiliation{Louisiana State University, Baton Rouge, LA 70803, USA}
% ...
 }{
 \author{\LVKCollabAuthors}
}
}

\date[\relax]{Compiled: \today}

\begin{abstract}

In this second of three papers on tests of \GR applied to the compact binary coalescence signals
    in the fourth Gravitational-Wave Transient Catalog (GWTC-4.0),
    we present the results of the parameterized tests of \GR and constraints on line-of-sight acceleration.
We include events up to and including the first part of the fourth observing run (O4a)
    of the LIGO--Virgo--KAGRA detectors.
As in the other two papers in this series,
    we restrict our analysis to the \TGRIINUMEVENTS confident signals,
    measured by at least two detectors,
    that have false alarm rates \TGRFARTHRESH from O4a,
    in addition to the \TGRNUMEVENTSPREV such events from previous observing runs.
This paper focuses on the \TGRIINUMTESTS tests that constrain parameterized deviations from the expected \GR (or unaccelerated) values.
These include modifications of post-Newtonian (PN) parameters,
    spin-induced quadrupole moments different from those of a binary black hole,
    and possible dispersive or birefringent propagation effects. 
Overall, we find no evidence for physics beyond \GR,
    for spin-induced quadrupole moments different from those of a Kerr black hole in \GR,
    or for line of sight acceleration,
    with more than 90\% of the events including the null result (no deviation) within their 90\% credible intervals.
We discuss possible systematics affecting the other events and tests,
    even though they are statistically not surprising, given noise.
The increased number of events analyzed allow us to improve the constraints on deviations from GR.
For instance, for the PN coefficients, we improve the constraints
    by factors of 
     $\TGRFTIBoundImprovement{min}$--$\TGRFTIBoundImprovement{max}$,
     though some of this improvement is due to allowing the PN coefficient deviations to affect more of the waveform.
We also provide illustrative translations to some modified theories.
Additionally, we update the bound on the mass of the graviton, at $90\%$ credibility, to $m_g \le$ \TGRMDRGravitonBound{gwtc4}.
Many of the bounds on possible deviations derived from our events are the best to date.

\end{abstract}

\subpapersection{Overview}\label{sec:paper II intro}

This paper is the second of three papers examining the nature of gravity through tests of \acf{GR}---and more generally of physics beyond the usual assumption of quasi-circular, isolated
    binaries
    composed of \acp{BH} and/or \acp{NS}---performed on the \ac{GW} signals
    as reported by the \ac{LVK} in the fourth Gravitational-Wave Transient Catalog \citep[GWTC-4.0;][]{GWTC:Introduction, GWTC:Results}.
Paper~I \citep{GWTC:TGR-I} gives an introduction and overview of our suite of events, tests, and shared methods,
    and also presents the results of the consistency tests.
Paper~III \citep{GWTC:TGR-III} focuses on tests of the remnants,
    considering the ringdown and possible echoes.
This paper describes the various parameterized tests performed to place quantitative bounds
    on possible deformations of the \ac{GW}s observed from \ac{CBC} sources,
    relative to those calculated in \GR. Specifically, we apply these tests to \BBH, \NSBH, and \BNS systems.
Where applicable, we describe the improved bounds relative to previous \ac{GW} catalogs and to existing bounds in the literature.

Specifically, we first examine deviations in \PN coefficients. The \PN approximation \citep[e.g.,][]{EIH1938,Chandrasekhar:1965gcg,Maggiore:2007ulw,2014grav.book.....P}
is the standard framework for computing and testing corrections to Newtonian gravity. Tests of gravity in the \PN framework have commonly been carried out in the weak field \citep[e.g., Solar System tests;][]{Will:2014kxa,2018tegp.book.....W}.
    However, for \acp{CBC}, the \PN approximation has been iterated to an impressively high order in perturbation theory \citep{Blanchet:2013haa}, so one can even use it to test highly relativistic effects during the end of inspiral
    \citep[e.g.,][]{Blanchet:1994ex,Blanchet:1994ez, Arun:2006hn, Arun:2006yw}.
For a binary system, the \PN approximation describes corrections to the binary's dynamics as a series in powers of $v/c$, where $v$ is the binary's orbital velocity.
The coefficients we test appear in the \ac{GW} phase, and thus describe a combination of the binary's conservative and dissipative dynamics.

The lowest few \PN parameters can be constrained with binary-pulsar observations \citep[e.g.,][]{Taylor:1982,Taylor:1989,Kramer:2021jcw},
    while higher \PN orders can only be constrained meaningfully with \ac{GW} observations,
    since these allow one to probe the strong-field, dynamical portion of the binary's evolution close to merger.
We use two frameworks
\clearpage
\startlongtable
\begin{deluxetable*}{c l cccc ccc c cc c cc}
\tabcolsep=0.06cm
\centerwidetable
\tablecaption{Event selection table for the analyses in this paper, from \ac{O4a}
       and the previous observing runs O1--O3}

\tablehead{
\colhead{} & \colhead{} & \colhead{} & \colhead{} & \colhead{} & \colhead{} &
    \multicolumn{3}{c}{Parameterized} &
    & \multicolumn{2}{c}{SIM}
    & \colhead{}
    & \multicolumn{2}{c}{PRP}    \\
\cline{7-9}\cline{11-12}
\cline{14-15}
\colhead{Run} &
 \colhead{Event Name} & 
 \colhead{SNR} & 
\colhead{$(1\!\!+\!\!z)\mathcal{M}/\Msun$} &
 \colhead{$\chi_{\mathrm{eff}}$} & 
 \colhead{$q$} & 
\colhead{TIGER} &
\colhead{FTI} &
\colhead{PCA} &
    &
\colhead{Phenom} &
\colhead{EOB} &
 \colhead{LOSA} & 
\colhead{MDR} &
\colhead{SSB}
 } 

\startdata
 O1 &
 \cellcolor{lightgray}{GW150914} & 
 \cellcolor{lightgray}{$\networkmatchedfiltersnrgwtctwopointoneuncert{GW150914_095045}$} & 
 \cellcolor{lightgray}{$\chirpmassdetgwtctwopointoneuncert{GW150914_095045}$} & 
 \cellcolor{lightgray}{$\chieffgwtctwopointoneuncert{GW150914_095045}$} & 
 \cellcolor{lightgray}{$\massratiogwtctwopointoneuncert{GW150914_095045}$} & 
 \cellcolor{lightgray}{\checkmark} & 
 \cellcolor{lightgray}{\checkprevlvk} & 
 \cellcolor{lightgray}{\checknodata} & 
 \cellcolor{lightgray}{} &
 \cellcolor{lightgray}{\checknodata} & 
 \cellcolor{lightgray}{\checknodata} & 
 \cellcolor{lightgray}{\checknodata} & 
 \cellcolor{lightgray}{\checkmark} & 
 \cellcolor{lightgray}{\checkprevext} \\
 & 
 GW151012* & 
 $\networkmatchedfiltersnrgwtctwopointoneuncert{GW151012_095443}$ & 
 $\chirpmassdetgwtctwopointoneuncert{GW151012_095443}$ & 
 $\chieffgwtctwopointoneuncert{GW151012_095443}$ & 
 $\massratiogwtctwopointoneuncert{GW151012_095443}$ & 
 \checkprevlvk & 
 \checknodata & 
 \checknodata & 
 &
 \checknodata & 
 \checknodata & 
 \checknodata & 
 \checkprevlvk &
  \checkprevext \\
 & 
 \cellcolor{lightgray}{GW151226} & 
 \cellcolor{lightgray}{$\networkmatchedfiltersnrgwtctwopointoneuncert{GW151226_033853}$} & 
 \cellcolor{lightgray}{$\chirpmassdetgwtctwopointoneuncert{GW151226_033853}$} & 
 \cellcolor{lightgray}{$\chieffgwtctwopointoneuncert{GW151226_033853}$} & 
 \cellcolor{lightgray}{$\massratiogwtctwopointoneuncert{GW151226_033853}$} & 
 \cellcolor{lightgray}{$\checkmark_\tigerinsp$} & 
 \cellcolor{lightgray}{\checkprevlvk} & 
 \cellcolor{lightgray}{\checknodata} & 
 \cellcolor{lightgray}{} &
 \cellcolor{lightgray}{\checkmark} & 
 \cellcolor{lightgray}{\checknodata} & 
 \cellcolor{lightgray}{\checknodata} & 
 \cellcolor{lightgray}{\checkmark} & 
 \cellcolor{lightgray}{\checkprevext} \\
\enddata

\tablecomments{
    Event selection table for the analyses in this paper, from \ac{O4a} \citep{GWTC:Results}
       and the previous observing runs O1 \citep{LIGOScientific:2016dsl},
       O2 \citep{GWTC1},
       O3a \citep{GWTC2,GWTC2p1},
       and O3b \citep{GWTC3},
       with binary parameters from each observing run's catalog paper.
       TIGER, FTI, and MDR results for GW170817 were published in \citet{LIGOScientific:2018dkp}.
       The starred events GW151012 and GW170729 had previously been used for the indicated tests of GR
    in \citet{LIGOScientific:2016dsl,LIGOScientific:2019fpa},
        but do not meet the current selection criteria.
       The O3a event GW190425 does meet the current selection criteria,
        but was previously excluded from most tests as a binary neutron star.
        A \checkmark{} indicates an event meeting our selection criteria for an analysis and thus included in our results;
        only these have been used to generate the bounds on \GR deviations presented in this paper.
    A \checkprevlvk{} indicates previous results by the LVK exist, but are not used for the new bounds here,
    while a \checkprevext{} indicates existing results in external works.
    A \checknodata indicates events not meeting selection criteria.
    For the TIGER tests, further subscripts $\tigerinsp / \tigerpostinsp$
        indicate where the test is only applied to the inspiral/post-inspiral stages, respectively.
    A few events have $\checkminus$ signs, indicating they had originally met the selection criteria,
        but technical difficulties in their analysis prevented obtaining meaningful results.
}

\label{tab:selectionII}

\end{deluxetable*}

\vspace{-22pt}
\noindent (TIGER and FTI, both defined and described below) to place bounds on deviations in individual \PN coefficients,
    and also illustrate how the posteriors on the PN coefficient deviations can be translated (with significant caveats) to the parameters in some alternative theories.

We also test for deviations in phenomenological post-inspiral coefficients with TIGER,
    in combinations of multiple \PN coefficients using principal component analysis (PCA),
    and in the spin-induced multipole (SIM) \PN coefficients.

Additionally, we place bounds on the line-of-sight acceleration (LOSA) of the source systems
    and on modifications to the propagation (PRP) of GWs to Earth,
    including in the GWs' dispersion relation (MDR),
    and of spacetime symmetry breaking (SSB) leading to anisotropic birefringence.
The LOSA analysis is not a test of \ac{GR}, but rather tests for astrophysics not included in the standard analyses
    performed in \citet{GWTC:Results}.
However, the LOSA analysis is included in this paper as such effects could appear as a \ac{GR} deviation in other tests.
Similarly, the LOSA analysis is also sensitive to other deviations from the assumption of quasi-circular isolated binaries,
    including deviations from \ac{GR},
    and provides a test of deviations starting at $-4$PN (see Section~\ref{sec:LOSA}) that is complementary to other analyses.

We apply these tests to the GW events from GWTC-4.0 that were observed with at least two detectors,
    and with a false-alarm rate of \TGRFARTHRESH,
    as described in Paper~I.
These include \TGRIINUMEVENTS events from \ac{O4a},
    which are new, as well as some events from previous runs,
    O1 \citep{LIGOScientific:2016dsl,LIGOScientific:2016lio},
    O2 \citep{GWTC1,LIGOScientific:2018dkp,LIGOScientific:2019fpa},
    O3a \citep{LIGOScientific:2020tif,GWTC2p1},
    and O3b \citep{GWTC3,LIGOScientific:2021sio},
    which have been used for a subset of the tests, in particular in creating combined results.
Table~\ref{tab:selectionII} details which events were analyzed by each test.
Table~\ref{tab:selectionII} also provides each binary's redshifted chirp mass $(1+z)\mathcal{M}$,
    effective inspiral spin $\chi_\mathrm{eff}$, and mass ratio $q$,
    as convenient combinations of the masses and spins \citep{GWTC:Introduction}.%

The rest of this paper is organized as follows:
Section~\ref{sec:generation} describes the parameterized tests of modifications in the generation of the GWs from the CBC systems,
    with the constraints on deviations in individual \PN and post-inspiral
    coefficients given in Section~\ref{sec:par} and the PCA multiparameter analysis given in Section~\ref{sec:pca},
    while Section~\ref{sec:SIM} presents the SIM test
    and Section~\ref{sec:LOSA} the LOSA test.
The propagation tests are given in Section~\ref{sec:propagation} (MDR in Section~\ref{sec:MDR} and SSB in Section~\ref{sec:SSB}).
We provide final conclusions along with a summary of the bounds and some of their implications in Section~\ref{sec:paper II conclusions}.
In Appendix~\ref{app:tiger_GW190728}, we consider the effects of real detector noise on the TIGER reanalysis of GW190728\_064510.
In Appendix~\ref{app:tiger_s231028}, we discuss the check of waveform systematics
    (i.e., systematic biases due to waveform modeling uncertainties) for the TIGER analysis of \FULLNAME{GW231028_153006},
    while in Appendix~\ref{app:mdr_s231028}, we discuss how prior effects lead to the apparent \GR deviation found in the MDR analysis of that event.
Finally, in Appendix~\ref{app:mdr_s231123},
    we discuss the waveform systematics found in the MDR analysis of \FULLNAME{GW231123_135430}
    (henceforth shortened to \COMMONNAME{GW231123}).

\subpapersection{Tests of \ac{GW} Generation}\label{sec:generation}

\subpapersubsection{FTI and TIGER: Single-parameter tests} \label{sec:par}

\newcommand{\dphi}[1]{\ensuremath{\delta\hat{\varphi}_{#1}}}

Alternative theories of gravity may lead to modifications of a binary's binding energy and angular momentum and the corresponding fluxes, which lead to changes in the orbital motion of the binary~\citep[e.g.,][]{Sotiriou:2006pq, Yagi:2011xp, Mirshekari:2013vb, Lang:2013fna, Lang:2014osa, Julie:2018lfp, Bernard:2022noq, Shiralilou:2021mfl}. This means that the inspiral portion of the \ac{GW} signal emitted by the binary differs from \GR. Such deviations from \GR can be captured by introducing extra parameters in the waveform. In this section, we focus on constraining deviations from \GR by introducing parametric deviations in the inspiral, post-inspiral, and merger--ringdown part of a waveform model. We will focus on deviations in the frequency domain \ac{GW} phase, since \ac{GW} observations are less sensitive to deviations in the amplitude \citep[see, e.g.,][]{Cutler:1994ys}.

The \ac{GW} signal from the early inspiral of compact binaries can be well described using the \PN formalism~\citep{Blanchet:2013haa}. In the \PN approximation, the waveform is expanded in powers of dimensionless velocity $v/c$, where $\mathcal{O}([v/c]^n)$ relative to the leading order is referred to as the $(n/2)$\PN order. Given the intrinsic parameters of a binary, the coefficients at each \PN order are uniquely determined in \GR. Measuring these coefficients by allowing them to differ from their \GR value can therefore be used as a null test of \GR~\citep{Blanchet:1994ex, Blanchet:1994ez, Arun:2006hn, Arun:2006yw, Yunes:2009ke, Mishra:2010tp, Li:2011cg, Li:2011vx}. Although many alternative theories of gravity admit a \PN expansion and can therefore be captured in this form, this is not always the case. For example, abrupt changes in the waveform such as dynamical scalarization do not admit a \PN expansion~\citep{Sampson:2013jpa, Khalil:2019wyy}. However, such a null test may still be able to detect such deviations from \GR.

The frequency domain phase of the \ac{GW} signal during early inspiral can be obtained from the \PN expanded time domain waveform using the stationary phase approximation~\citep{Sathyaprakash:1991mt, Cutler:1994ys, Buonanno:2009zt}. For quasi-circular, spin-aligned binaries, it has the form
\begin{multline}
    \Psi_{\ell m}(f) = 2\pi f t_c - \phi_c - \frac{\pi}{4} \\
		+ \frac{3 c^5}{128\eta v^5} \frac{|m|}{2} \sum_{n=0}^7\left( \psi_n + \psi_{nl} \log \frac{v}{c} \right) \frac{v^n}{c^n},
\end{multline}
where $v = \left[ 2\pi G (1+z)M f / |m| \right]^{1/3}$ with $f$ the \ac{GW} frequency, $M$ the total mass, $\eta$ the symmetric mass ratio, $z$ the redshift, and $t_c, \phi_c$ the time and phase at coalescence \citep{GWTC:Introduction}. The coefficients $\psi_n$ and $\psi_{nl}$ (where the $l$ denotes the coefficients of the logarithmic terms) are the $(n/2)$\PN coefficients and depend on the masses and spins of the binary. The subscript $\ell m$ denotes the $(\ell, m)$ multipole in the spherical harmonic decomposition of the waveform. The \PN coefficients are fully known up to 4.5\PN~\citep{Blanchet:2023bwj}, but we only include the ones up to 3.5\PN here since we do not test for higher orders.

Deviations from \GR can then be modeled by introducing parametric deviations to the non-spinning part of the \PN coefficients in \GR $\psi_i^\mathrm{GR, NS}$, i.e., of the form
\begin{equation}
    \psi_i \rightarrow (1+\dphi{i})\psi_i^\mathrm{GR, NS} + \psi_i^\mathrm{GR, S},
    \label{eq:frac_PN_dev}
\end{equation}
where $\dphi{i}$ is the deviation parameter and $\psi_i^\mathrm{GR, S}$ is the spin-dependent part of the \PN coefficient in \GR. This is the same parameterization that was previously used~\citep{LIGOScientific:2016lio, LIGOScientific:2018dkp, LIGOScientific:2019fpa, LIGOScientific:2020tif, LIGOScientific:2021sio, Sanger:2024axs}, and avoids possible singularities when the spin-dependent terms cancel with the non-spinning terms.
Since the spin terms are still included in the \GR part of the model, excluding the spin terms from the normalization of the deviations does not lead to any problems even for highly spinning systems. The full set of inspiral parameters that we constrain is then
\begin{equation}
    \left\{\dphi{-2}, \dphi{0}, \dphi{1}, \dphi{2}, \dphi{3}, \dphi{4}, \dphi{5l}, \dphi{6}, \dphi{6l}, \dphi{7}\right\}.
    \label{eq:frac_PN_terms}
\end{equation}
The non-logarithmic 2.5\PN term is degenerate with the coalescence phase, so it cannot be constrained. In \GR, the $-1$\PN and 0.5\PN terms vanish, so we instead treat $\dphi{-2}$ and $\dphi{1}$ as absolute deviations with a normalization factor equal to the 0\PN coefficient. All other deviation parameters are fractional deviations relative to the \GR values.

We use two different frameworks to constrain these inspiral deviation parameters. They use a different approach for how exactly the corrections are added to the \GR waveform, and differ in the underlying \GR waveform model used. Using two different implementations of the parameterized inspiral test with two different \GR waveform models makes the test more robust against waveform systematics coming from approximations made in the models.

The first framework is called the Flexible Theory-Independent (FTI) method~\citep{Mehta:2022pcn} and can use any frequency-domain, aligned-spin \GR waveform as a baseline model. In this work, we use the \SEOBNRFIVEHMROM~\citep{Pompili:2023tna} waveform approximant for \acp{BBH}. For \acp{NSBH} (only \FULLNAME{GW230518_125908} in O4a), we use \SEOBNRFOURNRtidalTWONSBH~\citep{Dietrich:2019kaq, Matas:2020wab}. Previous papers used \SEOBNRFOURROM~\citep{Bohe:2016gbl} for both \acp{BBH} and \acp{NSBH}, as well as \SEOBNRFOURHMROM~\citep{Cotesta:2020qhw} for a few \acp{BBH}, and additionally used \SEOBNRFOURNRTidalTWO~\citep{Dietrich:2019kaq} for the \acp{BNS} GW170817.
In FTI, the beyond \GR corrections are added to the phase of an inspiral--merger--ringdown frequency-domain waveform. The frequency-domain phase is computed for the inspiral part and then smoothly tapered off to zero towards merger--ringdown so that the post-inspiral part of the signal remains the same as in \GR. The tapering happens using a windowing function that is centered around a tapering frequency, which is a free parameter of the model together with the window width. We choose the tapering frequency to be the frequency at the peak of the $(2,2)$ multipole $f_\mathrm{tape} = f_\mathrm{peak}^{22}$ \citep[using the fit from][]{Bohe:2016gbl} and the window width to be one \ac{GW} cycle $\mathcal{N}_\mathrm{GW} = \TGRFTITaperingCycles$.

The settings used for FTI in previous papers~\citep{LIGOScientific:2016lio, LIGOScientific:2018dkp, LIGOScientific:2019fpa, LIGOScientific:2020tif, LIGOScientific:2021sio} were different than those employed here.
Specifically, the choice of the tapering frequency changed from $f_\mathrm{tape} = \TGRFTITaperingFreqOld f_\mathrm{peak}^{22}$ in GWTC-3.0 to $f_\mathrm{tape} = \TGRFTITaperingFreqNew f_\mathrm{peak}^{22}$ in this work. The motivation for increasing the tapering frequency is that this captures more of the late inspiral, which is especially important for higher \PN orders and more massive systems~\citep{Mehta:2022pcn}. Since the bounds obtained depend on the tapering frequency used, it is not possible to directly compare the GWTC-3.0 and O4a bounds. Typically, the increase in tapering frequency leads to tighter bounds on higher \PN coefficients (up to a factor of $\sim 3$) since more of the late inspiral is considered in the analysis, as illustrated in \citet{Mehta:2022pcn}. For low total-mass systems like \acp{BNS}, the tapering frequency lies outside the sensitive band of the detectors and therefore does not significantly influence the results. Only new events from O4a were analyzed with the updated FTI model, so any results obtained by combining events only use the events marked with a \checkmark in Table \ref{tab:selectionII}.

In the second framework, the Test Infrastructure for GEneral Relativity~\citep[TIGER;][]{Agathos:2013upa, Meidam:2017dgf, Roy:2025gzv}, the parameterized deviations are directly added to the \PN phase coefficients in the \IMRPhenomXAS~\citep{Pratten:2020fqn} waveform model. This means that it can only be used by waveforms in that family, and the corrections are propagated to the spin-precessing waveform with higher order multipole moments \IMRPhenomXPHMST~\citep[\IMRPhenomXPHM for brevity;][]{Pratten:2020ceb, Colleoni:2024knd}, which is used in this work. However, the corrections do not affect the spin precession.
By implementing the corrections directly into the \PN coefficients of the model, a smooth connection to the post-inspiral part of the waveform is guaranteed by construction. Previous versions of TIGER used the same construction for the \IMRPhenomPVTWO~\citep{Hannam:2013oca, Bohe:PPv2, Husa:2015iqa,Khan:2015jqa} and \IMRPhenomPVTHREEHM~\citep{Khan:2018fmp, Khan:2019kot} waveform models.

The TIGER framework also allows for parameterized deviations in the post-inspiral part of the waveform, similarly to~\citet{Cornish:2011ys}.
The \IMRPhenomXAS waveform has analytical expressions for the \ac{GW} phase in the intermediate and merger--ringdown regimes which contain coefficients that can be parameterized much like the inspiral \PN coefficients.
The main difference is that the post-inspiral coefficients come from calibrating the waveform to \NR simulations in \GR instead of from analytical results.
The parameters ${\delta \hat{b}_1, \delta \hat{b}_2, \delta \hat{b}_3, \delta \hat{b}_4}$ capture deviations in the \NR-calibrated coefficients in the intermediate regime, and the parameters ${\delta \hat{c}_1, \delta \hat{c}_2, \delta \hat{c}_4, \delta \hat{c}_\ell}$ capture deviations in the merger--ringdown calibration coefficients~\citep{Pratten:2020fqn, Roy:2025gzv}.

\begin{figure*}%
  \centering
  \includegraphics[width=\TGRFigureWidthPage]{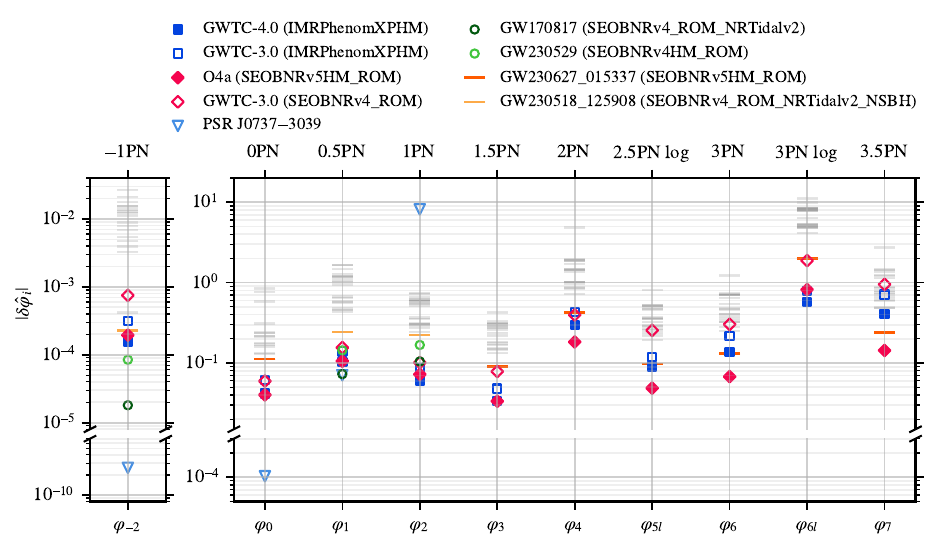}
  \caption{%
  Results for the magnitudes of the inspiral deviation coefficients (from $-1$PN to $3.5$PN) in terms of 90\% upper limits, compared with previous combined results from GWTC-3.0 \citep{LIGOScientific:2021sio}, including the updates to use the \IMRPhenomXPHM{} waveform model for the TIGER results.
  The horizontal stripes mark the results from individual events obtained with the \SEOBNRFIVEHMROM waveform model using the FTI analysis.
  Marginalizing over all results from O4a (GWTC-3.0), while assuming that all events share the same values for the violation coefficients, results in the filled (unfilled) red diamonds for the \SEOBNRFIVEHMROM (\SEOBNRFOURROM) waveform model.
  These bounds cannot be directly compared due to changes in the tapering frequency and waveform model used in the analysis.
  Marginalizing over all results from GWTC-4.0 (GWTC-3.0) for the TIGER analysis using the \IMRPhenomXPHM{} waveform model results in the filled (unfilled) blue squares.
  The dark and light orange stripes highlight \BBH \FULLNAME{GW230627_015337} and \NSBH \FULLNAME{GW230518_125908}, respectively, when they give the best individual event constraints on the inspiral parameters.
  The green circles show the bounds obtained with \BNS GW170817 \citep{LIGOScientific:2018dkp} and \NSBH \COMMONNAME{GW230529} \citep{Sanger:2024axs} when competitive; those two events are not included in the combined bounds.
  The blue upside-down triangles indicate bounds obtained with the double pulsar J0737$-$3039 \citep{Kramer:2021jcw} when competitive. Starting from 1.5PN the pulsar bounds get much worse than the bounds from \acp{GW} and are therefore not shown.
  }
  \label{fig:par:insp_bounds}
\end{figure*}

In this section, we vary only one deviation parameter
\begin{equation}
\delta \hat{p}_i \in \left\{ \underbrace{\dphi{-2}, \dphi{0}, \ldots, \dphi{7}}_{\text{Inspiral}}, \underbrace{\delta \hat{b}_1, \ldots, \delta \hat{b}_4}_{\text{Intermediate}}, \underbrace{\delta \hat{c}_1, \delta \hat{c}_2, \delta \hat{c}_4, \delta \hat{c}_\ell}_{\text{Merger-ringdown}} \right\}
\end{equation}
at a time. This is sufficient to detect and constrain deviations from \GR, even when the deviations are present in multiple \PN orders~\citep{Sampson:2013lpa, Meidam:2017dgf, Perkins:2022fhr}. Testing for multiple deviation parameters at the same time would lead to less informative posteriors due to correlations between the parameters~\citep{LIGOScientific:2016lio, Gupta:2020lxa}. In Section~\ref{sec:pca}, such a multi-parameter test is performed using the principal component analysis approach. For both FTI and TIGER, we use priors that are uniform in $\delta\hat{p}_i$ and symmetric around zero. For the TIGER analyses, we adopt the following prior ranges: $\dphi{-2} \in [-0.1, 0.1]$, $\dphi{0,1,2,3} \in [-5, 5]$, $\dphi{4, 5\ell, 6} \in [-20, 20]$, $\dphi{6\ell, 7} \in [-30, 30]$, and for the post-inspiral parameters, $[-20, 20]$. We use the following prior ranges for FTI: $\dphi{-2} \in [-1,1]$ ($[-0.01, 0.01]$ for \FULLNAME{GW230518_125908}) and $\dphi{i} \in [-20, 20]$ for all other parameters. These broad priors are chosen to avoid prior-railing issues (i.e., significant posterior probability density right up to at least one prior boundary), which often arise from degeneracies between the deviation parameters and the \GR parameters. Since the priors we use are wide enough to avoid truncating the posterior, the exact choice of range does not impact the constraints on deviation parameters, though it does impact the values of Bayes factors (quoted below for TIGER).

The FTI test is only applied to events that have a significant inspiral signal. The inspiral part of the signal is in this case defined in the frequency domain as the frequencies up to the tapering frequency used. The selection criteria are then a \SNR of at least \TGRFTIMinimumSNR{} in the inspiral and at least \TGRFTIMinimumCycles{} \ac{GW} cycles in the inspiral to ensure there is enough inspiral signal. The inspiral \SNR and cycles are calculated using the maximum likelihood parameters from the \GR analysis~\citep{GWTC:Results} with \IMRPhenomXPHM. In the TIGER analysis, we impose an \SNR threshold of greater than $6$ in the inspiral part of the signal to select an event for testing inspiral phasing coefficients. An identical criterion is applied in the post-inspiral regime to determine whether an event qualifies for the analysis of post-inspiral phasing coefficients. Following the prescription of the \IMRPhenomXAS model, we use the $(2,2)$ multipole frequency at the minimum energy circular orbit \citep[\acsu{MECO};][]{Cabero:2016ayq} to divide the waveform between the inspiral and post-inspiral regimes. (There are small corrections to the \ac{MECO} used to obtain the exact end-of-inspiral frequency in \IMRPhenomXAS, but these are less than a \TGRTIGERMaxFracDiffEndInspiralvsMECO{} correction for the range of mass ratios and spins allowed by our priors.) This prescription gives an end-of-inspiral frequency that is between \TGRFTIEndInspiraltoTIGEREndInspiralRatioMin{} and \TGRFTIEndInspiraltoTIGEREndInspiralRatioMax{} times smaller than the one used by FTI, with the largest differences occurring for negative effective inspiral spins \cite[defined in][]{GWTC:Introduction}. The median \SNR values are computed using posterior samples obtained from the standard \GR analysis with the \IMRPhenomXPHM model.

\begin{figure*}%
  \centering
  \includegraphics[width=\TGRFigureWidthPage]{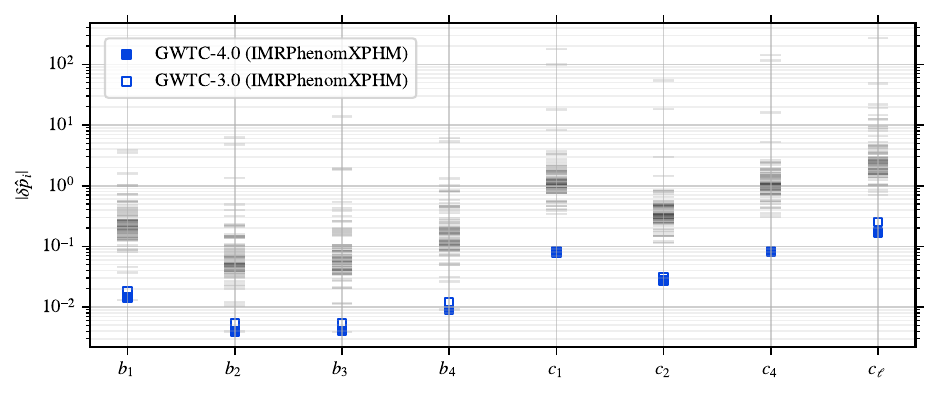}
  \caption{%
  Results for the magnitudes of the post-inspiral deviation coefficients using TIGER.
  The horizontal stripes mark the results from individual events obtained with the \IMRPhenomXPHM waveform model.
  Marginalizing over all analyzed events in GWTC-4.0 (GWTC-3.0), while assuming that all events share the same values for the violation coefficients, results in the filled (unfilled) blue squares.
  }
  \label{fig:tiger:post_bounds}
\end{figure*}

\begin{figure*}%
  \centering
\includegraphics[keepaspectratio, width=\TGRFigureWidthPage]{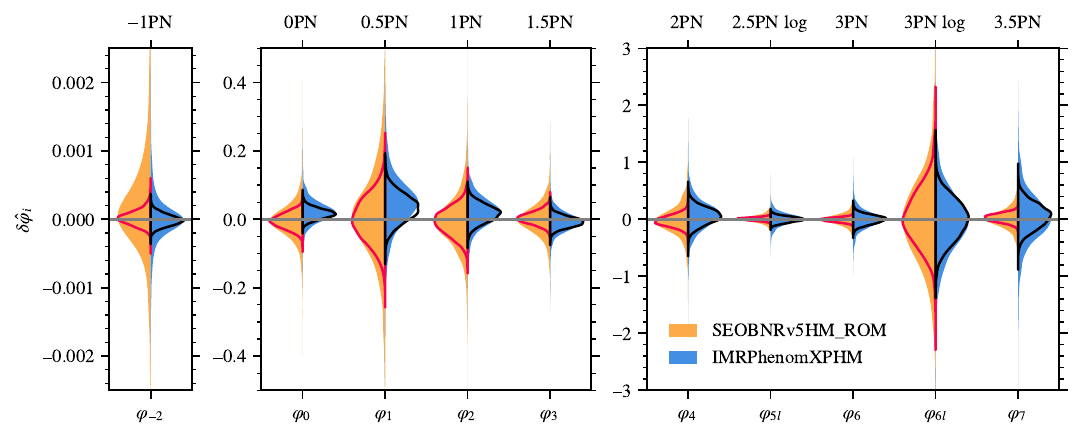}
\caption{%
The constraints on the \GR violation parameters in the inspiral, marginalized over all analyzed events,
  in orange on the left for the \SEOBNRFIVEHMROM waveform model (FTI analysis)
  and in blue on the right for the \IMRPhenomXPHM waveform model (TIGER analysis).
Consistency of \GR corresponds to vanishing parameters (the horizontal line).
Assuming independent violation parameters between events, using a hierarchical analysis, results in the filled probability densities. 
Instead, assuming that the violation parameters share the same values between different events leads to the unfilled distributions.
}
\label{fig:par:insp_combined}
\end{figure*}

For the TIGER part, we also analyze the GWTC-3.0 events using the new \IMRPhenomXPHM-based framework \citep{Roy:2025gzv}. We keep the same event-selection criteria as in the GWTC-2.0 tests of \GR{} paper \citep{LIGOScientific:2020tif}. GW170809 was not included in the GWTC-2.0 TIGER inspiral analysis. This event does not meet the selection criteria when using \IMRPhenomPVTWO, but it passes them when using the \IMRPhenomXPHM{} posterior samples from GWTC-2.1 \citep{GWTC2p1}. However, we leave its possible inclusion in the TIGER inspiral analysis to a future publication. GW191109\_010717 was excluded from the combined analysis despite meeting the post-inspiral selection criteria. We found apparent deviations for this event, similar to the findings reported in the GWTC-3.0 tests of \GR{} paper~\citep{LIGOScientific:2021sio}, with the apparent tension explained as being driven by data-quality issues in both LIGO detectors rather than by genuine departures from \GR.  For GW190728\_064510, we find apparent tension in the lower-\PN{} coefficients when analyzing it using the \IMRPhenomXPHM{} version of TIGER. However, we also find similar behavior when analyzing simulated signals added to real-noise and conclude that the apparent tension is likely caused by noise-induced fluctuations that the analysis including higher multipole moments is more sensitive to rather than evidence for a departure from \GR. This is further discussed in Appendix~\ref{app:tiger_GW190728}.

The results for the inspiral deviation parameters $\dphi{i}$ are consistent with \GR for all events analyzed.
There are two events where \GR is outside the $90\%$ credible interval for some parameters for FTI.
For the parameters with the largest deviations, \GR is still found within the $\TGRFTICredibleLevel{GW230628_231200}\%$ credible interval for \FULLNAME{GW230628_231200} and the $\TGRFTICredibleLevel{GW231110_040320}\%$ credible interval for \FULLNAME{GW231110_040320}.
However, for \FULLNAME{GW230628_231200}, TIGER finds the \GR value within the 90\% credible interval for the inspiral parameters, though it is found outside that interval for some post-inspiral parameters, but still within the $\TGRTIGERCredibleLevel{GW230628_231200}\%$ credible interval. For \FULLNAME{GW231110_040320}, TIGER finds that \GR lies outside the 90\% credible interval for all the inspiral parameters except $-1$PN, and for all of the intermediate parameters. However, \GR is still found within the $\TGRTIGERCredibleLevel{GW231110_040320}\%$ credible interval. Additionally, the TIGER Bayes factors prefer \GR for both events for all the deviation parameters. In general, we expect to find \GR outside the 90\% credible interval for a few events due to noise fluctuations, given the number of events under consideration.

\begin{figure}
\centering
\includegraphics[width=\TGRFigureWidth]{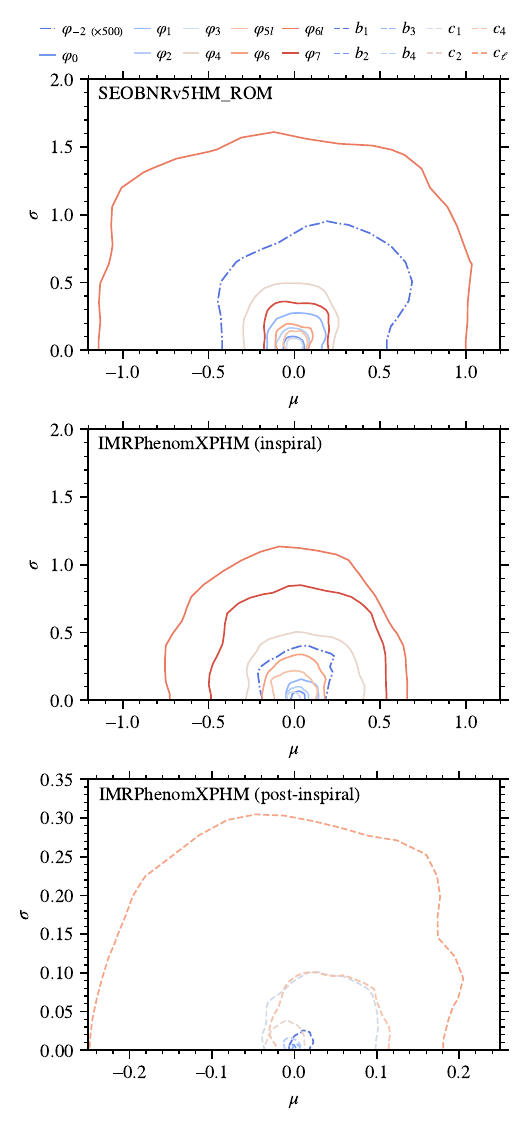}
\caption{%
Inferred hyperparameters $\mu$ and $\sigma$ of the \GR violation coefficients.
For each testing coefficient, the contours mark 90\% credible regions.
The inspiral parameters for the two waveform models \SEOBNRFIVEHMROM (using the FTI analysis) and \IMRPhenomXPHM (using the TIGER analysis) are considered in the top and middle panel respectively. The bottom panel shows the post-inspiral parameters for the \IMRPhenomXPHM waveform model.
All contours enclose the \GR expectation of $\mu=\sigma=0$;
the $\dphi{-2}$ values are rescaled $\times 500$ to improve visibility.
}
\label{fig:par:hier_contour}
\end{figure}

For four of the events, namely \FULLNAME{GW230518_125908}, \FULLNAME{GW231020_142947}, \FULLNAME{GW231104_133418}, and \FULLNAME{GW231113_200417}, we were unable to obtain constraints on the $0$\PN deviation parameter for FTI.
These are all low-mass events with low \SNR where strong correlations between $\dphi{0}$ and the chirp mass lead to problems in the analysis~\citep{Sanger:2024axs}. The higher tapering frequency in FTI, compared to the lower end-of-inspiral frequency in TIGER, implies that this degeneracy is stronger for FTI.
The TIGER results do not show this issue for these events, except for \FULLNAME{GW230518_125908} and one O3b event, GW200115\_042309. This is likely because the lower cutoff frequency leaves a few unmodified \ac{GW} cycles in band that break the degeneracy. The exceptional behavior in these two events is specifically due to a strong degeneracy between $\dphi{0}$ and the chirp mass, leading to a significantly larger chirp mass compared to the \GR case and, consequently, to an apparent deviation in the 0PN term with a positive value, as seen in~\citet{Sanger:2024axs}. The FTI analysis of GW200115\_042309 in \citet{LIGOScientific:2021sio} also exhibits a degeneracy with the chirp mass, resulting in a bias in the posterior distribution of the $0$\PN deviation parameter. Since a lower tapering frequency was used in that analysis, the bias was less severe, and the event was not excluded from the combined analysis.

The deviation parameters at other PN orders can also show significant correlations with the intrinsic parameters. How strong these correlations are depends on the PN order and where in parameter space the system is. In particular, the correlations are strongest at the order a given parameter first appears, so $\dphi{2}$ for the mass ratio and $\dphi{3}$ for spins in addition to $\dphi{0}$ and the chirp mass. However, these correlations are not as strong as is the case for the $\dphi{0}$--chirp mass degeneracy for low mass systems, and higher SNR events usually show smaller correlations. We therefore do not experience any problems with sampling for other PN deviation parameters or higher mass systems.

TIGER finds the \GR value near the boundary of $90\%$ credibility for the high-mass event \FULLNAME{GW231028_153006}, where the \GR analysis~\citep{GWTC:Results} finds significant systematic modeling uncertainties. As discussed in Appendix~\ref{app:tiger_s231028}, we find that the largest shifts away from GR (for the intermediate coefficients) are likely attributable to waveform modeling uncertainties, but the smaller shifts for the merger--ringdown coefficients seem instead to be attributable primarily to a prior effect.

\begin{figure*}%
  \centering
\includegraphics[keepaspectratio, width=\TGRFigureWidthPage]{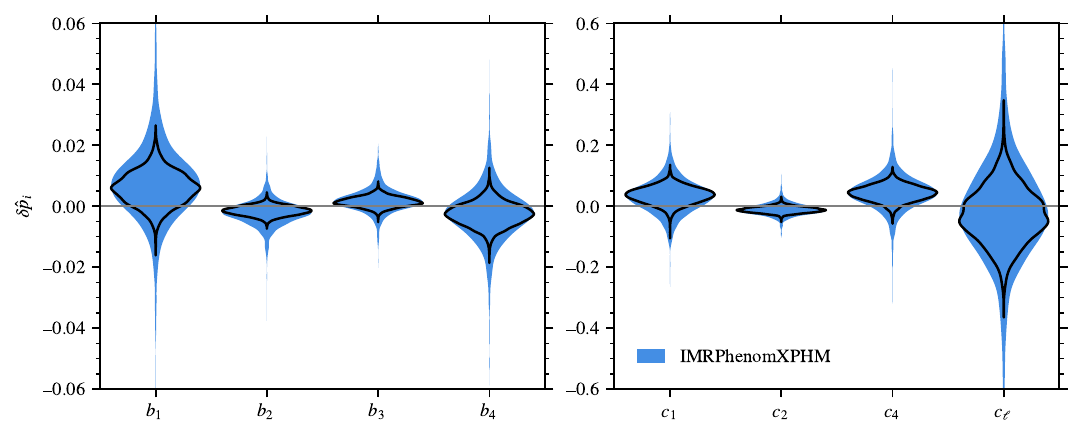}
\caption{%
The constraints on the \GR violation parameters in the post-inspiral using TIGER, marginalized over all analyzed events in GWTC-4.0.
Consistency of \GR corresponds to vanishing parameters (the horizontal line).
Assuming independent violation parameters between events, using a hierarchical analysis, results in the colored probability densities. 
Instead, assuming that the violation parameters share the same values between different events leads to the unfilled distributions.
}
\label{fig:tiger:post_combined}
\end{figure*}

In Figure \ref{fig:par:insp_bounds}, we show the 90\% upper bounds on the magnitude of the \PN deviation parameters in the inspiral; the analogous plot for the TIGER post-inspiral coefficients is Figure~\ref{fig:tiger:post_bounds}.
The bounds for individual events obtained with FTI are shown as grey stripes.
We highlight the event \FULLNAME{GW230627_015337}, which is the \BBH signal that gives the best constraints. This event is particularly good for inspiral tests of \GR because its low masses means there is a long inspiral while it also has a relatively high \SNR of $\networkmatchedfiltersnruncert{GW230627_015337}$.
We also show the bounds for the \BNS GW170817~\citep{LIGOScientific:2018dkp} and \NSBH \FULLNAME{GW230529_181500} \citep[abbreviated to \COMMONNAME{GW230529};][]{Sanger:2024axs} for comparison when their bounds are competitive.
\COMMONNAME{GW230529} does not meet the selection criteria for inclusion in this paper because it was detected by only one detector and therefore is excluded from the combined results we present here.
The high-\SNR single-detector event \FULLNAME{GW230814_230901}~\citep{GW230814} is not shown, however, since its bounds do not stand out compared to those of other events, though the more recent even higher-\SNR event GW250114~\citep{GW250114} from the \ac{O4b} does provide even better constraints on its own~\citep{GW250114_TGR}.

We also show the combined bounds for both FTI and TIGER obtained under the assumption that deviations have the same value for each event.
When combining events, we exclude the \BNS GW170817 due to the different nature of its source. For comparison, we also include the combined FTI and TIGER results from previous catalogs. For TIGER, we plot the updated results from the reanalysis of GWTC-3.0 using the \IMRPhenomXPHM{} version of TIGER \citep{Roy:2025gzv}, though there are no TIGER results in the GWTC-3.0 paper \citep{LIGOScientific:2021sio}.
 All combined bounds improve with the largest improvements seen for FTI for the high-\PN parameters.
As discussed before, the change in tapering frequency is contributing to this improvement.
The list of events included in the combined bounds is different between FTI and TIGER, making it impossible to directly compare the combined bounds between the two.
The differences between the bounds for individual events from FTI and TIGER should be seen as an estimate for the systematic uncertainties that can be expected from inspiral tests, with FTI pushing the transition from inspiral to post-inspiral to the highest possible frequency and TIGER providing more conservative bounds.
The best bound from \acp{GW} at $-1$PN is still from GW170817, and is an order of magnitude better than the combined constraints from O4a (FTI) or GWTC-4.0 (TIGER).

It is also possible to put constraints on the PN deviation parameters using binary pulsars~\citep{Yunes:2010qb,Nair:2020ggs}.
The best overall bound on dipole radiation comes from the double pulsar PSR J0737$-$3039A/B~\citep{Kramer:2021jcw}, with a 90\% upper bound corresponding to $|\dphi{-2}| \leq \TGRFTIPulsarBound{dchiMinus2}$.
This double pulsar also puts the most stringent constraint on deviations from \GR at Newtonian order with $|\dphi{0}| \leq \TGRFTIPulsarBound{dchi0}$. At $0.5$\PN, the double pulsar bound is $|\dphi{1}| \leq \TGRFTIPulsarBound{dchi1}$, which is on par with the bound from GW170817. From $1$\PN onwards the bounds from the double pulsar are less constraining than the bounds from \acp{GW}.
The bounds from PSR J0737$-$3039A/B are shown as blue upside-down triangles in Figure~\ref{fig:par:insp_bounds}.

\begin{table*}[t!]
  \caption{\label{tab:par}
  Results from parametrized tests of \ac{GW} generation
}
  \begin{center}
  \begin{center}
\begin{tabular}[t]{cccrrrrrrrr}
\toprule
 $p_i$ & pipeline & waveform & \phantom{X} & \multicolumn{4}{c}{Hierarchical} & \phantom{X} & \multicolumn{2}{c}{Restricted} \\
\cline{5-8} \cline{10-11}
 & & &  & \multicolumn{1}{c}{$\mu$} & \multicolumn{1}{c}{$\sigma$} & \multicolumn{1}{c}{$\delta\hat{\varphi}_i$} & \multicolumn{1}{c}{$Q_\mathrm{GR}$} &  & \multicolumn{1}{c}{$\delta\hat{\varphi}_i$} & \multicolumn{1}{c}{$Q_\mathrm{GR}$} \\
\midrule
$\varphi_{-2}~\scriptstyle{(\times 500)}$ & FTI & EOB &    & $0.06^{+0.47}_{-0.27}$ & $< 0.66$ & $0.05^{+0.81}_{-0.57}$ & $40\%$ &   & $0.01^{+0.10}_{-0.08}$ & $44\%$ \\
& TIGER & Phenom &   & $-0.00^{+0.17}_{-0.13}$ & $< 0.31$ & $-0.00^{+0.32}_{-0.28}$ & $52\%$ &   & $-0.01^{+0.07}_{-0.07}$ & $57\%$ \\
$\varphi_0$ & FTI & EOB &    & $0.00^{+0.05}_{-0.04}$ & $< 0.08$ & $0.00^{+0.09}_{-0.09}$ & $54\%$ &   & $-0.01^{+0.04}_{-0.04}$ & $59\%$ \\
& TIGER & Phenom &    & $0.02^{+0.03}_{-0.03}$ & $< 0.05$ & $0.02^{+0.06}_{-0.05}$ & $22\%$ &   & $0.02^{+0.03}_{-0.02}$ & $14\%$ \\
$\varphi_1$ & FTI & EOB &    & $0.01^{+0.13}_{-0.12}$ & $< 0.22$ & $0.01^{+0.24}_{-0.24}$ & $46\%$ &   & $0.00^{+0.11}_{-0.11}$ & $50\%$ \\
 & TIGER & Phenom &    & $0.05^{+0.07}_{-0.07}$ & $< 0.12$ & $0.05^{+0.13}_{-0.14}$ & $23\%$ &   & $0.03^{+0.07}_{-0.07}$ & $20\%$ \\
$\varphi_2$ & FTI & EOB &    & $-0.01^{+0.07}_{-0.07}$ & $< 0.13$ & $-0.01^{+0.13}_{-0.13}$ & $55\%$ &   & $-0.01^{+0.06}_{-0.07}$ & $57\%$ \\
& TIGER & Phenom &    & $0.02^{+0.05}_{-0.05}$ & $< 0.07$ & $0.02^{+0.08}_{-0.08}$ & $30\%$ &   & $0.02^{+0.04}_{-0.04}$ & $26\%$ \\
$\varphi_3$ & FTI & EOB &    & $0.00^{+0.04}_{-0.03}$ & $< 0.07$ & $0.00^{+0.08}_{-0.08}$ & $51\%$ &   & $0.00^{+0.03}_{-0.03}$ & $43\%$  \\
& TIGER & Phenom &    & $-0.01^{+0.03}_{-0.03}$ & $< 0.05$ & $-0.01^{+0.05}_{-0.05}$ & $61\%$ &   & $-0.01^{+0.03}_{-0.03}$ & $69\%$ \\
$\varphi_4$ & FTI & EOB &    & $-0.02^{+0.21}_{-0.19}$ & $< 0.40$ & $-0.02^{+0.45}_{-0.43}$ & $54\%$ &   & $-0.03^{+0.16}_{-0.15}$ & $62\%$  \\
& TIGER & Phenom &   & $0.06^{+0.25}_{-0.25}$ & $< 0.40$ & $0.07^{+0.45}_{-0.45}$ & $37\%$ &   & $0.06^{+0.24}_{-0.25}$ & $34\%$ \\
$\varphi_{5l}$ & FTI & EOB &    & $0.01^{+0.05}_{-0.05}$ & $< 0.11$ & $0.01^{+0.13}_{-0.12}$ & $45\%$ &   & $0.01^{+0.04}_{-0.05}$ & $40\%$  \\
& TIGER & Phenom &    & $0.00^{+0.09}_{-0.11}$ & $< 0.17$ & $0.01^{+0.18}_{-0.21}$ & $46\%$ &   & $0.00^{+0.08}_{-0.09}$ & $52\%$ \\
$\varphi_6$ & FTI & EOB &    & $0.00^{+0.09}_{-0.08}$ & $< 0.15$ & $0.00^{+0.18}_{-0.18}$ & $51\%$ &   & $-0.01^{+0.06}_{-0.06}$ & $58\%$  \\
& TIGER & Phenom &    & $0.00^{+0.14}_{-0.13}$ & $< 0.26$ & $0.00^{+0.28}_{-0.29}$ & $50\%$ &   & $0.00^{+0.13}_{-0.14}$ & $50\%$ \\
$\varphi_{6l}$ & FTI & EOB &    & $-0.04^{+0.79}_{-0.81}$ & $< 1.23$ & $-0.01^{+1.45}_{-1.40}$ & $51\%$ &   & $0.01^{+0.83}_{-0.75}$ & $49\%$  \\
& TIGER & Phenom &    & $-0.04^{+0.54}_{-0.54}$ & $< 0.86$ & $-0.04^{+0.94}_{-1.00}$ & $54\%$ &   & $0.00^{+0.55}_{-0.60}$ & $50\%$ \\
$\varphi_7$ & FTI & EOB &    & $0.01^{+0.15}_{-0.14}$ & $< 0.29$ & $0.01^{+0.33}_{-0.34}$ & $47\%$ &   & $0.03^{+0.11}_{-0.12}$ & $38\%$  \\
& TIGER & Phenom &    & $0.03^{+0.38}_{-0.39}$ & $< 0.66$ & $0.02^{+0.77}_{-0.78}$ & $48\%$ &   & $0.06^{+0.35}_{-0.35}$ & $40\%$ \\
\midrule
$b_1~\scriptstyle{(\times 10)}$ & TIGER & Phenom &   & $0.08^{+0.11}_{-0.10}$ & $< 0.19$ & $0.07^{+0.22}_{-0.18}$ & $23\%$ &   & $0.06^{+0.09}_{-0.09}$ & $13\%$ \\
$b_2~\scriptstyle{(\times 10)}$ &  TIGER & Phenom &  & $-0.02^{+0.03}_{-0.03}$ & $< 0.06$ & $-0.02^{+0.06}_{-0.07}$ & $76\%$ &   & $-0.01^{+0.02}_{-0.03}$ & $83\%$ \\
$b_3~\scriptstyle{(\times 10)}$ & TIGER & Phenom &   & $0.02^{+0.04}_{-0.03}$ & $< 0.06$ & $0.02^{+0.07}_{-0.06}$ & $26\%$ &   & $0.01^{+0.03}_{-0.03}$ & $18\%$ \\
$b_4~\scriptstyle{(\times 10)}$ & TIGER & Phenom &   & $-0.03^{+0.07}_{-0.08}$ & $< 0.12$ & $-0.02^{+0.13}_{-0.14}$ & $65\%$ &   & $-0.03^{+0.06}_{-0.06}$ & $78\%$ \\
$c_1$ & TIGER & Phenom &   & $0.03^{+0.05}_{-0.06}$ & $< 0.08$ & $0.03^{+0.09}_{-0.09}$ & $25\%$ &   & $0.04^{+0.04}_{-0.05}$ & $9\%$ \\
$c_2$ & TIGER & Phenom &   & $-0.01^{+0.02}_{-0.02}$ & $< 0.03$ & $-0.01^{+0.03}_{-0.03}$ & $75\%$ &   & $-0.01^{+0.02}_{-0.01}$ & $89\%$ \\
$c_4$ & TIGER & Phenom &   & $0.04^{+0.05}_{-0.06}$ & $< 0.08$ & $0.04^{+0.08}_{-0.09}$ & $17\%$ &   & $0.04^{+0.04}_{-0.04}$ & $5\%$ \\
$c_\ell$ &  TIGER & Phenom &  & $-0.02^{+0.17}_{-0.16}$ & $< 0.24$ & $-0.02^{+0.29}_{-0.27}$ & $56\%$ &   & $-0.03^{+0.15}_{-0.14}$ & $64\%$ \\
\bottomrule
\end{tabular}
\end{center}

  \end{center}
  \tablecomments{Combined constraints on the deviation parameters obtained by marginalizing over all analyzed events in GWTC-4.0
    using the analyses FTI (which uses the \SEOBNRFIVEHMROM waveform)
    and TIGER (which uses the \IMRPhenomXPHM and \IMRPhenomPVTWO waveforms).
    Hierarchical (restricted) constraints are obtained under the assumption that deviation coefficients can (cannot) vary across the observed events.
    The one-sided quantile corresponding to the \GR value for the distributions plotted in Figures~\ref{fig:par:insp_combined} and~\ref{fig:tiger:post_combined} is indicated by $Q_{\rm GR}$.
    For hierarchical constraints, we also provide the mean $\mu$ and standard deviation $\sigma$ of the inferred hyperdistribution.
    For $\dphi{i}$ and $\mu$, we report the median as well as the 90\%-credible intervals, while for $\sigma$ we only present 90\% upper bounds. We scale the $\varphi_{-2}$ and $b_i$ entries
    to display the bounds with only two decimal places (for $b_1$ and $b_4$ without rounding up). For $\varphi_{-2}$, we use the same scaling as in Figure~\ref{fig:par:hier_contour}.
  }  

\end{table*}

The combined posteriors for the inspiral deviation parameters are shown in Figure \ref{fig:par:insp_combined} (unfilled violins) and are consistent with \GR at the 90\% credible level for all parameters.
The combined posteriors are dominated by \FULLNAME{GW230627_015337} because of its tight constraints.
The posteriors for this event are centered around zero, a trait inherited by the displayed combined posteriors. This is reflected in the \GR quantiles shown in Table \ref{tab:par} being between 40\% and 60\%.
Here the \GR quantile is defined as the fraction of samples with $\dphi{i}<0$, so a posterior perfectly centered on \GR gives a quantile of 50\%.
We also combined the posteriors from the different events using a hierarchical approach to get the combined posteriors shown in Figure \ref{fig:par:insp_combined} (filled violins). This approach allows the deviation parameters to take independent values for each event, modeling their distribution as a Gaussian described by a mean $\mu$ and standard deviation $\sigma$.
We again see that the posteriors are consistent with \GR at the 90\% credible level for all \PN deviation parameters. The 90\% contours for the hyperparameters $\mu$ and $\sigma$ of the hierarchical approach are shown in Figure \ref{fig:par:hier_contour} for each PN deviation parameter.
\GR corresponds to $\mu=\sigma=0$, which is enclosed by the contours for all inspiral parameters. The medians, 90\% credible intervals, and \GR quantiles for all $\dphi{i}$ and both approaches to combine results are listed in Table~\ref{tab:par}, as well as the medians and 90\% credible intervals for $\mu$ and the 90\% upper limits on $\sigma$.

Figure \ref{fig:tiger:post_bounds} shows the 90\% upper bounds on the magnitude of the post-inspiral deviation parameters obtained with TIGER for each O4a event (grey stripes). The blue squares are the bounds obtained by combining all events assuming that deviations have the same value for each event. Figure \ref{fig:tiger:post_combined} shows the combined posteriors for the post-inspiral parameter obtained with both the restricted (unfilled violins) and hierarchical (filled violins) approaches. The 90\% contours for the hyperparameters of the hierarchical analysis are shown in the bottom panel of Figure~\ref{fig:par:hier_contour}. Table~\ref{tab:par} gives the median and 90\% credible intervals on the post-inspiral deviation parameters for the combined constraints using both approaches, as well as the medians and 90\% credible intervals for $\mu$ and the 90\% upper limits on $\sigma$. All results for the TIGER post-inspiral parameters are consistent with \GR at the 90\% credible level.

The improvement in the combined post-inspiral TIGER constraints from GWTC-3.0 to GWTC-4.0 is not as significant as that found for the inspiral coefficients. The combined post-inspiral constraints are mostly dominated by the O3a events GW190412 and GW190814, which provide the strongest bounds among the analyzed events for most coefficients. Thus, the addition of O4a events does not lead to a large improvement, even though 31 out of the 66 events in the combined analysis are from O4a. For the TIGER inspiral coefficients, the improvement is instead largely driven by the O4a events \FULLNAME{GW230518_125908} and \FULLNAME{GW230627_015337}, which enable significantly tighter constraints on several \PN{} coefficients. If these two events are removed from the combined inspiral analysis, the improvement becomes comparable to what is found for the post-inspiral analysis.

\begin{table*}[t!]
\caption{\label{tab:map_to_tgr}
Illustrative mappings of agnostic constraints on PN coefficients onto parameters of selected modified gravity theories for the event that gives the best result for each mapping
  }
  \begin{center}
  \begin{tabular}{cccccc}
\toprule
	Theory & $\varphi_i$ & Parameter  & FTI mapping & TIGER mapping & Existing constraint/mapping \\
\midrule
	Scalar--tensor  & $\varphi_{-2}$ & $|\dot{\phi}| \ [\rm{s}^{-1}]$  & $1.1 \times 10^3$  & $1.1 \times 10^3$  & $10^{-6} $ \citep{Horbatsch:2011ye} \\
	Einstein--dilaton--Gauss--Bonnet  & $\varphi_{-2}$ & $\sqrt{\alpha_{\rm{EdGB}}} \ [\rm{km}]$ & 0.98 & 1.07   & 0.28  \citep{Sanger:2024axs} \\
	Einstein--Maxwell  & $\varphi_{-2}$ & $\displaystyle \zeta=\left|\frac{q_1}{m_1}-\frac{q_2}{m_2}\right|$ & 0.047  & 0.052 & 0.024  \citep{Gao:2024rel} \\
	pseudo-complex GR $n=2$  & $\varphi_{4}$ & $b_c/b$ & \reviewed{0.34} & \reviewed{0.15}  & \reviewed{0.10} (FTI), \reviewed{0.07} (TIGER)${^\dagger}$ \\
	pseudo-complex GR $n=3$  & $\varphi_{6}$ & $b_c/b $ & \reviewed{0.25} & \reviewed{0.07}  & \reviewed{0.05} (FTI), \reviewed{0.05} (TIGER)${^\dagger}$ \\
\bottomrule
\end{tabular}

  \end{center}
   \tablecomments{
The values shown are mappings of posteriors on agnostic PN
deformation parameters (from FTI and TIGER analyses) into 90\% credibility results for parameters in 
modified gravity theories for the event that gives the best result for each mapping.
These mappings are performed independently for each O4a event using posterior samples for the intrinsic binary parameters along with the agnostic PN deformation parameter, and
includes consistent prior reweighting. The results therefore reflect how agnostic 
constraints project onto theory parameters under simplifying assumptions, rather than a 
joint or theory-specific inference. Specifically the binary parameters used for the translations involving $\varphi_{-2}$ are those of
GW230518\_125908, which provide the best results, while for
pseudo-complex \GR, GW230627\_015337 provides the best results for all but TIGER $\varphi_6$, where GW230919\_215712 provides the best result, so those events' parameters are used.
Additionally, for pseudo-complex \GR, we give the best illustrative translations for O3b events, for comparison, using the results from \citet{LIGOScientific:2021sio} for FTI and our reanalysis in this paper for TIGER. The O3b translations come from GW191204\_171526, except for FTI $\varphi_4$, which comes from GW191216\_213338. We mark those entries as a dagger to call attention to their status as only illustrative translations, compared to the bounds given for other theories.
All these translations should not be interpreted as statistical constraints on the underlying
theories, nor as exclusions. In particular, real modified gravity theories generically 
induce correlated deviations across multiple PN coefficients and modify the 
merger--ringdown signal, effects not included in this mapping.
The table is therefore intended only to provide intuition about the agnostic PN bounds we have obtained by showing how they would translate to the parameters of some commonly considered alternative theories, together with constraints on those parameters from the literature.
For pseudo-complex \GR, larger values correspond to better constraints and 
a $b_c/b$ bound above 1 rules out horizonless objects for that $n$ in this model \citep{Maimon:2025hsi}.
  }
\end{table*}

A useful framework for interpreting agnostic PN deformation tests is the parameterized 
post-Einsteinian (ppE) formalism, which provides a systematic description of leading 
deviations from general relativity in the gravitational waveform phase and amplitude 
\citep{Yunes:2009ke,Cornish:2011ys}. In this approach, deviations are introduced as 
power-law corrections at specific PN orders, allowing one to connect agnostic constraints 
on PN coefficients to a broad class of modified gravity theories.

To provide intuition for the meaning of the agnostic constraints on deviations in the PN coefficients we have obtained,
we thus follow the ppE formalism and
perform event-by-event illustrative mappings of the inferred PN deformation posteriors 
onto parameters appearing in selected modified gravity theories, with the translation for the event that provides the best result for each mapping given in Table~\ref{tab:map_to_tgr}.
The example theories considered are: scalar--tensor theories with an evolving scalar field \citep{Jacobson:1999vr,Horbatsch:2011ye}, Einstein--dilaton--Gauss--Bonnet \citep[EdGB;][]{Gross:1986mw,Kanti:1995vq}, dynamical Chern--Simons \citep[dCS;][]{Alexander:2009tp}, Einstein--Maxwell with black hole charge, and dirty black holes or pseudo-complex \GR \citep{Hess:2008wd, Caspar:2012ux, Hess:2016gmh,Maimon:2025hsi}.
 
These mappings should not be interpreted as theory-specific statistical constraints, 
but rather as illustrative translations of agnostic deviations into the language of 
particular models, obtained under simplifying assumptions 
(e.g., leading-order modifications and no deviations in the merger-ringdown phase).

Importantly, each mapping is performed independently for each \ac{GW} event using the 
corresponding posterior samples of both intrinsic binary parameters and the agnostic PN 
deformation parameter, with consistent reweighting to account for prior differences.

In order to map a \PN parameter constraint to a parameter in a given theory, we map use the posteriors of the PN coefficient constraint along with the intrinsic parameters for each analyzed O4a event and the PN expressions derived to leading
order in \citet{Yagi:2011xp} for EdGB, \citet{Yagi:2011xp,Yagi:2012vf} for dCS gravity,
\citet{Khalil:2018aaj} for Einstein--Maxwell with black hole charge, \citet{Nielsen:2017lpd} for pseudo-complex \GR, and \citet{Horbatsch:2011ye,Yunes:2016jcc}
for scalar--tensor theories with an evolving scalar field.

The mapping on theories we quote in Table~\ref{tab:map_to_tgr} are the 90\% result for the O4a \ac{GW} event that gives the tightest translation on a given parameter.
For comparison, we also provide existing bounds or illustrative translations on each theory.
Current constraints on scalar--tensor theories come from observations of 
the quasar OJ~287~\citep{Horbatsch:2011ye}.
For EdGB gravity and Einstein--Maxwell,
the strongest bound arises from the low-mass, long-inspiral signal GW230529 
\citep[not included here because it is a single-detector event]{Sanger:2024axs}.
For dCS gravity, the current bound of $8.5$~km comes from combining
NICER observations of 
a pulsar with the NS tidal deformability inferred from the binary NS event
GW170817~\citep{Silva:2020acr}.

The expression from \citet{Horbatsch:2011ye} used to obtain the translation to the evolution of the scalar field in the scalar--tensor case requires $\epsilon_\mathrm{ST} = 2Gm_1|\dot{\phi}|/c^3 \ll 1$.
While $|\dot{\phi}|$ in our translation is large in units of $\mathrm{s}^{-1}$, the $90\%$ upper bound on $\epsilon_\mathrm{ST}$ is $0.077$ and $0.070$ for TIGER and FTI respectively, which are smaller than the value of
$0.13$ for the primary of OJ~287 used to obtain the constraints in \citet{Horbatsch:2011ye}. The much better constraints on $|\dot{\phi}|$ from OJ~287 are due to its much larger total mass (its primary's mass is $\sim 10^{10}M_\odot$). Additionally, the assumptions made in obtaining the frequency-domain dephasing used in the translation assume
that the dipole radiation is a small perturbation to the quadrupole radiation \citep[as discussed in, e.g.,][]{Yunes:2009ke}, but this is also the case here: The $90\%$ upper bounds on the ratio of the
leading dipole to quadrupole energy fluxes are $0.062$ and $0.053$ for TIGER and FTI at the low-frequency cutoff of $20$~Hz and decrease at higher frequencies. The GW event that gives rise to the tightest value for $|\dot{\phi}|$ is GW230518\_125908, whose source is likely an \NSBH. We assume that the \ac{NS} secondary has a vanishing scalar charge, maximizing the amount of dipolar radiation emitted by the system.

Similarly, the PN expressions used for dCS and EdGB
gravity are derived under the perturbative assumption that the dimensionless coupling
$\zeta_{\mathrm{dCS},\mathrm{EdGB}} = 16 \pi \alpha_{\mathrm{dCS}/\mathrm{EdGB}}^{2}/ \left(G m_2/c^2\right)^{4} \ll 1$ throughout the inspiral~\citep{Yagi:2011xp}, where we have evaluated this quantity using $m_2$, since that sets the smallest length scale in the binary. Evaluating this condition for each theory, we find that although the translation for
EdGB gravity satisfies the small-coupling requirement within the 90\% credible region with a value for $\zeta_{\mathrm{EdGB}}$ of 
$\mathcal{O}(10^{-3})$, the dCS translation does not. Indeed, the dCS mapping gives a
value $\sim 4 $ times larger than the best current bounds on the theory \citep{Silva:2020acr}, but the 90\% upper bound on small-coupling parameter is $\mathcal{O}(10^4)$.
As a result, the perturbative waveform is not self-consistent for the systems
considered, and any inferred PN-to-theory translation for dCS gravity would lie outside 
the regime of validity of the effective field theory expansion.
We therefore do not report a dCS entry in Table~\ref{tab:map_to_tgr}. 
In order to obtain constraints for dCS that are within small-coupling approximation from
\ac{GW} observations, one would need a low-mass binary with at least one
rapidly spinning object.

The values given in Table~\ref{tab:map_to_tgr} are for the purposes of illustration, and there are several caveats to them being directly taken as constraints on the specific theories. To begin with, a given modified gravity theory will give deviations in multiple \PN parameters, while the analyses consider each \PN parameter deviation in isolation. In addition, these analyses make use of the merger--ringdown signal, but without taking into account how a specific theory would modify that portion of the \ac{GW} signal from \GR. 
For instance, \citet{Johnson-McDaniel:2021yge} analyzes phenomenological non-\GR signals 
with FTI and TIGER, finding that these methods can yield constraints on a PN deviation 
parameter that are much smaller than the true deviation.
Thus, these mappings will not be equivalent to those obtained with an analysis that assumes a specific theory.

\subpapersubsection{FTI and TIGER: Multiparameter \PN coefficient tests with PCA}
\label{sec:pca}

\begin{figure*}[hbt!]
    \centering
\includegraphics{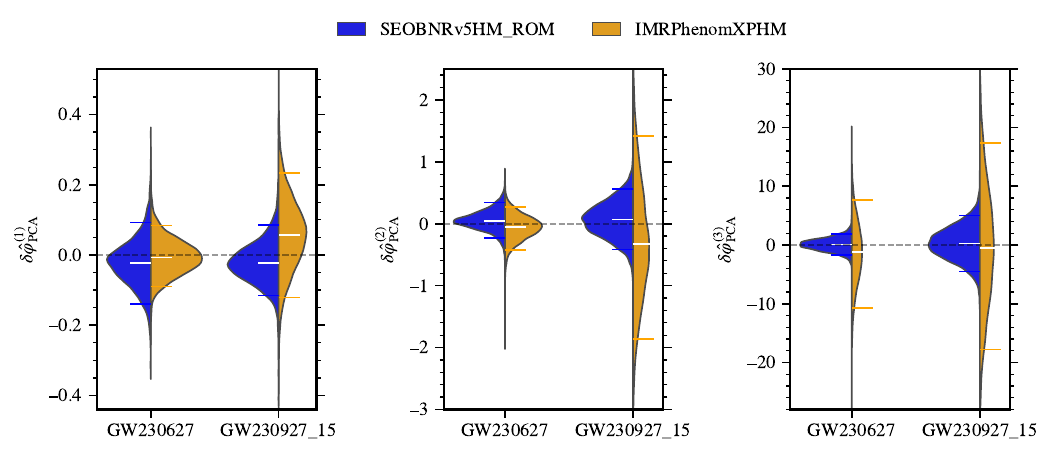}
    \caption{%
    Violin plots showing the posterior probability distribution of the first three leading PCA parameters, from the O4a events listed in the PCA column of Table~\ref{tab:selectionII} (\FULLNAME{GW230627_015337} and \FULLNAME{GW230927_153832}, abbreviated as \MINIMALNAME{GW230627_015337}{} and \MINIMALNAME{GW230927_153832}{}), passing the selection criteria described in Section~\ref{sec:pca}. In each violin plot, the colored horizontal bars and the horizontal white solid line denote the 90\% credible intervals and the posterior median, respectively. We mark the \GR value of zero with dashed grey lines.
    }
    \label{fig:pca_individuals}
\end{figure*}

\begin{figure}[hbt!]
    \centering
\includegraphics{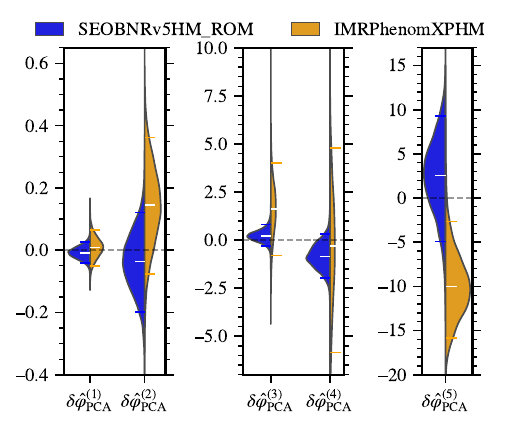}
    \caption{%
    The joint posterior probability distributions on the first five leading PCA parameters from all the selected GWTC-4.0 events are shown. The joint bounds are obtained through the marginalized-likelihood multiplication technique. The markers have the same meaning as in Figure~\ref{fig:pca_individuals}. All PCA parameters are statistically consistent with \GR at the 90\% credible level, except for the fifth parameter in the TIGER framework. This deviation may stem from instrumental noise features or from waveform-modeling systematic uncertainties.
    }
    \label{fig:pca_joints}
\end{figure}

The parametrized tests of \GR aim to identify potential deviations by modifying the GW phasing with dimensionless deviation parameters. This approach is implemented in the FTI and TIGER frameworks, as discussed before.
Focusing on the inspiral part of the GW signal, the null parametrization $\delta \hat{\varphi}_i$ is introduced at each \PN order, with $i$ denoting the \PN index, in Equation~(\ref{eq:frac_PN_dev}).
Considering the inspiral phase up to the $3.5$\PN order with non-vanishing \PN coefficients in \GR gives rise to eight deviation parameters, i.e., by dropping $\delta \hat{\varphi}_{-2}$ and $\delta \hat{\varphi}_{1}$ in Equation~(\ref{eq:frac_PN_terms}). However, all the deviation parameters are not measured simultaneously since such {\em multi-parameter} tests give rise to uninformative posteriors. This is caused by the correlations between the deviation parameters and other \GR parameters, which can be mitigated by increasing detector network sensitivity~\citep{Gupta:2020lxa, Datta:2020vcj} or by basing the parametrization on the radiative multipole moments~\citep{Kastha:2018bcr, Kastha:2019brk, Mahapatra:2023hqq, Mahapatra:2023ydi, Mahapatra:2023uwd}. While deviation from \GR can occur at any \PN order and hence the eight deviation parameters should be measured simultaneously, to avoid the aforementioned problem with uninformative posteriors, the FTI and TIGER analyses each perform eight {\em single parameter} tests, where only one deviation parameter is measured with the others being consistent with \GR.
However, we will instead use principal component analysis \citep[PCA;][]{Arun:2013bp, Pai:2012mv, Saleem:2021nsb, Shoom:2021mdj, Datta:2022izc, Mahapatra:2025cwk} to identify the best-measured linear combinations of a set of \PN deviation parameters. The bounds so obtained are more constraining than the single parameter tests since they contain the information from multiple \PN orders.

\begin{table}[bt!]
    \caption{\label{tab:pca_combined}
    Combined constraints on the leading five PCA parameters from all selected GWTC-4.0 events
    }
    \begin{center}
    \begin{tabular}{crrrrrr}
\toprule
$\delta\hat{\varphi}_{\mathrm{PCA}}^{(k)}$ & \phantom{X} & \multicolumn{2}{c}{Median \& 90\% errors} & \phantom{X} & \multicolumn{2}{c}{GR quantile}\\
\cline{3-4} \cline{6-7}
&  & \multicolumn{1}{c}{FTI} & \multicolumn{1}{c}{TIGER} & & \multicolumn{1}{c}{FTI} & \multicolumn{1}{c}{TIGER}\\
\midrule
$\delta\hat{\varphi}_{\mathrm{PCA}}^{(1)}$ & & $-0.01^{+0.04}_{-0.03}$ & $0.01^{+0.06}_{-0.06}$ & & 66\% & 41\%\\
$\delta\hat{\varphi}_{\mathrm{PCA}}^{(2)}$ & & $-0.04^{+0.16}_{-0.16}$ & $0.14^{+0.22}_{-0.22}$ & & 64\% & 14\%\\
$\delta\hat{\varphi}_{\mathrm{PCA}}^{(3)}$ & & $0.21^{+0.60}_{-0.52}$ & $1.63^{+2.39}_{-2.43}$ & & 24\% & 14\%\\
$\delta\hat{\varphi}_{\mathrm{PCA}}^{(4)}$ & & $-0.85^{+1.15}_{-1.10}$ & $-0.32^{+5.12}_{-5.53}$ & & 88\% & 54\%\\
$\delta\hat{\varphi}_{\mathrm{PCA}}^{(5)}$ & & $2.56^{+6.72}_{-7.48}$ & $-10.03^{+7.36}_{-5.82}$ & & 28\% & 98\%\\
\bottomrule
\end{tabular}
  \end{center}
 \tablecomments{We give the median, 90\% credible intervals, and the \GR quantile, i.e., $P(\delta \hat{\phi}_{\rm PCA}^{(i)}<0)$, of the combined PCA posteriors.
    }
\end{table}

In a multi-parameter test, along with the 15 \GR parameters, the six deviation parameters, introduced from $1.5$\PN to $3.5$\PN, are included: 
\begin{equation}
    \{\delta \hat{\varphi}_{3}, \delta \hat{\varphi}_{\rm 4},\delta \hat{\varphi}_{\rm 5\ell}, \delta \hat{\varphi}_{\rm 6}, \delta \hat{\varphi}_{\rm 6\ell}, \delta \hat{\varphi}_{\rm 7}\}\,,
    \label{eq:dev_params}
\end{equation}
with the assumption that $\{\delta \hat{\varphi}_{0}, \delta \hat{\varphi}_{\rm 1},\delta \hat{\varphi}_{\rm 2}\}$, the leading-order \PN deviation parameters, are consistent with \GR. While current GW detectors are sensitive enough to distinguish lower-order fractional deviations from intrinsic binary parameters in single-parameter tests for some events, in particular distinguishing 0PN and 1PN deviations from chirp mass and mass ratio, this is not the case for multi-parameter tests. Moreover, the $-1$PN and 0PN deviations are already well-constrained by binary pulsar measurements~\citep{Yunes:2010qb, Nair:2020ggs, Kramer:2021jcw}. Of course, constraints on the $-1$PN coefficient from \acp{BNS} are not directly applicable to \ac{BBH} signals, since there are theories in which \acp{BH} scalarize (leading to dipole radiation at $-1$PN) and \acp{NS} do not \citep{Yagi:2011xp, Yagi:2015oca}. However, \COMMONNAME{GW230529}, whose source has at least one \BH, has also placed fairly tight constraints on the $-1$PN coefficient \citep{Sanger:2024axs}. Thus, our analysis focuses on a six-dimensional test with fractional deviation parameters ranging from 1.5PN to 3.5PN~\citep{Mahapatra:2025cwk}.
This results in a 21-dimensional parameter space; hence, these analyses are computationally expensive. After obtaining the marginalized posteriors on the six deviation parameters, we post-process the posteriors using PCA to transition to a new basis. This new basis is a linear combination of the original deviation parameters with suitable weights, expressed as
\begin{equation}
    \delta \hat{\varphi}^{(k)}_{\rm PCA} = \sum_{i}\alpha^{ki} \delta \hat{\varphi}_i\,,
\end{equation}
where the summation index runs over the six deviation parameters in Equation~(\ref{eq:dev_params}).
The coefficient $\alpha^{ki}$ corresponds to elements of the transformation matrix that diagonalizes the covariance matrix \citep{Arun:2013bp,Pai:2012mv,Saleem:2021nsb,Datta:2022izc,Mahapatra:2025cwk}. Hence, shifting to the PCA basis reduces the correlations and provides better constraints. The PN coefficients that contribute relatively more to the observed inspiral signal have better-constrained fractional deviation parameters and therefore contribute dominantly to the leading informative PCA components, making the PCA method sensitive to the relative normalization of the different deviation parameters. PCA has inherent sign ambiguity because the direction of eigenvectors is not uniquely determined; their signs can be flipped without affecting the variance they explain. To resolve this, we adopt a sign convention where each eigenvector and each principal component is adjusted so that the largest element is always positive \citep{Mahapatra:2025cwk}.

We implement the PCA on GW events observed during the O4a run that meet specific selection criteria and constrain the deviations from \GR. Due to the high computational cost of multi-parameter runs, we use strict selection criteria of inspiral $\text{\SNR} \geq \TGRPCAFTITIGERInspiralSNR{}$ and detector-frame chirp mass $(1+z)\mathcal{M} < \TGRPCAFTITIGERChirpMass{}$ to ensure longer inspiral signals. The inspiral SNR in the FTI and TIGER frameworks is defined in Section~\ref{sec:par}. For chirp mass, we use the median of the chirp mass posterior from the \GR analysis with \IMRPhenomXPHM.
Imposing this criterion results in \TGRPCAFTITIGEREvents{} qualifying events from O3a (GW190412 and GW190814),
    \TGRPCAFTITIGEREvents{} from O3b (GW191204\_171526 and GW191216\_213338),
    and \TGRPCAFTITIGEREvents{} from O4a (\FULLNAME{GW230627_015337} and \FULLNAME{GW230927_153832}) selected for PCA analysis in both the FTI and TIGER frameworks.

We perform a multi-parameter run on the selected events using the parametrized extension of the \SEOBNRFIVEHMROM waveform for FTI and the \IMRPhenomXPHM waveform for TIGER, as in the single-parameter analyses in Section~\ref{sec:par}. Both waveforms give consistent results in the PCA basis.
In Figure~\ref{fig:pca_individuals}, we show the violin plots of the posterior probability distributions for the first three leading PCA parameters corresponding to the selected O4a events. The results from PCA analysis on the O3a and O3b events are presented in \citet{Mahapatra:2025cwk}. We restrict our attention to the first three PCA parameters since they carry the most information.
The higher-order PCA parameters receive relatively larger contributions from those original deviation parameters that have uninformative posteriors (i.e., are dominated by the priors). Consequently, the higher-order PCA parameters, which have wide posteriors, are not considered. We find that for both events, the posteriors are consistent with \GR at the 90\% credible level for all PCA parameters within both the FTI and TIGER frameworks.

To combine information from multiple events, we use the marginalized-likelihood multiplication technique applied to the original deviation parameters \citep{Saleem:2021nsb, Mahapatra:2025cwk}, where the likelihoods are constructed by sampling from six-dimensional kernel density estimator fits using a Gaussian mixture model \citep[as implemented in \soft{scikit-learn};][]{scikit-learn}. Subsequently, we apply PCA to the combined posteriors of the original deviation parameters in order to obtain the PCA parameters for multiple events.
We find that higher PCA parameters for multiple events contain useful information, even though they do not for the individual events: The leading three PCA parameters in the FTI framework and the leading two in the TIGER framework are informative in all individual event analyses, as the Jensen–Shannon (JS) divergence~\citep{JSdiv} between the posterior and prior distributions exceeds $0.1$~bit.
An exception is \FULLNAME{GW230627_015337} in the TIGER framework, where the JS divergence for the third PCA parameter also exceeds $0.1$~bit. However, in the joint analysis, the leading five PCA parameters are informative, with the JS divergence exceeding $0.1$~bit in both frameworks.

The combined posterior probability distributions on the first five leading PCA parameters from all the GW events passing the selection criteria are displayed in Figure~\ref{fig:pca_joints}. The median values, 90\% credible intervals, and \GR quantiles for combined PCA posteriors are listed in Table~\ref{tab:pca_combined}. The combined posterior distributions of the PCA parameters are statistically consistent with \GR within the 90\% credible interval, except for the fifth PCA parameter in the TIGER framework. The fifth PCA parameter in the TIGER framework exhibits a significant offset from zero, with a GR quantile of \FifthPCATIGERGRquantile{}.
Furthermore, the fourth PCA parameter in the FTI framework, as well as the second and third PCA parameters in the TIGER framework, are offset from zero. These offsets are all driven by the event GW190814~\citep{GW190814}, for which we observed a similar offset in the PCA posteriors \citep{Mahapatra:2025cwk}. If GW190814 is excluded from the combined results, they do not have any notable offsets from zero.
The deviations found for GW190814 could possibly be caused by noise artifacts present in the data or by systematic errors from the waveform model and parameterization framework \citep{Mahapatra:2025cwk}.
The difference between the FTI and TIGER results is an estimate of the latter.
We deduce a joint bound (at 90\% credibility) on the leading PCA parameter to be \FirstPCAFTIjointbound{} (\FirstPCATIGERjointbound) in the FTI (TIGER) framework, consistent with \GR predictions. The loud \ac{O4b} event GW250114~\citep{GW250114} also provides tight constraints on the first two PCA parameters~\citep{GW250114_TGR}.

\subpapersubsection{Spin-induced multipole moment coefficient null test}
\label{sec:SIM}

\begin{figure*}
\includegraphics{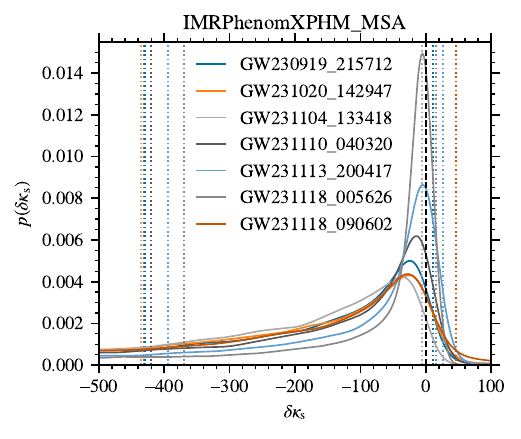}
\hfill
\includegraphics{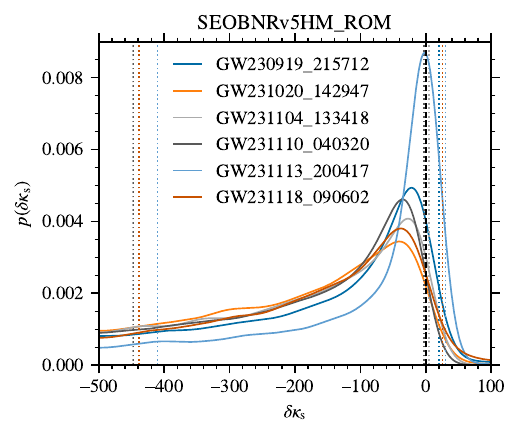}
\caption{
  The posterior probability distribution on the spin-induced quadrupole moment parameter $\delta \kappa_{\mathrm{s}}$ using the \IMRPhenomXPHMMSA (left panel) and  \SEOBNRFIVEHMROM (right panel) waveform models.
The black dashed vertical line indicates the BBH value ($\delta \kappa_{\mathrm{s}}=0$). The colored vertical lines show the 90\% symmetric bounds on $\delta \kappa_{\mathrm{s}}$ calculated from the individual events assuming a uniform prior ranging between $[-500, 500]$ on $\delta \kappa_{\mathrm{s}}$.}
\label{fig:sim_sym}
\end{figure*}

\begin{figure}
\includegraphics{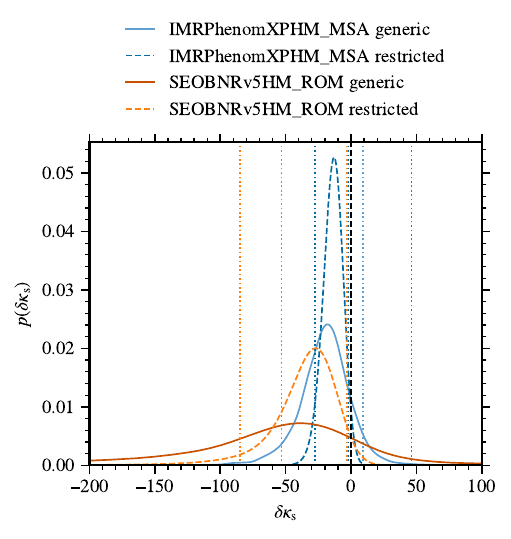}
\caption{
  Joint posterior probability distribution on the spin-induced quadrupole moment parameter $\delta \kappa_{\mathrm{s}}$ from the GWTC-4.0 events. Bounds obtained by multiplying the likelihoods (restricted) and by hierarchically combining events (generic) are shown. The analysis is performed assuming a uniform prior ranging between $[-500, 500]$ on $\delta \kappa_{\mathrm{s}}$. The dashed line indicates the BBH value of $\delta \kappa_{\mathrm{s}}$, while the dotted lines mark the 90\% symmetric credible intervals. The BBH value is found slightly outside of the $90\%$ credible interval for the restricted results, due to correlations with $\chi_\mathrm{eff}$.
}
\label{fig:sim_combined}
\end{figure}

The gravitational field of a spinning object includes contributions from the object's quadrupole and higher multipole moments, which arise from rotationally induced deformations. These higher-order contributions leave unique imprints on gravitational waveforms, encoding the values of the spin-induced multipole moments of the binary's constituent objects. For black holes, these moments take specific values that are uniquely determined by the mass and spin, as dictated by the no-hair conjecture~\citep{Carter,Hansen:1974zz}.
Inferring the values of these moments from their imprints on gravitational waveforms can allow one to distinguish exotic compact objects from black holes. In this analysis, we model the spin-induced quadrupole moment of each compact object using the parameterization
\begin{equation}\label{eq:sim} 
Q_{i}= -\, \kappa_{i}\, \chi_{i}^2 m_{i}^3,
\end{equation}
where $Q_{i}$, representing the quadrupole-moment scalar, appears in the leading-order 2PN term in the \ac{GW} phase and $\kappa_i$ encodes the response of the object’s shape to its spin for object $i=1, 2$.  Here $m_{i}$ and $\chi_{i}$ are the mass and the dimensionless spin of the compact object $i$. Along with the 2PN leading-order term, we also include the spin-induced quadrupole-moment terms at 3PN in the binary inspiral phase~\citep{Arun:2008kb, Mishra:2016whh}. For a given family of exotic compact objects, the value of $\kappa_i$, the dimensionless spin-induced quadrupole-moment coefficient,  is primarily determined by the mass of the compact object, with a subdominant dependence on the object's spin, but the value can be significantly different for different classes of exotic compact objects.

For Kerr black holes, the value is $\kappa_{\rm{BH}}=1$~\citep{Carter,Hansen:1974zz}; for other compact stars, the value of $\kappa$ is different due to the stars' internal structure. For models of spinning neutron stars, $\kappa$ can vary between ${\sim}2$ and ${\sim}14$~\citep{Pappas:2012qg,Pappas:2012ns,Harry:2018hke}. For available models of spinning boson stars, $\kappa$ can have values of ${\sim}10\text{--}150$~\citep{FDRyan1997}, and $\kappa$ can even be negative for exotic stars such as gravastars~\citep{Uchikata:2015yma}.
One expects any compact object that has a spin-induced quadrupole different from that of a black hole to have a nonzero tidal deformability \citep[e.g.,][]{Sennett:2017etc, Johnson-McDaniel:2018uvs, Pacilio:2020jza,Narikawa:2021pak} as well as differences in the tidal heating~\citep{Cardoso:2017cfl,Maselli:2017cmm,Datta:2019euh,Datta:2019epe,Chia:2024bwc} and modifications to the merger--ringdown part of the signal compared to that from an \ac{BBH}. However, including these effects is not essential for the effectiveness of the null test presented here. While we do not explicitly test signals from non-black hole objects, the measurement of the spin-induced quadrupole coefficient \(\kappa\) from \ac{GW} data serves as a physics-motivated null test: deviations from the Kerr prediction \(\kappa = 1\) may indicate the presence of exotic compact objects or deviations from \ac{GR}.

The coefficients $\kappa_i$ represent the primary and secondary components' spin-induced quadrupole moment parameters for a spinning compact binary system. Although we include the 3PN spin-quadrupole contributions in our waveform model, the estimation of individual spin-induced quadrupole parameters \(\kappa_1\) and \(\kappa_2\) remains challenging due to their strong correlations with the component masses and spins. In principle, higher-order corrections beyond 3PN could help disentangle these parameters. Given the current detector sensitivities, at least the additional $3.5$PN term \citep[where the spin-induced octupole first enters][]{Marsat:2014xea} does not significantly improve constraints, which remain broad. For this reason, we limit our analysis to the leading and next-to-leading spin-quadrupole terms.
However, with a combination of $\kappa_i$ values one can proceed further~\citep{Krishnendu:2017shb, Krishnendu:2018nqa, Krishnendu:2019tjp,LIGOScientific:2020tif,Divyajyoti:2023izl}. We consider the symmetric and anti-symmetric combinations of $\kappa_i$,
\begin{align}
    \kappa_{\mathrm{s}} &= \frac{\kappa_1+\kappa_2}{2},\\
    \kappa_{\mathrm{a}} &= \frac{\kappa_1-\kappa_2}{2}.
\end{align}
For binary black holes in general relativity, \(\kappa_{\mathrm{s}} = 1\) and \(\kappa_{\mathrm{a}} = 0\). In this analysis, we assume \(\kappa_{\mathrm{a}} = 0\) in order to construct a more constraining null test for the Kerr nature of compact objects by measuring deviations from the expected value via \(\kappa_{\mathrm{s}} = 1 + \delta\kappa_{\mathrm{s}}\). This assumption simplifies the parameter space and enhances our ability to detect deviations from the black hole prediction, though it does not capture possible asymmetries in the spin-induced quadrupole moments of the two components. However, for the loud \ac{O4b} event GW241011 \citep{GW241011/GW241110}, whose source was an unequal-mass binary, we have been able to constrain deviations in $\kappa_1$, allowing it and $\kappa_2$ to vary independently.

We introduce the parameterized deviations $\delta\kappa_{\mathrm{s}}$ to the inspiral phase for both the \IMRPhenomXPHMMSA and \SEOBNRFIVEHMROM waveform models, similar to the introduction of generic deviations of PN coefficients for these waveform models in Section~\ref{sec:par}.
Here \IMRPhenomXPHMMSA denotes that we use the version of this model in which precession is modeled using a multiscale analysis~\citep{Pratten:2020ceb}, which was previously studied on GWTC-3.0 events~\citep{Divyajyoti:2023izl}, instead of the SpinTaylor variant used throughout the rest of this paper.

We apply selection criteria based on the measurability of \(\delta\kappa_{\mathrm{s}}\), requiring that the posterior distribution is sufficiently constrained to distinguish potential deviations from the black hole value \(\kappa_{\mathrm{s}} = 1\).
For the \IMRPhenomXPHMMSA analysis, we require (i) an inspiral network SNR of at least \TGRSIMPhenomMinimumSNR{} and (ii) \(\chi_{\mathrm{eff}} > 0\)
at the 90\% credible level.
To compute the combined bounds, we include events reported in \citet{LIGOScientific:2020tif} that satisfy both selection criteria. This results in a total of \TGRSIMPhenomNumberOFouraEvents{} events from O4a and \TGRSIMPhenomNumberGWTCThreeEvents{} events from GWTC-3.0. The event selection criterion on \(\chi_{\mathrm{eff}}\) is stricter than the one imposed for GWTC-3.0, so GW200129\_065458 is no longer selected for this test.

Besides the spin-induced quadrupole-moment test previously used in~\citet{LIGOScientific:2020tif}, we also employ a new test based on the FTI framework used for parameterized inspiral tests (see Section~\ref{sec:par}). For this version of the test, the corrections due to $\delta\kappa_{\mathrm{s}}$ are added at 2PN and 3PN in the frequency-domain phase during inspiral~\citep{Mehta:2022pcn}. In order to only apply the corrections to the inspiral portion of the waveform, the corrections are tapered off towards the merger-ringdown, which is left the same as in \ac{GR}, as described in Section~\ref{sec:par}. This test uses the \SEOBNRFIVEHMROM waveform model and is applied only to events that satisfy the FTI event selection criteria of having an inspiral SNR of at least \TGRFTIMinimumSNR{} and at least \TGRFTIMinimumCycles{} \ac{GW} cycles in the inspiral. Additionally, as for the \IMRPhenomXPHMMSA analysis, we require that zero effective inspiral spin $\chi_\mathrm{eff}=0$ is outside the 90\% credible interval of the \ac{GR} posterior. Overall, \TGRSIMEOBNumberOFouraEvents{} events from O4a pass these selection criteria.

Figure~\ref{fig:sim_sym} shows the posterior distributions of $\delta \kappa_{\mathrm{s}}$ for individual events. They are derived under the assumption of a uniform prior on $\delta \kappa_{\mathrm{s}}$ between $[-500,500]$. As the parameter $\delta \kappa_{\mathrm{s}}$ is correlated with $\chi_\mathrm{eff}$, individual events constrain positive values of  $\delta \kappa_{\mathrm{s}}$ more strongly for positive effective inspiral spin and often have a long tail for negative $\delta \kappa_{\mathrm{s}}$~\citep{Krishnendu:2019tjp}. Since most events observed have small but positive $\chi_{\mathrm{eff}}$~\citep{GWTC:AstroDist}, it is expected that the combined posterior and the 90\%  bounds will have more stringent constraints on positive values of $\delta \kappa_{\mathrm{s}}$ than negative values.

The combined posterior distribution on \(\delta \kappa_{\mathrm{s}}\) from all selected GW events is shown in Figure~\ref{fig:sim_combined}.
The 90\% credible interval on $\delta\kappa_{\mathrm{s}}$ from the hierarchical analysis is $\delta\kappa_{\mathrm{s}} = \TGRSIMPhenomCombinedCI{HIER_POP}$ for \IMRPhenomXPHMMSA and $\delta\kappa_{\mathrm{s}} = \TGRSIMEOBCombinedCI{HIER_POP}$ for \SEOBNRFIVEHMROM. When restricted to the positive prior region, this analysis places a $90\%$ credibility constraint of $\delta\kappa_{\mathrm{s}} < \TGRSIMPhenomCombinedCI{HIER_POP_POS}$ for \IMRPhenomXPHMMSA and $\delta\kappa_{\mathrm{s}} < \TGRSIMEOBCombinedCI{HIER_POP_POS}$ for \SEOBNRFIVEHMROM.
The combined \IMRPhenomXPHMMSA bounds are tighter than the \SEOBNRFIVEHMROM ones because the GWTC-3.0 events have not yet been included in the \SEOBNRFIVEHMROM analysis, due to computational constraints. However, the \IMRPhenomXPHMMSA and \SEOBNRFIVEHMROM results for individual events in Figure~\ref{fig:sim_sym} agree quite well.

The distribution hyperparameters are just consistent with the null hypothesis ($\mu=\sigma=0$), with $\mu = \TGRSIMPhenomCombinedCI{HIER_MU}$ and $\sigma < \TGRSIMPhenomCombinedCI{HIER_SIGMA}$ for \IMRPhenomXPHMMSA, and with $\mu = \TGRSIMEOBCombinedCI{HIER_MU}$ and $\sigma < \TGRSIMEOBCombinedCI{HIER_SIGMA}$ for \SEOBNRFIVEHMROM.
Both $\mu$ and the population-marginalized posterior in Figure~\ref{fig:sim_combined} inherit the asymmetry of the individual events, which tend to be skewed toward $\delta\kappa_{\mathrm{s}} < 0$. This explains why the results are marginally consistent with the null hypothesis and suggests that negative values of $\delta\kappa_{\mathrm{s}}$ are harder to constrain.

The dashed curves in Figure~\ref{fig:sim_combined} shows the joint-likelihood posterior obtained under the assumption that all events share a common value of $\delta\kappa_{\mathrm{s}}$. Within this framework, we obtain $\delta\kappa_{\mathrm{s}} = \TGRSIMPhenomCombinedCI{RESTRICTED_POP}$ for \IMRPhenomXPHMMSA and $\delta\kappa_{\mathrm{s}} = \TGRSIMEOBCombinedCI{RESTRICTED_POP}$ for \SEOBNRFIVEHMROM.
The combined results of the restricted analysis are not consistent with $\delta\kappa_{\mathrm{s}} = 0$ at the 90\% credible level due to the correlation with $\chi_\mathrm{eff}$, just like the hierarchical results.
When restricted to positive values, the constraint becomes $\delta\kappa_{\mathrm{s}} < \TGRSIMPhenomCombinedCI{RESTRICTED_POP_POS}$ for \IMRPhenomXPHMMSA and $\delta\kappa_{\mathrm{s}} < \TGRSIMEOBCombinedCI{RESTRICTED_POP_POS}$ for \SEOBNRFIVEHMROM.

Overall, the results are consistent with BBHs described by \ac{GR}.

\subpapersubsection{LOSA: Line-of-Sight Acceleration}\label{sec:LOSA}

Motion of a CBC relative to a GW detector will cause the signal to be Doppler shifted. Uniform (constant velocity) motion will cause a constant Doppler shift, the measurement of which is perfectly degenerate with the intrinsic mass of the CBC following the mass--redshift degeneracy in GW signals \citep{GWTC:Introduction}. However, a CBC moving with a \textit{time-varying} relative velocity caused, e.g., by a constant LOSA $a$ such as due to a nearby supermassive black hole,  will cause a $-4 {\rm PN}$ modulation \citep{Vijaykumar:2023tjg, Tiwari:2025aec} at the leading order: the strength of this modulation is proportional to the magnitude of the LOSA. \citet{Tiwari:2025aec} discusses how LOSA compares to other effects like repeated lensing by a supermassive black hole.

The modulated inspiral GW waveform in the frequency domain can be written as
\begin{equation}
    \label{eq: WF_losa}
    \tilde{h}(f) = \tilde{h}_{\rm non-acc}(f) \left(1 + \frac{\Delta \mathcal{A}_{\rm LOSA}}{\mathcal{A}_{\rm non-acc}} \right) e^{i \Delta \Psi_{\rm LOSA}},
\end{equation}
where $\tilde{h}_{\rm non-acc}(f)$ is the waveform without the effects of LOSA, $\Delta \mathcal{A}_{\rm LOSA} / \mathcal{A}_{\rm non-acc} \propto a/v^8$ is the modulation in the amplitude, and $\Delta \Psi_{\rm LOSA} \propto a/v^{13}$ is the phase correction due to LOSA.
Here $v = \left[ \pi G (1 + z) M f \right]^{1/3}$ is the binary's orbital velocity, where $z$ is the binary's redshift and $M$ is its total mass. The phase correction contains contributions from point-particle, aligned spin, and tidal effect terms \citep{Vijaykumar:2023tjg,Tiwari:2025aec}.

\begin{figure}
    \centering
    \includegraphics[width=\TGRFigureWidth]{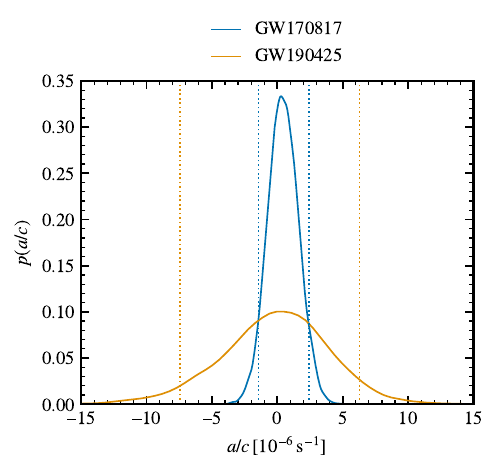}
    \caption{The posterior probability distributions for events satisfying the LOSA selection criteria. The colored dotted lines represent the 90\% credible levels.}
    \label{fig: losa_post_preO4}
\end{figure}

The LOSA corrections have been derived assuming $\vert a/c \vert t_{\rm sd} \ll 1$,
    where $t_{\rm sd}$ is the signal duration in the detector's sensitive band,
    so we can linearize in $\vert a/c \vert t_{\rm sd}$.
Additionally, the current setup of phase corrections due to LOSA only has corrections to the dominant $(2,\pm 2)$ multipoles of the waveform and does not include contributions from precessing spins.
For a precessing binary, the aligned spin LOSA corrections are computed using the spin components at the reference frequency.

We impose a set of selection criteria~\citep{Tiwari:2025aec} to identify events to which to apply the LOSA analysis,
    where we evaluate these criteria using the intrinsic parameters of the binary based on the GR analyses \citep{GWTC:Results},
    in addition to using the LOSA results themselves in the final two a posteriori criteria:
\begin{enumerate}
    \item We ensure that the median of the redshifted total mass $(1 + z) M$ of the CBC is $\leq \TGRLOSAmaxMtdet$, since lower-mass binaries have longer signals in the detectors sensitive band, with many cycles at relatively small $v$, which give the best constraints on the LOSA.
    \item We ensure the median value of mass ratio $q \geq \TGRLOSAminq$, to avoid biased LOSA recovery due to higher multipole content.
    \item We ensure the median value of $\chi_{\rm p} \leq \TGRLOSAmaxchip$, to avoid biased LOSA recovery due to precession, since the LOSA corrections used are only for aligned spins.
    \item After we perform LOSA inference on the events passing the above selection criteria, we calculate the quantity $\vert a/c \vert t_{\rm sd}$ to ensure it is $\leq \TGRLOSAtsdprodmax$ at the 90\% credible level. Here we compute $t_{\rm sd}$ using the Newtonian-order expression for the time to coalescence, $5G(1+z)Mc^5/(256\eta v^8)$ \citep[e.g.,][]{Buonanno:2009zt}, starting from the low frequency used in the analysis.
\end{enumerate}
We use precessing waveforms \IMRPhenomXPMSA \citep{Pratten:2020ceb} and \IMRPhenomXPNRTidalTWO \citep{Colleoni:2023ple} as the base waveforms for BBHs and BNSs, respectively, while we use \IMRPhenomNSBH \citep{Thompson:2020nei}, which is non-precessing, for NSBHs.

Among the O4 events, we find \FULLNAME{GW230518_125908} to have a total redshifted mass very close to the cutoff, so the first selection criteria is only satisfied for the results using some waveform models. We thus restrict consideration to the \IMRPhenomNSBH analysis, where we found that the median total mass of \FULLNAME{GW230518_125908} is slightly above the threshold. However, three pre-O4 events fulfilled the first three selection criteria regardless of the waveform model used in the analysis, namely GW170817, GW190425, and GW200115\_042309. We find GW170817 and GW190425 to also fulfill the fourth criterion and to be non-accelerating (see Figure~\ref{fig: losa_post_preO4}): $a/c$ is constrained to be $\TGRLOSAResults{GW170817} \, {\rm s}^{-1}$ for GW170817 and $\TGRLOSAResults{GW190425} \, {\rm s}^{-1}$ for GW190425. We also find that GW200115\_042309 is consistent with zero LOSA at slightly less than $90\%$ credibility. However, the GW200115\_042309 results fail criterion four, so we cannot claim a reliable constraint on LOSA, and thus do not provide quantitative results for that event.

\subpapersection{Tests of GW Propagation}\label{sec:propagation}

This section is dedicated to plausible modification (beyond \GR) of the \ac{GW} signal as it propagates between the source and the detector.  As discussed in \citet{LIGOScientific:2019fpa}, we assume that \GR accurately models the generation of \acp{GW} near the source in its local wave zone~\citep{Thorne:1980ru}, or that the deviations from GR during the generation or due to non-linearity of the field in the near zone
are not observable. This is an excellent assumption in many cases. For instance, for the massive graviton, the combined constraints from propagation mean that the Yukawa length scale $\lambda_g$ that determines the corrections to the Newtonian potential is $\lambda_g/r \gtrsim10^{\TGRMDRGravitonOrbitScaleOoM}$ times the maximum separation of the binary $r$ in the sensitive band of ground-based detectors, thus leading to corrections to the binary's \GR dynamics from a massive graviton that are of order $r^2/(2\lambda_g^2) \lesssim 10^{\TGRMDRGravitonOrbitEoMCorrectionScaleOoM}$ (bounds on $m_g$ from individual events would lead to corrections at most an order of magnitude higher). We also assume that the \ac{GW} signal is described sufficiently accurately by two
(tensorial) polarizations similar to \GR.  In Section~\ref{sec:MDR}, we consider a phenomenological model where the  modification in the propagation of \ac{GW} is caused by a modification to the dispersion relation.
In Section~\ref{sec:SSB}, we consider modifications in \ac{GW} signal caused by the difference in propagation of each polarization.
For both of the above tests, the modification to the GR waveform scales with the distance to the source. As such, GWTC-4.0, containing many events at $>1 \,\mathrm{Gpc}$~\citep{GWTC:Results}, is well suited to improve constraints on the modified propagation theories.

The companion paper focusing on the inference of cosmological parameters with GWTC-4.0~\citep{GWTC:Cosmology}, also constrains a broad class of theoretical models that allow for violations of \ac{GR}. In that work, we concentrate on a specific effect often referred to as \ac{GW} friction, which manifests itself as a modification of the \ac{GW} amplitude and thus alters the inferred luminosity distance. We place constraints on the ratio of gravitational-wave to electromagnetic luminosity distances by using the distribution of source-frame \ac{BBH} masses and assuming luminal \ac{GW} propagation, finding results consistent with \ac{GR}.

\subpapersubsection{Tests of a modified GW dispersion relation}
\label{sec:MDR}

GR predicts that the gravitational interaction (graviton) propagates with the speed of light, and the corresponding energy-momentum relation
takes a simple form $E^2 = p^2c^2$. However, in the case of massive graviton and Lorentz invariance violating theories of gravity, we have a modified dispersion relation (MDR) with an extra term:

\begin{equation}\label{eq:mdr:dispersion}
E^2 = p^2 c^2 + A_{\alpha} p^{\alpha} c^{\alpha},
\end{equation}
where the phenomenological parameter $A_\alpha$ governs the strength of the violation and $\alpha$ controls the frequency dependence of the dispersion~\citep{Will:1997bb, Mirshekari:2011yq}.  The modified dispersion causes different frequency components of the signal to travel with different group velocities, changing the morphology of the signal, which can be expressed as a frequency-dependent phase modification of the GR waveform $\tilde{h}(f) = \tilde{h}_\mathrm{GR}(f) \exp(i \delta \Psi(f))$, with the modification~\citep{Ezquiaga:2022nak}:
\begin{equation}
  \label{eq:mdr:correction_group}
  \delta \Psi(f) = -\frac{\pi D_\alpha h^{\alpha-2}(1+z)^{\alpha-1}}{c}A_{\alpha}f^{\alpha-1} \, .
\end{equation}
Here $h$ is the Planck constant, $z$ is the redshift of the source, and $D_\alpha$ is a modified distance parameter, defined by:
\begin{equation}
  D_{\alpha} = \frac{c(1+z)^{1-\alpha}}{H_0}\int_0^z \frac{(1+\bar{z})^{\alpha-2}}{\sqrt{\Omega_\mathrm{m}(1+\bar{z})^3+\Omega_\Lambda}} \mathrm{d}\bar{z}\, ,
\end{equation}
where $H_0$ is the Hubble constant, $\Omega_\mathrm{m}$ is the matter density parameter, and $\Omega_\Lambda$ is the dark energy density parameter. Here, we use the $\Lambda$CDM cosmological model given in~\citet{GWTC:Introduction}.  As such, we neglect the radiation density in the expression for our distance parameter.

Previous tests of MDR performed by the LVK~\citep{Abbott:2017vtc,LIGOScientific:2019fpa,LIGOScientific:2020tif, LIGOScientific:2021sio} quoted the results using the particle velocity, for which the constraints on $A_\alpha$ are $1-\alpha$ times larger than those obtained with the group velocity for $\alpha \neq 1$; for $\alpha = 1$ the particle velocity yields a deviation with a different frequency dependence than the group-velocity results. \citet{Ezquiaga:2022nak} showed that the phase modification obtained using the group velocity is consistent with that obtained through the Wentzel--Kramers--Brillouin approach~\citep{Jimenez:2019xxx}, which is why we use the group-velocity expressions here.

The MDR given in Equation~\eqref{eq:mdr:dispersion} provides a simple phenomenological parametrization to a broad class of modified theories. Different values of $\alpha$ describe the leading dispersive effect in different non-GR theories. Particular predictions correspond to $\alpha=0$~\citep[massive graviton;][]{Will:1997bb}, $\alpha=2.5$~\citep[multi-fractal spacetime;][]{Calcagni:2009kc}, $\alpha=3$~\citep[doubly special relativity;][]{AmelinoCamelia:2002wr}, and $\alpha=4$~\citep[Ho{\v{r}}ava--Lifshitz and extra dimensional theories;][]{Horava:2009uw,Sefiedgar:2010we}.

Following the phenomenological approach, we test $\alpha \in \{ -3, -2, -1, 0, 0.5, 1.5, 2.5, 3, 3.5, 4\}$. The case  $\alpha=2$ is excluded, as this choice is equivalent to changing the overall speed of GW propagation and results in no dispersion. Compared with GWTC-3.0, we drop the test of $\alpha=1$; in this case the GW group velocity is equal to the speed of light at leading order and the amplitude $A_{\alpha=1}$ cannot be constrained~\citep{Baka:2025drk}. However, we now also include negative values of $\alpha$, which can result from a dark energy phase transition~\citep{deRham:2018red,LISACosmologyWorkingGroup:2022wjo,Harry:2022zey}.

We have introduced two novelties in the analysis after GWTC-3.0~\citep{LIGOScientific:2021sio}. First, we use \IMRPhenomXPHM for the underlying \GR waveform.
Second, for each value of $\alpha$, we sample in the effective amplitude parameter $A_{\rm{eff}}$ (which decouples the sampling parameter from the choice of cosmology) and reweight the final result to a prior flat in $A_\alpha$, instead of sampling separately for the positive and negative values of $A_\alpha$ using the logarithm of an effective wavelength parameter~\citep{LIGOScientific:2019fpa,Baka:2025drk}. For $\alpha=0$, we additionally sample in the effective graviton mass \citep{Baka:2025drk}. For the combined bounds on the graviton mass, we opt to transform the $A_0$ posterior to a prior flat in graviton mass \citep{Baka:2025drk}, to ensure consistency between between our bounds on $A_0$ and $m_g$.

Contrary to the tests of \ac{GW} generation, where the deviation from \ac{GR} is expected to depend on the parameters of each individual source, we assume that the modification in the propagation of the gravitational waves themselves is universal to all events, and independent of their \ac{SNR} or parameters. Therefore, we do not perform a hierarchical analysis as in Section~\ref{sec:generation}. Instead, we just apply kernel density estimation to the $A_\alpha$ posteriors and combine the posteriors from different events by multiplying together the individual likelihoods (which are proportional to the posteriors as the priors were chosen to be flat) and renormalizing the result. Finally, we estimate bounds on the amplitude parameters using the combined results.

\begin{figure*}
  \centering \includegraphics[width=\TGRFigureWidthPage]{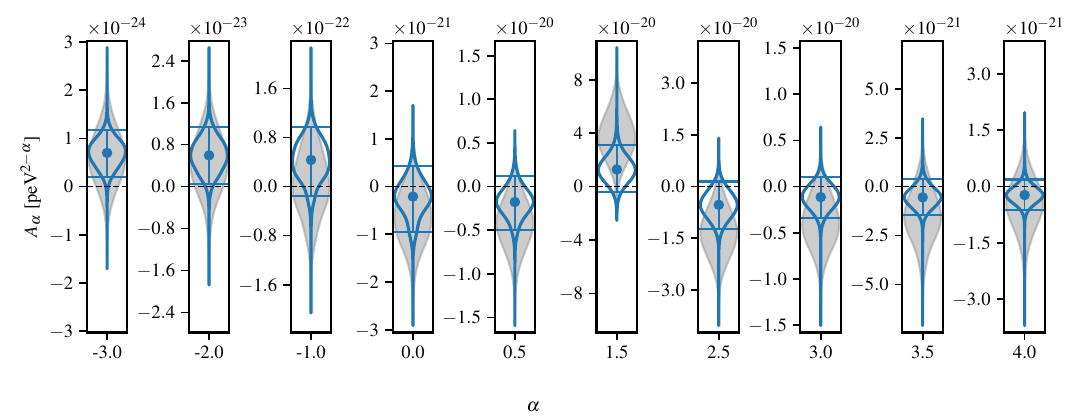}
  \caption{%
  Posteriors on the MDR amplitude parameters $A_\alpha$. Results from GWTC-3.0 \citep{LIGOScientific:2021sio,Baka:2025drk} are indicated by a shaded light-gray area, while the new results in GWTC-4.0 are represented by blue curves. The error bars indicate 90\% credible intervals. The significant shifts away from zero for $\alpha=-3,-2$ are driven by prior effects for \FULLNAME{GW231028_153006}.}
  \label{fig:mdr:amplitudes}
 \end{figure*}

 \begin{table*}
   \caption{Combined results for the MDR analysis}
   \label{tab:mdr:combined_amplitudes}
   \centering
   \setlength\tabcolsep{3pt}
 \begin{tabular}{cc c cc c cc c cc c cc c cc}
 \toprule
 & $\bar{m}_g$  & & \multicolumn{2}{c}{$\bar{A}_{-3.0}$}  & & \multicolumn{2}{c}{$\bar{A}_{-2.0}$}  & & \multicolumn{2}{c}{$\bar{A}_{-1.0}$}  & & \multicolumn{2}{c}{$\bar{A}_{0.0}$}  & & \multicolumn{2}{c}{$\bar{A}_{0.5}$} \\
 \cline{4-5}
 \cline{7-8}
 \cline{10-11}
 \cline{13-14}
 \cline{16-17}
  & $90\%$ & & $90\%$ & $Q_\text{GR}$ & & $90\%$ & $Q_\text{GR}$ & & $90\%$ & $Q_\text{GR}$ & & $90\%$ & $Q_\text{GR}$ & & $90\%$ & $Q_\text{GR}$ \\
 & $[10^{-23}]$  & & \multicolumn{1}{c}{$[10^{-84}]$} &  \% & & \multicolumn{1}{c}{$[10^{-71}]$} &  \% & & \multicolumn{1}{c}{$[10^{-58}]$} &  \% & & \multicolumn{1}{c}{$[10^{-45}]$} &  \% & & \multicolumn{1}{c}{$[10^{-39}]$} &  \%\\
 \midrule
GWTC-3.0 & \reviewed{2.23} &  & $\reviewed{-0.38}$ -- $\reviewed{1.59}$ & $\reviewed{15.7}$ &  & $\reviewed{-0.53}$ -- $\reviewed{1.42}$ & $\reviewed{22.3}$ &  & $\reviewed{-0.86}$ -- $\reviewed{1.04}$ & $\reviewed{45.5}$ &  & $\reviewed{-1.63}$ -- $\reviewed{0.41}$ & $\reviewed{84.1}$ &  & $\reviewed{-9.40}$ -- $\reviewed{0.66}$ & $\reviewed{92.2}$ \\
GWTC-4.0 & \reviewed{1.92} &  & $\reviewed{0.19}$ -- $\reviewed{1.18}$ & $\reviewed{1.1}$ &  & $\reviewed{0.06}$ -- $\reviewed{1.13}$ & $\reviewed{3.4}$ &  & $\reviewed{-0.15}$ -- $\reviewed{0.97}$ & $\reviewed{10.3}$ &  & $\reviewed{-0.96}$ -- $\reviewed{0.42}$ & $\reviewed{71.7}$ &  & $\reviewed{-4.94}$ -- $\reviewed{1.20}$ & $\reviewed{84.8}$ \\
\toprule
 &   & & \multicolumn{2}{c}{$\bar{A}_{1.5}$}  & & \multicolumn{2}{c}{$\bar{A}_{2.5}$}  & & \multicolumn{2}{c}{$\bar{A}_{3.0}$}  & & \multicolumn{2}{c}{$\bar{A}_{3.5}$}  & & \multicolumn{2}{c}{$\bar{A}_{4.0}$} \\
 \cline{4-5}
 \cline{7-8}
 \cline{10-11}
 \cline{13-14}
 \cline{16-17}
 &   & & \multicolumn{1}{c}{$[10^{-26}]$} &  \% & & \multicolumn{1}{c}{$[10^{-14}]$} &  \% & & \multicolumn{1}{c}{$[10^{-9}]$} &  \% & & \multicolumn{1}{c}{$[10^{-3}]$} &  \% & & \multicolumn{1}{c}{$[10^{3}]$} &  \%\\
 \midrule
GWTC-3.0 &  &  & $\reviewed{0.95}$ -- $\reviewed{6.83}$ & $\reviewed{1.3}$ &  & $\reviewed{-2.50}$ -- $\reviewed{-0.11}$ & $\reviewed{96.4}$ &  & $\reviewed{-8.53}$ -- $\reviewed{0.95}$ & $\reviewed{90.4}$ &  & $\reviewed{-3.89}$ -- $\reviewed{0.85}$ & $\reviewed{85.0}$ &  & $\reviewed{-1.69}$ -- $\reviewed{0.64}$ & $\reviewed{75.3}$ \\
GWTC-4.0 &  &  & $\reviewed{-0.40}$ -- $\reviewed{3.11}$ & $\reviewed{10.5}$ &  & $\reviewed{-1.22}$ -- $\reviewed{0.15}$ & $\reviewed{90.0}$ &  & $\reviewed{-3.42}$ -- $\reviewed{1.08}$ & $\reviewed{79.8}$ &  & $\reviewed{-1.44}$ -- $\reviewed{0.38}$ & $\reviewed{84.2}$ &  & $\reviewed{-0.62}$ -- $\reviewed{0.19}$ & $\reviewed{82.5}$ \\
\bottomrule
\end{tabular}

   \tablecomments{We compare the results for GWTC-4.0 with those for GWTC-3.0 \citep{LIGOScientific:2021sio,Baka:2025drk}. The table shows 90\% credible intervals for the dimensionless graviton mass $\bar{m}_g = m_g/ (\text{eV}/c^2)$ and the dimensionless amplitude parameters $\bar{A}_\alpha = A_\alpha/\text{eV}^{2-\alpha}$. We also include the quantiles of the \GR hypothesis $Q_{\mathrm{GR}} = P(A_\alpha <0)$. The significant shifts away from zero for $\alpha=-3,-2$ are driven by prior effects for \FULLNAME{GW231028_153006}.}
 \end{table*}

We use all \TGRMDREventNumber{gwtc3}{} events analyzed for MDR in GWTC-3.0~\citep{LIGOScientific:2021sio}, as well as all new \BBH events in GWTC-4.0 that pass the general selection criteria described in Section~\ref{sec:paper II intro}. \BNS and \NSBH events are excluded due to computational constraints, as these signals have long durations. Additionally, these relatively nearby events provide limited information on dispersion, with only loosely constrained posteriors, as demonstrated for GW170817 \citep{LIGOScientific:2018dkp}.
We exclude \FULLNAME{GW230518_125908} due to its likely \NSBH nature.
Although we analyzed \COMMONNAME{GW231123}, we omit it from our combined analysis due to significant waveform systematic uncertainties \citep{GW231123}, which led to an apparent \ac{GR} violation when analyzed with the \IMRPhenomXPHM waveform model. We give details of our studies of the systematic errors from waveform modeling for this event and MDR in Appendix~\ref{app:mdr_s231123}.

Therefore, we include \TGRMDREventNumber{gwtc4new}{} new events from O4a in our combined analysis. In~\citet{Baka:2025drk}, the authors reanalyze the GWTC-3.0 data using all of the improvements described above and provide a more detailed explanation of the methods.
All comparisons to the GWTC-4.0 results are made with respect to these reanalyzed GWTC-3.0 results, to ensure that any differences are driven only by the inclusion of results from the new mergers, and not by changes in methodology.

Figure~\ref{fig:mdr:amplitudes} shows the violin plots describing the combined bounds on the amplitude parameters $A_\alpha$. The shaded gray region shows the result for \TGRMDREventNumber{gwtc3}{} events from GWTC-3.0~\citep{LIGOScientific:2021sio,Baka:2025drk}, while the blue posteriors show the result for the \TGRMDREventNumber{gwtc4cumulative}{} cumulative events in GWTC-4.0 (i.e., after adding the events from O4a). Overall, the new results are more constraining
than those from GWTC-3.0 and more consistent with the GR prediction (zero)
for positive values of $\alpha$. On average, the new posteriors are narrower by a factor \TGRMDRBoundImprovement{amplitude_mean}{} compared to the factor of \TGRMDRExpectedBoundImprovement{} improvement expected from the multiplication of Gaussian distributions. Although for $\alpha=-3$ and $\alpha=-2$ cases, GR lands outside the central 90\% credible interval, this is driven primarily by \FULLNAME{GW231028_153006} for which prior effects lead to the apparent GR violations. The deviations disappear if this event is excluded and therefore we find no evidence for dispersion of GWs. See Appendix~\ref{app:mdr_s231028} for details about the prior effects for \FULLNAME{GW231028_153006}.

In Table~\ref{tab:mdr:combined_amplitudes} we summarize the data from Figure~\ref{fig:mdr:amplitudes}. For each tested value of $\alpha$, we report the $5\%$ and $95\%$ quantiles together with the quantile at which GR was found, $Q_{\mathrm{GR}} = P(A_\alpha <0)$. These values show the difference in width of the marginalised posteriors and consistency with  GR between GWTC-3.0 and its extension to GWTC-4.0.

\begin{figure*}
  \centering \includegraphics[width=\TGRFigureWidthPage]{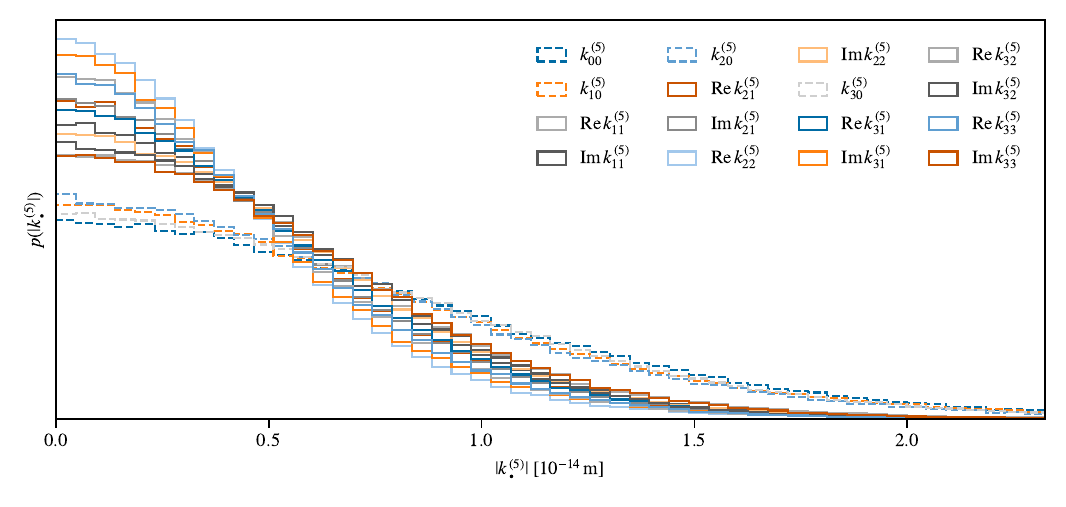}
  \caption{
  The combined posterior probability on the 16 anisotropic dispersion and  birefringence coefficients $k_{\ell m}^{(5)}$ from O4a.
  We highlight the $m = 0$ coefficients, which have weaker constraints that the others.} 
  \label{fig:ssb:kv5_plot}
\end{figure*}

We applied a special treatment for $\alpha=0$, which can be interpreted as giving the dispersion due to a massive graviton, and 
obtained the 90\% upper bound on the graviton mass, \TGRMDRGravitonBound{gwtc4}.  The new result gives a factor of \TGRMDRBoundImprovement{mg}{} improvement compared to the GWTC-3.0 bound \TGRMDRGravitonBound{gwtc3}~\citep{LIGOScientific:2021sio,Baka:2025drk}. The bound on the graviton mass from planetary ephemerides is \num{1.01e-24} $\mathrm{eV}/c^2$ at the $99.7\%$ confidence level~\citep{Mariani:2023ubf}, which is an order of magnitude better than our constraint. However, these results are complementary, since they use different set of observations (GW versus astrometry) and focus on different phenomenologies (wave propagation versus orbital motion).

\subpapersubsection{Tests of anisotropic birefringence from the Standard Model Extension}
\label{sec:SSB}

The effective field theory referred to as the Standard Model Extension (SME) is a phenomenological framework for deriving observational constraints of
  SSB, such as violations of Lorentz invariance and charge--parity--time (CPT) symmetry~\citep{Colladay:1998fq, Kostelecky:2003fs}.
For GWs, Lorentz and CPT violation can result in anisotropic and dispersive propagation.
One particular prediction of SME is birefringence, which can cause the two polarizations to travel at different speeds in vacuum. 

Within the SME's gravity sector, all symmetry-breaking operators that appear in the Lagrangian are associated with a certain mass dimension $d$.
Operators at higher mass dimensions  are suppressed by larger powers of the natural energy scale~\citep{Kostelecky:2016kfm}.
For gauge-invariant theories that allow SSB, the leading order manifestation of anisotropic birefringence appears at mass-dimension $d=5$.
This introduces a frequency-dependent birefringent dispersion relation~\citep{Kostelecky:2016kfm,Mewes:2019dhj}, which changes the frequency-domain waveform,
\begin{align}
\tilde{h}_{+}(f) &= \tilde{h}_{+}^{\rm GR}(f) \cos{\left(\beta f^2\right)}   -  \tilde{h}_{\times}^{\rm
GR}(f) \sin{\left(\beta f^2\right)}   \,,\\
\tilde{h}_{\times}(f) &= \tilde{h}_{+}^{\rm GR}(f) \, \sin{\left(\beta f^2\right)}   + \tilde{h}_{\times}^{\rm GR}(f) \cos{\left(\beta f^2\right)}  \,.
\end{align}
Here $\tilde{h}_{+}^{\rm GR}(f)$ and $\tilde{h}_{\times}^{\rm GR}(f)$ are the polarization components in GR and $\beta$ is the deviation parameter directly related to the symmetry breaking operators that appear in the SME Lagrangian at $d=5$:
\begin{align}
\beta &= \frac{4\pi^{2}}{c} \left| k^{(5)}_{{\rm eff}} \right| \tau(z) \,, \;\;\; k^{(5)}_{{\rm eff}} = \sum_{\ell, m} Y_{\ell m}(\boldsymbol{\hat{n}}) \, k^{(5)}_{{\ell m}} \,, \\
\tau(z) &= \int_{0}^z \frac{1+\bar{z}}{H_0 \sqrt{\Omega_\mathrm{m}(1+\bar{z})^3+\Omega_\Lambda}} \, {\rm d}\bar{z} \,.
\end{align}
where the superscript (5) denotes parameters related to dimension $d=5$ and the parameters $k^{(5)}_{{\ell m}}$ often appear as $k^{(5)}_{(V)\,{\ell m}}$ in other papers~\citep{Kostelecky:2016kfm,ONeal-Ault:2021uwu,Haegel:2022ymk}.
Here $\ell = 0, 1, \ldots, d-2 = 3$ and $-\ell \leq m \leq \ell$, while $\boldsymbol{\hat{n}}$ is the direction of the source in spherical polar coordinates centered on the Earth and $\tau(z)$ is the effective propagation time associated with cosmological redshift $z$, where we use the cosmological model given in \citet{GWTC:Introduction}.
Similarly to MDR, we only consider the propagation effect and neglect any modifications in the \acp{GW} that could arise during their generation.

Since CBC sources are isotropically distributed over the sky \citep{Essick:2022slj},
  we have sufficiently many GW detections to 
  extract all $16$ $k^{(5)}_{{\ell m}}$ parameters to determine all the degrees of freedom possible in $d=5$ symmetry breaking operators in the Lagrangian.

We place a uniform prior on $k^{(5)}_{{\rm eff}} \tau(z)$ and 
  use \IMRPhenomXPHM as the baseline \GR waveform model.
We then derive the posteriors on $k^{(5)}_{{\rm eff}}$ and perform the singular value decomposition of the matrix of spherical harmonics to obtain the posteriors of the 16 $k^{(5)}_{{\ell m}}$ parameters we are interested in.
While we do not have uniform priors on the individual $k^{(5)}_{{\ell m}}$ parameters, we have checked that we obtain almost identical results when placing a uniform prior on $k^{(5)}_{{\rm eff}}$.
These post-processing details can be found in \cite{Haegel:2022ymk},
  which provides the limits obtained from 45 events in GWTC-3.0.
However, there were some typographic errors in their calculations, and the actual constraint is $\left(Y_{00}(\boldsymbol{\hat{n}})\right)^2$ times the original constraint, which is about one order of magnitude tighter.
We do not provide results on GWTC-3.0, awaiting re-analysis with the updated \BILBY implementation.

\begin{table}
  \caption{Combined results for the SSB analysis}
  \label{tab:ssb:kij_coeffs}
  \begin{center}
  \setlength{\tabcolsep}{12pt}
\centering
\begin{tabular}{ c  c c  }
\hline \hline
$|k^{(5)}_{ij}| \; [10^{-14} \mathrm{m}]$  & 90\%  \\ 
  \hline 
$|k^{(5)}_{00}|$  & $\reviewed{1.52}$ \\
$|k^{(5)}_{10}|$  & $\reviewed{1.46}$ \\
$|{\rm Re} \,k^{(5)}_{11}|$  & $\reviewed{1.05}$ \\
$|{\rm Im} \,k^{(5)}_{11}|$  & $\reviewed{0.98}$ \\
$|k^{(5)}_{20}|$  & $\reviewed{1.39}$ \\
$|{\rm Re} \,k^{(5)}_{21}|$  & $\reviewed{0.95}$ \\
$|{\rm Im} \,k^{(5)}_{21}|$  & $\reviewed{0.94}$ \\
$|{\rm Re} \,k^{(5)}_{22}|$  & $\reviewed{0.81}$ \\
$|{\rm Im} \,k^{(5)}_{22}|$  & $\reviewed{0.99}$ \\
$|k^{(5)}_{30}|$  & $\reviewed{1.42}$ \\
$|{\rm Re} \,k^{(5)}_{31}|$  & $\reviewed{0.95}$ \\
$|{\rm Im} \,k^{(5)}_{31}|$  & $\reviewed{0.85}$ \\
$|{\rm Re} \,k^{(5)}_{32}|$  & $\reviewed{0.87}$ \\
$|{\rm Im} \,k^{(5)}_{32}|$  & $\reviewed{0.98}$ \\
$|{\rm Re} \,k^{(5)}_{33}|$  & $\reviewed{0.88}$ \\
$|{\rm Im} \,k^{(5)}_{33}|$  & $\reviewed{1.05}$ \\
  \hline \hline
\end{tabular}
\label{tab:kij_GWTC3_GWTC4}
  \end{center}
  \tablecomments{We give the 90\% upper bounds for the magnitudes of all 16 $k_{\ell m}^{(5)}$ SSB coefficients from the O4a marginalized posteriors in Figure~\ref{fig:ssb:kv5_plot}.}
\end{table}

Figure~\ref{fig:ssb:kv5_plot} shows the marginalized posteriors of the absolute values of the 16 $k_{\ell m}^{(5)}$ coefficients from $40$ events in O4a.
Here we omit GW231123 from the combined results due to the waveform modeling uncertainties discussed in \citet{GW231123} and also illustrated for MDR in Appendix~\ref{app:mdr_s231123}.
For \FULLNAME{GW231028_153006}, the SSB analysis finds the same prior-driven shifts away from \ac{GR} found by MDR (see Appendix~\ref{app:mdr_s231028}),
  and we thus include this event in the combined results, as does MDR.
The posteriors for all parameters peak at zero, which corresponds to GR.
In Table~\ref{tab:ssb:kij_coeffs} we summarize the data from Figure~\ref{fig:ssb:kv5_plot}.
Compared to the original GWTC-3.0 results in \cite{Haegel:2022ymk}, the constraints from just O4a are about a factor of two tighter, after accounting for the correction mentioned above.
These are thus the current best constraints on these coefficients---see Table~D51 in \citet{SME_data_tables_2026}, which also shows results from other studies that have obtained more
constraining illustrative bounds by varying the coefficients one at a time, though this procedure is only valid for the purposes of illustration.

\subpapersection{Conclusions}\label{sec:paper II conclusions}

We have presented the results of \TGRIINUMTESTS tests of \GR or physics beyond isolated, quasi-circular binaries, \reviewed{four} of which are new
    (and one, TIGER, has been significantly updated),
    all testing various parameterized modifications to \GR, and placing constraints on such deviations.

Overall, our tests find that the signals conform to our expectations from \GR.
We provide highlights of the bounds and improvements in Table 1 of Paper I.
In particular, we improve the constraints on deviations in the \PN coefficients compared to GWTC-3.0 by up to a factor of $\TGRFTIBoundImprovement{max}$
    (where some of this improvement is due to an increase in the tapering frequency used by FTI, as discussed in Section~\ref{sec:par}).
We also provide illustrative translations (with caveats) of the \PN coefficient constraints to constraints on a variety
    of alternative theories in Table~\ref{tab:map_to_tgr}.
The loudest event during O4a, \FULLNAME{GW230814_230901},
    can by itself place a tight bound on some PN deviations \citep{GW230814},
    but it is excluded from this paper's analysis because it is a single-detector event;
    even without it, the PN coefficient constraints obtained from the full catalog are better.
However, the loud \ac{O4b} event GW250114 \citep{GW250114}
    provides even better constraints than the full catalog \citep{GW250114_TGR}.
Additionally, the MDR analysis updates the bound on the graviton mass to $m_g \le$ \TGRMDRGravitonBound{gwtc4}{} at $90\%$ credibility.

Since we apply these tests to \TGRNUMEVENTSPREVPLUSOFOURA events in GWTC-4.0,
    statistically a few should display apparent deviations from \GR expectations.
Of the \TGRIINUMEVENTS O4a events covered in this paper, this occurred for four events,
    namely \FULLNAME{GW230628_231200}, \FULLNAME{GW231028_153006}, \FULLNAME{GW231110_040320}, and \COMMONNAME{GW231123},
        which showed some deviations.
The last three of these produced results with \GR outside the $90\%$ credible interval for the MDR test,
    though for \FULLNAME{GW231028_153006} this is due to priors,
    as discussed in Appendix~\ref{app:mdr_s231028}.
The deviations for \FULLNAME{GW231110_040320} are only slight, and the Bayes factors still favor \GR,
    while the deviations for \COMMONNAME{GW231123} can be attributed to waveform systematics,
    as discussed in Appendix~\ref{app:mdr_s231123},
    though alternative explanations such as wave-optics lensing have also been raised \citep{GW231123}.
TIGER also finds deviations for \FULLNAME{GW231028_153006} for most post-inspiral coefficients
    but the Bayes factors still favor \GR (FTI does not analyze this event),
    while both FTI and TIGER find deviations for
    \FULLNAME{GW231110_040320}.
Both FTI and TIGER find \GR outside the 90\% credible interval for \FULLNAME{GW230628_231200}.

For events before O4,
    the tests in this paper find \GR inside the $90\%$ credible interval for all but GW190814~\citep{GW190814},
    where the TIGER and PCA analyses find \GR to be outside the $90\%$ credible interval for some parameters.
    FTI also finds minor shifts away from zero for some parameters (though \GR is still in the $90\%$ credible interval),
        which studies in \citet{Mehta:2022pcn} attribute to properties of the noise around the time of the event.
    Inclusion of GW190814 in the PCA combined results for TIGER also leads to shifts away from zero for some parameters, with the
    largest shift in the fifth PCA parameter, with a $98\%$ \GR quantile, while there are no notable shifts for any PCA parameters
    if GW190814 is excluded from the combined results.

Our analyses find consistency with GR in these tests for all other events.
For all other tests, we find such consistency for the entire GWTC-4.0 catalog,
    though the SIM analyses find that \GR is slightly outside the $90\%$ credible interval in the combined results,
    due to correlations with the effective inspiral spin.
The few events that met the criteria for the LOSA test are consistent with zero LOSA.

In summary, almost all events analysed were found to be consistent with the expectations of \GR.
In those few cases where the results were not seen to be consistent,
    likely causes of the inconsistency were identified which
    are likely sufficient to explain these apparent deviations, given the large number of events analysed,
    without requiring violations of \GR.
Continued improvement in
    waveform modeling and detector sensitivities
    as well as the analysis of additional events will further clarify the situation.

We thus expect to place even better constraints on deviations from \GR with the analysis
    of additional events from the remainder of the fourth observing run and from future
    observing runs \citep{Aasi:2013wya}.\footnote{LVK observing run plans \url{https://observing.docs.ligo.org/plan}}
Additionally, improvements in analyses will also lead to better constraints,
    particularly performing tests for specific alternative theories,
    where inspiral--merger--ringdown waveform models have started to be developed and applied to data \citep{Julie:2024fwy,Liu:2024atc},
    improving on analyses that place constraints using either the inspiral \citep{Nair:2019iur,Perkins:2021mhb,Lyu:2022gdr,Perkins:2022fhr} or ringdown \citep{Silva:2022srr,Chung:2025wbg,Maenaut:2024oci} alone.

All strain data analyzed in this paper are available from the Gravitational Wave Open Science Center \citep{OpenData}. The data and scripts used to prepare the figures and tables are available at \citet{PaperII_DCC_release}.

\section*{Acknowledgements}

This material is based upon work supported by NSF's LIGO Laboratory, which is a
major facility fully funded by the National Science Foundation.
The authors also gratefully acknowledge the support of
the Science and Technology Facilities Council (STFC) of the
United Kingdom, the Max-Planck-Society (MPS), and the State of
Niedersachsen/Germany for support of the construction of Advanced LIGO 
and construction and operation of the GEO\,600 detector. 
Additional support for Advanced LIGO was provided by the Australian Research Council.
The authors gratefully acknowledge the Italian Istituto Nazionale di Fisica Nucleare (INFN),  
the French Centre National de la Recherche Scientifique (CNRS) and
the Netherlands Organization for Scientific Research (NWO)
for the construction and operation of the Virgo detector
and the creation and support  of the EGO consortium. 
The authors also gratefully acknowledge research support from these agencies as well as by 
the Council of Scientific and Industrial Research of India, 
the Department of Science and Technology, India,
the Science \& Engineering Research Board (SERB), India,
the Ministry of Human Resource Development, India,
the Spanish Agencia Estatal de Investigaci\'on (AEI),
the Spanish Ministerio de Ciencia, Innovaci\'on y Universidades,
the European Union NextGenerationEU/PRTR (PRTR-C17.I1),
the ICSC - CentroNazionale di Ricerca in High Performance Computing, Big Data
and Quantum Computing, funded by the European Union NextGenerationEU,
the Comunitat Auton\`oma de les Illes Balears through the Conselleria d'Educaci\'o i Universitats,
the Conselleria d'Innovaci\'o, Universitats, Ci\`encia i Societat Digital de la Generalitat Valenciana and
the CERCA Programme Generalitat de Catalunya, Spain,
the Polish National Agency for Academic Exchange,
the National Science Centre of Poland and the European Union - European Regional
Development Fund;
the Foundation for Polish Science (FNP),
the Polish Ministry of Science and Higher Education,
the Swiss National Science Foundation (SNSF),
the Russian Science Foundation,
the European Commission,
the European Social Funds (ESF),
the European Regional Development Funds (ERDF),
the Royal Society, 
the Scottish Funding Council, 
the Scottish Universities Physics Alliance, 
the Hungarian Scientific Research Fund (OTKA),
the French Lyon Institute of Origins (LIO),
the Belgian Fonds de la Recherche Scientifique (FRS-FNRS), 
Actions de Recherche Concert\'ees (ARC) and
Fonds Wetenschappelijk Onderzoek - Vlaanderen (FWO), Belgium,
the Paris \^{I}le-de-France Region, 
the National Research, Development and Innovation Office of Hungary (NKFIH), 
the National Research Foundation of Korea,
the Natural Sciences and Engineering Research Council of Canada (NSERC),
the Canadian Foundation for Innovation (CFI),
the Brazilian Ministry of Science, Technology, and Innovations,
the International Center for Theoretical Physics South American Institute for Fundamental Research (ICTP-SAIFR), 
the Research Grants Council of Hong Kong,
the National Natural Science Foundation of China (NSFC),
the Israel Science Foundation (ISF),
the US-Israel Binational Science Fund (BSF),
the Leverhulme Trust, 
the Research Corporation,
the National Science and Technology Council (NSTC), Taiwan,
the United States Department of Energy,
and
the Kavli Foundation.
The authors gratefully acknowledge the support of the NSF, STFC, INFN and CNRS for provision of computational resources.

This work was supported by MEXT,
the JSPS Leading-edge Research Infrastructure Program,
JSPS Grant-in-Aid for Specially Promoted Research 26000005,
JSPS Grant-in-Aid for Scientific Research on Innovative Areas 2402: 24103006,
24103005, and 2905: JP17H06358, JP17H06361 and JP17H06364,
JSPS Core-to-Core Program A.\ Advanced Research Networks,
JSPS Grants-in-Aid for Scientific Research (S) 17H06133 and 20H05639,
JSPS Grant-in-Aid for Transformative Research Areas (A) 20A203: JP20H05854,
the joint research program of the Institute for Cosmic Ray Research,
University of Tokyo,
the National Research Foundation (NRF),
the Computing Infrastructure Project of the Global Science experimental Data hub
Center (GSDC) at KISTI,
the Korea Astronomy and Space Science Institute (KASI),
the Ministry of Science and ICT (MSIT) in Korea,
Academia Sinica (AS),
the AS Grid Center (ASGC) and the National Science and Technology Council (NSTC)
in Taiwan under grants including the Science Vanguard Research Program,
the Advanced Technology Center (ATC) of NAOJ,
and the Mechanical Engineering Center of KEK.

Additional acknowledgements for support of individual authors may be found in the following document: \\
\href{https://dcc.ligo.org/LIGO-M2300033/public}{https://dcc.ligo.org/LIGO-M2300033/public}.

For the purpose of open access, the authors have applied a Creative Commons Attribution (CC BY)
license to any Author Accepted Manuscript version arising.
We request that citations to this article use `A. G. Abac {\it et al.} (LIGO-Virgo-KAGRA Collaboration), ...' or similar phrasing, depending on journal convention.

\clearpage
\emph{The following open-source software has been used:}

Calibration of the \ac{LIGO} strain data was performed with a \GSTLAL{}-based
    calibration software pipeline~\citep{Viets:2017yvy}.
    Calibration of the Virgo strain data is performed with C-based software~\citep{VIRGO:2021kfv}.
Data-quality products and event-validation results were computed using the
    \soft{DMT}{}~\citep{DMTdocumentation}, \soft{DQR}{}~\citep{DQRdocumentation},
    \soft{DQSEGDB}{}~\citep{Fisher:2020pnr}, \soft{gwdetchar}{}~\citep{gwdetchar-software},
    \soft{hveto}{}~\citep{Smith:2011an}, \soft{iDQ}{}~\citep{Essick:2020qpo},
    \soft{Omicron}{}~\citep{Robinet:2020lbf} and
    \soft{PythonVirgoTools}{}~\citep{pythonvirgotools} software packages and contributing
    software tools.
Analyses in this catalog relied upon the \LALSUITE{} software library~\citep{lalsuite-software,Wette:2020air}.
The detection of the signals and subsequent significance evaluations in this catalog were performed with the
    \GSTLAL{}-based inspiral software pipeline~\citep{Messick:2016aqy,Sachdev:2019vvd,Hanna:2019ezx,Cannon:2020qnf},
    with the \MBTA{} pipeline~\citep{Adams:2015ulm,Aubin:2020goo}, and with the
    \PYCBC{}~\citep{Usman:2015kfa,Nitz:2017svb,Davies:2020tsx} and the
    \CWB{}~\citep{Klimenko:2004qh,Klimenko:2011hz,Klimenko:2015ypf} packages.
Estimates of the noise spectra and glitch models were obtained using
    \BAYESWAVE{}~\citep{Cornish:2014kda,Littenberg:2015kpb,Cornish:2020dwh,Gupta:2023jrn}.
Noise subtraction for one candidate was also performed with \soft{gwsubtract}{}~\citep{Davis:2022ird}.
Source-parameter estimation was performed with the \BILBY{}
    and \PBILBY{} libraries~\citep{Ashton:2018jfp,Romero-Shaw:2020owr,Smith:2019ucc} using the
    \DYNESTY{} nested sampling package~\citep{Speagle:2020spe}.

FTI, TIGER, SIM, LOSA, MDR, and SSB waveforms used for testing \GR were generated using \BILBYTGR~\citep{ashton_2025_15676285}.
\PESUMMARY{} was used to postprocess and collate parameter-estimation results~\citep{Hoy:2020vys}.
The various stages of the parameter-estimation analysis were managed with the \ASIMOV{} library~\citep{Williams:2022pgn}
    together with \soft{CBCFlow}~\citep{cbcflow}.
Plots were prepared with \soft{Matplotlib}~\citep{Hunter:2007ouj},
    \SEABORN{}~\citep{Waskom:2021psk}, and \GWPY{}~\citep{Macleod:2021goi}.
\NUMPY{}~\citep{Harris:2020xlr}, \soft{scikit-learn}~\citep{scikit-learn},  and \SCIPY{}~\citep{Virtanen:2019joe} were used
    in the preparation of the manuscript.

\appendix

In these Appendices, we discuss apparent \ac{GR} deviations found for various events and either the TIGER or MDR analyses. Specifically, we discuss the effects of noise on GW190728\_064510 in the TIGER
    reanalysis with \IMRPhenomXPHM{} (Appendix~\ref{app:tiger_GW190728}), as well as systematic errors from waveform modeling for the events \FULLNAME{GW231028_153006}
    for the TIGER and MDR analyses (Appendices~\ref{app:tiger_s231028} and~\ref{app:mdr_s231028}) and \COMMONNAME{GW231123} for the MDR analysis (Appendix~\ref{app:mdr_s231123}).
    See \citet{GW231123} for studies of waveform-driven systematic uncertainties in standard parameter estimation for \COMMONNAME{GW231123}.\\

\section{\label{app:tiger_GW190728} TIGER and GW190728\_064510}

\begin{figure}
    \centering \includegraphics[width=\TGRFigureWidth]{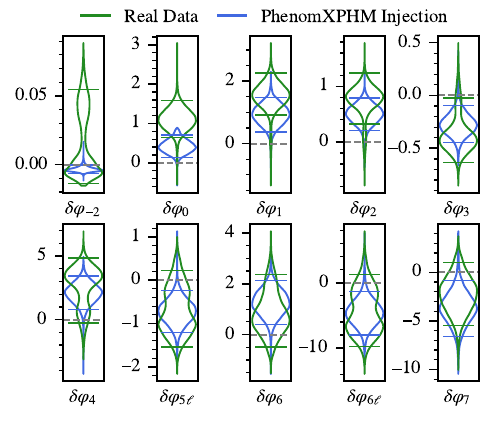}
    \caption{%
    Results from the TIGER test on the inspiral part of the signal GW190728\_064510, including the real data and  real-noise injection studies, one with \IMRPhenomXPHM, based on the maximum-likelihood sample from the \GR{} analysis. All TIGER analyses were carried out using the same waveform model as the baseline.}
    \label{fig:tiger:GW190728_posteriors}
\end{figure}

In the TIGER analysis of GW190728\_064510 using the \IMRPhenomXPHM{} model, we find that the \GR{} value lies outside the 90\% credible interval for the lower \PN{} terms, in particular from 0\PN{} to 1.5\PN, as shown by the orange violin plots in Figure~\ref{fig:tiger:GW190728_posteriors}. The posterior distributions of the deviation parameters also exhibit bimodal features. However, the $\log_{10}$ Bayes factors in favor of the beyond \GR{} hypothesis over the \GR{} hypothesis are negative for all deviation parameters, except for the 0\PN{} and 0.5\PN{} cases, where the values are positive but small, with $\log_{10}\mathcal{B}<0.2$.

Also, the TIGER analysis of this event using the \IMRPhenomPVTWO{} model was reported in the GWTC-2.0 tests of \GR~\citep{LIGOScientific:2020tif}. In that analysis, no apparent deviation was found, although strong bimodal features were present for several deviation parameters, with one mode consistent with \GR and another mode shifted away from the \GR value. To understand whether the differences with the \IMRPhenomXPHM{} results are driven primarily by the improved treatment of precession or by the inclusion of higher-order modes, we additionally analyze this event using the \soft{IMRPhenomXP\_SpinTaylor} model (henceforth just \IMRPhenomXP), which includes the updated precession treatment but not higher-order modes. We find a similar bimodal structure. However, the posteriors obtained with \IMRPhenomPVTWO{} and \IMRPhenomXP{} are not fully consistent. Since the \IMRPhenomXP{} result still shows a \GR-consistent mode, while this mode is significantly suppressed in the \IMRPhenomXPHM{} analysis, the comparison suggests that the apparent deviation in the lower \PN{} coefficients is mainly associated with the inclusion of higher-order modes.

To investigate whether the apparent tension with \GR{} is driven by waveform systematics, we first perform a zero-noise injection analysis (i.e., we analyze simulated observations with no noise added) using \IMRPhenomXPHM{} and the event PSDs released with GWTC-2.1~\citep{GWTC2p1}. The injection waveform is generated using the maximum-likelihood sample from a new \IMRPhenomXPHMST{} analysis of this event. (Here we do not abbreviate the name of the waveform model to emphasize the difference compared to the \IMRPhenomXPHMMSA{} version used in the GWTC-2.1 parameter-estimation analysis.) We do not observe any comparable deviation or bias in the posterior distributions of the deviation parameters.

We then perform a set of injections in real detector noise around the event time. The median detector-frame component masses of GW190728\_064510 are $15.1\,M_\odot$ and $9.1\,M_\odot$, and the standard parameter-estimation analysis was performed using a $16$~s data segment. To avoid overlap with the data segment used to analyze the event, we choose nearby integer-second offsets somewhat randomly, injecting the signal $20$~s before and $18$~s after the merger time. These choices place the injections close to the event time while keeping the corresponding data segments non-overlapping with the event segment. As described above, we use the same injection parameters and waveform model for these injections. For the injection performed $20$~s before the event, the posteriors of the deviation parameters are consistent with the \GR{} value.

\begin{table}[ht]
    \label{tab:tiger:gw190728_qgr}
    \caption{
Comparison of the \GR{} quantiles, $Q_\mathrm{GR}=P(\delta\hat{\varphi}_i<0)$, for the TIGER analysis of GW190728\_064510 and the corresponding real noise injection}
    \begin{center}
    \begin{tabular}{lrr}
\toprule
Parameter & Real event & Injection \\
\midrule
$\delta \varphi_{-2}$ & 40.3\% & 94.0\% \\
$\delta \varphi_{0}$ & 0.3\% & 0.7\% \\
$\delta \varphi_{1}$ & 1.4\% & 0.6\% \\
$\delta \varphi_{2}$ & 3.8\% & 0.7\% \\
$\delta \varphi_{3}$ & 93.5\% & 99.2\% \\
$\delta \varphi_{4}$ & 9.2\% & 1.1\% \\
$\delta \varphi_{5\ell}$ & 87.0\% & 99.1\% \\
$\delta \varphi_{6}$ & 14.7\% & 1.2\% \\
$\delta \varphi_{6\ell}$ & 87.4\% & 98.1\% \\
$\delta \varphi_{7}$ & 87.6\% & 97.6\% \\
\bottomrule
\end{tabular}
    \end{center}
\tablecomments{The injection was performed $18$~s after the event merger time and generated using the maximum-likelihood sample from the \IMRPhenomXPHM{} analysis. The TIGER analyses were carried out with the same waveform.}
\end{table}

However, for the injection performed $18$~s after the event, we find qualitatively similar biases to those seen in the real event analysis, with the \GR{} value lying outside the 90\% credible intervals for several deviation parameters, as shown in Figure~\ref{fig:tiger:GW190728_posteriors}. The shifts are not identical to those in the real event, and in some lower \PN{} cases the real event posteriors show larger apparent deviations. This is expected, since the injection probes only one nearby realization of real detector noise and is not intended to reproduce the exact noise fluctuation present at the event time. Instead, it demonstrates that real detector noise for GW190728\_064510 like event can produce apparent deviations with comparable character.

Table~\ref{tab:tiger:gw190728_qgr} lists the \GR{} quantiles, $Q_\mathrm{GR}=P(\delta\hat{\varphi}_i < 0)$, for the \PN{} deviation coefficients considered in this analysis. Except for the 0\PN{} coefficient, the real noise injection shows \GR{} quantiles that are comparable to, or more extreme than, those of the real event, supporting the interpretation that such apparent deviations can arise from detector noise fluctuations. We also find that the $\log_{10}$ Bayes factors are positive for the 0.5\PN, 1\PN, and 3\PN{} logarithmic deviation parameters, but all remain below $0.4$.

A similar injection study was performed in previous work using NR waveforms injected into 21 different stretches of real detector noise and recovered with the previous version of TIGER \citep[based on \IMRPhenomPVTWO;][]{Meidam:2017dgf}. Figure~8 of \citet{Meidam:2017dgf} shows the 90\% credible intervals of the deviation parameters. In several cases, the \GR{} value lies outside, or close to the edge of, the 90\% credible interval. In particular, simulation number 20 shows behavior that is qualitatively consistent with our findings for GW190728\_064510.

We therefore conclude that the apparent deviations observed for GW190728\_064510 are most likely caused by a particular realization of detector noise rather than by waveform systematics or evidence for a genuine violation of \GR. The weak Bayes-factor support for the beyond-\GR{} hypotheses further supports this interpretation. Thus, while GW190728\_064510 shows interesting behavior in the lower \PN{} coefficients, the result is consistent with noise-induced fluctuations and does not provide evidence for a departure from \GR.

\section{\label{app:tiger_s231028} TIGER and GW231028\_153006}

\begin{figure}
    \centering \includegraphics[width=\TGRFigureWidth]{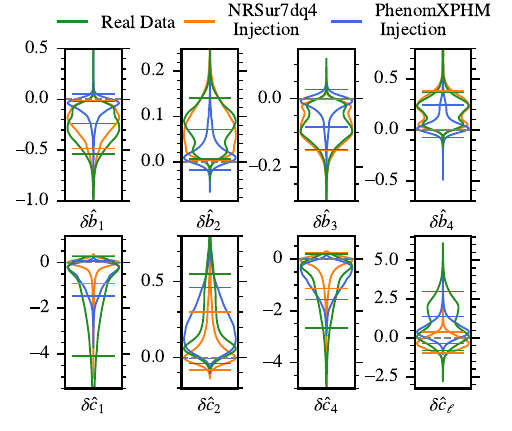}
    \caption{%
    Results from the TIGER test on the intermediate and merger–ringdown portion of the signal \FULLNAME{GW231028_153006}, including the real data and two zero-noise injection studies, one with \SURSEVENDQFOUR and one with \IMRPhenomXPHM, both based on the maximum-likelihood sample from the \ac{GR} analysis. All TIGER analyses were carried out using the \IMRPhenomXPHM waveform model as the baseline.}
    \label{fig:tiger:s231028_posteriors}
\end{figure}

As the \ac{GR} analysis of \FULLNAME{GW231028_153006} found potential waveform systematic uncertainties \citep{GWTC:Results}, we highlight the TIGER results for this event. This event only passes the criteria for TIGER analysis with post-inspiral parameters. For all intermediate parameters and two merger--ringdown parameters, the \ac{GR} value lies at the edge of the 90\% credible interval, as shown in the orange violin plot in Figure~\ref{fig:tiger:s231028_posteriors}. However, the Bayes factor in favor of the beyond-\ac{GR} model against the \ac{GR} model is negative for all deviation parameters; specifically, all $\log_{10}$ Bayes factor values are less than $-1.2$.

To investigate whether the apparent tension with \ac{GR} is driven by waveform uncertainties, we perform a zero-noise injection analysis using the \SURSEVENDQFOUR model \citep{Varma:2019csw}, which directly interpolates numerical-relativity simulations, avoiding the approximations used to construct \IMRPhenomXPHM. The injection waveform is generated from the maximum-likelihood sample of the \SURSEVENDQFOUR \ac{GR} analysis for this event \citep{GWTC:Results}. The same injection study for the MDR analysis is presented in Appendix~\ref{app:mdr_s231028}. As shown in Figure~\ref{fig:tiger:s231028_posteriors}, the posteriors of the deviation parameters from the injection analysis are consistent with those from the real data. In particular, the \ac{GR} value for the posterior of intermediate parameters is also found at the edge of the 90\% credible interval. Since the TIGER baseline model is \IMRPhenomXPHM, the apparent tension may be driven by the choice of waveform model used in the injection simulations.

To further investigate the degeneracy between the deviation parameter and the \ac{GR} parameters, we performed a zero-noise injection study using the maximum-likelihood sample from the \IMRPhenomXPHM \ac{GR} analysis of this event. This approach avoids systematics arising from the choice of waveform models. The blue violin plots in Figure~\ref{fig:tiger:s231028_posteriors} show the posterior distributions of the deviation parameters. For the intermediate parameters, the \ac{GR} values are recovered within the 90\% credible intervals. However, for the merger--ringdown parameters, the \ac{GR} values lie at the edge of the intervals. We find that the masses and spins are strongly correlated with the deviation parameter. In particular, the spin magnitude of the injection parameters is nearly unity, while the recovery with the TIGER model shows reduced consistency at high spin values for some deviation parameters. Moreover, the posteriors of the mass-related parameters differ significantly from those obtained with the \ac{GR} model. Therefore, we conclude that while the
shifts away from GR seen in the intermediate coefficients are primarily due to waveform uncertainties, the results for the merger--ringdown parameters are partly due to a strong degeneracy between the deviation parameter and the \ac{GR} parameters, not by physics beyond \ac{GR}.

\section{\label{app:mdr_s231028} MDR and GW231028\_153006}

\begin{figure}
    \centering \includegraphics[width=\TGRFigureWidth]{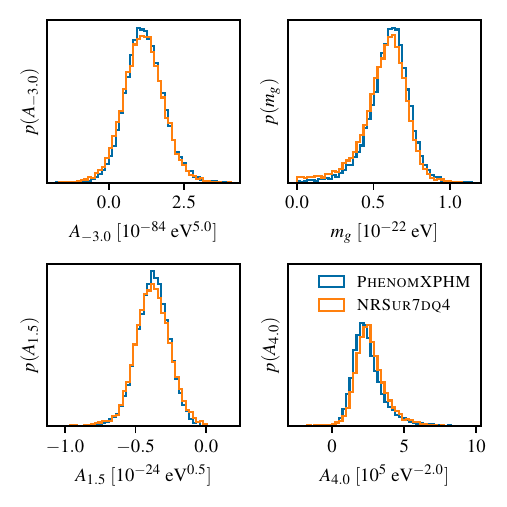}
    \caption{%
    Representative  posteriors for the MDR amplitude parameters $A_\alpha$ and the graviton mass $m_g$ for \FULLNAME{GW231028_153006}. Inference using both the \IMRPhenomXPHM and \SURSEVENDQFOUR waveform models does not show any significant systematic uncertainties. We instead find that this shift away from GR is due to a prior effect.}
    \label{fig:mdr:s231028_posteriors}
\end{figure}

\begin{figure}
    \centering \includegraphics[width=\TGRFigureWidth]{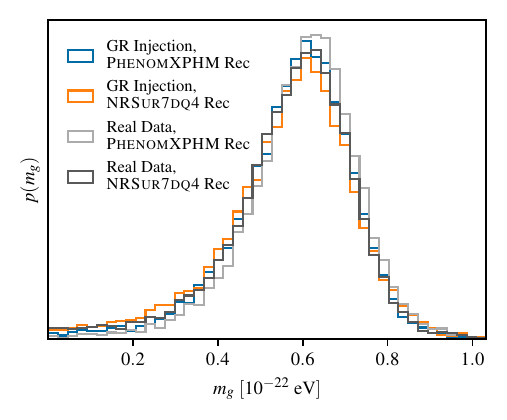}
    \caption{%
    Graviton-mass posterior for \FULLNAME{GW231028_153006} both for real data and \SURSEVENDQFOUR injections into zero noise based on the maximum-likelihood sample from \ac{GR} analysis. Both are analyzed with \IMRPhenomXPHM and \SURSEVENDQFOUR waveform models.
    We thus find that this shift away from GR is due to a prior effect.}
    \label{fig:mdr:s231028_injection}
\end{figure}

Of the \TGRMDREventNumber{gwtc4new}{} new events contributing to the improved bounds on the dispersion amplitudes $A_\alpha$, \FULLNAME{GW231028_153006} is the least consistent with \GR. Since the \ac{GR} analysis identified potential waveform systematic uncertainties for this event, we performed an additional analysis using the more accurate \SURSEVENDQFOUR waveform model. Representative posteriors are shown in Figure~\ref{fig:mdr:s231028_posteriors}. The posteriors are consistent across waveform models, indicating that our initial analysis with \IMRPhenomXPHM is reliable.

To better understand the apparent tension with \ac{GR}, we analyzed a \SURSEVENDQFOUR injection into zero noise, based on the maximum-likelihood sample from the \ac{GR} analysis of this event. In Figure~\ref{fig:mdr:s231028_injection}, we show the inferred posterior distribution of the graviton mass $m_g$ for this injection, compared to the result from the analysis of the actual data. The posteriors agree and both exhibit a peak in $m_g$ away from zero, even though the injected signal is consistent with \GR.

This can be attributed to a degeneracy between $m_g$ and other binary parameters: the inferred parameters differ from those of the injection, yet the corresponding waveforms remain similar. We quantify this similarity using the mismatch between two templates $h_1$ and $h_2$, defined as~\citep{Cutler:1994ys}:
\begin{equation}
	\mathrm{MM} = 1 - \left|\frac{\ip{h_1}{h_2}}{\sqrt{\ip{h_1}{h_1}\ip{h_2}{h_2}}}\right|,
\end{equation}
where $\ip{\cdot}{\cdot}$ denotes the noise-weighted inner product and a mismatch of $0$ corresponds to identical templates. Comparing the injected signal with the waveform corresponding to the maximum-posterior sample, we find a low mismatch of $\mathrm{MM} = \TGRMDRInjectionRecoveryMismatch$. While the likelihood near both the injection point and the maximum-posterior sample is similar, the prior density at the maximum-posterior location is higher by a factor of $\TGRMDRInjectionRecoveryPriorFactor$. Consequently, samples near the maximum-posterior point dominate, and GR-consistent solutions lie at the edge of the posterior distribution.

\begin{figure}
    \centering \includegraphics[width=\TGRFigureWidth]{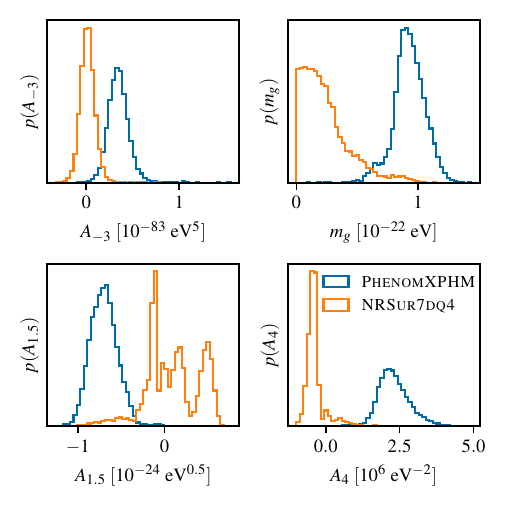}
    \caption{%
    Representative  posteriors for the MDR amplitude parameters $A_\alpha$ and the graviton mass $m_g$ for \COMMONNAME{GW231123}.
    We find a large discrepancy in the results between those obtained using \IMRPhenomXPHM and \SURSEVENDQFOUR.}
    \label{fig:mdr:s231123_posteriors}
\end{figure}

\section{\label{app:mdr_s231123} MDR and GW231123}

\begin{table}[ht]
    \caption{\label{tab:mdr:combined_s231123_bayes}
    $\mathcal{B}^\mathrm{MDR}_\mathrm{GR}$ comparison for \COMMONNAME{GW231123}
    }
    \begin{center}
    \begin{tabular}{lrr}
\toprule
 & \textsc{XPHM} $\log_{10}\mathcal{B}^\mathrm{MDR}_\mathrm{GR}$ & \textsc{NRSur} $\log_{10}\mathcal{B}^\mathrm{MDR}_\mathrm{GR}$ \\
\midrule
$\alpha=-3.0$ & $\reviewed{-0.5}$ & $\reviewed{-2.5}$ \\
$\alpha=-2.0$ & $\reviewed{-0.3}$ & $\reviewed{-2.5}$ \\
$\alpha=-1.0$ & $\reviewed{0.1}$ & $\reviewed{-2.5}$ \\
$m_g$ & $\reviewed{1.0}$ & $\reviewed{-1.3}$ \\
$\alpha=0.0$ & $\reviewed{0.2}$ & $\reviewed{-2.4}$ \\
$\alpha=0.5$ & $\reviewed{0.3}$ & $\reviewed{-2.2}$ \\
$\alpha=1.5$ & $\reviewed{0.8}$ & $\reviewed{-1.2}$ \\
$\alpha=2.5$ & $\reviewed{1.0}$ & $\reviewed{-0.6}$ \\
$\alpha=3.0$ & $\reviewed{1.0}$ & $\reviewed{-0.5}$ \\
$\alpha=3.5$ & $\reviewed{1.1}$ & $\reviewed{-0.2}$ \\
$\alpha=4.0$ & $\reviewed{1.8}$ & $\reviewed{0.1}$ \\
\bottomrule
\end{tabular}

    \end{center}
    \tablecomments{We give the values of the log Bayes factor $\mathcal{B}^\mathrm{MDR}_\mathrm{GR}$ in favor of MDR over \GR for \COMMONNAME{GW231123} analyzed with the \IMRPhenomXPHM and \SURSEVENDQFOUR waveform models and consider a selection of different dispersion models (different values of $\alpha$). The $m_g$ row denotes the analysis with a prior uniform in the graviton mass. Each entry compares analyses performed with the same waveform model.}
\end{table}

While \COMMONNAME{GW231123} passed the MDR selection criteria, this event shows significant issues with waveform systematic uncertainties \citep{GW231123}. As the MDR analysis with \IMRPhenomXPHM showed an apparent \ac{GR} violation, we reanalyzed the data with the \SURSEVENDQFOUR waveform model to check if waveform uncertainties are responsible. \SURSEVENDQFOUR is a surrogate model that shows better agreement than \IMRPhenomXPHM with numerical-relativity simulations for systems with similar total mass to \COMMONNAME{GW231123}~\citep{GW231123}.
The results show significant discrepancies in the inferred posteriors of the amplitude parameters $A_\alpha$ and the graviton mass $m_g$; we show a few examples in Figure~\ref{fig:mdr:s231123_posteriors}. The use of the \SURSEVENDQFOUR waveform model shifts the posteriors so that the \ac{GR} value of $0$ is included in the posterior mass. For $\alpha\leq0$, the posteriors are unimodal and centered near zero. For $\alpha \in \{0.5,1.5,2.5\}$, the posteriors have multiple peaks but still show consistency with \ac{GR}. For $\alpha\geq3$, the posteriors are almost unimodal, with a small peak near $0$ and a larger peak further away. The inferred amplitude parameter $A_\alpha$ is closer to $0$ and with the opposite sign compared to the \IMRPhenomXPHM results.

In Table~\ref{tab:mdr:combined_s231123_bayes}, we compare how the Bayes factors $\mathcal{B}^\mathrm{MDR}_\mathrm{GR}$ in favor of MDR over \ac{GR} vary depending on the waveform used. For \IMRPhenomXPHM, the Bayes factor shows a preference for MDR for $\alpha>-2$, despite its larger prior volume compared to \ac{GR} (due to one additional parameter and an expanded prior range for the chirp mass to avoid railing). In the \SURSEVENDQFOUR analysis, \ac{GR} is generally preferred over MDR, except for the $\alpha = 4$ case, where there is a small preference for MDR, though the uncertainty on the Bayes factors has a standard deviation of $\sim 0.2$.

Due to these issues with the waveform systematic uncertainties, we have excluded \COMMONNAME{GW231123} from the combined bounds on GW dispersion.

\bibliography{}

\iftoggle{endauthorlist}{
 \let\author\myauthor
 \let\affiliation\myaffiliation
 \let\maketitle\mymaketitle
 \iftoggle{fullauthorlist}{
  \title{Authors}
  \pacs{}
  
  \newpage
  \maketitle
 }{}
}{}

\end{document}